\documentclass[11pt%
]{tumDiss}
\usepackage{scrhack}
\usepackage{hyphenat}
\usepackage{graphicx}
\usepackage{pgfplots}
\usepackage{pgfplotstable}
\tikzset{>=stealth}
\usetikzlibrary{patterns}
\usetikzlibrary{pgfplots.statistics}

\usepackage{moreverb}
\usepackage{listings}

\usepackage{amsmath}
\usepackage{amssymb}
\usepackage{amsfonts}
\usepackage{upgreek}
\usepackage{multicol}
\usepackage{enumerate}
\usepackage[utf8]{inputenc}
\usepackage{tcolorbox}
\usepackage{empheq}
\usepackage{blindtext}
\usepackage{multicol}
\usepackage{adjustbox}
\usepackage{booktabs}
\usepackage{longtable}
\usepackage{bm}
\usepackage{amsthm}
\usepackage{etaremune}
\usepackage{wrapfig}
\usepackage{float}

\usepackage{setspace}

\usepackage{listings}
\usepackage[all]{nowidow}
\usepackage{multirow}
\usepackage{lipsum}
\usepackage{url}
\usepackage[
  hidelinks,
  bookmarksnumbered
]{hyperref}
\hypersetup{%
    pdftoolbar   = true,         
    pdfmenubar   = true,         
    pdffitwindow = false,        
    pdfstartview = {FitH},       
    pdftitle     = {Test},    
    pdfauthor    = {Stavros Nousias},   
    pdfsubject   = {PhD Thesis}, 
    pdfcreator   = {Stavros Nousias},   
    pdfnewwindow = true,         
}

\usepackage{caption}
\usepackage{subcaption}
\usepackage[
  toc,
  acronym,
  style = long
]{glossaries}

\usepackage{float}
\usepackage{graphicx}
\usepackage{multirow}
\usepackage{amsmath}
\usepackage{amsfonts}
\usepackage{hhline}
\usepackage{amssymb}
\usepackage{enumerate}
\usepackage{multirow}
\usepackage{adjustbox}
\usepackage{color}
\usepackage{xcolor}
\usepackage{cite}
\usepackage{colortbl}
\usepackage{url}
\usepackage{tabularx}
\usepackage{enumitem}
\usepackage{amsmath}
\usepackage{cite}
\usepackage{textcomp}
\usepackage{framed}
\usepackage{epigraph}
\usepackage{algorithm}
\usepackage[noend]{algpseudocode} 

 \algnewcommand\algorithmicforeach{\textbf{for each}}
 \algdef{S}[FOR]{ForEach}[1]{\algorithmicforeach\ #1\ \algorithmicdo}
 \algnewcommand{\Initialize}[1]{%
   \State \textbf{Initialize:}
 }

\usepackage[bottom=1.3in]{geometry}
\definecolor{color1}{HTML}{4F81BD}
\definecolor{color2}{RGB}{220,0,0}
\definecolor{bblue}{HTML}{4F81BD}
\definecolor{rred}{HTML}{C0504D}
\definecolor{ggreen}{HTML}{9BBB59}
\definecolor{ppurple}{HTML}{9F4C7C}
\definecolor{lgreen} {RGB}{180,210,100}
\definecolor{dblue}  {RGB}{20,66,129}
\definecolor{ddblue} {RGB}{11,36,69}
\definecolor{lred}   {RGB}{220,0,0}
\definecolor{nred}   {RGB}{224,0,0}
\definecolor{norange}{RGB}{230,120,20}
\definecolor{nyellow}{RGB}{255,221,0}
\definecolor{ngreen} {RGB}{98,158,31}
\definecolor{dgreen} {RGB}{78,138,21}
\definecolor{nblue}  {RGB}{28,130,185}
\definecolor{jblue}  {RGB}{20,50,100}
\definecolor{LightPink}			{RGB}{255,178,178}
\definecolor{LightSteelBlue} 	{RGB}{178,178,255}
\definecolor{Wheat}				{RGB}{236,217,198}
\definecolor{GreenYellow}       {RGB}{217, 229, 6} 	    
\definecolor{Yellow}            {RGB}{254, 223, 0} 	    
\definecolor{Goldenrod}         {RGB}{249, 214, 22} 	
\definecolor{Dandelion}         {RGB}{253, 200, 47} 	
\definecolor{Apricot}           {RGB}{255, 170, 123} 	
\definecolor{Peach}             {RGB}{255, 127, 69} 	
\definecolor{Melon}             {RGB}{255, 129, 141} 	
\definecolor{YellowOrange}      {RGB}{240, 171, 0} 	    
\definecolor{Orange}            {RGB}{255, 88, 0} 	    
\definecolor{BurntOrange}       {RGB}{199, 98, 43} 	    
\definecolor{Bittersweet}       {RGB}{189, 79, 25} 	    
\definecolor{RedOrange}         {RGB}{222, 56, 49} 	    
\definecolor{Mahogany}          {RGB}{152, 50, 34} 	    
\definecolor{Maroon}            {RGB}{152, 30, 50} 	    
\definecolor{BrickRed}          {RGB}{170, 39, 47} 	    
\definecolor{Red}               {RGB}{255, 0, 0}        
\definecolor{BrilliantRed}      {RGB}{237, 41, 57} 	    
\definecolor{OrangeRed}         {RGB}{231, 58, 0} 	    
\definecolor{RubineRed}         {RGB}{202, 0, 93}       
\definecolor{WildStrawberry}    {RGB}{203, 0, 68} 	    
\definecolor{Salmon}            {RGB}{250, 147, 171} 	
\definecolor{CarnationPink}     {RGB}{226, 110, 178} 	
\definecolor{Magenta}           {RGB}{255, 0, 144} 	    
\definecolor{VioletRed}         {RGB}{215, 31, 133} 	
\definecolor{Mulberry}          {RGB}{163, 26, 126} 	
\definecolor{RedViolet}         {RGB}{161, 0, 107} 	    
\definecolor{Fuchsia}           {RGB}{155, 24, 137} 	
\definecolor{Lavender}          {RGB}{240, 146, 205} 	
\definecolor{Thistle}           {RGB}{222, 129, 211} 	
\definecolor{Orchid}            {RGB}{201, 102, 205} 	
\definecolor{DarkOrchid}        {RGB}{153, 50, 204} 	
\definecolor{Purple}            {RGB}{182, 52, 187} 	
\definecolor{Plum}              {RGB}{79, 50, 76} 	    
\definecolor{Violet}            {RGB}{75, 8, 161} 	    
\definecolor{RoyalPurple}       {RGB}{82, 35, 152} 	    
\definecolor{BlueViolet}        {RGB}{33, 7, 106} 	    
\definecolor{Periwinkle}        {RGB}{136, 132, 213} 	
\definecolor{CadetBlue}	  	    {RGB}{95, 158, 160} 	
\definecolor{CornflowerBlue}  	{RGB}{99, 177, 229} 	
\definecolor{MidnightBlue}	  	{RGB}{0, 65, 101} 	    
\definecolor{NavyBlue}          {RGB}{0, 70, 173}       
\definecolor{RoyalBlue}         {RGB}{0, 35, 102}       
\definecolor{Blue}              {RGB}{0, 24, 168}       
\definecolor{Cerulean}          {RGB}{0, 122, 201}      
\definecolor{Cyan}              {RGB}{0, 159, 218}      
\definecolor{ProcessBlue}       {RGB}{0, 136, 206}      
\definecolor{SkyBlue}           {RGB}{91, 198, 232}     
\definecolor{Turquoise}         {RGB}{0, 255, 239} 	    
\definecolor{TealBlue}          {RGB}{0, 124, 146} 	    
\definecolor{Aquamarine}        {RGB}{0, 148, 179} 	    
\definecolor{BlueGreen}         {RGB}{0, 154, 166} 	    
\definecolor{Emerald}           {RGB}{80, 200, 120} 	
\definecolor{JungleGreen}       {RGB}{0, 115, 99} 	     
\definecolor{LightGreen}{RGB}{204, 255, 204} 	
\definecolor{SeaGreen}          {RGB}{0, 176, 146} 	    
\definecolor{Green}             {RGB}{0, 173, 131} 	    
\definecolor{ForestGreen}       {RGB}{0, 105, 60} 	    
\definecolor{PineGreen}         {RGB}{0, 98, 101} 	    
\definecolor{LimeGreen}         {RGB}{50, 205, 50} 	    
\definecolor{YellowGreen}       {RGB}{146, 212, 0} 	    
\definecolor{SpringGreen}       {RGB}{201, 221, 3} 	    
\definecolor{OliveGreen}        {RGB}{135, 136, 0} 	    
\definecolor{RawSienna}         {RGB}{149, 82, 20} 	    
\definecolor{Sepia}             {RGB}{98, 60, 27} 	    
\definecolor{Brown}             {RGB}{134, 67, 30}      
\definecolor{Tan}               {RGB}{210, 180, 140}	
\definecolor{Gray}              {RGB}{139, 141, 142} 	
\definecolor{Black}		  	    {RGB}{30, 30, 30}       
\definecolor{White}		  	    {RGB}{255, 255, 255}    

\definecolor{color1}{RGB}{200, 200, 200}
\definecolor{color2}{RGB}{255, 255, 255}

\newlist{inlinelist}{enumerate*}{1}
\setlist*[inlinelist,1]{%
  label=\arabic*),
}

\newlist{inlineitemlist}{enumerate*}{1}
\setlist*[inlineitemlist,1]{%
  label=\null,
}

\usepackage{nicefrac}
\usepackage{mathtools}

\DeclarePairedDelimiterX{\Set}[2]\{\}{%
  \, #1 \;\delimsize\vert\; #2 \,
}

\DeclarePairedDelimiter\floor{\lfloor}{\rfloor}

\DeclareMathOperator*{\argmin}{arg\,min}

\def\showFigures{}

\newcommand{\norm}[1]{\left\lVert#1\right\rVert} 
\usepackage{ragged2e}

\makeatletter
\let\@texttop\@textbottom
\makeatother

\faculty{Department of Electrical and Computer Engineering}

\degree{Doctor of Philosophy}

\title{%
Patient-specific modelling, simulation and real time processing for respiratory diseases 
}

\author{Stavros Nousias}
\vorsitz{Prof. Dr.-Ing. Vorname Nachname}
\erstpruef{Prof. Dr.-Ing. Vorname Nachname}

\zweitpruef[University of Patras]{Prof. Dr.-Ing. Vorname Nachname}

\date{15.05.2022}

\dateaccepted{05.07.2022}

\makeglossaries

\definecolor{grey60} {RGB} {102, 102, 102} 

\newcommand{\figureHeight}{0.5625} 
\pgfplotsset{
  compat           = 1.13,
  grid             = major,
  enlarge x limits = 0,
  cycle list name  = tum,
  major grid style = {dotted},
  minor grid style = {dotted},
  legend style     = {
    at     = {(0.98,0.96)},
    anchor = north east,
  },
  width            = \hsize * 0.9,
  height           = \hsize * 0.9 * \figureHeight,
}

\newacronym{cpu}{CPU}{Central Processing Unit}
\newacronym{gpu}{GPU}{Graphics Processing Unit}
\newacronym{lstm}{LSTM}{Long Short Term Memory}
\newacronym{cnn}{CNN}{Convolutional neural networks}
\newacronym{cfd}{CFD}{Computational Fluid Dynamics}
\newacronym{gmm}{GMM}{Gaussian Mixture Models}
\newacronym{pmdi}{PMDI}{Pressurized Metered Dose Inhalers}
\newacronym{rnn}{RNN}{Recurrent Neural Networks}
\newacronym{acq}{ACQ}{Asthma Control Questionaire}
\newacronym{act}{ACT}{Asthma Control Test}
\newacronym{copd}{COPD}{Chronic Obstructive Pulmonary Disorder}
\newacronym{ct}{CT}{Computed tomography}
\newacronym{sdf}{SDF}{Shape Diameter Function}
\newacronym{gfv}{GFV}{Gradient Vector Flow}
\newacronym{asm}{ASM}{Airway Smooth Muscle}
\newacronym{hrct}{HRCT}{High Resolution Computed Tomography}
\newacronym{vfb}{VFB}{Volume Filling Branching}
\newacronym{mdct}{MDCT}{Multi-Detector Computed Tomography}
\newacronym{bc}{BC}{Boundary Conditions}

\newacronym{lll}{LLL}{Left Lower Lobe}
\newacronym{lul}{LUL}{Left Upper Lobe}

\newacronym{rll}{RLL}{Right Lower Lobe}
\newacronym{rul}{RUL}{Right Upper Lobe}
\newacronym{rml}{RML}{Right Middle Lobe}
\newacronym{zcr}{ZCR}{Zero Cross Rate}
\newacronym{qda}{QDA}{Quadratic Discriminant Analysis}
\newacronym{mfcc}{MFCC}{Mel Frequency Ceptral Coefficients}
\newacronym{dft}{DFT}{Discrete Fourier Transform}
\newacronym{dct}{DCT}{Discrete Cosine Transform}
\newacronym{svm}{SVM}{Support Vector Machine}
\newacronym{rf}{RF}{Random Forest}
\newacronym{gui}{GUI}{Graphical User Interface}
\newacronym{loso}{LOSO}{Leave One Subject Out}

\glsadd{loso}
\glsadd{gui}
\glsadd{rf}
\glsadd{svm}
\glsadd{dct}
\glsadd{dft}
\glsadd{mfcc}
\glsadd{qda}
\glsadd{zcr}
\glsadd{zcr}
\glsadd{lul}
\glsadd{rml}
\glsadd{rul}
\glsadd{lll}
\glsadd{rll}
\glsadd{bc}
\glsadd{mdct}
\glsadd{hrct}
\glsadd{vfb}
\glsadd{asm}
\glsadd{cpu}
\glsadd{gpu}
\glsadd{lstm}
\glsadd{cnn}
\glsadd{cfd}
\glsadd{gmm}
\glsadd{pmdi}
\glsadd{rnn}
\glsadd{acq}
\glsadd{act}
\glsadd{copd}
\glsadd{ct}
\glsadd{sdf}
\glsadd{gfv}

\begin{document}
\frontmatter
\maketitle

\newpage

\noindent \textbf{\huge Certification} 

\vspace{2em}

\noindent It is certified that, the doctoral dissertation entitled

\vspace{1em}

\begin{center}	\Large
\textbf{Patient-specific modelling, simulation and real time processing for respiratory diseases}
\end{center}

\vspace{1em}

\noindent was elaborated by Stavros Nousias, Dipl.-ing Electrical and Computer Engineer, was publicly presented to the University of Patras's Department of Electrical Engineering on 4th of July 2022, was examined and accepted by the following committee:

\vspace{1em}
\begin{multicols}{2}
\committeeMember{Prof. Konstantinos Moustakas}{Dept. Electrical \& Computer Engineering}{University of Patras, Greece}
\committeeMember{Associate Prof. Evangelos Dermatas}{Computer Engineering \& Informatics Dept.}{University of Patras, Greece}
\committeeMember{Assistant Prof. Sotirios Fouzas}{School of Medicine}{University of Patras, Greece}
\committeeMember{Prof. Pavlos Peppas}{Dept. Electrical \& Computer Engineering}{University of Patras, Greece}
\vfill\null
\columnbreak
\committeeMember{Associate Prof. Kyriakos Sgarbas}{Dept. Electrical \& Computer Engineering}{University of Patras, Greece}
\committeeMember{Researcher Grade B Aris Lalos}{Industrial Systems Institute}{Athena Research Institute}
\committeeMember{Associate Prof. Christos Makris}{Computer Engineering \& Informatics Dept.}{University of Patras, Greece}

\end{multicols}

\begin{textblock*}{\defaultwidth}(22mm, 200mm)
\raggedright
Patras 04, 07, 2022 
\end{textblock*}

\begin{textblock*}{\defaultwidth}(-20mm, 220mm)
\centering
Supervisor\\
Professor\\
Konstantinos Moustakas
\end{textblock*}

\begin{textblock*}{\defaultwidth}(60mm, 220mm)
\centering
Department Chair\\
Professor\\
Odyseas Koufopavlou
\end{textblock*}

\newpage 

\ 

\newpage

\noindent \textbf{\huge Dedication} 

\vspace{10cm}

\noindent \textbf{\LARGE To my wife and daughter}

\chapter{Summary}

Asthma is a common chronic disease of the respiratory system causing significant disability and societal burden. According to estimates, it affects more than 300 million people worldwide, while more than 100 million people will likely have asthma by 2025 \cite{nunes2017asthma}. According to the National Health Interview Survey (NHIS)-2012, approximately 13 percent of the USA population, nearly 40 million Americans, had asthma at some point in their lives, and 8 percent, approximately 26 million Americans, had asthma in 2017. The price of asthma varies greatly from nation to nation. It is possible to estimate that the mean cost per patient per year in Europe is calculated to \texteuro 1900 while in the United States it is calculated to \$ 3,100 (USD) taking into account patients with all severity levels, ranging from intermittent and mild to severe \cite{nunes2017asthma}.

Managing asthma involves controlling symptoms, preventing exacerbations, and maintaining lung function. Improving asthma control affects the daily life of patients and is associated with a reduced risk of exacerbations and lung function impairment, reduces the cost of asthma care and indirect costs associated with reduced productivity. 
Despite the widespread availability of therapies in randomized controlled trials, different levels of asthma control have been observed in several studies using well-validated self-assessment questionnaires, such as the Asthma Control Questionnaire (ACQ) and the Asthma Control Test (ACT).

Traditionally the primary pillar of asthma control is monitoring and intervention. This route entails asynchronous spirometry data, questionnaires, environmental parameters and medication monitoring.
However, in an optimized asthma control approach, improvements in multiple stages take place, including data management, intelligent data processing, data mining predictive models fused with computational models of the pulmonary system. A primary layer of assessing the concentrated data corresponds to a system-level solution that is forwarded and visualized via innovative interfaces to the medical personnel. A second level includes simplistic data mining and handmade decision trees that generate simple notifications to the doctor and the patient. Complex artificial intelligence models process the data to extract biomarkers and predict exacerbations, hospital visits, and conditions at a third level.

Understanding the complex dynamics of the pulmonary system and the lung's response to disease, injury, and treatment is fundamental to the advancement of Asthma treatment. Computational models of the respiratory system seek to provide a theoretical framework to understand the interaction between structure and function. Well-designed computational models can make sense of highly complex systems, but they can also become quite complicated. A model will never wholly represent reality, and simplifying assumptions are always required for tractability.
Integrative computational models can bring together two or more aspects of lung function to study their combined contribution to overall function.
Their application can improve pulmonary medicine by introducing a patient-specific approach to medicinal methodologies optimizing the delivery given the personalized geometry and personalized ventilation patterns while introducing a patient-specific technique that maximizes drug delivery.

A three-fold objective, addressed within this dissertation, becomes prominent at this point. The first part refers to the comprehension of pulmonary pathophysiology and the mechanics of Asthma and subsequently of constrictive pulmonary conditions in general. The second part refers to the design and implementation of tools that facilitate personalized medicine to improve delivery and effectiveness. Finally, the third part refers to the self-management of the condition, meaning that medical personnel and patients have access to tools and methods that allow the first party to easily track the course of the condition and the second party, i.e. the patient to easily self-manage it alleviating the significant burden from the health system.

More specifically, we \textbf{i)} introduce validated variations of space-filling algorithms that regenerate and predict the structure of the entire bronchial tree.
\textbf{ii)} We generate 3D digital twins of pulmonary structures that allow the application of computational fluid dynamics. 
\textbf{iii)} We investigate and validate principal component analysis as a method to predict maximal space allocation lung bronchial tree growing.
\textbf{iv)} We introduce geometry processing methods that facilitate the simulation of bronchoconstriction and
\textbf{v)} perform CFD studies that quantify the airflow in normal and asthmatic patient-specific image-based 3-dimensional bronchial tree representations.
\textbf{vi)} Furthermore, we focus on the macroscopic monitoring of constrictive pulmonary diseases investigating monitoring medication adherence through content-based audio classification for pressurized metered-dose inhalers.  
\textbf{i)} To this end, we compile two datasets containing inhaler use audio and
\textbf{ii)} present a comparative study investigating the classification accuracy of multiple machine learning and deep learning classifiers for a series of features.
\textbf{iii)} We propose a Gaussian Mixture Model-based method that adequately differentiates inhaler events and, in certain cases, outperforms them.
\textbf{iv)} Finally, we present CNN based approaches that adequately capture features and allow for computationally inexpensive real-time embedded deployments.

\tableofcontents
\listoffigures
\listoftables
\printglossary[type=\acronymtype, nonumberlist]

\mainmatter

\chapter{Introduction}

Asthma is a common chronic disease of the respiratory system causing significant disability and societal burden. According to estimates, it affects more than 300 million people worldwide, while more than 100 million people will likely have asthma by 2025 \cite{nunes2017asthma}. According to the National Health Interview Survey (NHIS)-2012, approximately 13 percent of the USA population, nearly 40 million Americans, had asthma at some point in their lives, and 8 percent, approximately 26 million Americans, had asthma in 2017. The price of asthma varies greatly from nation to nation. It is possible to estimate that the mean cost per patient per year in Europe is calculated to \texteuro 1900 while in the United States it is calculated to \$ 3,100 (USD) taking into account patients with all severity levels, ranging from intermittent and mild to severe \cite{nunes2017asthma}.

Specifically, Asthma is a complex condition encompassing multiple underlying pathological conditions that interact on multiple scales of the pulmonary system (e.g. cell level, tissue level, organ level). It falls under the category of constrictive pulmonary diseases characterized by an inflammatory reaction that generates airway remodelling airflow limitation leading to a diminished lung function. 
Asthma encompasses pathophysiological changes across various biological, spatial and time domains. The diminished lung function is characterized by an underlying inflammatory response that induces airway remodelling, airflow limitation, and increased ventilation-perfusion (V/Q) mismatch. The hallmark of Asthma is variable airflow obstruction and airway hyper-responsiveness of the airway smooth muscle (ASM), primarily referred to as exacerbations. The mechanism of exacerbations in Asthma remains unknown. and the pathogenesis of exacerbations remains poorly understood. 

There is currently no cure for Asthma, but treatment can help control the symptoms so patients can live an everyday, active life. The primary means to address the effects of Asthma is the use of inhaled corticosteroids through devices known as inhalers. Tablets and other treatments may also be needed if your Asthma is severe.
Traditionally the control of Asthma consists of two main steps: a) the initial diagnosis of the disease and b) followed by a continuous cycle of readjustments to fit the patient's physiology and lifestyle, followed by weekly to monthly doctor visits.
Diagnosis and severity classifications are primarily based on global lung measurements of pulmonary function tests (PFTs). PFTs provide measurements only for global lung function and can be used to attribute future risk of disease progression and future exacerbations and mortality in populations.

Although asthma deaths appear to decrease due to the consistent use of inhaled corticosteroids and broader recognition of markers associated with adverse risk, many patients with Asthma continue to have persistent and poorly controllable symptoms. Current therapies are still ineffective due to insufficient understanding of pathophysiology and the disease heterogeneity, complexity and multiscale nature. Thus a patient-specific approach is required.
Computational models of the pulmonary system can help medical experts to evaluate the effectiveness of pulmonary medicine. Since Asthma is associated with multiple inflammatory phenotypes, targeting these different phenotypes in specific ways using existing or novel asthma therapies reduces asthma exacerbations. Characterizing the Asthma heterogeneities on a patient-specific approach can improve the treatment plan for a certain patient or even produce patient-specific medicine.
Capturing the physics of the pulmonary system requires detailed structural models and models predicting alterations in mechanical properties of the tissue during disease and the dynamics of fluid flows.
Medical imaging is the basis for deriving personalized digital twins of a patient's pulmonary systems. However, small airways are not visible with current imaging modalities. For this reason, a combined imaging and modelling approach is required to obtain novel information about the respiratory system. Imaging and image processing techniques are critical in constructing, parameterizing, and validating patient-specific computational models.

The issues in the definition of patient-specific models refer to the definition of structural and functional pulmonary models interlinked with a feedback process. Imaging and image processing techniques are critical in constructing, parameterizing, and validating patient-specific computational models. 
Since small airways are not visible with current imaging modalities, a combined imaging and modelling approach is required to obtain novel information about the respiratory system. 
Furthermore, understanding the flow characteristics within the airways could facilitate the improvement of inhaled pulmonary medicine since characteristics of airway flow determine particle (either noxious or therapeutic) transport and its deposition. Penetration and deposition of particles within the airways depend on airway size and branching patterns, which vary between species and age. Therefore, variation in individual airway geometry makes subject-specific models essential for studying pulmonary airflow and drug delivery. It has also been confirmed that a strong interaction exists between lung geometry and gas properties, which significantly affects gas delivery to and clearance from the lung periphery during ventilation.

A three-fold objective addressed within this dissertation becomes prominent at this point. The first part refers to comprehending pulmonary pathophysiology and the mechanics of Asthma and subsequent constrictive pulmonary conditions in general. The second part refers to designing and implementing tools that facilitate personalized medicine to improve delivery and effectiveness. Finally, the third part refers to the self-management of the condition, meaning that medical personnel and patients have access to tools and methods that allow the first party to easily track the course of the condition and the second party, i.e. the patient to easily self-manage it alleviating the significant burden from the health system.

This dissertation is organized into six (6) separate sections. Section 1 introduces the addressed topics to the reader. Section 2 presents the novel components and contributions. Section 3 outlines current state-of-the-art tools and techniques related to i) pulmonary image capturing and ii)current state-of-the-art patient-specific computational modelling of the lung, and iii) methods and preliminaries that facilitate patient adherence to asthma medication. Section 4 presents and analyses this dissertation's novelties and contributions related to patient-specific structural modelling and simulation for constrictive respiratory conditions.
Section 5 presents the novelties and contributions related to this dissertation with respect to asthma medication adherence monitoring. Finally, section 6 concludes this document.

\chapter{Novelty and contribution}
\label{section:novelty-and-contribution}
The main goal of asthma management is to control symptoms, prevent exacerbations, and improve and maintain lung function while minimizing the side effects of asthma medication. Improvements in asthma control do not only impact the patients' daily life but are also associated with a reduced risk of exacerbations and lung function impairment. Additionally, gaining and maintaining asthma control is expected to be cost-effective by reducing the overall cost of asthma care and indirect costs related to decreased productivity.

Despite the widespread availability of therapies reported as highly effective in randomized controlled trials, variable levels of asthma control have been shown in several studies\cite{chapman2008suboptimal,boonsawat2015survey}  using well-validated self-assessment questionnaires, such as the Asthma Control Questionnaire (ACQ) and the Asthma Control Test (ACT). One study performed in the Netherlands in 200 adult asthma patients \cite{van2010weekly} reported that 35.5\% of patients had partially controlled Asthma and 27 \% had uncontrolled Asthma, respectively, as measured by the ACQ. In contrast, observational studies performed in Italy revealed that approximately 35\% of patients with suboptimal asthma control. In population-based surveys conducted across Europe, prevalence rates of unsatisfactory asthma control ranged from 56.6\% to 80.0\%. 
The discrepancies between studies conducted through randomized controlled trials and observational studies arise from the different methodologies applied. The former required stricter eligibility criteria to select a highly selected population and closely monitor each participant. Reality-driven studies are more varied in terms of patient populations, levels of treatment adherence, and methods of disease management based on a variety of factors, such as the qualification of the treating physician, the availability of healthcare resources, or the patient's education regarding healthy living.

\section{Traditional and optimized asthma management}

The traditional approach to the control of Asthma consists of two main steps: a) the initial diagnosis of the disease, b) followed by a continuous cycle of readjustments to fit the patient's physiology and lifestyle, followed by weekly to monthly doctor visits. The diagnosis of Asthma is mainly based on clinical examination and questionnaires, which add to the patient's medical history and allow the identification of the disease and the assessment of its severity. After the diagnosis, the patient and the doctor try to better assess the patient's lifestyle, physiology and Asthma severity through a cycle of monitoring and controlling the disease and the readjustment of the suggested treatment. The suggested treatment consists of a written action plan describing the timing and type of medicines used. Additionally, the doctor tries to educate the patients on how to use the inhaler, control the disease's symptoms, and avoid possible exacerbations. 
The patient tries to use a combination of signs related to the disease by logging diaries and information found online to adjust the treatment based on the doctor's generic guidelines.
Regular visits to the doctor allow the identification of essential changes that will lead to the readjustment of the action plan based on the patient's experience and additional clinical examination.

On the other hand, monitoring, intervention and personalization of treatment pave the way toward optimized Asthma Control. Continuous monitoring of one's condition has been proven to improve control and decline symptoms. Van Sickle et al.\cite{van2013remote,van2013monitoring,kim2016using} aimed to investigate the use of a device to monitor the time and location of inhaler use as a measure of asthma control objectively and to determine whether information about inhaler use and feedback on asthma control via weekly email reports was associated with improved scores on composite measures of asthma control. Research endeavours within myAirCoach programme \cite{honkoop2017myaircoach,myaircoach2020effectiveness} aimed to enhance asthma diagnosis and control mechanisms by increasing their accuracy and decreasing the time response of all the steps involved.
Optimized control entails the asynchronous collection of spirometry data, questionnaires, environmental parameters and medication monitoring. A primary layer of assessing the concentrated data corresponds to a system-level solution where they are forwarded and visualized via innovative interfaces to the medical personnel. A second level includes simplistic data mining and handmade decision trees that generate simple notifications to the doctor and the patient. On a third level, complex artificial intelligence models process the data to extract biomarkers and predict exacerbations and hospital visits.

Novel patient-specific lung function models support the progress monitoring of the disease and form the basis for developing a prediction engine that also helps prevent asthma attacks. Traditional inhalers are integrated with modern devices with sensing capabilities which, combined with a smartphone and individual action plans, will help the patient learn and respond fast and accurately to any symptoms and environmental conditions. 
Improved sensing capabilities can widen the spectrum of inputs and allow the monitoring of the patients' environment and behavioural profile in addition to crucial physiological characteristics. The information base created by all the previous steps can guide the patients in the real-time control of their disease and will also support the decisions of doctors who will have up-to-date, objective and accurate information about their patients' condition.

To elaborate further, a modern asthma control approach \textbf{i)} empowers patients to modify their treatment towards personalized preset goals and guidelines either automatically or driven by a healthcare professional \textbf{ii)} incorporates visualization of the overall personalized clinical status of their health condition, with calculated risks for disease exacerbation (personalized patient models and predictions), \textbf{iii)} supports patients in receiving appropriate and prompt feedback on their asthma management process remotely, increases patients' awareness of Asthma and the effects that affect its progression and provide an optimized treatment.
\textbf{iv)} Furthermore, patients should have an overall view of their health status and lifestyle and assist them in evaluating the true overall benefits of specific interventions and the benefits of personal lifestyle management.
\textbf{v)} Providing to patients efficient knowledge and education about the asthma triggers and how to avoid them in order to be able to respond quickly to their worsening asthma control procedure.
\textbf{vi)} Thus, individuals can self-manage their condition and maintain good health according to their medical recommendations.
\textbf{vii)} Additionally, interactions among patients having similar health issues, exchanging disease and health-related information and co-operative management of Asthma.
Concerning medical personnel, \textbf{viii)} they should be allowed to make decisions based on more accurate patient-specific tailor-made computational models, allowing for improved and more accurate decision support as to when and how a particular patient should change its treatment, intervention (action plan, lifestyle, protocol). Allowing medical personnel to monitor the asthma progress facilitates an overall visual understanding of the potential disease interactions and progression pathways. Medical experts can review patient profiles and objective data on inhaled prescription adherence. \textbf{ix)} Furthermore, alerts for increased risk for acute disease episodes for a particular patient could allow for immediate intervention

\section{Computational models}
Computational models can be used to take a deeper insight into asthma pathophysiology and causes. It can be potentially employed to investigate interaction across spatial scales and assess normal physiology or pathophysiology. They also have the potential to explain how physiology is affected by sex and age. Numerous experimental studies seek to understand the precise function of specific mechanisms, and many clinical studies seek to document their change with pathology. Well-designed computational models can make sense of highly complex systems but can also become quite complicated. A model will never completely represent reality, and simplifying assumptions is always required for model tractability.

Integrative computational models have the distinct advantage of allowing detailed analysis of their parts. Their subcomponents can be evaluated individually, especially when actual experiments are not feasible or in cases where experiments cannot be easily repeated. Each parameter or variable within a model can be precisely controlled. Ideally, a comprehensive integrative model for the respiratory system should include all critical structures and functional mechanisms known or considered necessary. The components should be parameterized using experimental data, informed by physiological data, and the interaction between them should be based on the laws of physics. Unfortunately, experimental data for model parameterization is often lacking. This situation is continually improving with the generation of vast amounts of experimental, physiological, and clinical data, some feeding into models and continuing to refine them.

Integrative computational models of the respiratory system seek to provide a theoretical framework to understand the interaction between structure and function.
Their utilization can improve pulmonary medicine, introducing a patient-specific approach to medicinal methodologies and optimizing the delivery given the personalized geometry and personalized ventilation patterns while introducing a patient-specific technique that maximizes drug delivery.

\begin{figure}
\centering
\includegraphics[width=0.8\textwidth]{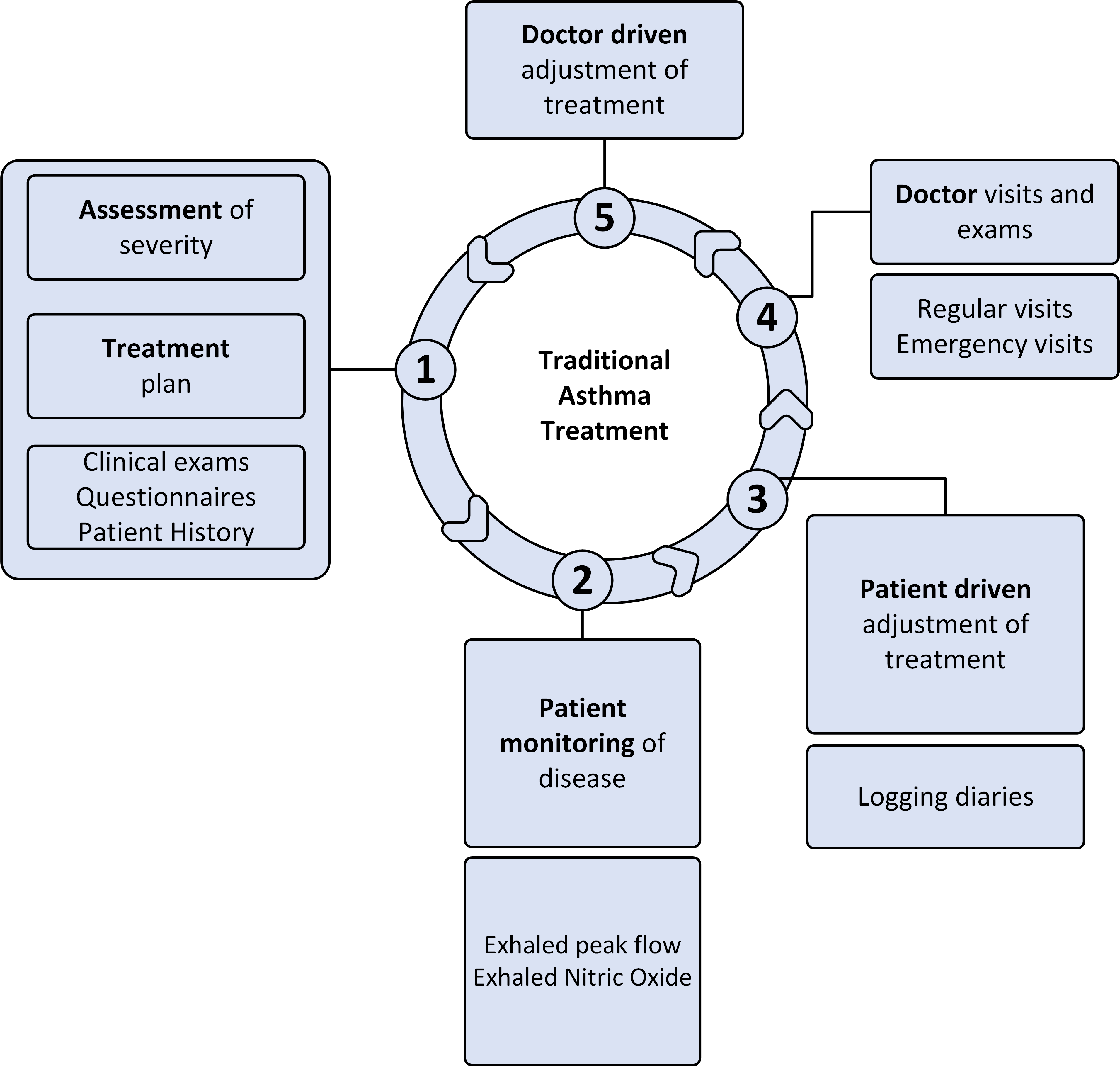}
\caption{The traditional approach towards the control of Asthma consists of two main steps: a) the initial diagnosis of the disease b) followed by a continuous cycle of readjustments to fit the patient's physiology and lifestyle, followed by weekly to monthly doctor visits. The diagnosis of Asthma is mainly based on clinical examination and questionnaires, which add to the patient's medical history and allow the identification of the disease and the assessment of its severity.}
\label{fig:contribution_traditional}
\end{figure}

\section{Contribution}
\begin{figure}
\centering
\includegraphics[width=0.7\textwidth]{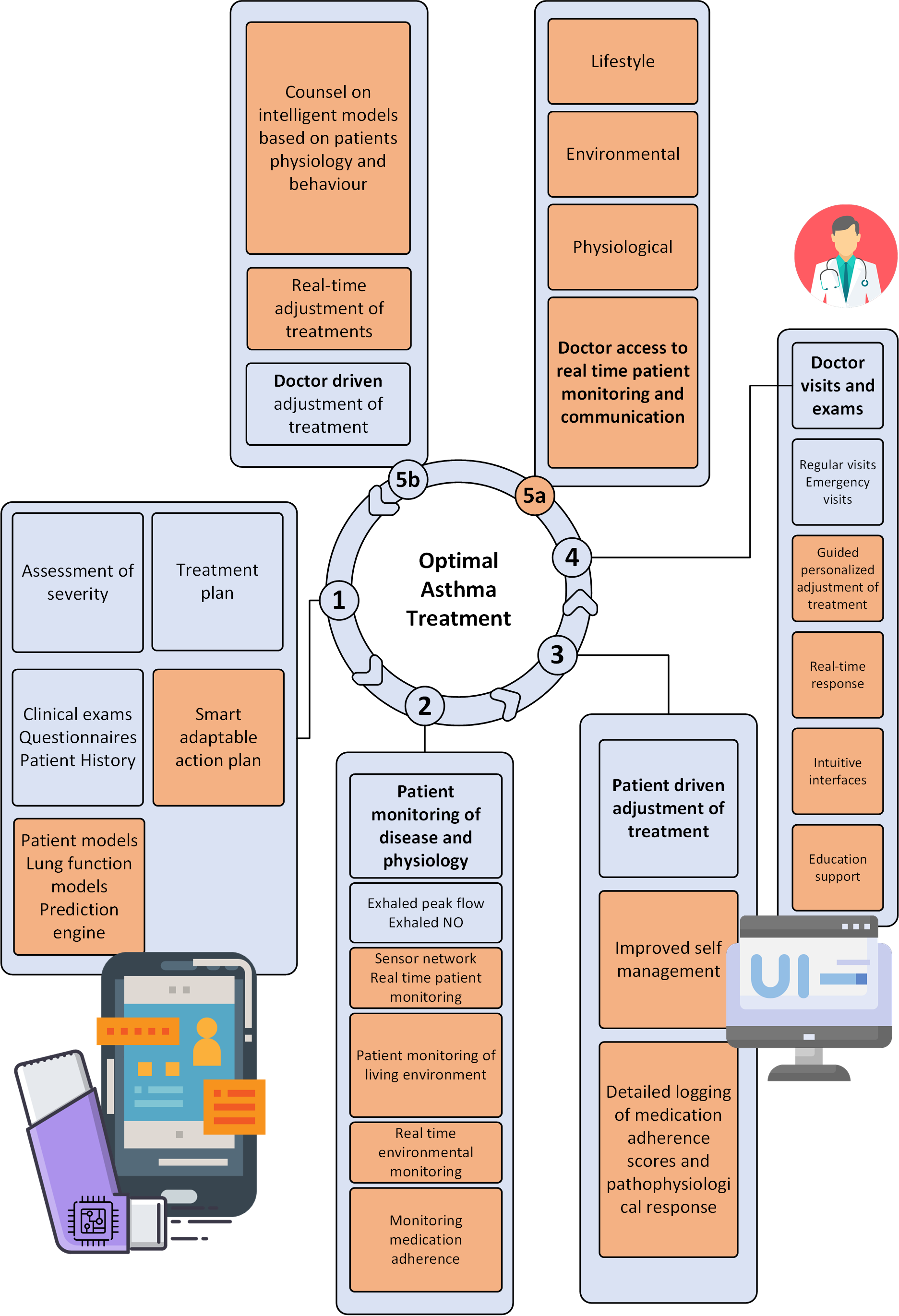}
\caption{Best practices in asthma treatment. The modern Asthma control approach entails the asynchronous collection of spirometry data, questionnaires, environmental parameters and medication monitoring. Primarily the data are forwarded and visualized via innovative interfaces to the medical personnel. At a second level, simplistic data mining and handmade decision trees are utilized. On a third level, complex artificial intelligence models process the data to extract biomarkers and predict exacerbations, hospital visits and conditions. Novel patient and lung function models support the progress monitoring of the disease and form the basis for the development of a prediction engine that also helps the prevention of asthma attacks. Modern devices integrate traditional inhalers with sensing capabilities which, combined with a smartphone and individual action plans, will help the patient learn and respond fast and accurately to any symptoms and environmental conditions.}
\label{fig:contribution_optimal}
\end{figure}

\begin{figure}
\centering
\includegraphics[width=\textwidth]{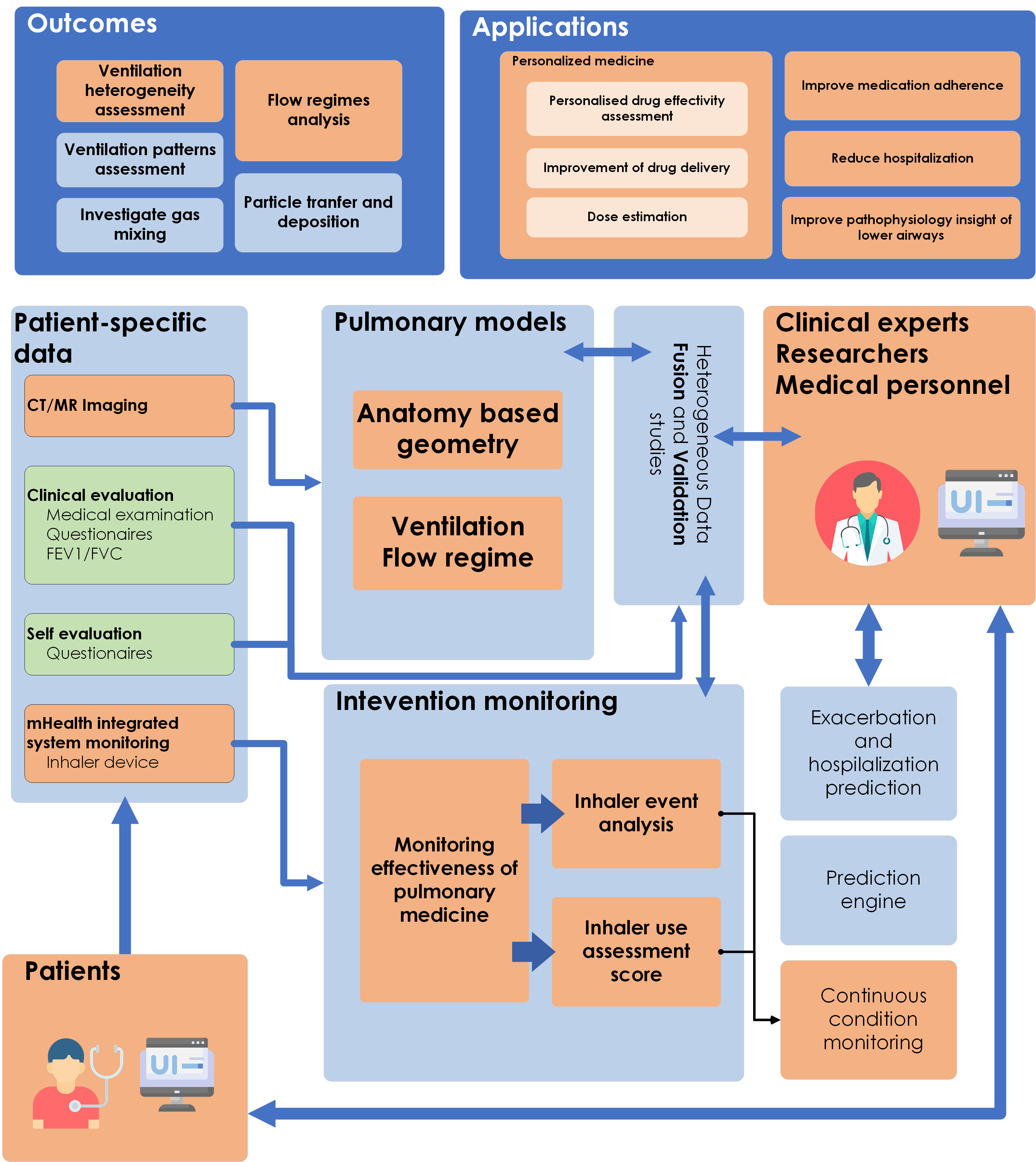}
\caption{Visualisation of the overall pipeline of our approach. The combination of pulmonary models and improved intervention monitoring guides medical personnel and the patient towards improved management of constrictive pulmonary conditions.}
\label{fig:Overall}
\end{figure}

This dissertation contributes to optimized asthma management, providing tools that describe the personalized pulmonary function and intelligent tools that allow for fast and efficient medication adherence monitoring, empowering patients to self-manage their condition through innovative interfaces and medical personnel to keep track of patient adherence to medication. 
The novelties introduced by this dissertation lie within the generation of anatomy-driven geometric models also used to study ventilation regimes in normal and asthmatic lungs. Furthermore, medication adherence tools, whose results are visualized through innovative interfaces, allow for improved patient-driven self-management and improved doctor-driven condition monitoring and intervention. Figures \ref{fig:contribution_traditional}, \ref{fig:contribution_optimal} present the traditional and optimal asthma control pipelines. Figure \ref{fig:Overall} presents the interconnect of integrative pulmonary models combined with optimal asthma control to tackle suboptimal asthma management holistically. The boxes highlighted in orange correspond to the open issues addressed by this dissertation.
Specifically, we \textbf{i)} introduce validated variations of space-filling algorithms\cite{tawhai2000generation,tawhai2004ct,lin2013multiscale} that regenerate and predict the structure of the entire bronchial tree. 
\textbf{ii)} We generate 3D digital twins of pulmonary structures that allow the application of computational fluid dynamics. 
\textbf{iii)} We investigate and validate principal component analysis as a method to predict maximal space allocation lung bronchial tree growing and 
\textbf{iv)} introduce geometry processing methods that facilitate the simulation of bronchoconstriction.
\textbf{v)} Additionally, we perform CFD studies that quantify the airflow in normal and asthmatic patient-specific image-based 3-dimensional bronchial tree representations. Furthermore, we focus on the macroscopic monitoring of such diseases. To this end,
\textbf{i)} we investigate monitoring medication adherence through content-based audio classification for pressurized metered-dose inhalers.  
\textbf{ii)} we compile two datasets containing inhaler use audio and present a comparative study investigating the classification accuracy of multiple machine and deep learning classifiers for a series of features.
\textbf{iv)} We propose a GMM-based method that adequately differentiates inhaler events and, in certain cases, outperforms them.
\textbf{v)} Furthermore, we present an approach driven by convolutional neural networks that adequately capture features and allow for computationally inexpensive real-time embedded deployments.

\chapter{Related work, trends, challenges and applications}
\label{chapter:trends}

\section{Patient-specific pulmonary structural and anatomical functional modelling}
\label{section:patient-specific}

With the advancements in computing power and the current medical imaging capabilities, the interest in the simulation of lung function based on personalized geometric models has significantly increased \cite{tawhai2011computational}.
Many studies have proposed the development and adoption of mathematical and geometrical models to study the structure of the airways and pulmonary physiology. Some address the problem of airway tree segmentation from CT images, while others analyze the branching patterns and bifurcations through airway morphometry or mathematical modelling. Airway segmentation, bronchial morphometry and mathematical models of bifurcating distributing systems are required to derive patient-specific structural and functional modelling approaches.

\textbf{Early studies on airway morphometry} \cite{haefeli1988morphometry,phillips1994diameter,pisupati1995central} used casts of human lungs to study branching patterns and the relation between airway lengths and diameters. The most commonly used conducting airway model has been Weibel's symmetric model "A" \cite{weibel1963morphometry}. 
The airway position has also been described by Horsfield order\cite{horsfield1986morphometry} and Strahler order \cite{huang1996morphometry}.

\textbf{Segmentation of lungs, airway trees, vessels and lobes in chest volumetric computed tomography (CT)} is discussed in an early review study \cite{sluimer2006}. Segmentation plays an essential role in analyzing pulmonary diseases such as emphysema, pulmonary embolism, lung cancer, asthma, and COPD, and the segmentation of the airway tree is employed to measure airway lumen and wall dimensions. 

A comparative study of automated and semi-automated segmentation methods of the airway tree from CT images was presented in \cite{lo2012extraction}. Overall, segmentation approaches can be classified into methods based on morphology \cite{irving20093d}, morphological aggregation \cite{fetita2009morphological}, voxel classification \cite{lo2010vessel}, adaptive region growing with  constraints\cite{pinho2009robust,feuerstein2009adaptive,fabijanska2009results,mendoza2009maximal,van2009automatic,weinheimer2008fully,wiemker2009simple,tschirren2009airway}, tube similarity \cite{bauer2009airway,smistad2014gpu} and gradient vector flow \cite{bauer2009segmentation}. Several implementations of the aforementioned approaches are available in the literature. The tube segmentation framework \cite{smistad2014gpu} utilizing gradient vector flow \cite{bauer2009segmentation} and the FAST heterogeneous medical image computing and visualization framework \cite{smistad2015fast} utilizing the seeded region growing approach.

Furthermore, \textbf{mathematical models of the airway structure} were formulated to derive branching and structural rules. Deterministic mathematical models of bifurcating distributing systems were examined by Wang et al. \cite{wang1992bifurcating} defined the basis for modelling bronchial tree branching as a function of available lung space\cite{tawhai2000generation}. 
Deriving airway diameter as a relation of branching features allows the complete determination of the geometry given one-dimensional skeletal representations.
Several studies mention scaling properties \cite{hagmeijer2018critical,pepe2017optimal,varner2017computational,florens2010anatomical} for the airway diameters. Kamiya et al.\cite{kamiya1974theoretical} validated the relationship between airway diameter and branching angles, and Kitaoka et al.\cite{kitaoka1999three} proposed a branching model allowing the prediction of the relationship between branching angle and flow rate and between airway length and diameter. 
For the surface reconstruction of airway surface, Tawhai et al.\cite{tawhai2003developing,tawhai2004ct,tawhai2006imaging} employed fitting cubic Hermite surfaces as described in \cite{fernandez2004anatomically}. In a similar direction, Hegedus et al. \cite{hegedHus2004detailed} generated surface models of idealized bifurcation through mathematical modelling, rigorously extending the previous definitions\cite{heistracher1995physiologically}.

Towards \textbf{patient-specific structural and functional modelling}, Tawhai et al. and Lin et al. \cite{tawhai2004ct,lin2009multiscale} studied the imposition of patient-specific boundary conditions to generate one-dimensional and three-dimensional computational models taking into consideration also the effects of turbulence. Towards the same direction, a review article \cite{tawhai2011computational} provides insight into multiscale finite element models of lung structure and function, aiming toward a computational framework for bridging the spatial scales from molecular to the whole organ. 
Bordas et al. \cite{bordas2015development} developed image analysis and modelling pipeline applied to healthy and asthmatic patient scans to produce complete personalized airway models to the acinar level incorporating CT acquisition, lung and lobar segmentation, airway segmentation and centerline extraction, algorithmic generation of distal airways and zero-dimensional models. Their implementation and results were included in the Chaste framework \cite{mirams2013chaste}, an open-source framework to facilitate computational modelling in heart, lung and soft tissue simulations. Towards the same direction, Montesantos et al.\cite{montesantos2016creation} presented a detailed algorithm for the generation of a personalized 3D deterministic model of the conducting part of the human tracheobronchial tree.

Using as basis patient-specific structural modelling,  \textbf{computational fluid dynamics} are subsequently employed to investigate flow regimes in the human lung taking into account CT-based patient specific geometries\cite{vial2005airflow,de2008flow,luo2008modeling,yin2010simulation,srivastav2011computational,swan2012computational}. Computational fluid dynamics have also been further deployed to investigate particle deposition\cite{tian2015validating,miyawaki2016effect,soni2013large,stylianou2016direct},
constrictive pulmonary diseases\cite{verbanck2007small,lalas2017substance,nousias2016computational,sul2014computational,kim2015dynamic}, micro-airway flow regimes\cite{verbanck2007small}, turbulence modelling\cite{lin2007characteristics}, four-dimensional (space and time) dynamic simulations\cite{miyawaki20164dct}, ventilation heterogeneity\cite{verbanck2007small},airflow in the acinar region \cite{kumar2009effects}. \textbf{Validation studies} conducted by Montesantos et al.\cite{montesantos2013airway} include morphometric studies on healthy and asthmatic patients providing, among others, measurements of branching angles, length and diameter of airways as a function of Generation. This dissertation employs such measurements for macroscopic validation of the generated trees.

\subsection{Studies on airway morphometry}

\subsubsection{Morphometric estimation of pulmonary diffusion capacity}

\begin{wrapfigure}{r}{0.5\textwidth}
\includegraphics[width=\linewidth]{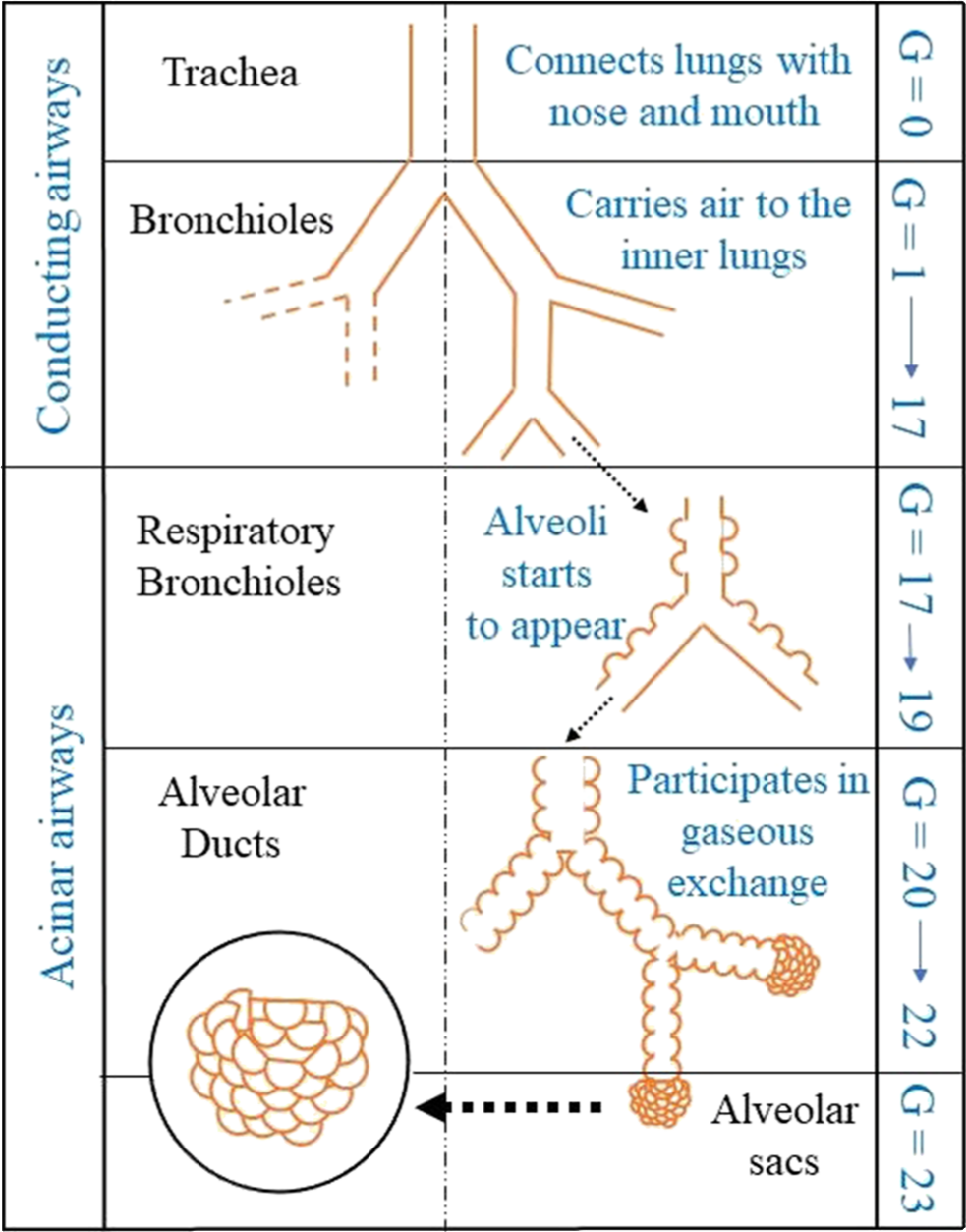}
\caption{A schematic of the human lung morphology as proposed by Weibel \cite{weibel1963morphometry} showing the conducting and acinar airways. The conducting airways transport air from the nose and mouth to the lung through the trachea, whereas the acinar airways consisting of alveolar ducts, participate in gas exchange with the blood.\cite{mallik2020experimental}}
\label{fig:human_lung_morphology}
\end{wrapfigure}
Weibel's models \cite{weibel1963morphometry} are an extremely influential work, entailing detailed measurements of the dimensions of both the conducting and the respiratory airways in the human lung. Using these measurements, Weibel constructed two different models of lung geometry in order to provide more accurate information on airway dimensions than was available as a basis for investigating the physical mechanisms of lung function.
The most widely adopted of these models has been his symmetrical model (Model A). The model was intended to represent regular features of the branching structure from the trachea right down to the alveoli. \textbf{Generation of an airway is defined as the number of bifurcations between the trachea and the airway. }

Generation classified the airways, and each generation was assigned an average diameter and length. The dimensions of the airways in the resulting symmetrical tree were deduced from measurements of a partial cast of the first ten(10) generations of a human lung, which was complete as far as the $5^{th}$ generation. Beyond this, various extrapolation procedures were used to estimate the dimensions. For generations 6-10, the frequency distribution of diameters within each generation was estimated from the observed frequencies for diameters of 2.5 mm or more by fitting to a hypothetical binomial distribution. Diameters were assigned to the remaining conducting airways using measurements of histological sections of respiratory bronchioles and alveolar ducts.

By estimating the total number of these structures and assuming symmetrical branching throughout the lung, it was deduced that they should occupy generations 17-23, and average diameters and lengths for these generations were obtained directly from the measurements of histological sections. Then, based on the diameters assigned to generations 4-10 and 17-23, a single smooth curve was deduced and was used to define airway diameters for all generations, including the intermediate range (generations 11-16) not covered by either the cast or the histological sections. A similar procedure was used to obtain an average length for each generation. 

The recognition of successive asymmetrical branching resulted in a large departure from the Model A symmetrical structure. Weibel constructed a second model (Model B) reflecting the resulting variability of the path lengths along different pathways in the bronchial tree. He suggested that the probability distribution of the path lengths required to reach airways of each given diameter was Gaussian and deduced mean and variance from his measurements using similar to Model A assumptions. In contrast to the symmetrical model, the asymmetrical one has rarely been adopted by other authors\cite{phillips1994diameter}. Both Weibel's models are highly idealized and rest on a relatively small amount of morphometric data. Nevertheless, his symmetrical model has been almost universally adopted for theoretical studies of transport in the human lung, with the dimensions sometimes even being treated as the result of direct measurement rather than extrapolation.

\subsubsection{Symmetrical and asymmetrical models of bronchial architecture}

The most widely adopted asymmetrical models of bronchial architecture have been those produced by Horsfield \cite{horsfield1986morphometry} and his colleagues. Their original paper applied this classification to a nearly complete cast of the conducting airways of a human lung, choosing the peripheral reference site to be the first airway of diameter \textbf{0.7 mm} or less. In the Horsfield ordering system, each airway is assigned an order equal to the most significant number of bifurcations between it and some chosen peripheral site. 
\begin{figure}[t!]
\centering
\includegraphics[width=0.5\linewidth]{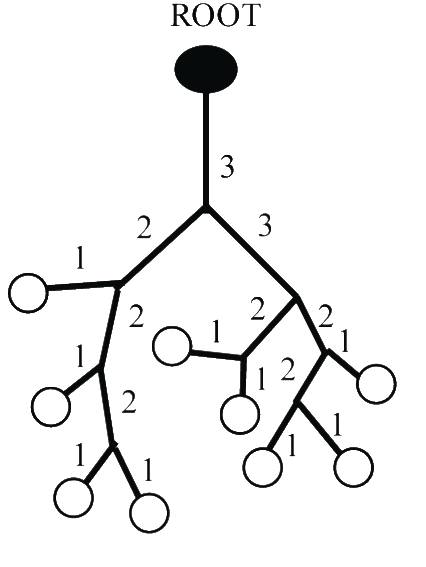}
\caption{A schematic of the Horsfield order. The first edge starts from the terminal point in Horsfield's ordering system. At a junction where edges of order j and order k cross, the parent edge is assigned to the order greater than that of j and k, or $j + 1$ if $j = k$. \cite{min2012use}}
\label{fig:horsfield_order}
\end{figure}
For each order, they counted the total number of airways and measured the average diameter, the average length and an estimate of the average number of distal respiratory bronchioles supplied (see the section on model flow velocity). Based on these measurements, they constructed a model of bronchial geometry as follows. 
For each order, the total number of observed airways was given, together with approximations for the average values of the other measured properties. \textbf{The model value of the average number of distal respiratory bronchioles was taken from a regression line drawn through a logarithmic plot of the measured average values against the order.} A similar procedure was used for diameter, with the data first being separated into three ranges of orders. 
Measurements were also made of the average length of airways of each diameter, and these allowed a length to be deduced from the diameter assigned to each order. Horsfield and his colleagues produced a series of models based on this ordering technique and Strahler's system.

Within Strahler's system, the basic unit consists of a Strahler branch, which may be composed of several successive generations of airways\cite{phillips1994diameter}. The main advantage of Horsfield ordering is its ability to characterize economically many of the properties of asymmetric ally branching trees. The principal practical drawback in using the technique to summarise morphometric data is that the assignment of orders requires the structure being measured to be virtually complete down to some chosen peripheral level. 

Applying Horsfield's original model to theoretical work is challenging because it is difficult to know how to combine airways of different orders into one whole lung. The order of daughters within each lobe should differ by a constant amount\cite{horsfield1986morphometry,phillips1994diameter}, defined as equal to six(6). In order to set this value to three, a regression line was drawn through a logarithmic plot of the number of airways observed versus order.

\subsubsection{Detailed measurements of tracheobronchial geometry}

The measurements of Raabe et al.\cite{raabe1976tracheobronchial} are the most detailed published morphometric data on the human bronchial tree. The authors measured the conducting airways from casts of 7 pairs of lungs from 4 mammalian species, including two humans. For the tree parts measured, the airways were coded so the branching structure could be recorded. The diameter, length and angles were measured for each airway. The angle was measured from the parent airway towards the direction of gravity. It was noted whether the airway was terminal, broken or otherwise anomalous. Raabe et al.'s data include two sets of measurements of the bronchial tree of one individual. The first set includes all the airways with a diameter of 3 mm or more and a more extensive set with several morphometric sampling techniques.

\subsubsection{Diameter-based reconstruction of the branching pattern of the human bronchial tree}

Philips et al.\cite{phillips1994diameter} present predictions for the total number of airways of each diameter, for the distribution of bronchial surface area and volume between airways of different diameters, and the probability distribution of the lengths of different pathways through the bronchial tree. With some additional assumptions, they also present a calculation of the proportion of the alveoli supplied by airways of each diameter and derive a simple model for the average flow velocity as a function of diameter. Their results are based on the measurements of Raabe et al. comparing them with measurements of Weibel's symmetrical model and Horsfield and Cumming's order-based model.
It must be borne in mind that these three sets of results are based on measurements of casts of three different lungs made with different materials and techniques. Comparing the average diameters in each of the first four generations, we find that they are between a tenth and a third more minor in the Weibel model and about a quarter smaller in the cast used for the Horsfield and Cumming model than in the Raabe data \cite{raabe1976tracheobronchial}. 

In order to allow approximately for the gross effects of lung size, they adopt the crude expedient of rescaling all the linear dimensions of the Weibel and Horsfield and Cumming models so that their tracheal diameters are the same in the Raabe data. It should be stressed that quantitative comparisons based on such a rescaling can only be very tentative, especially for the Weibel model. In the latter, the dimensions of the large airways, relative to the corresponding values in the Raabe data, are variable. Additionally, the anatomy of the smallest airways can be expected to be independent of the lung's size, so the rescaling's applicability to the smallest conducting airways is limited.

\begin{figure}[!ht]
\centering
\includegraphics[width=\textwidth]{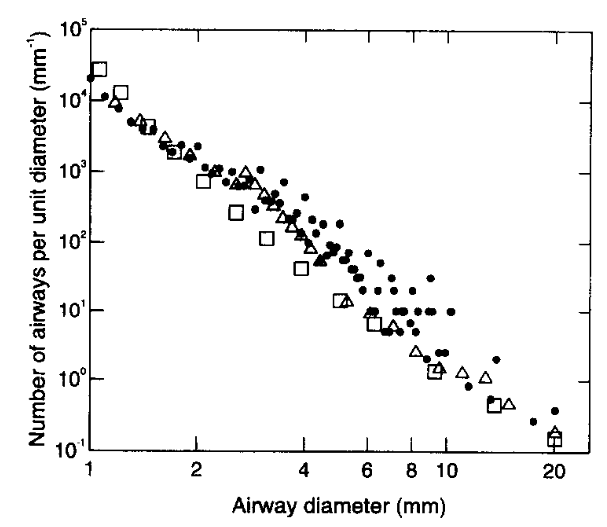}
\caption{Number of airways per unit diameter, plotted against diameter. \textbf{a)} $\bullet$ Reconstruction based on Raabe data\cite{raabe1976tracheobronchial}. \textbf{b)} $\boldsymbol{\square}$ Data based on Horsfield and Cumming model. The linear dimensions are rescaled with respect to tracheal diameter. \textbf{c)} $\boldsymbol{\Delta}$ Data based on symmetrical Weibel model. Linear dimensions are rescaled with respect to tracheal diameter}
\label{fig:philips}
\end{figure}

Figure \ref{fig:philips} shows a doubly logarithmic plot, the number of airways per unit diameter, as a function of diameter. For an airway diameter of 3 mm, the authors predict that the number of airways per unit diameter is about $10^3$ $mm^{-1}$. In order to obtain the absolute number of airways with diameters in a range near this value, one should multiply this quantity by the width of the range. Thus the number of airways we should expect in the range $2.8-3.2 mm$ is about $0.4 x 10^3 = 400$. It is clear from the figure that the authors' predictions and those based on the Weibel and Horsfield and Cumming models exhibit broadly similar behaviour, with the frequency per unit diameter decreasing steadily as the diameter increases.

\subsection{Segmentation of bronchial tree in chest volumetric computed tomography (CT)}

With the advancement of medical imaging techniques, the extraction of airway structure and lung volume from imaging started to play an important role in analyzing pulmonary diseases. A literature review on the analysis of lung CTs, including segmentation of the various pulmonary structures, can be found in \cite{sluimer2006}. In contrast, a comparative study of automated and semi-automated segmentation methods of the airway tree from CT images was presented in \cite{lo2012extraction}. Overall, segmentation approaches can be classified into methods based on morphology \cite{irving20093d}, morphological aggregation \cite{fetita2009morphological}, voxel classification \cite{lo2010vessel}, adaptive region growing with  constraints\cite{pinho2009robust,feuerstein2009adaptive,fabijanska2009results,mendoza2009maximal,van2009automatic,weinheimer2008fully,wiemker2009simple,tschirren2009airway}, tube similarity \cite{bauer2009airway,smistad2014gpu} and gradient vector flow \cite{bauer2009segmentation}. Several implementations of the approaches mentioned above are available in the literature. The tube segmentation framework \cite{smistad2014gpu} utilizes gradient vector flow \cite{bauer2009segmentation} and the FAST heterogeneous medical image computing and visualization framework \cite{smistad2015fast} utilizes the seeded region growing approach. In this dissertation, we employ airway segmentation\cite{nousias2020avatree} as a first step to obtain the personalized structure in the first generations. In contrast, the more advanced generations are simulated based on a tree extension algorithm.

\subsection{Mathematical models of the airway structure}

\textbf{Wang et al.}\cite{wang1992bifurcating} investigated the mathematical modelling of bifurcating distributive systems in biological organisms, such as veins on leaves, the lung airways, and the arterial (or venous) system for blood. The need for such models is justified since the experimental evaluation of the function and dynamics of distributive systems may be either impossible or too costly. The modelling principle needs to ensure that any location in a given region should be reachable by a path through the system. Previous modelling fractal-based and area halving solutions such as Mandelbrot solutions were highly tedious because, for N generations of branching, there are $2N - 1$ computations of areas of various irregular shapes. 

\noindent The Monte Carlo method proved its ability to solve the branching tree problem efficiently. Specifically, given a collection of points on a plane sampled under a particular distribution (e.g. uniform or Gaussian), the following algorithm would construct a bifurcating branching system which serves these points:

\begin{enumerate}
\item The center of mass of the points is found by averaging the individual coordinate positions. 
\item From a given starting point, a dividing line connecting the center of mass is constructed.
\item A branch from the starting point lies on this line and has a length of a certain fraction of the distance to the center of mass.
\item The dividing separates two subcollections of points with new starting points at the tip of the branch just found.
\item The process is repeated until the sub-collection contains only one point. Then the branch ends at that point.
\end{enumerate}

 \begin{figure}[!ht]
  \centering
  \includegraphics[width=0.8\textwidth]{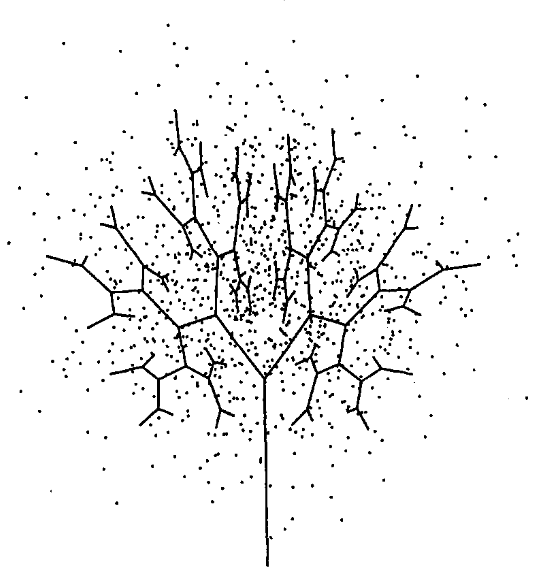}
  \caption{Gererated distributive structure on a randomly sampled set of points as presented in Wang et al. \cite{wang1992bifurcating}}
\end{figure}

\subsubsection{Relationship Between diameter, branching angle and Flow Rate}

Kitaoka et al.\cite{kitaoka1999three} investigate the optimal relationship between the flow rate through and diameter of a segment in a living organ \cite{murray1926physiological, kamiya1974theoretical}. 
Given that each branch is a circular, rigid tube with a constant diameter, d is diameter, Q is the flow rate, $n$ is a constant called the diameter exponent, and C a constant that relies on the organ in question and the fluid then:
\begin{equation}
    Q=Cd^n
    \label{eq:qcdn}
\end{equation}
This relationship is independent of the generation index of the branch. It relies on the idea that the conduit has to optimally overcome the energy expenses associated with fluid flow friction through the segment and power dissipation due to the biological maintenance of the segment. When their values normalize flow rate and diameter in the parent branch, equation \eqref{eq:qcdn} can be rewritten as follows. 
According to Kitaoka et al.\cite{kitaoka1999three}, assuming that each branch is a circular, rigid tube with a constant diameter, the optimal relationship between the flow rate through and diameter of a segment in a living organ \cite{murray1926physiological, kamiya1974theoretical} can be defined as 
\begin{equation}
    Q'=\left(d'\right)^n
    \label{eq:qdn}
\end{equation}
where Q and d are the normalized flow and diameter, respectively. Equation \eqref{eq:qdn} is assumed to be \textbf{common to all organs containing any fluid}. When the flows are conserved before and after branching, we can relate the diameter of the parent $d_0$ to the diameters of the daughters $d_1$ and $d_2$ 
\begin{equation}
    d^n_0=d^n_1+d^n_2
\end{equation}
The flow-dividing ratio (r) as the ratio of flow in the daughter branch receiving the smaller flow to that in the parent. 
\begin{align}
    d_1=&d_0r^\frac{1}{n}\\
    d_2=&d_0(1-r)^\frac{1}{n}
\end{align}
\begin{figure}[!ht]
  \centering
  \includegraphics[width=0.5\textwidth]{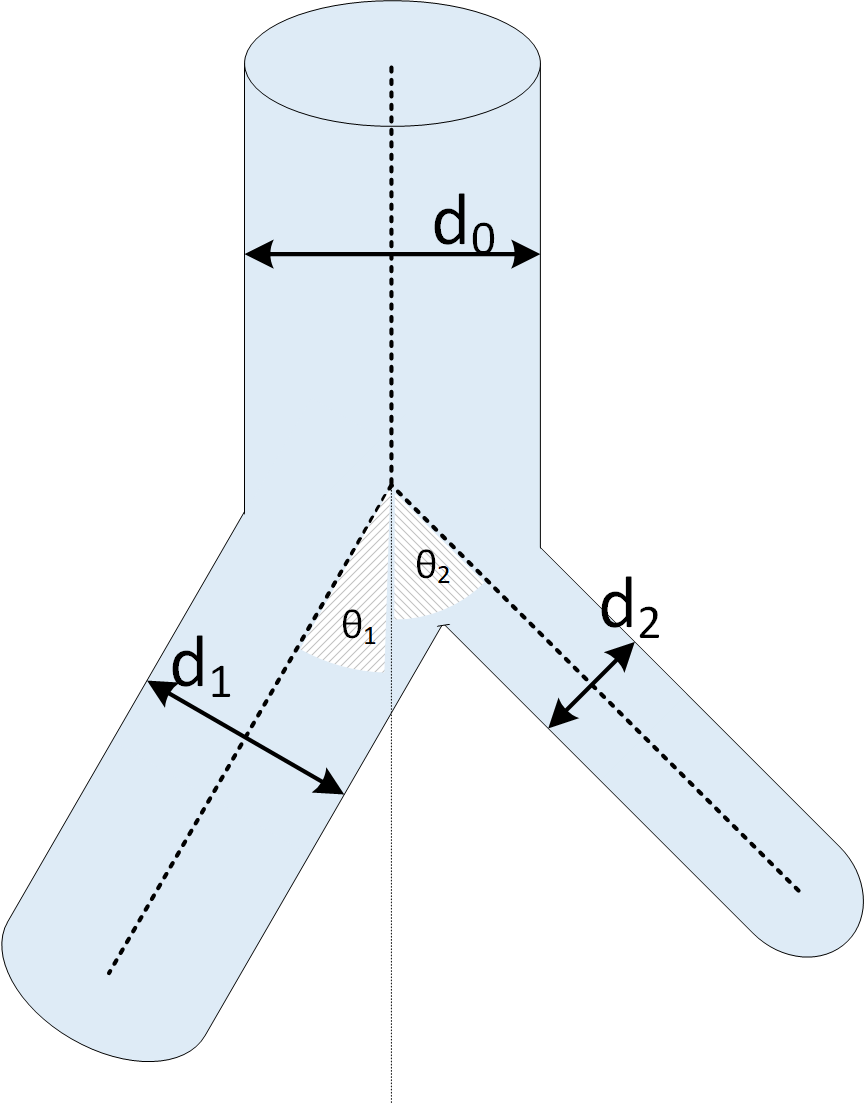}
  \caption{Visualization of power law to describe the relationship of diameters between successive generations, $d^n_0=d^n_1+d^n_2$}
\end{figure}

Based on a minimum energy loss principle, the
value of n was suggested to be three (3)\cite{kamiya1974theoretical}.
In the case of several (4) mammal airways, the value was $2.4$ –$2.9$. Given morphometric data of four mammalian airway trees published by Raabe et al. (25) and Wang et Kitaoka et al. found $n=2.8$ for human lungs.
Using the assumption that the total volume of the three branches in a bifurcation is minimized, Kamiya et al.\cite{kamiya1974theoretical} derived the following equation between the branching angles and the diameters using used $n=3$.
\begin{equation}
    \frac{d^n_0}{sin(\theta_1+\theta_2)}=\frac{d^n_1}{sin\theta_1}+\frac{d^n_2}{sin\theta_2}
\end{equation}
Uylings \cite{uylings1977optimization} argued that the value of n depends on the flow condition in the conduit: for laminar flow $n = 3$, whereas for turbulent flow $n = 2.333$.

\subsubsection{Design principles of branching duct system}

Additionally, \textbf{Kitaoka et al.}\cite{kitaoka1999three} proposed a three-dimensional model of branching ducts that supply fluid evenly to a given organ, such as the human tracheobronchial tree. The authors assumed two basic principles: A) The amount of fluid delivered through a branch is proportional to the volume of the region it supplies. Therefore, each bifurcation should result in a proportional volume distribution between parent and daughter regions and a proportional flow rate distribution between parent and daughter branches. B) The terminal branches of the tree are homogeneously arranged within the organ until they become daughter branches. This division proceeds until the daughter branches become terminal branches. The terminal branches correspond to the terminal bronchioles for the human airway tree because they are the airway segments beyond which convective airflow dominates diffusion.
\textbf{The pulmonary acini, being regions of the lung distal to the terminal branchioles, are the terminal structural units that receive almost identical flows per unit volume and sample the space uniformly.
}
The authors establish the following nine basic rules to generate two daughter branches given a parent branch and its volume region by repeatedly applying these rules to successive daughter branches. 
\begin{enumerate}
\item Branching is dichotomous. 
\item The parent branch and its two daughter branches lie in the same plane, referred to as "the branching plane". 
\item The volumetric flow rate through the parent
branch is maintained after branching. The sum of the flows in the daughter branches is equal to the flow in the parent branch. 
\item The region supplied by a parent branch is divided into two daughter regions by a space-dividing plane. This plane is perpendicular to the branching plane and extends out to the border of the parent region. 
\item The flow-dividing ratio r is set to be equal to the volume-dividing ratio, defined as the ratio of the volume of the smaller daughter region to that of its parent. 
\item Diameters and branching angles of the two daughter branches are computed given value $r$.
\item The length of each daughter branch is assigned a value three times its diameter. There is a supplementary rule (rule 7a) for correcting the length according to the region's shape. 
\item If branching continues in a given direction, the daughter branch becomes the new parent branch. The associated branching plane is set perpendicular to the branching plane of the old parent (Fig. 2). We call the angle between the two successive branching planes the "rotation angle of the branching plane." A supplementary rule (rule 8a) for correcting the rotation angle according to the region's shape. 
\item The branching process in a given direction
stops whenever the flow rate becomes less than a specified threshold or the branch extends beyond its region.
\end{enumerate}
Despite the success of the Kitaoka model, the resulting model was markedly more asymmetric than the human lungs. The algorithm was susceptible to the assumed geometry of the thoracic cavity and parameters fundamental to the method. As an extension of the Kitaoka model, Tawhai et al. developed a CT/MRI driven method combining both Wang et al.\cite{wang1992bifurcating} and Kitaoka\cite{kitaoka1999three} approaches.   

\subsubsection{Anatomically Based Three-Dimensional Model of the Conducting Airways}

Tawhai et al.\cite{tawhai2000generation} extended the 2D algorithm of Wang et al.\cite{wang1992bifurcating} into 3D space introducing a series of novel features. The authors defined the lung surface for the host volume using a surface mesh derived from magnetic resonance imaging (MRI) data. The surface mesh describes only the surface of the lungs, as more detailed information was not visible on the MRI slices. The horizontal and oblique fissures were determined by examination of slices through the torso from the Visible Human project \cite{ackerman1999visible}. These torso slices are 1 mm sections through a frozen cadaver, on which the lung fissures are identifiable.
The central airways from the trachea to the lobar bronchi are based mainly on the study of Horsfield and Cumming. However, they have been adjusted to fit the MRI data and model from Bradley et al. \cite{bradley1997geometric}. The end of each lobar bronchus provides a starting point for model generation into each corresponding lobe. 
Instead of the random point approach, a fine uniform grid of points was fit inside the three-dimensional host spaces. A uniform grid was computationally less expensive than generating random points. The irregular shape of the host spaces means that meshes generated using the uniform grid will be grown in response to host geometry rather than according to a defined distribution of the random points.

\begin{figure}[!ht]
\label{fig:tawhai2000}
  \centering
  \includegraphics[width=\textwidth]{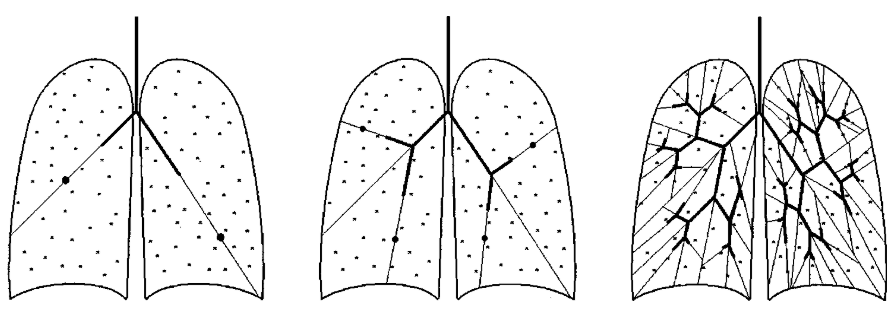}
  \caption{Volume filling method introduced by Tawhai et al. \cite{tawhai2000generation} using a point distribution across the filled space and a successive division scheme.}
\end{figure}

The vector in the direction of the parent branch and the coordinates of the centre of mass were used to define the splitting plane. For consistency between the three-dimensional and two-dimensional algorithms, the splitting plane in the three-dimensional algorithm should include the line from the starting point through the centre of mass. However, the choice of a splitting plane that includes this line is infinite.

The two-dimensional algorithm aims to supply each random point in the hosting space with a terminal branch. Terminal branches occur in the three-dimensional model when their length is less than a defined limit or the subcollection of points the branch supplies is less than a defined number. Hence termination of an airway path occurs when a subcollection of random points contains only one point. In contrast, it is not appropriate to have a terminal branch supplying each uniform grid point in the three-dimensional host volumes. The terminal airways are not necessarily uniformly distributed nor as numerous as the number of grid points.

Branch lengths decrease with the order in the conducting airway tree. A limit on the length is imposed such that generated branches with a length less than or equal to the length limit are considered terminal bronchioles. The terminal bronchioles are on the order of 1–1.5 mm \cite{tawhai2000generation} therefore, a limit of 1.2 mm was imposed to allow some variation in the terminal bronchiole length.

As a result, the three-dimensional volume filling algorithm was formulated as follows
\begin{enumerate}
    \item Initially, the host volume was filled with a fine uniform grid of points 
    \item The centre of mass of the points contained by a single host lobe is found by averaging the individual coordinate positions. 
    \item The vector in the direction of the corresponding lobar bronchus and the centre of mass coordinates are used to define a splitting plane. The splitting plane is extended to the host boundaries, and points on either side of the plane are assigned into two sub-collections of points.
    \item The centre of mass of each sub-collection of points is calculated.
    \item An imaginary line is constructed from the end of the lobar bronchus to each centre of mass. A branch is generated from the end of the lobar bronchus, lying on the imaginary line, extending a defined fractional distance from the branching fraction toward the centre of mass for each sub-collection of points. 
    \item The branching angle is defined as the angle between the projection of the parent branch and the new generated branch calculated. Suppose the branch angle is larger than an angle limit. In that case, the angle is set equal to the limit, so the resulting branch continues to lie in the plane of its parent branch and the imaginary dividing line. 
\item The length of the branch is calculated. The branch is a terminal airway if the length is less than or equal to a length limit. 
    \item The position of the branch end is checked to ensure it is inside the hosting space. If the branch end is outside the hosting space, its length is reduced until the endpoint lies within the host.
    \item The number of grid points in the sub-collection is compared with the point number limit. If the number of points is smaller than the limit, the branch is a terminal airway. 
    \item When all branching is completed for a single generation, grid points from terminal branches up to a generation limit are reassigned to the neighbouring branches.
    \item The process continues until a terminal airway terminates all pathways. Diameters are randomly assigned to branches in each Horsfield order using data from Horsfield as mean values and a coefficient of variation of 0.1.
\end{enumerate}

\subsection{Patient-specific structural and functional modelling}

Personalized models of the lung entail multiple spatial scales of interest. From the organ to the large airways, the smaller airways, the alveoli and blood vessels. Personalized image-based models allow us to shed light on intra-organ inter-scale relationships to predict function based on physical laws. Image-based models of the lung have been used, among other purposes, to study phenomena associated with flow characteristics in the most significant airways and the alveolated airways, blood flow in the largest pulmonary vessels, and perfusion of the whole lung, as well as soft-tissue deformation. Recent imaging studies are paving the way toward new models of alveolar structure that will provide insight into tissue mechanics at the alveolar level. In addition, four-dimensional CT has also been used to simulate lung tissue motion to account for respiratory motion in radiation therapy of thoracic tumours. However, in those studies, material properties of the lung parenchymal (including tumours) were assumed homogeneous, which may limit the accuracy of the models. These studies and others aim to provide a functional interpretation that complements imaging and experimental studies.

\subsubsection{Computed tomography imaging data}

According to Rawson\cite{rawson2020x}, CT imaging involves taking multiple  X-ray projections from various angles around a sample, either 360-wide or 180-wide. X-ray projections correspond to the absorption rate of the photons. Afterwards, the data are computationally reconstructed. The outcome is a greyscale virtual 3D reconstruction of the sample attenuation capability. Slices can be subsequently extracted for efficient inspection. Segmentation can be employed to distinguish certain parts facilitating volumetric quantification.

\begin{figure}
\label{fig:ctexample}
  \centering
  \includegraphics[width=\textwidth]{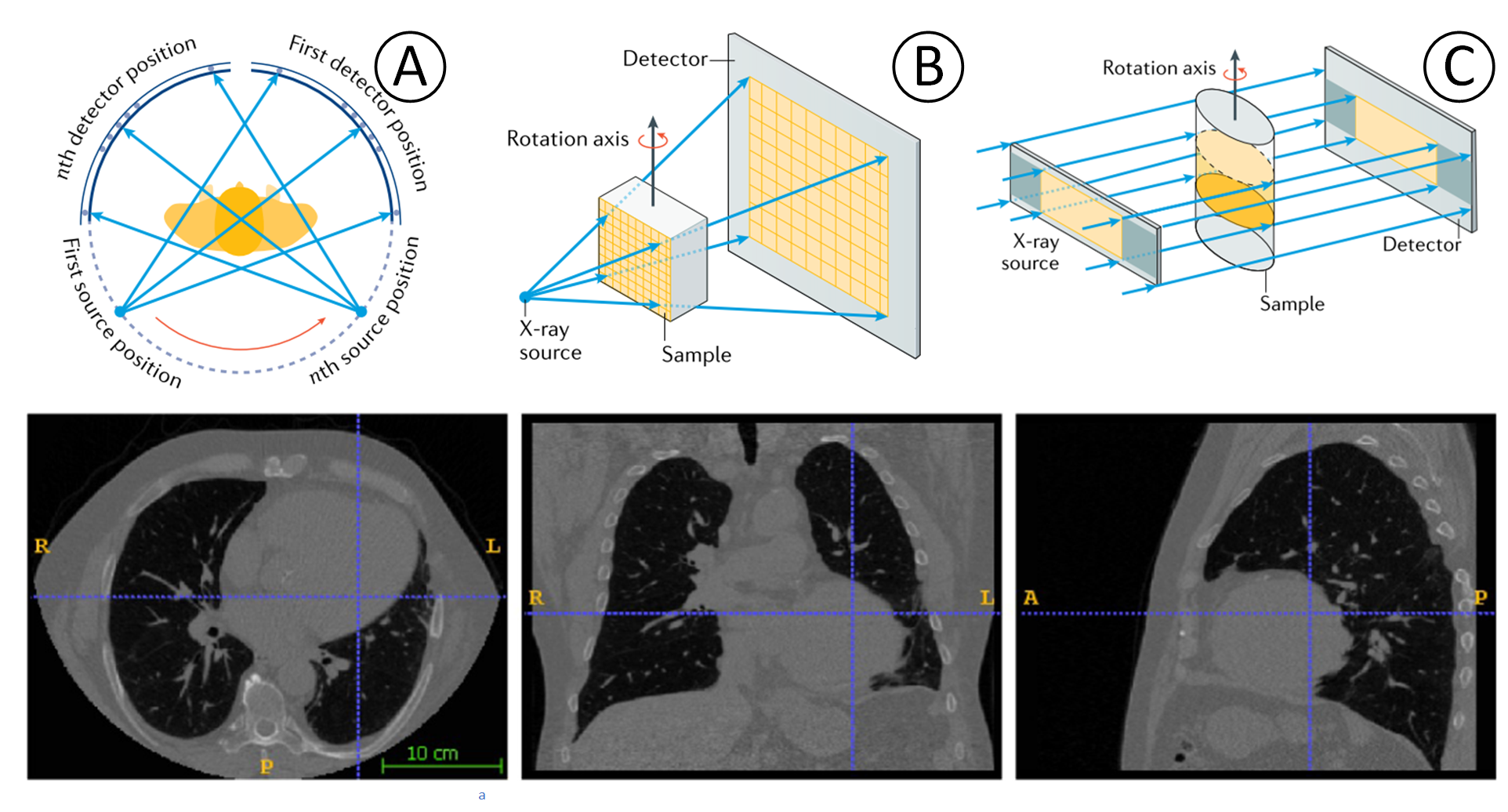}
  \caption{Common X-ray computed tomography configurations. Top row is reproduced from the work of Withers et al. \cite{withers2021x} (A) In Gantry system the source and the detector rotate in tandem around the patient\cite{withers2021x}. (B) Cone beam system typical of laboratory systems\cite{withers2021x}. (C) Parallel beam system geometry typical of synchrotron X- ray systems\cite{withers2021x}. Second row visualize virtual slices of human lung\cite{nousias2020avatree}}
\end{figure}

\subsubsection{The Human Atlas Project}

The Human Lung Atlas, elaborated by Tawhai et al.\cite{tawhai2009lung}, includes male and female subjects spanning several decades of life, integrating structural and functional information from MDCT imaging and spirometry. The human female lung is smaller in volume relative to body size than the male lung\cite{hopkins2004gender,harms2008sex}, has smaller mean airway diameters and a smaller diffusion surface. Functional gender differences include greater expiratory flow limitation and increased work of breathing in the female lung during heavy exercise. One goal of the Atlas is to establish the range of normality of an extensive set of structural and functional measures of the adult human lung. A second goal is to catalogue these normative values in a form that can then serve as the basis for detecting early pathology in an individual and a basis for tracking the changes in an individual lung over time. A further goal is to partner imaging-based measures with the computational model-based interpretation of physiological mechanisms to analyze structure-function relationships in lung disease.

\subsubsection{3D models of the bronchial tree}

The earliest studies of airway CFD used smooth cylindrical tubes merging in a single bifurcation \cite{balashazy2003local}, building to several bifurcations\cite{luo2007particle}, and then more recently incorporating imaging-defined airway surface data. Studies based on single and double bifurcation models have been insufficient for analyzing particle transport and deposition in the bronchial airways. Furthermore, several studies have demonstrated the importance of anatomically accurate geometry for accurate flow and particle distribution predictions. Recognizing the need for more accurate modelling of the airway bifurcation, Hegedus et al. \cite{hegedHus2004detailed} presented a mathematical description of a morphologically realistic airway bifurcation. The bifurcation was designed for CFD study and used their method to merge several bifurcations into multiple airway geometries. A critical feature was to enforce a smooth transition between the airways and rounding at the carina. Farkas et al. \cite{farkas2006characterization} used the model from Hegedus et al. \cite{hegedHus2004detailed} to piece together a model from generation 1 to generation 5 in the right upper lobe. They found that aerosol deposition highly depends on airway geometry, breathing parameters, and particle characteristics. Recognizing the importance of subject-specific geometry in predicting airflow, several groups have constructed geometric models based on in vivo-volumetric imaging. These models are typically derived by using commercial or in-house software to create a mesh of connected triangles that covers the surface of anatomical structures (the airways) that have been segmented from volumetric imaging. The triangulated surface is then filled with a connected mesh of tetrahedral elements. Other shaped elements can also be used, but tetrahedra are most frequently employed. The quality of the resulting geometric model depends on the image quality, segmentation accuracy, the algorithms employed for meshing, and the rigour applied to testing the mesh refinement for convergence of the numerical solution. For example, poor segmentation could lead to an irregular model surface that must be smoothed. This can be time-consuming, requiring subjective user input to decide which structures are actual and which are not.

\subsubsection{1D models}

Anatomically based models of the airway or pulmonary vascular trees can be derived using measurements from casts \cite{horsfield1986morphometry,weibel1963morphometry}. 
The models derived from such studies have been employed extensively in mathematical studies of the lung. However, the limitation of this approach is that the models are not subject-specific, are not the same as the lung in vivo, and do not have a spatial relationship with the tissue in which the model is embedded.
Tawhai et al. \cite{tawhai2010image} proposed a volume-filling branching (VFB) method for creating patient-specific imaging-based models of the airway tree that are geometrically compatible with morphometric studies. The models are generated within the lungs/lobes. Therefore, the spatial relationship between the airways and the lung tissue is intrinsic. Burrowes et al. later extended the method to the pulmonary arterial and venous trees, including specific definitions of the supernumerary vessels often overlooked.

This approach of combining in-vivo imaging with supplemental airways can produce a model specific to an imaged subject\cite{tawhai2010image}. The airway trees created with the VFB method are morphometrically consistent with cast and imaging studies measurements. Individual airway trees generated for human and ovine subjects have geometry appropriate to their species: the human airways form a relatively symmetric bifurcating tree, whereas the sheep airways branch in a monopodial approach and hence are far more asymmetric. The VFB algorithm reproduces these features in response to the lung or lobe boundary shape. This algorithm is the only method that produces complete airway or pulmonary vascular tree models within anatomically realistic lung shapes. It is the only method used to generate nonhuman airway trees.

The model generated by VFB is a 1D tree with branches distributed in 3D space\cite{tawhai2010image}. In particular, this feature becomes useful when considering interacting functions, such as how changes in tissue properties affect airway tethering and airway collapse or when comparing simulated with spatially distributed experimental results. The model must define diameters throughout the entire structure to be used for computation. The diameters of the imaging-based (uppermost) airways are assigned directly by measuring the cross-sectional area of the airway in a plane orthogonal to the central axis. Diameters of algorithm-based airways can be assigned using a Horsfield or Strahler ordering weighted against the ratio of the diameter of a child compared to that of its parent.

The image-based 1D model has been used in various lung function studies: inert gas mixing, fluid dynamics with thermal coefficients, ventilation distribution, oscillation mechanics, and the distribution of pulmonary perfusion. The advantage of the one-dimensional model for these studies is that it accounts for all airways from the trachea to the terminal bronchioles, includes realistic heterogeneity in-branch dimensions and connectivity, and produces realistic distributions of perfusion and ventilation concerning the direction of gravity.
The 1D model necessarily requires 1D equations to simulate function. These are derived from generalized (3D) equations by making simplifying assumptions. The 1D equations typically |average (integrate) the solution over the airway cross-section. Therefore, the 1D model cannot provide any details about the airflow's turbulent structures. Because it cannot correctly account for turbulence, there may be an error in estimating airway resistance.

\subsubsection{Subject specific 3D-1D models of the bronchial tree }

Lin et al.\cite{lin2013multiscale} have established a method to create a fully resolved three-dimensional (3D) mesh from the trachea to any terminal bronchioles of interest, allowing one to simulate fluid and particle transport from the model entrance to the level of lung parenchyma. The airway tree beyond CT resolution is generated by a volume filling method (VFM) developed by Tawhai et al.\cite{tawhai2000generation,tawhai2003developing,tawhai2006imaging,tawhai2011computational}. The VFM takes the skeleton of the 3D CT-resolved central airway tree in a human subject and then generates a tree to fill the entire volume within the subject's five lobes. The resulting airway trees are consistent with measurements from airway casts \cite{horsfield1986morphometry,phalen1978application} and imaging studies\cite{sauret2002study}. However, they are only specific to the subject's lung lobe shapes and orientation of their central airways.

\subsubsection{3D-1D transition}

The study of airflow in the lung must ultimately be able to span from the mouth to the alveolated airways in the lung periphery. Beyond approximate generations 6 to 9 in the human lung, the smaller airways cannot usually be visualized with current clinical imaging. 
Lin et al.\cite{lin2013multiscale} presented a method for creating subject-specific 3D and 1D coupled airway mesh structures with seamless transition between the 3D and 1D scales, incorporating the desired level of geometric detail wherever it is needed in the airway tree. 
Tawhai et al.\cite{tawhai2004ct} converted subject-specific 1D models for the entire conducting airway tree to a high-order (cubic Hermite) 2D surface mesh of the entire domain. The parent and child branches merge with a smooth, continuous surface at the bifurcations. The uppermost airways—for which there is MDCT (multidetector-row CT) surface data—are geometry fitted to enforce accurate airway surface geometry. The resulting 2D surface mesh is continuous with the surface of the volume filling algorithm-based airways, which are generally assumed to have a circular cross-section. Any portion of this surface mesh can be converted to a 3D CFD-ready mesh by selecting a region of interest to study in detail. The 3D mesh is created only within those airways, and the remainder of the domain is the original 1D tree. The 3D and 1D trees are a single continuous model with different dimensions in specified regions. An advantage of this approach is that transport can be studied in, for example, models that include successive generations to define the conditions under which a 1D model representation will suffice.

The 3D airway model allows detailed and accurate simulation of gas flow and particle transport in transition from turbulent flow in the central airways to laminar flow in the small airways in any region of interest. The 3D model includes all central airways segmented from MDCT imaging for this subject and five selected pathways. Each pathway corresponds to one in each of the five lobes. The MDCT-based airways extend from generation 0, referring to the larynx and trachea, reaching up to generation 5. The 1D airway model supplements the remainder of the airways to bridge from the 3D airways to the lung parenchyma, allowing the specification of subject-specific regional ventilation as described in the following section.

\subsubsection{Image Registration and regional ventilation}

Image registration can be applied to register consecutive CT image volumes and study human subjects' lung tissue expansion and contraction over multiple breathing cycles.
By registering two or more images, one image can be used as the reference image and the others as the floating images. The floating image is then transformed to match the reference image. The registration process generates a point-wise voxel-by-voxel displacement field between the two images for various applications. 
The displacement field between two lung volumes can be used to deform the conducting airway for breathing-lung simulation.
Local lung volume change is calculated using the Jacobian of the transformation field. The Jacobian measures local volume expansion and contraction, which can be used with CT numbers to assess regional ventilation and lung tissue mechanics. 
The regional ventilation can be overlapped with the entire airway tree to derive flow rate fractions for parenchymal units that enclose terminal bronchioles. These fractions can subsequently be used to determine subject-specific CFD boundary conditions at the 3D ending airway branches via tree connectivity and mass conservation \cite{yin2010simulation}. This boundary condition is referred to as an image-based BC. Yin et al.\cite{yin2013multiscale}  extended the same technique to three CT image volumes to account for the non-linearity of lung mechanics.

\subsubsection{Coupling of Structure and Function}

Yin et al.\cite{yin2010simulation} compare the distributions of outlet velocity and static pressure for three boundary conditions: a) image-based BC, b) uniform velocity BC, and c) uniform pressure BC. The outlet velocity and pressure distributions obtained from the image-based BC are more heterogeneous than those obtained from the uniform velocity BC and the uniform pressure BC. In particular, the image-based BC predicts a much greater pressure drop at the airways in the left lower lobe (LLL) and right lower lobe (RLL). On the other hand, the uniform velocity BC produces the greatest pressure drop in the right middle lobe (RML). In contrast, the uniform pressure BC fails to account for the pressure variation at different ending airways. 
The distributions predicted by the image-based BC agree well with the measurements. In contrast, both uniform pressure and uniform velocity BCs under-predicted the ventilation to the LLL and RLL, whereas the uniform velocity BC over-predicted the ventilation to the RML, the left upper lobe (LUL) and the right upper lobe (RUL).
The implication of the above BC study for regional deposition of particles is significant.

\subsection{Computational fluid dynamics to investigate flow regimes and particle transfer within the bronchial tree}

Flow and pressure in the airways change during the breathing cycle and cannot be measured directly. CFD studies compute flow in a domain by performing simulations through boundary conditions (BCs) that must be defined to find a unique solution to the equations. The boundary conditions typically specify the pressure, flow, or velocity at the inlet and outlets of the models. 
The approach taken in earlier CFD analysis has been to specify equal pressure or flow BCs at all outlets of the model, neglecting any variation due to geometry, downstream resistance, and gravitational effects on the regional volume expansion of the lung. Both stationary and pulsatile BCs have been investigated. The former yields a constant flow rate, whereas the latter adopts a breathing waveform. 
De Backer et al. \cite{de2008flow} approached this problem by specifying two different pressure values at the ends of the 3D CT-resolved left and right main bronchi to approximate a subject-specific BC. The authors produced a steady inspiratory flow with different proportions distributed to the left and right lungs. 
Other approaches to overcome the boundary condition problem have been proposed by Lin et al. \cite{lin2007characteristics,lin2009multiscale} and Tawhai et al. \cite{tawhai2010image}, using image registration and soft-tissue mechanics, respectively. 

Lin et al. proposed a multiscale CFD framework that utilizes 3D-1D coupled meshes and image-registration-derived regional ventilation and deformation for realistic simulation of pulmonary airflow that relates directly to an imaged subject. 
Tawhai et al. \cite{tawhai2011computational,tawhai2006imaging} developed a soft-tissue-mechanics-based model for elastic deformation of the compressible lung tissue, which can be used to provide flow and pressure boundary conditions for a 1D tissue-embedded airway model.

\subsection{Validation studies}

Montesantos et al. \cite{montesantos2013airway} extracted the morphometric information from the HRCT scans of seven healthy adults and six moderately persistent asthmatic patients through reconstructed three-dimensional models of their lungs. The results were then analyzed to determine whether there are differences in geometry and connectivity between the healthy and asthmatic populations that extend beyond the lumen diameter and wall thickness. The authors compared the data available in the literature to assess the consistency of our methods and the validity of some theoretical relationships used for airway tree models. Finally, certain aspects of intrasubject variability in different locations within the lung were investigated to evaluate the levels of geometrical non-uniformity at different lung regions in health and
asthma.

\section{Monitoring medication adherence in Obstructive Respiratory Diseases}
\label{section:monitoring-medication}

The respiratory system is a vital structure vulnerable to airborne infection and injury. Respiratory diseases are the leading cause of death and disability globally. Specifically, nearly 334 million people have asthma, the most common chronic disease of childhood, affecting 14\% of all children globally \cite{world2017global}. The effective management of constrictive pulmonary conditions lies mainly in proper and timely medication administration. However, as recently reported \cite{ngo2019inhaler}, a large proportion of patients misuse their inhalers. Studies have shown that possible technique errors can harm clinical outcomes for users of inhaler medication \cite{darcy2014method,jardim2019importance}. Incorrect inhaler usage and poor adherence were associated with high asthma test scores \cite{gupta2014copd}, long periods of hospitalization and high numbers of exacerbations.

Several methods have been introduced to monitor a patient's adherence to medication. As a series of studies indicate, effective medication adherence monitoring can be defined by successfully identifying actions performed by the patient during inhaler usage. Several inhaler types are available in the market, among which the pressurized metered-dose inhalers and dry powder inhalers are the most common\cite{schreiber2020inhaler}. Developing an intelligent inhaler setup that allows better monitoring and direct feedback to the user independently of the drug type is expected to lead to more efficient drug delivery, thereby becoming the main product used by patients.

The pMDI usage technique is successful if a specific sequence of actions is followed \cite{murphy2019help}. 
Specifically, \begin{inlinelist}
\item inhaler cap should be removed, and the subject should shake the inhaler to ensure consistent dose delivery. The device should be primed in case of first use for a certain canister.
\item Subject should breathe out completely.
\item Then, start breathing in slowly before pressing the drug activation button.
\item Upon inhalation, the patient should hold their breath for at least ten seconds.
\item Afterwards, the subject should slowly breathe out.
\end{inlinelist}
Common errors in this technique made by patients are summarized in the following points:
\begin{inlinelist}
\item Negligence of shaking the aerosol inhaler before use.
\item Negligence of priming the aerosol inhaler before first use.
\item Failing to identify empty inhaler before use.
\item Failing to breathe out fully before inhaling.
\item Incorrect coordination of MDI and pMDI actuation with inspiration.
\item Incorrect inspiration flow rate (breathing in too fast or too slow).
\item Failing to hold breath after inhalation.
\end{inlinelist}

Landmarks of relevant literature on monitoring methodologies are shortly presented in the following paragraph. 
The earliest monitoring methodologies encompass electronic or mechanical meters integrated into the device, activated with the drug delivery button. Howard et al. \cite{howard2014electronic} reported the existence of several such devices, able to record the time of each drug actuation of the total number of them. 
The use of audio analysis came up later as a method which can characterize the quality of inhaler usage while also monitoring the timings of each audio event. The classical audio analysis involves a transformation of the time-domain into a set of features, mainly, in the frequency domain, including Spectrogram, Mel-Frequency Cepstral Coefficients (MFCCs), Cepstrogram, Zero-Crossing Rate (ZCR), Power Spectral Density (PSD) and Continuous Wavelet Transform (CWT). Subsequently, audio-based evaluation employs the extracted features via classification approaches to locate and identify medication-related audio events. 
Holmes et al. \cite{holmes2012automatic, holmes2013acoustic, holmes2014acoustic} designed decision trees for blister detection and respiratory sound classification. This study includes detection of drug activation, breath detection and inhalation-exhalation differentiation and provides feedback regarding patient adherence. 
Taylor et al. \cite{taylor2014acoustic, taylor2016monitoring} used the continuous wavelet transform to identify pMDI actuations to quantitatively assess the inhaler technique, focusing only on the detection of inhaler actuation sounds. As a step forward, data-driven approaches learn from features and distributions found in the data by example. 
Furthermore, Taylor et al. in  \cite{taylor2018advances} compared Quadratic Discriminant Analysis (QDA) and Artificial Neural Network (ANN) based classifiers using MFCC, Linear Predictive Coding, ZCR and CWT features.

\subsection{Preliminaries on audio feature extraction}
\label{sec:audio-features-extraction}

\subsubsection{Spectrogram}
The spectrogram of each audio signal that represents the time-localized signal power at various frequencies.  In details, for a timeseries $x[n]$ with $n$ timepoints, where $n \in (1,\cdots,N)$, the spectrogram $\mathbf{S}$ is described as follows:
\begin{equation}
    S(m,\omega)=spectrogram\{x[n]\}(m,\omega)=\vert  X(m,\omega)\vert^2
\end{equation}
where $X(m,\omega)$ is the Short Time Fourier transform of $x[n]$,

\begin{equation}
\omega \in (1,\cdots,\frac{w}{2}+1)
\end{equation}
\begin{equation}
   m \in (1,\cdots,\floor*{\nicefrac{N-(w-h)}{h}}) 
\end{equation}
$w$ is the window size and $h$ is the frame increment.
For the sake of self-completeness according to the authors in \cite{esmaeilpour2020sound}, for a given continuous signal $a(t)$ which is distributed over time, its STFT using window function $w(\tau)$ can be computed using Eq.~\ref{STFT_contin}.
\begin{equation}
 \mathrm{STFT}\begin{Bmatrix} a(t) \end{Bmatrix}(\tau, \omega)=\int_{-\infty}^{\infty}a(t)w(t-\tau)e^{-j\omega t}dt
 \label{STFT_contin}
\end{equation}
\noindent where $\tau$ and $\omega$ are time and frequency axes, respectively. This transform is quite generalizable to discrete time domain for a discrete signal $a[n]$ as:
\begin{equation}
 \mathrm{STFT}\begin{Bmatrix} a[n] \end{Bmatrix}(m,\omega)=\sum_{n=-\infty}^{\infty}a[n]w[n-m]e^{-j\omega n}
 \label{STFT_disc}
\end{equation}
\noindent where $m\ll n$ and $\omega$ is a continuous frequency coefficient. In other words, for generating the STFT of a discrete signal, we need to divide it into overlapping shorter-length sub-signals and compute Fourier transform on it, which results in an array of complex coefficients. Computing the square of the magnitude of this array yields a spectrogram representation as shown in Eq.~\ref{STFT_spec}.

\begin{equation}
 \mathbf{Sp_{STFT}}\begin{Bmatrix} a[n] \end{Bmatrix}(m,\omega)=
 \left | \sum_{n=-\infty}^{\infty}a[n]w[n-m]e^{-j\omega n} \right |^2
 \label{STFT_spec}
\end{equation}

\noindent This 2D representation shows frequency distribution over discrete-time, and compared to the original signal $a[n]$, it has a lower dimensionality, although it is a lossy operation.


\subsubsection{Mel-Frequency Cepstral Coefficients (MFCC)}
\label{section:preliminaries:mfcc}
In "Speech and Processing in Mobile Environments"\cite{rao2014speech}, the MFCC feature extraction technique involves the following steps: 1) windowing the signal, 2)applying the DFT, 3) log of the magnitude, 4) warping the frequencies on a Mel scale, 4) applying the inverse DCT. Following is a description of the various steps involved in the MFCC feature extraction process. 
\begin{enumerate}
\item \textbf{Pre-emphasis} aims to remove some of the glottal effects from the vocal tract parameters 
\begin{equation}
H(z)=1-b z^{-1}
\end{equation}
\item \textbf{Frame blocking and windowing}. Short-time Fourier transform takes place with a window of 20ms and an overlap of 10ms.
According to the overlapping analysis, each speech sound of an input sequence will be approximately centred at some frame. A window is applied to each frame to taper the signal towards the frame boundaries. Typically, Hanning or Hamming windows are employed. As part of the DFT, the harmonics are enhanced, the edges are smoothed, and the edge effect is reduced.
\item DFT spectrum: Each windowed frame is converted into a magnitude spectrum by applying DFT
\begin{equation}
X(k)=\sum_{n=0}^{N-1} x(n) e^{\frac{-j 2 \pi n k}{N}} ; \quad 0 \leq k \leq N-1
\end{equation}
\item Mel spectrum: Mel spectrum is computed by passing the Fourier transformed signal through a set of band-pass filters known as Mel-filter bank. The approximation of Mel frequency can be expressed as
\begin{equation}
f_{Mel}=2595 \log _{10}\left(1+\frac{f}{700}\right)
\end{equation}
where f denotes the physical frequency in Hz, and $f_{Mel}$ denotes the perceived frequency. Filter banks can be implemented in both time domain and frequency domain. The most commonly used filter shaper is triangular, and in some cases the Hanning filter can be found. The triangular filter banks are presented in \ref{fig:Mel-Spaced-Filter-Bank}
\begin{figure}[!ht]
    \centering
    \includegraphics[width=\textwidth]{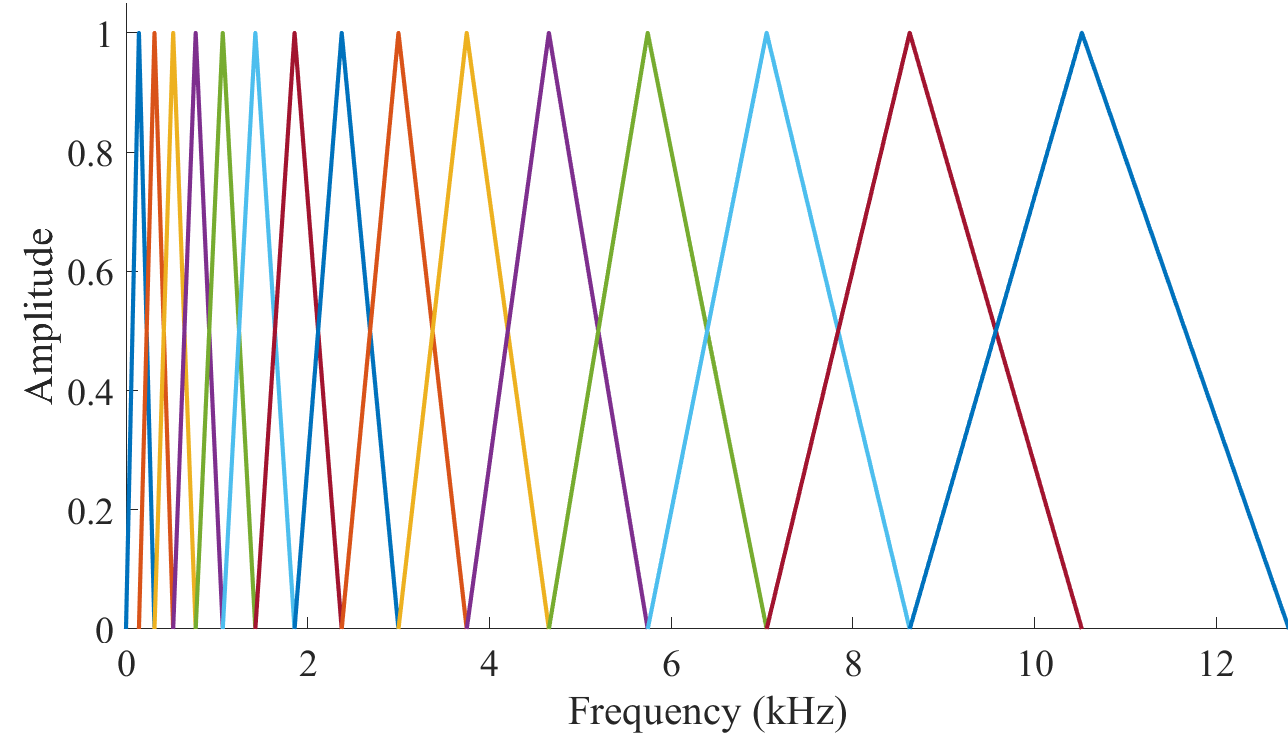}
    \caption{Mel-spaced filterbank triangular bandpass filters \cite{rao2014speech}}
    \label{fig:Mel-Spaced-Filter-Bank}
\end{figure}
The Mel spectrum of the magnitude spectrum $X(k)$ is computed by multiplying the magnitude spectrum by each of the of the triangular Mel weighting filters.
\begin{equation}
s(m)=\sum_{k=0}^{N-1}\left[|X(k)|^{2} H_{m}(k)\right] ; \quad 0 \leq m \leq M-1
\end{equation}
where M is total number of triangular Mel weighting filters. $H_{m}(k)$ is the weight given to the $k^{th}$ energy spectrum bin contributing to the $m^{th}$ output band and is expressed as:
\begin{equation}
H_{m}(k)=\left\{\begin{array}{cl}
0, & k<f(m-1) \\
\frac{2(k-f(m-1))}{f(m)-f(m-1)}, & f(m-1) \leq k \leq f(m) \\
\frac{2(f(m+1)-k)}{f(m+1)-f(m)}, & f(m)<k \leq f(m+1) \\
0, & k>f(m+1)
\end{array}\right.
\end{equation}
\item \textbf{Discrete cosine transform (DCT)}.The DCT is applied to the transformed Mel frequency coefficients and produces a set of cepstral coefficients. Before computing DCT, the Mel spectrum is usually represented on a log scale. This results in a signal in the cepstral domain with a quefrequency peak corresponding to the pitch of the signal and several components representing low quefrequency peaks. Since most of the signal information is represented by 
\begin{equation}
c(n)=\sum_{m=0}^{M-1} \log _{10}(s(m)) \cos \left(\frac{\pi n(m-0.5)}{M}\right) ; \quad n=0,1,2, \ldots, C-1
\end{equation}
where $c(n)$ is the cepstral coefficients, and $C$ is the number of MFCCs. Traditional MFCC systems use only 8–13 cepstral coefficients. The zeroth coefficient is often excluded since it represents the average log-energy of the input signal, which only carries little speaker-specific information.
\item  \textbf{Dynamic MFCC features} The cepstral coefficients are usually referred to as static features since they only contain information from a given frame. The extra information about the temporal dynamics of the signal is obtained by computing the first and second derivatives of cepstral coefficients. The first-order derivative is called delta coefficients, and the second-order derivative is called delta-delta coefficients. Delta coefficients tell about the speech rate, and delta-delta coefficients provide information similar to the acceleration of speech. The commonly used definition for dynamic computing parameters is
\begin{equation}
\Delta c_{m}(n)=\frac{\sum_{i=-T}^{T} k_{i} c_{m}(n+i)}{\sum_{i=-T}^{T}|i|}
\end{equation}
where $c_{m}(n)$ denotes the $m^{th}$ feature for the $n^{th}$ time frame, $k_{i}$ is the $i^{th}$ weight, and T is the number of successive frames used for computation. Generally, T is taken as 2. The delta-delta coefficients are computed by taking the first-order derivative of the delta coefficients.
\end{enumerate}

\subsubsection{Zero-cross rate} The average zero-crossing rate refers to the number of times speech samples change the algebraic sign in a given frame. The rate at which zero-crossings occur is a simple measure of the frequency content of a signal. It measures the number of times in a given time interval/frame that the amplitude of the speech signals passes through a value of zero. Unvoiced speech components typically have much higher ZCR.
Zero Crossing Rate is defined formally as the number of time-domain zero-crossings within a defined region of signal, divided by the number of samples of that region \cite{gouyon2000use}:
\begin{align*}
\mathcal{ZCR} = \frac{1}{T-1}\sum_{t-1}^{T-1} II\{S_{t}S_{t-1}<0\}
\end{align*}
where $S$ is a signal of length $T$ and the indicator function $II{A}$ is $1$ if its argument $A$ is true and zero $(0)$ otherwise.

\subsubsection{Discrete Wavelet Transform (DWT)}
Wavelet transform maps the continuous signal $a(t)$ into time and scale (frequency) coefficients similar to STFT using Eq.~\ref{dwt_contin}. 
\begin{equation}
 \mathrm{DWT}\begin{Bmatrix} a(t) \end{Bmatrix} = \frac{1}{\sqrt{\left | s \right |}}\int_{-\infty}^{\infty}a(t)\psi \begin{pmatrix} \frac{t-\tau}{s} \end{pmatrix}dt
 \label{dwt_contin}
\end{equation}
\noindent where $s$ and $\tau$ denote discrete scale and time variations, respectively, and $\psi$ is the core transformation function which is also known as mother function (see Eq.~\ref{morlet_func}). There are a variety of mother functions for different applications such as the complex Morlet which is given by Eq.~\ref{morlet_func}: 
\begin{equation}
 \psi(t)=\frac{1}{\sqrt{2\pi}}e^{-j\omega t}e^{-t^{2}/2}
 \label{morlet_func}
\end{equation}
\noindent Discrete time formulation for this transform is shown in Eq.~\ref{dwt_discrete}.
\begin{equation}
 \mathrm{DWT}\begin{Bmatrix} a[k,n] \end{Bmatrix}=\int_{-\infty}^{\infty}a(t)h\begin{pmatrix} na^{k}T-t \end{pmatrix}
 \label{dwt_discrete}
\end{equation}
\noindent where $n$ and $k$ are integer values for the continuous mother function of $h$. Spectral representation for this transformed signal is a 2D array which is computed by Eq.~\ref{dwt_spectrogram}:
\begin{equation}
 \mathrm{Sp_{DWT}}\begin{Bmatrix} a(t) \end{Bmatrix}=\left | \mathrm{DWT}\begin{Bmatrix} a[k,n] \end{Bmatrix} \right |
 \label{dwt_spectrogram}
\end{equation}

\subsection{Preliminaries on pattern classifiation}
Pattern classification in machine learning is described as the following problem \cite{kotsiantis2007supervised}. A set of entities is available, where an attribute vector describes each entity, and a special attribute is called the class. While the attributes of the attribute vector can be either discrete or continuous, the class attribute is discrete. Pattern classification is the issue of estimating a function f that maps the attributes vector to the class attribute. Such a function is referred to as a classification model. A classification model is helpful because it explains how the attributes vector distinguishes a set of entities into diverse classes. However, its most widespread objective is predictive. It can be applied to differentiate into classes new entities, which have known attributes but unknown classes. Pattern classification is the most frequently seen problem in supervised learning.

A classification algorithm or classifier is a systematic approach to constructing classification models from data. Each classifier applies a learning technique to identify the model that best fits the data (notably, identifying the dependencies between the attributes vector and the class). The resulted model must be able to fit the data optimally and predict entities with unknown classes. Thus, it is required to dispose of generalization properties. The following steps comprise a classification problem. Initially, a training dataset is given, where entities have known classes, and a classification algorithm is applied to produce a classification model. Then, the generated model is applied to a testing dataset, where entities have unknown classes, to predict their classes.
In some cases, there exists no testing dataset but only training. In such a case, a classification model is evaluated with k-fold cross-validation: the training dataset is split into k parts (usually 10). Each part sequentially forms the testing set, and all the others form the training set. Finally, the $k$ results are combined.

The following paragraphs describe three well-known, sophisticated, and recently invented classification algorithms: support vector machines, random forests, and AdaBoost. We will be referring to the vector of attributes of entity $i$ as $x_i$ (assuming cardinality p) and to its class value $c_i$ (taking integer values from 1 to K). Thus, $K$ is the number of classes.

\subsubsection{Support vector machines} 
Support vector machines \cite{burges1998tutorial} is a well-known and sophisticated method for supervised learning. In describing SVMs, we will assume that the number of classes equals two (K=2); that is $c_i=\pm 1$. In the end, we will generalize to more distinct class values. Furthermore, let us assume that all entities in the training dataset are linearly separable in their attributes (p-dimensional) space. This fact denotes that there exists a separating hyperplane and consequently a vector of weights w and number b that satisfy the following conditions: 
\begin{align}
    b+wx_i\geq 1,c_i&=1\\
    b+wx_i\leq-1,c_i&=-1
\end{align}
 
Equivalently, $c_i(b+wx_i ) \geq 1 $ for every entity of the dataset. In that simple situation of linearly separable entities, SVMs effort to satisfy the aforementioned condition while minimizing the quantity
\begin{equation}
    \frac{1}{2}\left\|w\right\|^2
\end{equation}
This is a curve quadratic programming problem that is solved approximately. If the separating hyperplane is recognized, its boundaries are referred to as support vectors, and the solution can be represented simply by a linear combination of those.
Unfortunately, in real-life situations, the entities of a dataset are rarely linearly separable. There are two ways to deal with the issue in such a case that can be used simultaneously. The first is that one can leave a margin $\xi$ for misclassified entities. In quadratic programming form, this equals adding a slack variable in the constraint equation: 
\begin{equation}
    c_i\left(b+wx_i\right)\geq 1-\xi
\end{equation}
and minimizing the quantity
\begin{equation}
    \frac{1}{2}\left\|w\right\|^2+C\xi
\end{equation}
where C is a trade-off constant. The second one attempts to separate the entities linearly in another space of higher dimensionality. For that purpose, a kernel function is used that transforms the inner product of any two entities to the new space. Common kernel functions are:
\begin{itemize}
	\item The polynomial of degree q:
	
	\begin{equation}
	K\left(x_i,x_j \right)=\left(x_i x_j+1\right)^q
	\end{equation}
	
	\item The RBF with parameter q:
	\begin{equation}
	K\left(x_i,x_j\right)=e^{-q\left\|x_i-x_j\right\|^2}
	\end{equation}
	
	\item The sigmoid of order q:
	\begin{equation}
	K\left(x_i,x_j\right)=\tanh{\left(\kappa x_1 x_2-\delta\right)}^q
	\end{equation}
	\end{itemize}
There are two ways to deal with the multi-class problems $\left(K\geq2\right)$. One way is the one versus all approach. Here, we create K SVM classifiers, and for each classifier, we attempt to distinguish one particular class from all the rest. To determine the optimal class to pick, we assign the class for which the observation produces the highest distance from the separating hyperplane, therefore lying farthest away from all other classes. An alternative approach is known as the one versus one approach. We create a classifier for all possible pairs of output classes. We then classify our observation with each of these classifiers and tally up the totals for every winning class. Finally, we pick the class that has the most votes.

\subsubsection{Bagging and boosting}

\paragraph{Random Forests} 
Random forests \cite{breiman2001random} are the most widespread paradigm for the concept of classification bagging. Bagging (bootstrap aggregating) is a machine learning ensemble meta-algorithm designed to improve the stability and accuracy of machine learning algorithms. It is based on combining classifications of randomly generated training sets. In the case of random forests, the individual classifiers used in parallel are small classification trees, and they are applied to different bootstrap samples of the training dataset. The number of trees is selected to be 500 or 1000 usually.
More specifically, random forests grow many classification trees. To classify a new entity with an attributes vector, put the attributes vector down each tree in the forest. Each tree gives a classification, and we say the tree "votes" for that class. The forest chooses the classification having the most votes (over all the trees in the forest). Each tree is growing as follows:
\begin{itemize}
	\item If the number of entities in the training dataset is n, sample n entities at random but with replacement. This sample will be the training set for growing the tree.
	\item If the cardinality of $x_i$ is $p$, a number of $\sqrt{p}$ attributes is selected at random for each node of the tree, and the best split on these is used to split the node.
	\item Each tree is grown to the largest extent possible. There is no pruning.
	\end{itemize}
The forest error rate (accuracy) depends on (a) the correlation between any two trees in the forest. Increasing the correlation increases the forest error rate. (b) the strength of each tree in the forest. Increasing the strength of the individual trees decreases the forest error rate.
An advantage of random forests is that they do not need k-fold cross-validation. Instead, the out-of-bag (OOB) error estimate can be computed. Specifically, each tree is built using a different bootstrap sample from the original data. About one-third of the cases are left out of the bootstrap sample and not used in constructing each tree. Put every entity left out in the construction of each tree down the tree to get a classification. This way, a test set is obtained for each entity in about one-third of the trees. The OOB error is estimated in that test set.

\paragraph{AdaBoost} 
AdaBoost \cite{zhu2009multi} is the most widely used classification boosting algorithm. Boosting answers the following question: Can a set of weak learners create a single strong learner? A weak learner is a classifier slightly correlated with valid classification (for example, it can label entities better than random guessing). In contrast, a strong learner is a classifier arbitrarily well-correlated with the valid classification. As in the case of random forests, in AdaBoost, the weak learners are simple classification trees.
The AdaBoost algorithm is an iterative process that attempts to combine a set of weak classifiers. Starting with the unweighted training dataset, the AdaBoost builds a classifier, for example, a classification tree, that produces class labels for the entities. Then, if an entity is misclassified, the weight of that entity is increased (boosted). Another classifier is built with the new weights, which are no longer equal. Once more, the misclassified entities have their weights boosted, and the process is repeated. Typically, one may build 500 or 1000 classifiers this way. A score is assigned to each classifier, and the final classifier is defined as the linear combination of the classifiers from each stage. Specifically, let T be a weak multi-class classifier (classification tree).
\begin{itemize}
	\item Initialize the weights for all the entities as $w_i=\frac{1}{n}$.
	\item Repeat the following for, say, 500 or 1000 times, indexed by j:
	\begin{itemize}
	\item Fit a classification tree $T^j\left(x\right)$ to the training data using weights $w_i$.
	\item Compute the error :
	\begin{equation}
	err^j=\frac{\sum_{i=1}^n w_i  I(c_i \neq T^j (x_i)  )}{\sum_{i=1}^n w_i}
	\end{equation}
	\item Compute the coefficient :
\begin{equation}	
a^j=\log \left(\frac{1-err^j}{err^j}\right)+\log\left(K-1\right))
\end{equation}
	\item Update the weights :
	\begin{equation}	
	w_i=w_i e^{a^j I\left(c_i \neq T^j \left(x_i \right)  \right) }
	\end{equation}
	\item Renormalize the weights.
	\end{itemize}
	\item Output $C(x)=argmax_k \sum_j{a^j I\left(T^j\left(x\right)=k\right)} $.
\end{itemize}

\chapter[Patient-specific structural modelling of the pulmonary system]{Novel patient-specific structural modelling and simulation processing pipeline of the pulmonary system}
\label{chapter:novel-patient-specific}



Early studies focused on airway morphometry generating the first human bronchial tree models\cite{horsfield1986morphometry}. These studies employed casts to decipher the relationship between bronchi lengths, branching angles and airway diameters\cite{kitaoka1999three}.
On this basis, researchers built and validated a simulation model of airway morphogenesis from generation $1$ to generation $23$ \cite{wang1992bifurcating,tawhai2000generation}. 
Deterministic parameterized bronchial tree generation algorithms are used as a single input for the location of the first one or two generations and the lung volume, thus constituting the core of patient-specific modelling \cite{tawhai2003developing,tawhai2004ct,bordas2015development}. Personalized boundary conditions based on diagnostic imaging were combined with generative approaches and lumped models of resistive trees \cite{ionescu2008parametric,ionescu2013respiratory} constituting state of the art in pulmonary system modelling. Later studies incorporated patient-specific boundary conditions into computational fluid dynamics to examine flow regimes, wall stresses and aerosol deposition. In the same direction, modelling the airflow in constrictive conditions, such as asthma and chronic obstructive pulmonary disease (COPD), became feasible with the approaches mentioned above. Wall constriction and remodelling combined with patient-specific boundary conditions allowed the quantification of breathing conditions for asthmatic patients. 
We are motivated by these advancements by introducing an end-to-end modelling approach that produces anatomically valid airway tree conformations. Such conformations are adapted to personalized geometry and boundary conditions derived from diagnostic imaging and well-established airway extraction methods. Specifically, this study aims to provide a simulation framework to \begin{inlinelist}
    \item exploit imaging data to provide patient-specific representations
    \item perform structural analysis
    \item extend the segmented airway tree to predict the airway branching across the whole lung volume
    \item visualize probabilistic confidence maps of generation data
    \item simulate bronchoconstriction to 
    \item access patient-specific airway functionality
    \item perform fluid dynamics simulation in patient-specific boundary conditions to access pulmonary function
\end{inlinelist}.

\begin{figure}[t]
    \centering
    \includegraphics[width=\textwidth]{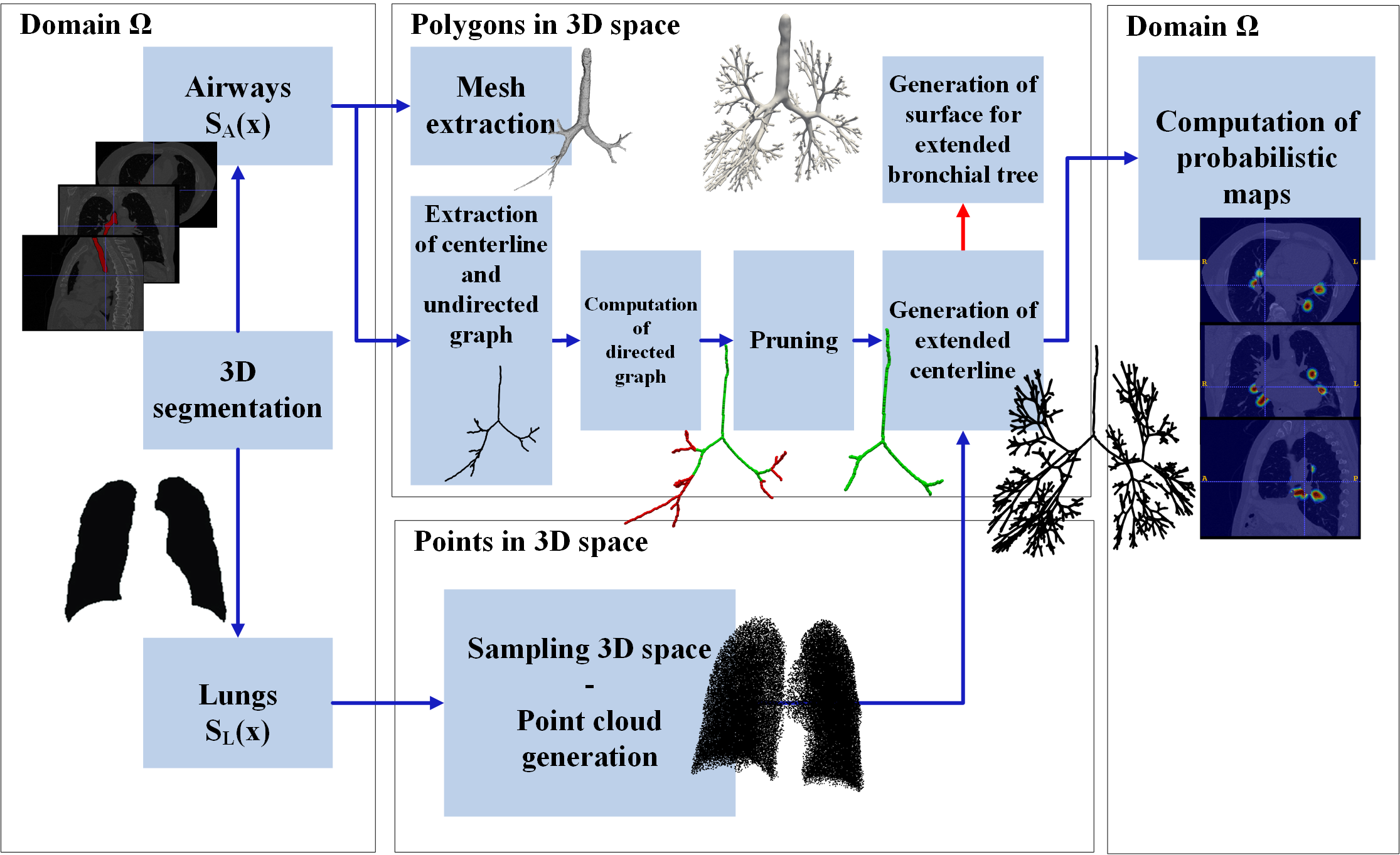}
    \caption{Processing pipeline of AVATREE}
  \label{fig:pipeline}
\end{figure}

The objective in this field of research is to enable the prediction of gas flow\cite{soni2013large,srivastav2011computational}, gas mixing\cite{hasler2019multi}, heat transfer\cite{karamaoun2018new}, particle deposition\cite{miyawaki2016effect,kolanjiyil2016computationally,kolanjiyil2017computationally,lalas2017substance}, and ventilation distribution \cite{whitfield2018modelling} in the pulmonary system. Lung ventilation patterns prediction \cite{pozin2017tree,pozin2018predicted}  can provide grounds for performance and fatigue estimation in high-frequency ventilation cases \cite{meyers2019high}, disease severity quantification, such as in asthma, and give insight into drug delivery or even in transfer of harmful particulates. Motivated by the recent advances in this field and building upon previous work \cite{lalas2017substance}, we developed an end-to-end approach facilitating pulmonary structural modelling that is based on the definition of the personalized boundary conditions required for fluid dynamics simulations. Specifically, in this work we
\begin{itemize}
    \item present an open-source simulation framework that utilizes imaging data to provide patient-based representations of the structural models of the bronchial tree,
    \item build and extend 1-dimensional graph representations of the bronchial tree and generate 3D surface models of extended bronchial tree models appropriate for computational fluid dynamics simulations
    \item generate probabilistic visualization of airway generations projected on the personalized CT imaging data,
    \item provide an open-source toolbox in C++ and a graphical user interface integrating modelling functionalities.
\end{itemize}

\noindent The processing pipeline uses as input CT images and is presented in Figure \ref{fig:pipeline}. Airway segmentation is applied to the imaging data to extract a 3D surface mesh and a 1-dimensional representation of the airways. 


\section{Segmentation and airways centerline extraction}
\label{subsection:segmentation}
The input of the presented approach is unlabeled CT scans required to extract bronchial tree and airways structural features. CT-based lung segmentation and annotation are required for the definition of the lung volume. For lung segmentation, we employ the FAST heterogeneous medical image computing, and visualization framework \cite{smistad2015fast} is employed. The result of the lung segmentation process is a binary mask visualized in Figure 1. We further process the segmentation result to distinguish the left and right lungs as a next step. The process is described below:

\begin{enumerate}
    \item A second region growing takes place starting from a single random point inside any segmented region only if all its immediate neighbours bare the same label.
    \item To advance the region growing front, all points neighbouring a candidate voxel must not include background voxel. This region is given a new label.
    \item Steps 1 and 2 are repeated for the other lung volume. The result is an image with three labels(background and two lung volumes). 
    \item To distinguish left or right, we employ the directed graph extracted from the main airways and follow the generic rule. The topological distance between the bifurcations of the first and the second generation is longer in the left lung. 
\end{enumerate}
The next step involves the segmentation of the first generations of the airways identifiable in the patient's CT image. However, any available airway tree segmentation method can also be applied. For this purpose, we investigated two algorithms. The first algorithm is the gradient vector flow \cite{bauer2009segmentation,bauer2009airway} which achieved high accuracy with low false-positive rate (only 1.44\%) in a comparative study \cite{lo2012extraction} in the context of the EXACT09 airway segmentation challenge. The second is a standard and stable approach based on seeded region growing \cite{adams1994seeded}. The former is included in the tube segmentation framework \cite{smistad2014gpu} and the latter in the FAST heterogeneous medical image computing and visualization framework \cite{smistad2015fast}.

\begin{figure}
    \centering
    \ifdefined\showFigures
    \includegraphics[width=\textwidth]{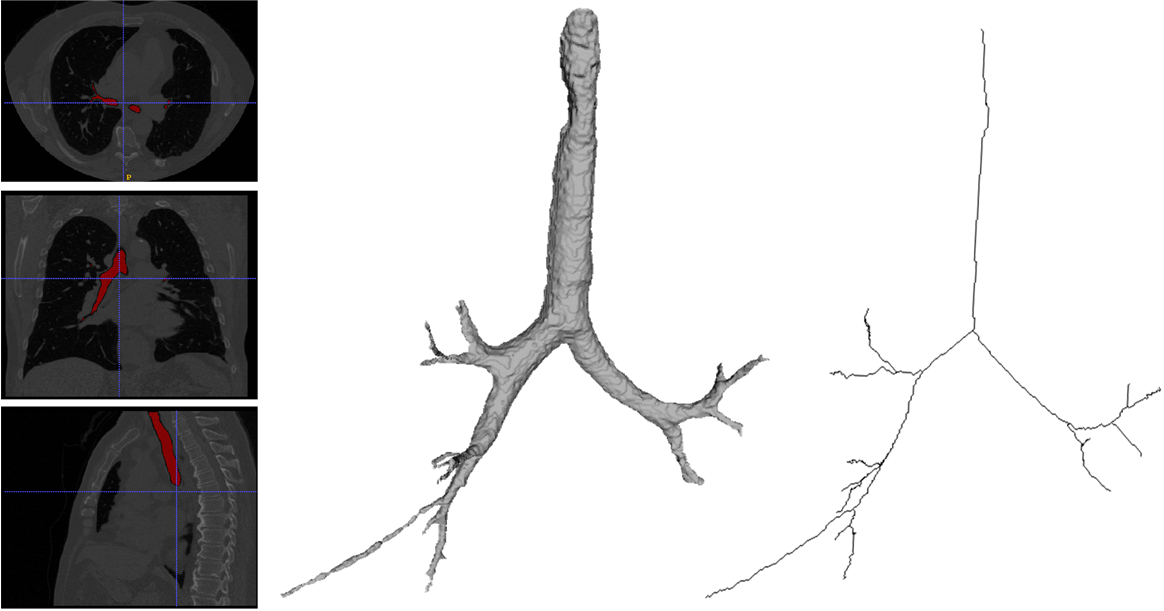}
    \fi
    \caption{Extraction of airway surface and centerline from CT scans using the FAST framework \cite{smistad2015fast}. Airway segmentation (left) and airway surface generation(center) and generation of one-dimensional representation through skeletonization.}
  \label{fig:extraction-of-airway-surface-n-centerline}
\end{figure}

\noindent Let's denote with 
\begin{equation}
  I(\mathbf{x}), I:\Omega \xrightarrow{}\mathcal{R},\mathbf{x}=(x,y,z),\mathbf{x} \in \Omega
\end{equation}
the gray level 3D medical image, where $\mathbf{x}=(x,y,z)$, $\mathbf{x} \in \Omega$ is a voxel in the spatial domain $\Omega \subset \mathcal{R}^3$ of the volumetric imaging data. The output of the segmentation algorithm for the airways is a binary image $S_A$ of equal size with $I$. Likewise, the output of the segmentation algorithm for the lung volumes is a binary image $S_L$ of equal size with $I$. 
The result is presented in Figure \ref{fig:extraction-of-airway-surface-n-centerline} and utilized to generate the prediction of complete bronchial tree structures based on personalized lung volumes. To derive the centerline from $S_A$, many methods are provided in the literature, including skeletonization or thinning. 
Figure \ref{fig:extraction-of-airway-surface-n-centerline} presents the up-to-four generations centerline of the airways.

\noindent The generation of the initial airway surface, lung volume and 1-dimensional representation are performed using the FAST framework\cite{smistad2015fast}.
This 1D representation of the bronchial tree is modelled by an undirected graph $\mathcal{G}=\{\mathcal{V},\mathcal{E}\}$ where $\mathcal{V}$ is the set of vertices and $\mathcal{E}$ is the set of edges. 
Each vertex, indexed by $i$, can be represented as a point $\mathbf{v}_i=(x_i,y_i,z_i)$. We denote the function $N(\mathbf{v}_i)$
yielding the set of vertices indices neighbouring vertex $i$. The undirected graph is extracted by the FAST framework \cite{smistad2015fast}
Moreover, converted into a directed graph with the following process. Initially, the graph starting point is defined as the one closest to the air inlet, i.e. the oral cavity or the trachea.

\begin{algorithm}
\caption{Automatic detection of inlet node in undirected graph}\label{alg:detection}
    \hspace*{\algorithmicindent} \textbf{Input : Graph $\mathcal{G}$ } \\
    \hspace*{\algorithmicindent} \textbf{Output : Index of graph inlet $y$} 
\begin{algorithmic}[1]
\Procedure{Derivation of graph inlet}{}
\State Initialize set $\mathcal{P}=\Set[\big]{}{}$
\ForEach {vertex $\mathbf{v}_i$}
\If {$|N(\mathbf{v}_i)| > 2 $} 
    \ForEach {$n \in  N(\mathbf{v}_i) $ }
        \State Initialize empty set $\mathcal{K}=\{i,n\}$
        \While {$N(\mathbf{v}_n) < 3$}
            \ForEach {$ m \in N(\mathbf{v}_n) $}
                \If{$m \not\in \mathcal{P}$ }
                    $\mathcal{K} \xleftarrow{} \mathcal{K} \cup m$
                \EndIf 
            \EndFor
        \EndWhile
        \State  $\mathcal{P} \xleftarrow{} \mathcal{P} \cup \mathcal{K} $
    \EndFor
\EndIf 
\EndFor
\State $\mathcal{K}_{max} = \max_{1 \leq i \leq |\mathcal{P}|} Length(\mathcal{K}_i) $ 
\State  $y \xleftarrow{} \mathcal{K}_{{max}_{|\mathcal{K}_{max}|}}$
\EndProcedure
\end{algorithmic}
\end{algorithm}

\noindent Given index $y$ the starting point for $\mathcal{G}$ we generate the directed tree $\mathcal{G}_D$. We define as distal point the vertex of the graph with no children and the distal branch as the edge containing a distal point.

\begin{figure}
    \centering
    \ifdefined\showFigures
    \includegraphics[width=0.7\textwidth]{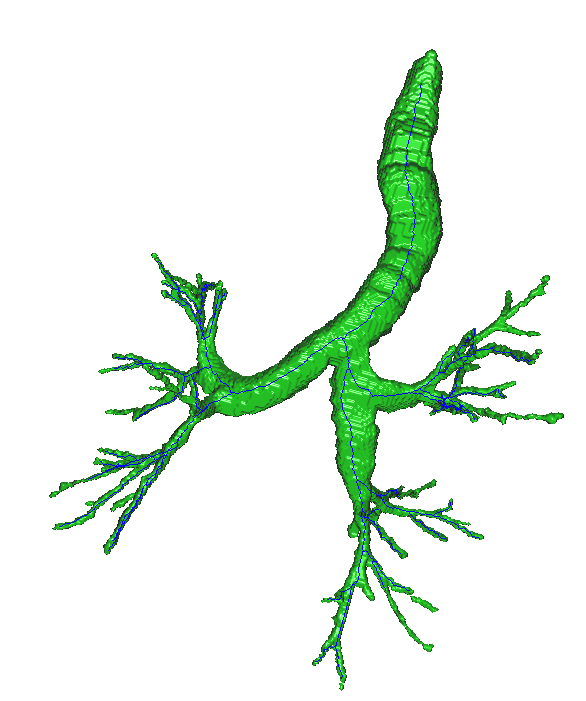}
    \fi
    \caption{Visualization of skeletonization and fitting process of one dimensional representation from extracted airway surface}
  \label{fig:generation1DSkeleton}
\end{figure}

\section{Generation of extended bronchial tree}
\label{subsection:generation}

\begin{figure}
  \begin{subfigure}[b]{0.42\linewidth}
    \includegraphics[width=\textwidth]{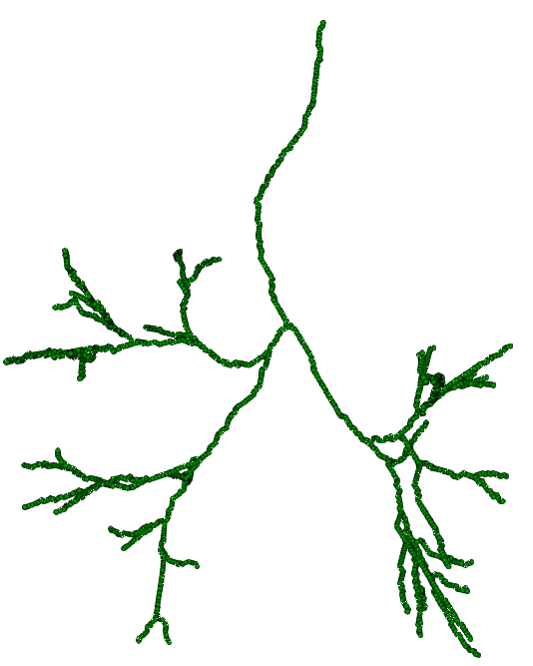}
    \caption{}
    \label{fig:pone:extractedCenterline}
  \end{subfigure}
  \hfill
  \begin{subfigure}[b]{0.42\linewidth}
    \includegraphics[width=\textwidth]{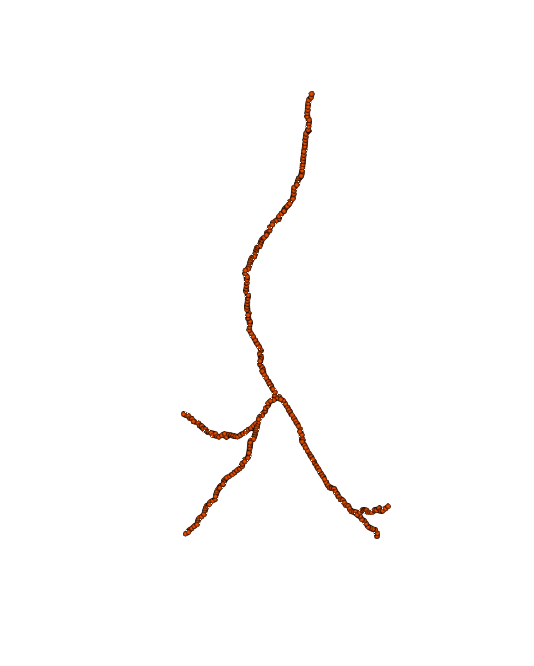}
    \caption{}
    \label{fig:pone:extractedCenterlineReduced}
  \end{subfigure}
  \caption{Reduction of bronchial tree to first two generations to avoid trifurcations in areas of the lung with high uncertainty}
  \label{fig:extractedCenterlineReduction}
\end{figure}

Since higher generations cannot be identified from the personal imaging data, we extend the bronchial tree based on population-wise empirical observations.
Initially, the directed graph generated by the procedure explained in subsection \ref{subsection:segmentation} is pruned. Specifically, the extracted 1-dimensional representation is processed to include all the bifurcations located at the end of a given generation to facilitate the volume filling algorithm. Figure \ref{fig:extractedCenterlineReduction} shows the result of pruning where all generations after the $n^{th}$ have been pruned.

\begin{figure}[t!]
    \centering
    \ifdefined\showFigures
    \includegraphics[width=0.4\textwidth]{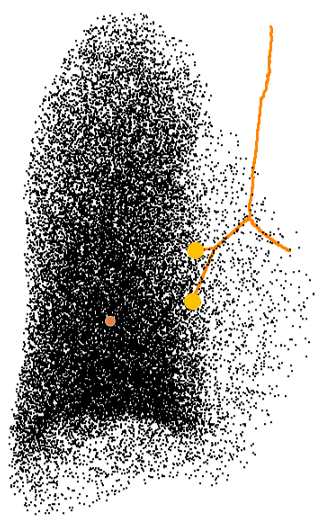}
    \fi
    \caption{Definition of center of mass (orange dot) and distal branches (yellow dots).}
  \label{fig:distal_branch}
\end{figure}

\begin{figure}[t!]
    \centering
    \ifdefined\showFigures
    \includegraphics[width=0.8\textwidth]{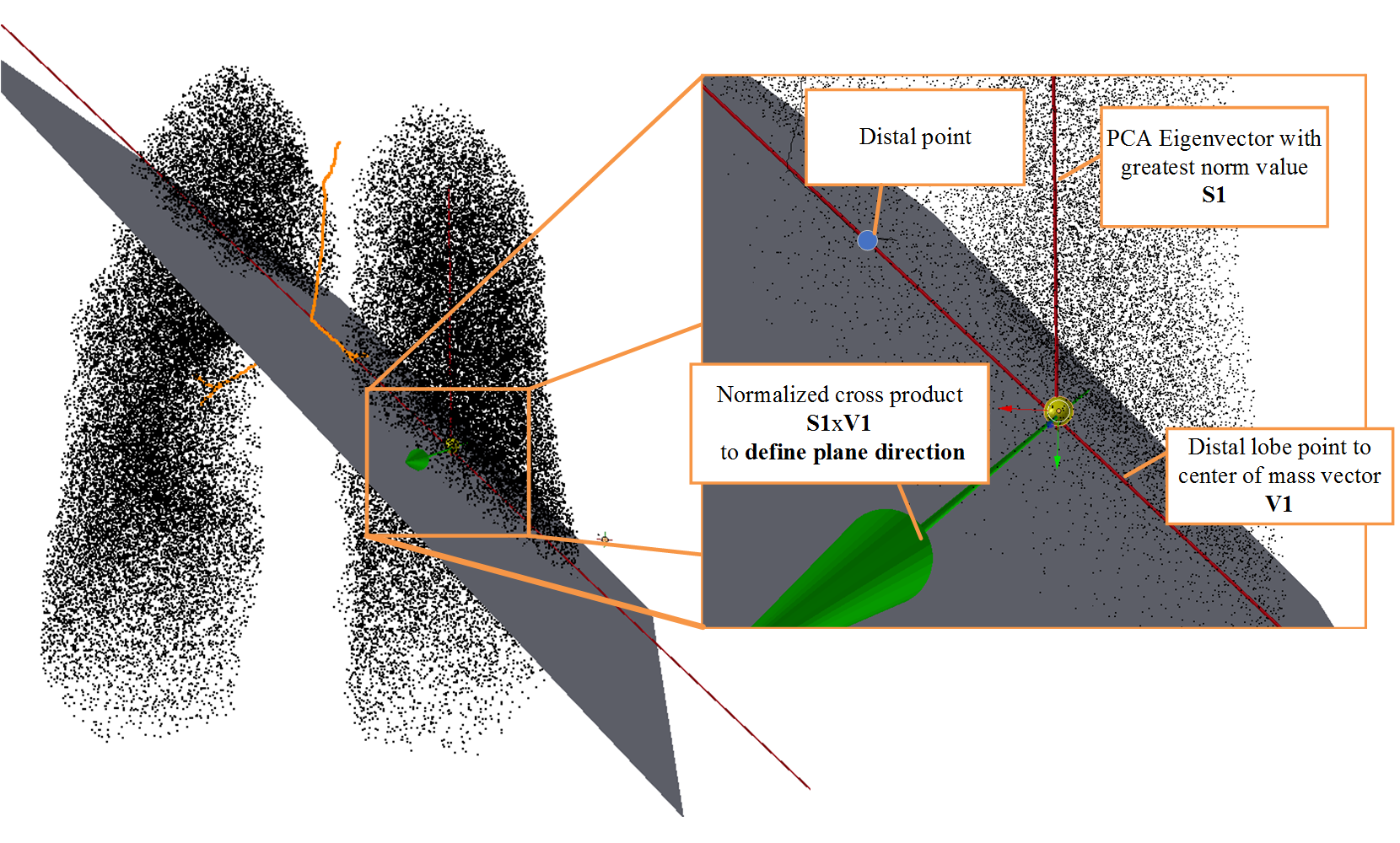}
    \fi
    \caption{Definition of splitting plane for bifurcating distributive structures.}
  \label{fig:plane_selection}
\end{figure}

The motivation behind pruning lies in that generations with lower index have greater certainty than generations with a higher index number.
The corrected tree is subsequently used for the bronchial tree extension.
The generation process utilizes the bronchial tree extension algorithm initially proposed by Tawhai et al. \cite{tawhai2000generation} and later improved by Bordas et al. \cite{bordas2015development}  while introducing a few safeguards to allow maximal space utilization. 
The following steps can describe the bronchial tree extension algorithm. 
\noindent For each lung subvolume $S_{L_L}$ and $S_{L_R}$:
\begin{enumerate}
    \item Generate a point cloud sampling the subvolume with a uniform random process. Figure \ref{fig:distal_branch} depicts the uniform sampling of each lung subvolume with a total number of $n=30000$ points \cite{tawhai2000generation,tawhai2004ct}.
    \item Assign a seed point to the closest distal branch
    as presented in Figure \ref{fig:distal_branch}.
    \item Calculate the center of mass of the sampled points as presented in Figure \ref{fig:distal_branch}.
    \begin{equation}
        \mathbf{c}=\frac{\sum_{\mathbf{p}_i\in\mathcal{P}} \mathbf{p}_i}{|\mathcal{P}|}
    \end{equation}
   
    \item 
    Employ principal component analysis (PCA) on the set of sampled points to define the splitting plane.
    The motivation for employing PCA is to address a space utilization aspect. The direction of the eigenvector with the greatest norm indicates the dimension of the data with the greatest variance denoting the direction where more space is available for the branches to grow. Picking a plane so that the resulting bounding box demonstrates the lowest possible variation inhibits the appearance of very long branches.
    Given data points 
    \begin{equation}
        \mathbf{D}=[\mathbf{p}_1\, \mathbf{p}_1\,\mathbf{p}_1\,\cdots\,\mathbf{p}_n]
    \end{equation}
    $\mathbf{A}=\mathbf{D}\mathbf{D}^T$ is the auto-correlation matrix. Direct singular value decomposition yields 
    \begin{equation}
    \mathbf{A}=\mathbf{U}\mathbf{\Sigma}\mathbf{U}^T
    \end{equation}
    where 
    \begin{equation}
       \mathbf{U}=[\mathbf{u}_1 \, \mathbf{u}_2\,\mathbf{u}_3]
    \end{equation}
    Then the largest eigenvector is defined as 
    \begin{equation}
    \mathbf{u}_m = \max_{1 \leq i \leq 3} \mathbf{u}_i    
    \end{equation}
    Given the vector $\mathbf{d}$ expressing the direction of the distal airway, the splitting plane is described by center of mass $\mathbf{c}$ and vector $\mathbf{d} \times [\mathbf{d} \times \mathbf{u}_m]$. The selected plane maximizes the available space for each new subdivision. Figure \ref{fig:plane_selection} presents a splitting plane splitting the set of points into two subdivisions. 
    \item Calculate the centroid of each new subdivision.
    \item For each centroid, define a line segment starting from the seed point extending \textbf{40\%} of the distance towards the centroid of the subdivision.
    \item If a newly created branch is more minor than $2 mm$, it is considered terminal.
    \item The process is repeated until no seed points remain.
    \item 
    Any branch found outside the lung volume is removed along with any apparent daughter branches.
\end{enumerate}
\noindent It is important to denote that the presented pipeline generates a tunable user-defined number of generations. If $n$ is the number of desired generations, we set stopping criteria in the extension of the bronchial tree until $2^{(n+1)}$ bifurcations have been reached.
The resulting 1D representation (Figure \ref{fig:extend-1D-rep}) predicts the location of the bifurcating distributive \cite{wang1992bifurcating} structure given the patient-specific available space. The outcome of the volume filling algorithm will be used later to create maps that express the probability of a voxel belonging to a certain generation. 
This information can be a very informative and powerful clinical decision support tool when projected on CT slices.

\begin{figure}[!t]
    \centering
    \ifdefined\showFigures
    \includegraphics[width=\textwidth]{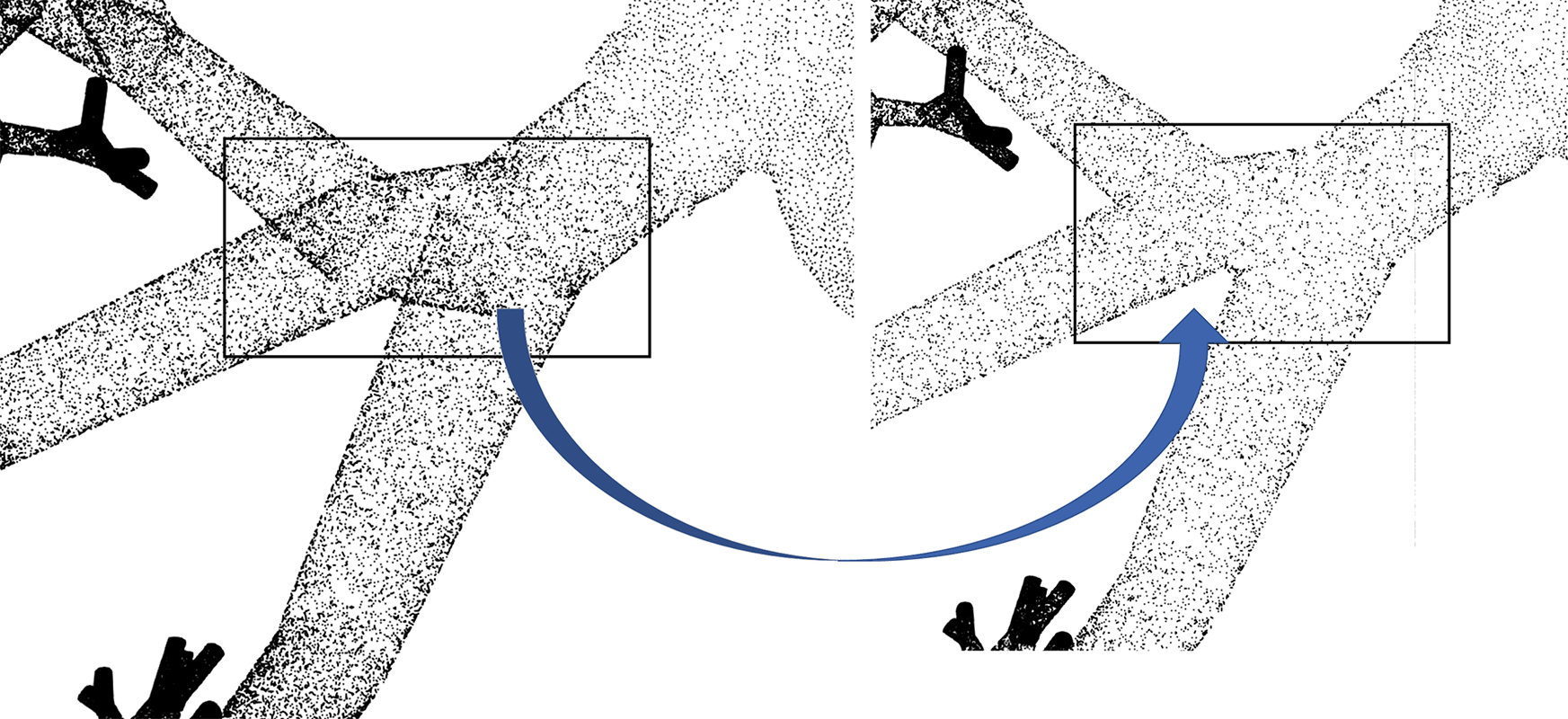}
    \fi
    \caption{Outcome of point cloud generation process. Extra refinements remove the inner points facilitating normal estimation for Poisson surface reconstruction.}
  \label{fig:point-cloud-correction}
\end{figure}

\section{Surface generation on predicted 1-dimensional representation}
The extracted 3D geometries are required to conduct studies on computational fluid dynamics, particle transfer and deposition, ventilation, stress analysis and deformation simulations. Marching cubes algorithm \cite{lorensen1987marching} is a very well established method implemented in FAST \cite{smistad2015fast} allowing the generation of 3D geometric models from airway segmentation label maps. The constriction simulation method aims to generate 3D tubular surface structures with smaller diameters. To this end, Laplacian surface contraction offers a solution that deforms the geometry pushing the vertices towards the direction of the inward normals. 

The extension of the extracted centerline generates a predictive representation of the bronchial tree given the available space. However, for the outcome of space-filling algorithms to be useful in fluid dynamics simulation, particle deposition simulation or stress finite element analysis based studies, the boundary conditions in the form of triangular 3D meshes need to be defined. Initially, as a simplified approach, to define the diameter of each generation, we can employ the power-law consistent with Murray's law of symmetric branching\cite{hagmeijer2018critical,pepe2017optimal}.
\begin{equation}
    d_z=d_0 \times 2^{-z/3}
\end{equation}
where $d_0$ denotes the branch diameter of the trachea and $d_z$ the branch diameter for generation $z$.

Furthermore, if we take into account that each branch demonstrates different branching angle and diameter properties, the relation between airway diameter ($d$) and branching angles ($\theta$) is based on the following rules validated by Kamiya et al. \cite{kamiya1974theoretical} and Kitaoka et al. \cite{kitaoka1999three}:

\begin{align} 
d_0^2&=d_1^2+d_2^2 \\
 \frac{d_0^2}{\sin(\theta_1+\theta_2)}&=\frac{d_1^2}{\sin\theta_1}=\frac{d_2^2}{\sin\theta_2}
\end{align}
where the index $0$ stands for the parent branch, and the indices 1 and 2 for the two children branches, respectively.

We employ a point cloud sampling approach as input for Poisson surface reconstruction to reconstruct the lung surface. The outcome is a smooth surface with smooth transitions instead of abrupt transitions in the intersection with the original tubular meshes. The tubular-shaped point cloud is sampled using a uniform random distribution. A clean-up step, visualized in Figure \ref{fig:point-cloud-correction}, ensures that no point can be found at a distance less than the prescribed diameter of every available branch. The resulting point cloud is used to compute normals. A bilateral normal smoothing \cite{zheng2010bilateral} function prepares the point cloud for Poisson surface reconstruction\cite{kazhdan2006poisson}. smoothing the point normals. This step facilitates the surface reconstruction in bifurcations and transitional parts. Furthermore, since the directed graph is extracted where each point on the centerline corresponds to a point on the lung surface, it is possible to further deform the surface with a custom function or pattern. The generated surface for seven and ten generations is presented in Figure \ref{fig:extended-surface}.

 \begin{figure}
  \centering
  \ifdefined\showFigures
  \includegraphics[width=\textwidth]{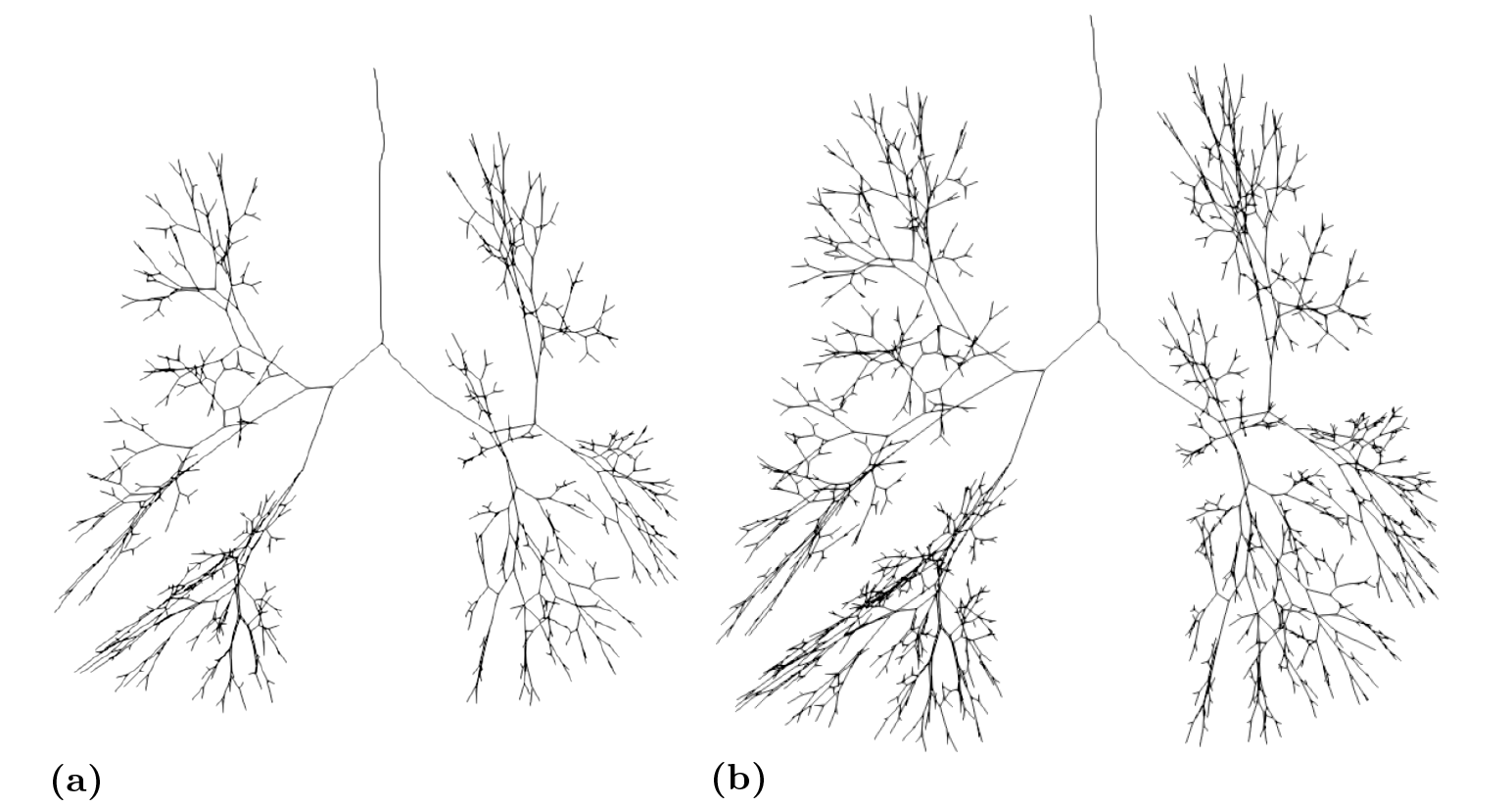}
    \fi
 \caption{Extended bronchial tree (a) 9 generations, (b) 12 generations.}
 \label{fig:extend-1D-rep}
\end{figure}

\begin{figure}
        \centering
        \ifdefined\showFigures
        \includegraphics[width=0.8\textwidth]{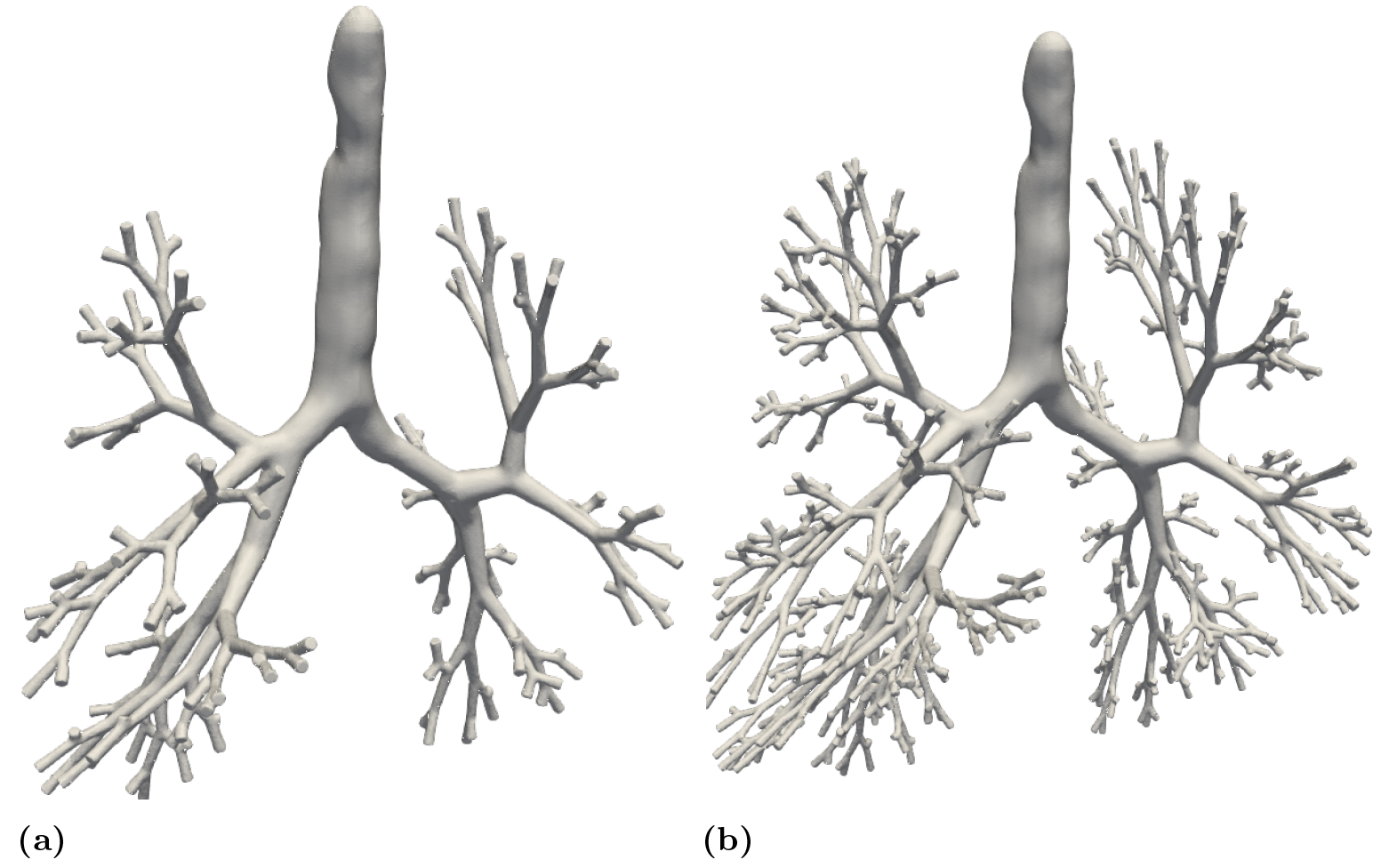}
        \fi
    \caption{%
Generation of surface from the extended bronchial tree centerline for a) 7 generations and b) 9 generations.}%
  \label{fig:extended-surface}
\end{figure}

\section{Simulation of constrictive pulmonary diseases' effect on airway tree}
\label{subsection:contraction}
This section aims to provide the methodology for simulation of bronchoconstriction, allowing to subsequently estimate the dynamic behaviour of the lung airways in the case of an exacerbation event. A bronchial tree 3D geometry is the input for this process yielding as output contracted airways. The proposed geometry contraction procedure is presented by Nousias et al.\cite{nousias2016computational} and Lalas et al.\cite{lalas2017substance} and is an extension of the work of Au et al. \cite{au2008skeleton} facilitating a Laplacian smoothing process that shifts vertices along the estimated curvature normal direction. 

\subsection{Structural modelling}

The airway geometric model consists of connected triangles forming the boundary conditions. Each triangular mesh $\mathcal{M}$ can be described as $\mathcal{M}=\{\mathcal{V},\mathcal{E},\mathcal{F}\}$ where $\mathcal{V}$ is the set of vertices, $\mathcal{E}$ is the set of edges, and $\mathcal{F}$ is the set of faces constituting the 3D surface. Each vertex $i$ can be represented as a point.

\begin{equation}
    \mathbf{v}_i=(x_i,y_i,z_i),\forall i=1,2,\cdots,N
\end{equation}

\noindent For each face $\mathbf{f}_i$, $\forall i=1,2,\cdots,l$ we denote the centroid

\begin{equation}
    \mathbf{m}_i=\frac{\mathbf{v}_{i_1}+\mathbf{v}_{i_2}+\mathbf{v}_{i_3}}{3},\forall i=1,2,\cdots,l
\end{equation}

\noindent  The plane corresponding to each face can also be represented by cendroid $\mathbf{m}_i$ and the outward unit normal
The outward unit normal $\mathbf{n}_{m_i}$ to the face $\mathbf{f}_i$ (located at the centroid $\mathbf{m}_i$) is calculated as
$\mathbf{n}_{m_i}$:
\begin{equation}
\mathbf{n}_{m_i}=\frac{(\mathbf{v}_{i_2}-\mathbf{v}_{i_1})\times(\mathbf{v}_{i_3}-\mathbf{v}_{i_1})}{\norm{(\mathbf{v}_{i_2}-\mathbf{v}_{i_1})\times(\mathbf{v}_{i_3}-\mathbf{v}_{i_1})}},\forall i=1,\cdots,l
\end{equation}
where $\mathbf{v}_{i_1},\mathbf{v}_{i_2},\mathbf{v}_{i_3}$ are the vertices corresponding to face $\mathbf{f}_i$. 

\noindent $\mathbf{L} \in \mathcal{R}^{N \times N}$ is the curvature flow Laplacian operator

\begin{equation}
\small
\mathbf{L}_{i,j}=
 \begin{cases}
      \omega_{i,j} = cot\ a_{ij}+ cot\ b_{ij},& \left(i,j\right) \in E \\
      \sum^k_{i,k\in E}-\omega_{ik},& i=j \\
			0 ,& \text{otherwise}
    \end{cases}
	\label{eq:laplacian}
\end{equation}

\begin{figure}
  \begin{subfigure}[b]{0.42\linewidth}
    \includegraphics[width=\textwidth]{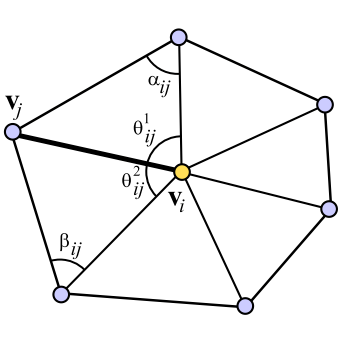}
    \caption{}
    \label{fig:bmc:angles}
  \end{subfigure}
  \hfill
  \begin{subfigure}[b]{0.42\linewidth}
    \includegraphics[width=\textwidth]{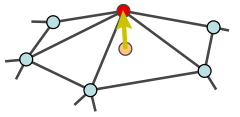}
    \caption{}
    \label{fig:bmc:delta}
  \end{subfigure}
  \caption{Addressing unwanted edge effects. The region between the narrowed and the unprocessed part is smoothed out using Taubin algorithm\cite{Taubin1995CurveAS} and Bilateral Normal filtering \cite{zheng2010bilateral}}
\end{figure}

The product 
\begin{equation}
   \mathbf{\delta}=\mathbf{L}\mathbf{V} 
\end{equation}
approximates the inward curvature flow normals \cite{sorkine2004laplacian}. 

\noindent The motivation for employing the curvature flow Laplacian operator\cite{nealen2006laplacian} on the mesh is that its output is not affected by mesh density. Specifically,
\begin{equation}
\mathbf{\delta}=\mathbf{L}\mathbf{V}=[\mathbf{\delta}_1^T,\mathbf{\delta}_2^T,\ldots,\mathbf{\delta}_N^T]^T,\delta_i=-4A_i\kappa_i\mathbf{n}_i
\end{equation}
where $A_i$ is the one-ring area, $\kappa_i$ is the local curvature and $\mathbf{n}_i$ is the inward curvature flow normal of the $i^{th}$ vertex. 

\noindent \textbf{The model vertices towards the inward normal leads to shrinkage of the 3D geometry.} The positions of the vertices satisfying $\mathbf{L}\mathbf{V}=0$ result in a zero volume string-like mesh and can be used to simulate mesh contraction. However, since such an optimization problem has more than one solution, further constraints are required\cite{au2008skeleton}. 
\noindent Introducing weighting matrices $\mathbf{W}_H \in \mathcal{R}^{N \times N}$ $\mathbf{W}_L \in \mathcal{R}^{N \times N}$ can smoothly move vertex positions $\mathbf{V} \in \mathcal{R}^{3 \times n}$ towards the direction of the inward unit normal by iteratively solving the following minimization problem
\begin{equation}
    \label{eq:new-vertex-positions}
    \hat{\mathbf{V}}=\argmin_\mathbf{V}
   \{
    \norm{\mathbf{W}_L\mathbf{L}\mathbf{V}}^2+\mathbf{W}_H\norm{\mathbf{V}-\mathbf{V}_a}
    \}
\end{equation}
where $\mathbf{V}_a \in \mathcal{R}^{3 \times N}$ corresponds to the vertex positions before the contraction at each iteration.

\noindent The weighting matrices $\mathbf{W}_H$ and $\mathbf{W}_L$ regulate the mesh contraction and mesh attraction, respectively. Utilization of proper weights $\mathbf{W}_L$ and $\mathbf{W}_H$ results in a step contraction of the 3D surface. 
Initially, we set them to

\begin{align}
\label{eqn:WLWH}
\begin{split}
\mathbf{W}_L &=k \cdot \sqrt{A} \cdot \mathbf{I},
\\
\mathbf{W}_H &=\mathbf{I} .
\end{split}
\end{align}
where $\mathbf{I} \in \mathcal{R}^{N \times N}$ is the identity matrix, $k$ a double constant experimentally tuned to $10^{-3}$ and $A$ the average face area of the model.

\noindent Equation \eqref{eq:new-vertex-positions} can be expressed as
\begin{equation}
    \begin{bmatrix} 
    \mathbf{W}_L\mathbf{L}
    \\ 
    \mathbf{W}_H
    \end{bmatrix}\cdot\mathbf{V}'
    =
    \begin{bmatrix} 
    \mathbf{0}
    \\ 
    \mathbf{W}_H\mathbf{V}
    \end{bmatrix}
\end{equation}
The analytical solution can be formulated as 
\begin{equation}
    \mathbf{V}'=(\mathbf{A}T\mathbf{A})^{-1}\mathbf{A}\mathbf{b}
\end{equation}
where matrices $\mathbf{A}$ and $\mathbf{b}$ are given by
\begin{equation}
\mathbf{A}= \begin{bmatrix} 
    \mathbf{W}_L\mathbf{L}
    \\ 
    \mathbf{W}_H
    \end{bmatrix}
    ,  
    \mathbf{b}= 
    \begin{bmatrix} 
    \mathbf{0}
    \\ 
    \mathbf{W}_H\mathbf{V}
    \end{bmatrix}
\end{equation}

After each iteration $t$ we update the contraction and inflation weights to be used in iteration $t+1$ so that:
\begin{equation}
\mathbf{W}^{t+1}_L=s_L\mathbf{W}^{t}_L
\end{equation}
 and 
\begin{equation}
 W^t_{H,i}=W^0_{H,i}\sqrt{\frac{A_{0_i}}{A_{t_i}}}
\end{equation}
where ${A_{0_i}}$ and ${A_{t_i}}$ are the original and the current one-ring area respectively. The Laplacian matrix for iteration $t+1$, $\mathbf{L}^{t+1}$ is also recomputed.

\subsection{Termination criteria}

On these grounds, to simulate bronchoconstriction, we must reduce the airway diameter to a predefined level of narrowing. This level is defined by certain termination criteria \cite{nousias2016computational,lalas2017substance}. Thus, a metric is required that measures the diameter of the bronchi under process. To estimate the degree of contraction of the airway's geometry after each iteration, we employ a shape diameter function (SDF) based scheme\cite {shapira2008consistent} implemented in \cite{cgal:eb-19a} that evaluates the local volume based on the estimated local diameter assigned to each face of the mesh, also known as raw SDF values. Measuring the volume before and after the Laplacian contraction iteration can set the termination criteria.

\paragraph{Computing Raw SDF Values}
Given a point on the surface mesh, a cone is used centred around its inward-normal direction (the opposite direction of its normal vector), and several rays are sent inside this cone to the other side of the mesh. For each such ray, we check the normal at the point of intersection and ignore intersections where the normal at the intersection points in the same direction as the point-of-origin (the same direction is defined as an angle difference less than $90^\circ$). This way, false intersections with the outer part of the mesh are removed. The SDF at a point is defined as the weighted average of all rays lengths which fall within one standard deviation from the median of all lengths. The weights used are the inverse of the angle between the ray to the centre of the cone. This is because rays with larger angles are more frequent and therefore have smaller weights.
\paragraph{Post-processing of Raw SDF Values}
After having calculated the raw SDF value for each facet, the SDF values that can be used in the segmentation algorithm are the result of several post-processing steps:
\begin{itemize}
\item Facets with no raw SDF values are assigned the average raw SDF value of their edge-adjacent neighbours.
\item A bilateral smoothing \cite{tomasi1998bilateral} is applied. This smoothing technique removes the noise while keeping fast changes on SDF values unchanged since they are natural candidates for segment boundaries. The bilateral smoothing \cite{tomasi1998bilateral} has three parameters that are set as follows:

\begin{figure}[t!]
  \centering
  \includegraphics[width=0.8\textwidth]{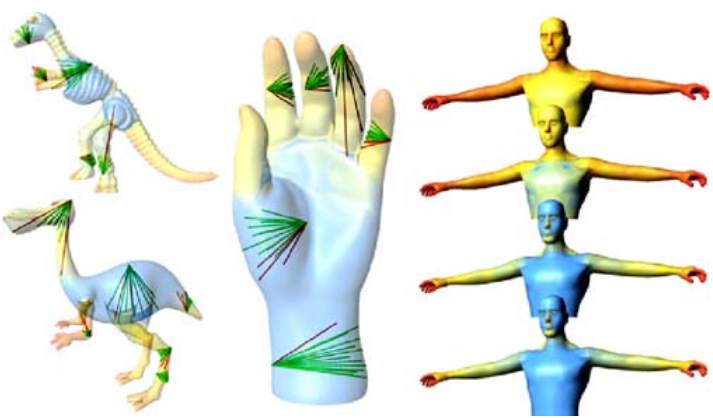}
 \caption{Demonstration of Shape Diameter Function based calculation of local diameter in other geometries \cite{shapira2008consistent}}
 \label{fig:SDFdemontration}
\end{figure}

\begin{itemize}

\item window size (i.e. maximum level for breadth-first neighbor selection)
\begin{equation}
w = \lfloor\sqrt{\frac{ |F|}{ 2000}}\rfloor + 1
\end{equation}, 
where $F$ denotes the set of facets
\item spatial parameter 
\begin{equation}
    \sigma_s = \frac{w}{2}
\end{equation}

\item range parameter set for each facet
\begin{equation}
    \sigma_{r_i} = \sqrt{\frac{1}{|w_i|} \sum_{f_j \in w_i}(SDF(f_j) - SDF(f_i))^2}
\end{equation}
$f_i$ and $w_i$ denotes the set of neighboring facets of $f_i$ collected using $w$ in the facet neighbor breadth-first search 
\end{itemize}
\item SDF values are linearly normalized between $[0,1]$.
\end{itemize}

\paragraph{Soft \& hard clustering}
Given a number k of clusters, the soft clustering is a Gaussian mixture model that consists in fitting $k$ Gaussian distributions to the distribution of the SDF values of the facets. It is initialized with \textbf{k-means++} \cite{arthur2007k}, and run multiple times with random seeds. Among these runs, the best result is used for initializing the expectation-maximization algorithm for fitting the Gaussian distributions.

The hard clustering yields the final segmentation of the input mesh and results from minimizing an energy function combining the probability mentioned above matrix and geometric surface features. The energy function minimized using alpha-expansion graph cut algorithm \cite{boykov2001fast} is defined as follows:

\begin{equation}
E(\bar{x}) = \sum\limits_{f \in F} e_1(f, x_f) + \lambda \sum\limits_{ \{f,g\} \in N} e_2(x_f, x_g)
\end{equation}

\begin{equation}
e_1(f, x_f) = -log(max(P(f|x_f), \epsilon))
\end{equation}

\begin{equation}
e_2(x_f, x_g) = \left \{ \begin{array}{rl} -log(\theta(f,g)/\pi) &\mbox{ $x_f \ne x_g$} \\ 0 &\mbox{ $x_f = x_g$} \end{array} \right \}
\end{equation}

where $F$ denotes the set of facets, $N$ denotes the set of pairs of neighboring facets, $x_f$ denotes the cluster assigned to facet $f$,
$P(f|x_p)$ denotes the probability of assigning facet $f$ to cluster $x_p$, $\theta(f,g)$ denotes the dihedral angle between neighboring facets $f$ and $g$: concave angles and convex angles are weighted by $1$ and $0.1$ respectively, $\epsilon$ denotes the minimal probability threshold, $\lambda \in [0,1]$denotes a smoothness parameter.

\paragraph{Deriving airway narrowing ratio}
The SDF values assigned to each face correspond to the local diameter of the shape. Thus, by summing up the SDF values of the faces of the processed part of the mesh, we derive a characteristic value for the local mean diameter of the processed segment. The ratio $r^t$  of the sum of the SDF values before the narrowing process to the sum of the SDF values after the narrowing process corresponds to a metric of the narrowing percentage of the airway. 

\begin{equation}
r^t=\frac{\sum_{i=0}^{N}SDF(f_i^t)}{\sum_{i=0}^{N}SDF(f_i^0)}
\end{equation}
where $r^t$ is the narrowing ratio after iteration $t$, $f_i^t$ is the face with index i after iteration $t$, $f_i^0$ is the initial face with index $i$ and $N$ is the number of processed faces.

\begin{figure}[t!]
\centering
 \includegraphics[width=\textwidth]{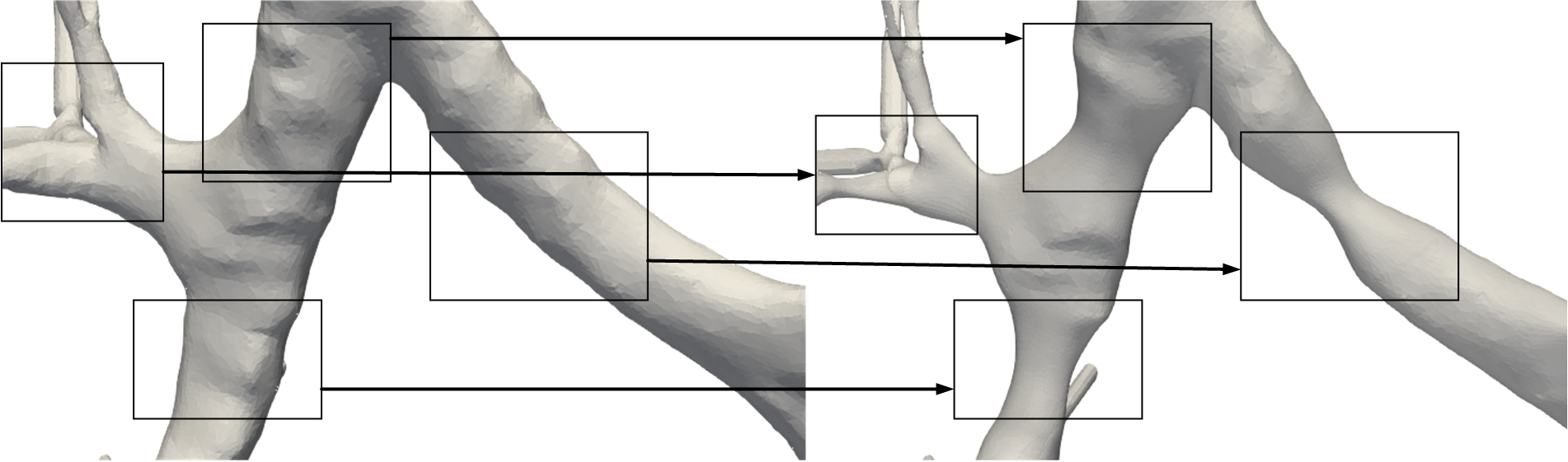}
    \caption{%
Simulation of constrictive pulmonary conditions. A selected region of the bronchial tree undergoes a controlled narrowing to the desired degree.}%
  \label{fig:bronchocostriction-simulation}
\end{figure}

Figure \ref{fig:bronchocostriction-simulation} presents a simulation of constrictive pulmonary conditions. Based on the geometry mentioned above processing schemes, we employ the following algorithm to simulate bronchoconstriction given the desired number of iterations $t$, user-defined weight $\omega$, and desired to narrow ratio $r$ and accepted narrowing ratio error $e$.

\begin{algorithm}[t]
\begin{enumerate}
    \item Get vertex positions \textbf{V}
    \item Initialize Laplacian matrix
\begin{equation}
\small
\mathbf{L}_{i,j}=
 \begin{cases}
      \omega_{i,j} = cot\ a_{ij}+ cot\ b_{ij},& \left(i,j\right) \in E \\
      \sum^k_{i,k\in E}-\omega_{ik},& i=j \\
			0 ,& \text{otherwise}
    \end{cases}
\end{equation}
\item Initialize $\mathbf{W}_L$ and $\mathbf{W}_H$ in the following manner: 
\begin{align}
\mathbf{W}_L & =k \cdot I \cdot \sqrt{A} \\
\mathbf{W}_H & =I
\end{align}
where I is a unitary matrix, k a double constant set to $10^{-3}$ and A the average face area of the model.
\item Solve
\begin{equation}
    \hat{\mathbf{V}}=\argmin_\mathbf{V}
   \{
    \norm{\mathbf{W}_L\mathbf{L}\mathbf{V}}^2+\mathbf{W}_H\norm{\mathbf{V}-\mathbf{V}_a}
    \}
\end{equation} 
for $\mathbf{V'}$.
\item Update $\mathbf{W}_L$ and $\mathbf{W}_H$ so that
\begin{eqnarray}
\mathbf{W}^{t+1}_L&=&{s}_L \cdot {W}^t_L \\ 
\mathbf{W}^{t+1}_{H,i}&=&{W}^0_{H,i} \cdot \sqrt{A^0_i/A^t_i}
\end{eqnarray}
where t denotes the iteration index.
\item Recompute L.
\item Compute narrowing ratio $r^t$
\item Repeat steps 4 to 7 for $r^t > r+e$ , where $r$ is the desired
\end{enumerate}
\caption{Bronchoconstriction simulation}\label{al:bronchoconstriction_simulation}
\end{algorithm}

\begin{figure}[t!]
  \begin{subfigure}[b]{0.42\linewidth}
    \includegraphics[width=\textwidth]{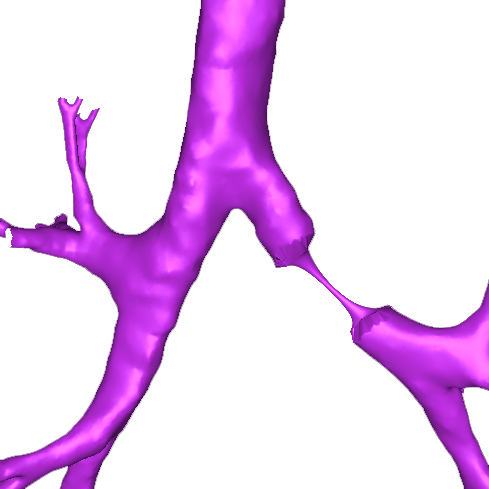}
    \caption{}
    \label{fig:edge_effects_a}
  \end{subfigure}
  \hfill
  \begin{subfigure}[b]{0.42\linewidth}
    \includegraphics[width=\textwidth]{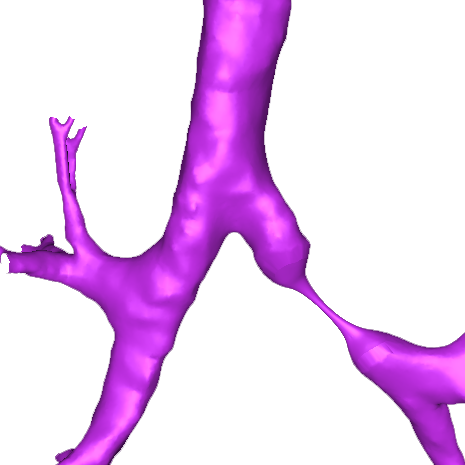}
    \caption{}
    \label{fig:edge_effects_b}
  \end{subfigure}
  \caption{Addressing unwanted edge effects. The region between the narrowed and the unprocessed part is smoothed out using Taubin algorithm\cite{Taubin1995CurveAS} and Bilateral Normal filtering \cite{zheng2010bilateral}}
\end{figure}

\noindent For the current study, we set $r=0.5$ , $e=0.05$ and $t=8$ and $\omega=0.8$. The area under investigation is segmented for more efficient processing and mesh contraction is performed.
As the processed part is reconnected with the rest of the mesh, discontinuities and edge effects may appear. To cope with this fact, we apply surface smoothing methods such as Taubin smoothing \cite{Taubin1995CurveAS} or bilateral surface denoising \cite{zheng2010bilateral} at the part of the mesh under process, including the edge points (anchor points) of the stable region of the mesh.
The bilateral technique estimates the new normals that can be used for vertex updating according to:

\begin{equation}
\label{eq:norm2}
\mathbf{\bar{n}}_{m} = \left(\mathbf{D}^{-1}\mathbf{C}_w \right)^\zeta \mathbf{\hat{n}}_{n}
\end{equation}

where $\zeta$ is an integer that represents a denoising factor and

\begin{align}
\mathbf{C}_w = \mathbf{W}_{m} \mathbf{W}_{n} \mathbf{C}_a. \label{W}
\end{align}
is the weighted bilateral adjacency matrix. Eq. \eqref{eq:norm2} shows that any averaging filter on the normals, is a frequency selective transform.
For every centroid $\mathbf{m}_i \forall \ i= 1,\cdots, n_f $ the weights of $\mathbf{W}_{m}, \mathbf{W}_{n} \in \Re^{n_f \times n_f}$ are estimated according to the following equations:
{\small
\begin{align}
\mathbf{W}_{m_{ij}} =
\left\{\begin{matrix}
exp(-\begin{Vmatrix}\mathbf{m}_{i}-\mathbf{m}_{j}\end{Vmatrix}^{2}/2\sigma^{2}_{m} ) & if \ \mathbf{C}_{a_{ij}} =1\\ 
0 & otherwise
\end{matrix}\right. 
\label{Wm}
\end{align}
\begin{align}
\mathbf{W}_{n_{ij}} =  
\left\{\begin{matrix}
exp(-\begin{Vmatrix}\mathbf{\hat{n}}_{mi}-\mathbf{\hat{n}}_{mj}\end{Vmatrix}^{2}/2\sigma^{2}_{n} )  & if \ \mathbf{C}_{a_{ij}} =1 \\ 
0 & otherwise
\end{matrix}\right.
\label{Wn}
\end{align}
}
\noindent where $\mathbf{C}_a$ is the binary adjacency matrix constructed using the k-nearest neighbours method. Finally, the vertices in each iteration $t$ are updated according to:
{\small
\begin{align}
\mathbf{v}^{(t+1)}_{i} = \mathbf{v}^{(t)}_{i} + \frac{ \sum_{j \in \mathbf{\Psi}_i} \mathbf{\bar{n}}^{(t)}_{mj} (<\mathbf{\bar{n}}^{(t)}_{mj} , (\mathbf{m}^{(t)}_{j} - \mathbf{v}^{(t)}_{i})>)}{|\mathbf{\Psi}_{i}|}. \label{vt}
\end{align}
\begin{align}
\mathbf{m}^{(t+1)}_{j} = (\mathbf{v}^{(t+1)}_{j1} + \mathbf{v}^{(t+1)}_{j2} +  \mathbf{v}^{(t+1)}_{j3} )/3 \ \ \forall \ j \in \mathbf{\Psi}_i. \label{mj}
\end{align}
}
where \(<\mathbf{a},\mathbf{b}>\) represents the inner product of \(\mathbf{a}\) and \(\mathbf{b}\), $(t)$ represents iteration number and \(\mathbf{\Psi}_{i}\) represents the set of the face vertices which are connected with \(\mathbf{v}_{i}\). Eqs. \eqref{eq:norm2}, \eqref{vt}, \eqref{mj} are executed iteratively for a given number of iterations until a certain reconstruction quality criterion is met.







\section{Datasets}
For evaluating the approaches mentioned above, lung CT scans are required. To this end, we employed the VESSEL12 dataset provided by VESSEL12 (VESsel SEgmentation in the Lung) challenge \cite{rudyanto2014comparing}, and EXACT09 \cite{lo2012extraction}. 

The VESSEL dataset comprises 20 anonymized scans in Meta (MHD/raw) format. The scans for this challenge were collected from the anonymized image repositories of three hospitals: University Medical Center Utrecht (Utrecht, The Netherlands), the University Clinic of Navarra (Pamplona, Spain), and Radboud University Nijmegen Medical Centre (Nijmegen, The Netherlands). The data included both clinical exams taken for a variety of indications and scans from two lung cancer screening trials: NELSON, the Dutch-Belgian randomized controlled lung cancer CT screening trial (van Klaveren, 2011) and I-ELCAP, the International Early Lung Cancer Action program. In the institutes where approval of the institutional ethics committee is required, written consent for retrospective studies had been previously obtained from each participant.

\begin{table}[t!]
\caption{VESSEL Dataset description, part A}
\label{table:vessel_dataset_description_A}
\resizebox{\textwidth}{!}{
\begin{tabular}{|l|p{2cm}|p{3cm}|p{3cm}|p{2cm}|p{2cm}|p{1.5cm}|p{1.5cm}|}
\hline
Scan & Image type  & Pathology                                 & Scanner and kernel                 & Spacing (mm) & Z-spacing (mm) & \# Of slices & kV/mAs   \\ \hline
01   & Angio-CT    & Alveolar inflammation                     & Siemens SOMATOM Sensation 64, B60f & 0.76         & 1              & 355          & 120/40   \\ \hline
02   & Chest CT    & Alveolar inflammation                     & Philips Mx8000 IDT 16, B Kernel    & 0.71         & 0.7            & 415          & 140/74   \\ \hline
03   & Chest CT    & ILD                                       & Philips Mx8000 IDT 16, B Kernel    & 0.62         & 0.7            & 534          & 120/77   \\ \hline
04   & LD Chest CT & ILD                                       & Toshiba Acquilion ONE, FC55        & 0.86         & 1              & 426          & 100/44   \\ \hline
05   & Chest CT    & ILD                                       & Philips Mx8000 IDT 16, B Kernel    & 0.72         & 0.7            & 424          & 140/73   \\ \hline
06   & Angio-CT    & ILD                                       & Siemens SOMATOM Sensation 64, B30f & 0.63         & 1              & 375          & 120/81   \\ \hline
07   & LD Chest CT & ILD                                       & Toshiba Acquilion ONE, FC55        & 0.69         & 1              & 461          & 100/23   \\ \hline
08   & Chest CT    & ILD                                       & Philips Mx8000 IDT 16, B Kernel    & 0.78         & 0.7            & 442          & 140/64   \\ \hline
09   & Angio-CT    & ILD                                       & Siemens SOMATOM Sensation 64, B25f & 0.68         & 1              & 543          & 100/150  \\ \hline
10   & Angio-CT    & ILD                                       & Toshiba Acquilion ONE, FC83        & 0.88         & 1              & 426          & 120/68   \\ \hline
11   & Angio-CT    & ILD and emphysema                         & Toshiba Acquilion ONE, FC83        & 0.77         & 1              & 421          & 100/120  \\ \hline
12   & Angio-CT    & Secondary pulmonary arterial hypertension & Toshiba Acquilion ONE, FC83        & 0.8          & 1              & 446          & 100/92   \\ \hline
\end{tabular}
}
\end{table}

\begin{table}[t!]
\caption{VESSEL Dataset description, part B}
\label{table:vessel_dataset_description_B}
\resizebox{\textwidth}{!}{
\begin{tabular}{|l|p{2cm}|p{3cm}|p{3cm}|p{2cm}|p{2cm}|p{1.5cm}|p{1.5cm}|}
\hline
Scan & Image type  & Pathology                                 & Scanner and kernel                 & Spacing (mm) & Z-spacing (mm) & \# Of slices & kV/mAs   \\ \hline
13   & Angio-CT    & Pulmonary thromboembolism                 & Toshiba Acquilion ONE, FC83        & 0.89         & 1              & 471          & 120/ 117 \\ \hline
14   & LD Chest CT & Pulmonary thromboembolism and emphysema   & Toshiba Acquilion ONE, FC83        & 0.71         & 1              & 386          & 100/33   \\ \hline
15   & Angio-CT    & Pulmonary thromboembolism                 & Siemens SOMATOM Sensation 64, B25f & 0.65         & 1              & 378          & 100/150  \\ \hline
16   & LD Chest CT & Small nodules                             & Toshiba Acquilion ONE, FC83        & 0.75         & 1              & 451          & 100/38   \\ \hline
17   & Angio-CT    & Nodules and diffuse abnormalities         & Siemens SOMATOM Sensation 64, B25f & 0.59         & 1              & 429          & 100/135  \\ \hline
18   & Chest CT    & Normal                                    & Philips Brilliance 16P, B Kernel   & 0.78         & 0.7            & 408          & 140/73   \\ \hline
19   & HR Chest CT & Small nodules                             & Toshiba Acquilion ONE, FC83        & 0.69         & 1              & 396          & 120/68   \\ \hline
20   & LD Chest CT & Emphysema                                 & Toshiba Acquilion ONE, FC55        & 0.75         & 1              & 406          & 100/32   \\ \hline
\end{tabular}
}
\end{table}

The variety of sources ensures that a wide range of clinical images typically used in diagnostic settings is present in the dataset: high and low resolution, standard or low-dose chest CT, and Angiography CT each with their scanning parameters and reconstruction kernels. Scanners from three major manufacturers were included: Philips, Siemens, and Toshiba. CT scan images were taken from individual patients diagnosed with a spectrum of lung pathologies, including diffuse interstitial disease, pulmonary thromboembolism, pulmonary hypertension, alveolar inflammation, lung nodules, and emphysema. To ensure that the images were as isotropic as possible, we selected only thin slice images having slice spacing between 0.59 mm and 0.89 mm, averaging at 0.74 mm. Accurate vessel segmentation requires thin slice data, as vessels are often blurred out in CTs with thicker slices. From this cohort of images, we finally selected 20 scans, described in Tables \ref{table:vessel_dataset_description_A} and \ref{table:vessel_dataset_description_B}. This heterogeneous data closely reflects the diversity of CT scans encountered in clinical practice.
All 20 scans were then anonymized and made available for download to registered VESSEL12 challenge participants in Meta (MHD/raw) format. The lung masks for each image were also provided to facilitate vessel segmentation only within the lung areas.

EXACT09\cite{lo2012extraction} consists of 75 wholly anonymized chest CT scans contributed by eight different institutions, acquired with several different CT scanner brands and models, using a variety of scanning protocols and reconstruction parameters. The conditions of the scanned subjects varied widely, ranging from healthy volunteers to patients showing severe abnormalities in the airways or lung parenchyma.

\section{Experimental evaluation}
\label{section:results}

\subsection{Structural modelling and validation}
\label{subsection:results_structural}

Our simulation framework processes the initial tree centerline and generates a structural estimation given the first three to four available generation and their morphometric characteristics, i.e., lengths and diameters. 
To facilitate the comparison with morphometric data, we employed a publicly available dataset provided by Montesantos et al.\cite{montesantos2016creation} labelled as \textit{pone.0168026.s001}. For the sake of self-completeness, the authors of \cite{montesantos2016creation}  provided morphometric data extracted from HRCT images acquired at the University Hospital Southampton NHS Foundation Trust as a part of the study described in \cite{fleming2015controlled,majoral2014controlled}. Data from seven healthy subjects and six patients with moderate or persistent asthma were included in the dataset. Asthmatic patients were diagnosed exacerbation-free for at least one month and were male non-smokers. 

\begin{figure}[t!]
    \centering
    \ifdefined\showFigures
     \includegraphics[width=\linewidth]{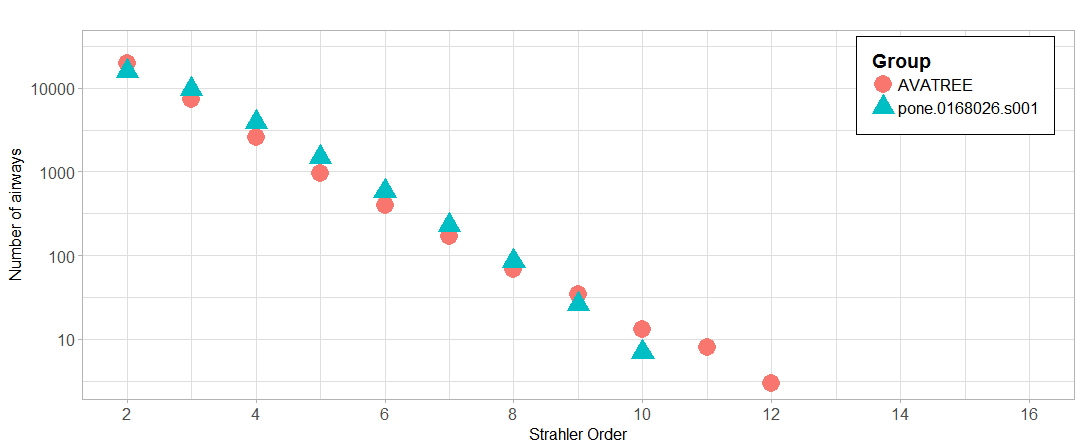}
     \fi
    \caption{
    Comparison in terms of the number of airways for each level of Strahler orders. This comparison confirms that our model comes into agreement with \textit{pone.0168026.s001}\cite{montesantos2016creation}.}
    \label{fig:noairwayshorsfield}
\end{figure}

\begin{figure}[t!]
    \centering
    \ifdefined\showFigures
     \includegraphics[width=\linewidth]{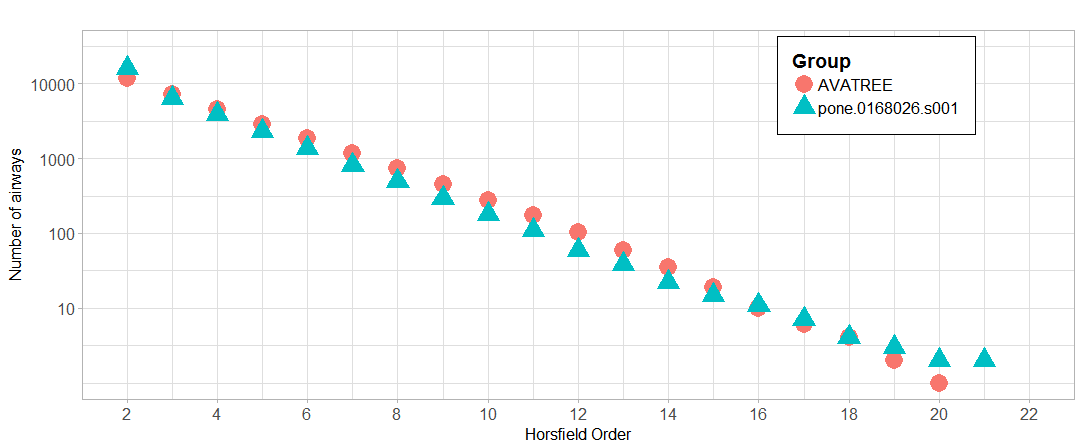}
     \fi
    \caption{Number of airways for each Horsfield order as predicted by our model AVATREE \cite{nousias2020avatree}, and \textit{pone.0168026.s001} \cite{montesantos2016creation}.}
    \label{fig:noairwaysstrahler}
\end{figure}

A \textit{"Sensation"} 64-slice HRCT scanner (Siemens, Enlargen, Germany) was utilized to capture 3D images from the mouth to the base of the lungs. Subjects were posed in the supine position and were instructed to perform slow exhalation. Groundtruth data for the development of bronchial tree models in \cite{montesantos2016creation} were extracted by Pulmonary Workstation 2 Software, including 3 to 4 generations in the upper lobes and 6 to 7 generations in the lower lobes. For each subject, a morphology file includes the total lung volume of the subject lung (in $cm^3$) and the percent volume per lobe, while a translation file contains the airway connectivity, starting from the trachea to the terminal nodes. We used the generated trees from \cite{montesantos2016creation} to validate our approach and compare our results with relevant literature findings. Specifically, we compared the distributions of diameters, airway lengths and branching angles for each generation and the total number of airways for Horsfield and Strahler orders.
In total, 31204 acini were calculated, agreeing with the results reported by \cite{tawhai2004ct,montesantos2016creation}. Figures \ref{fig:noairwayshorsfield} and \ref{fig:noairwaysstrahler} present a comparison in terms of the number of airways for each level of Strahler and Horsfield orders. This comparison confirms that our model comes into agreement with 
\textit{pone.0168026.s001}

\begin{figure}[t!]
    \centering
    \ifdefined\showFigures
     \includegraphics[width=\linewidth]{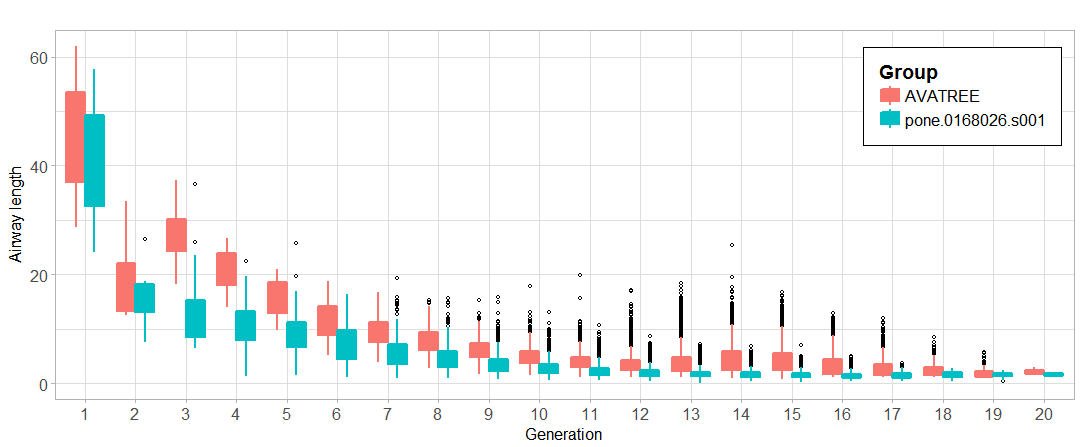}
     \fi
     \caption{Distribution of airway lengths for each generation for AVATREE and \textit{pone.0168026.s001} \cite{montesantos2016creation}}
    \label{fig:dist_lengths}
\end{figure}

\begin{figure}[t!]
    \centering
    \ifdefined\showFigures
     \includegraphics[width=\linewidth]{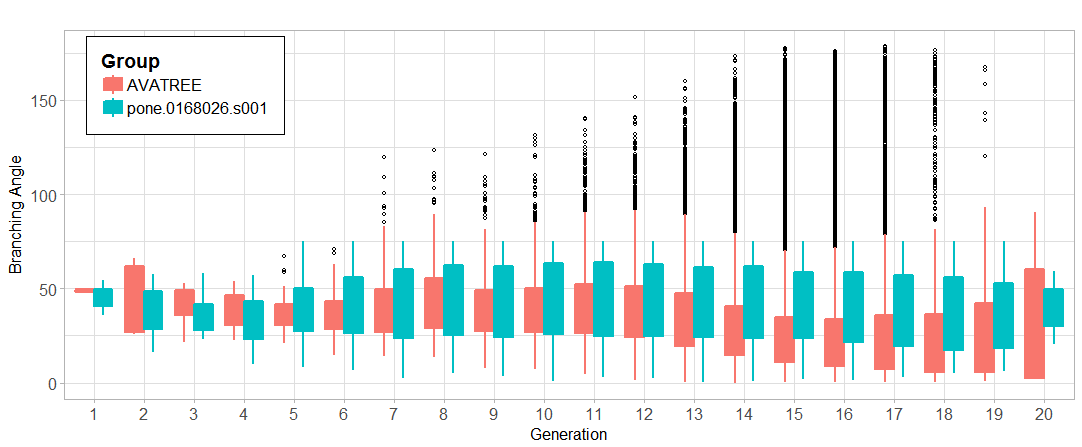}
     \fi
    \caption{Distribution of branching angles for each generation as predicted by our model AVATREE \cite{nousias2020avatree} and \textit{pone.0168026.s001}}
    \label{fig:dist_angles}
\end{figure}

\begin{figure}[t!]
    \centering
    \ifdefined\showFigures
     \includegraphics[width=0.8\linewidth]{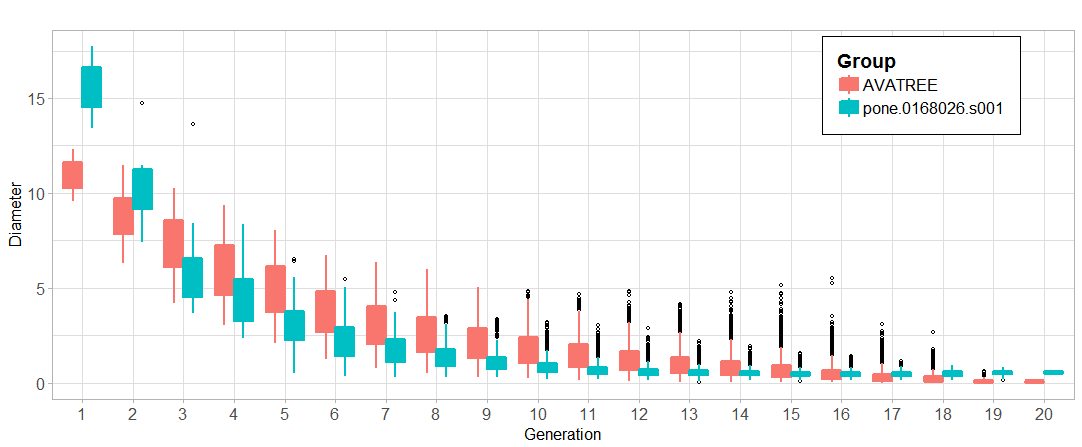}
     \fi
    \caption{Distribution of diameters for each generation for AVATREE and \textit{pone. 0168026. s001}}
    \label{fig:dist_diameters}
\end{figure}

\begin{figure}[t!]
    \centering
    \ifdefined\showFigures
    \includegraphics[width=0.8\linewidth]{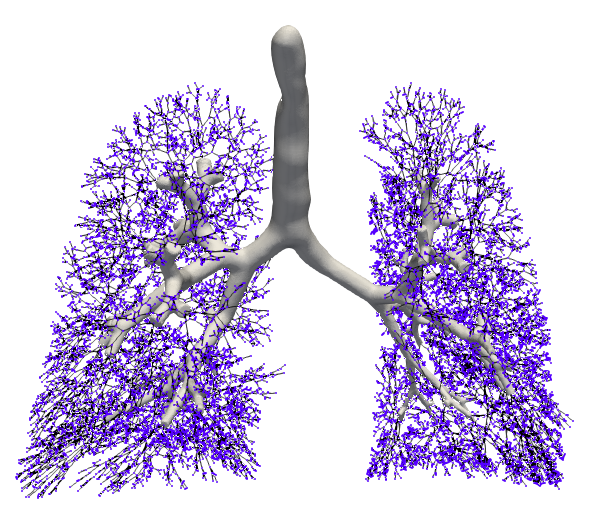}
    \fi
    \caption{Estimation of bronchial tree for 12 generations. Surface reconstruction was performed only for the first 7 generations}
    \label{fig:tree_final_12}
\end{figure}

\begin{figure}[t!]
    \centering
    \ifdefined\showFigures
    \includegraphics[width=0.8\linewidth]{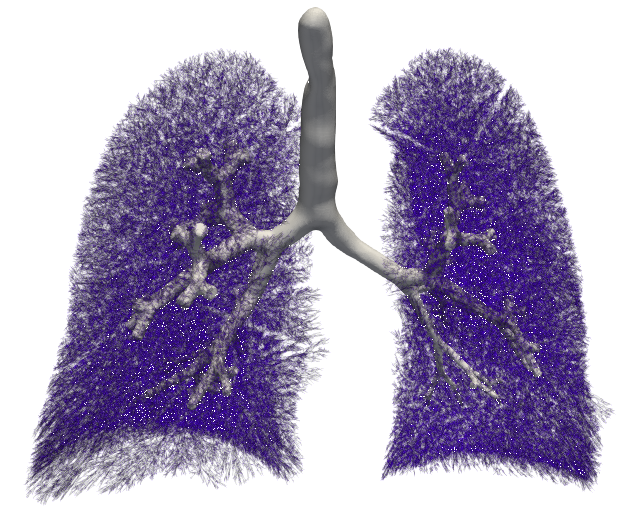}
    \fi
    \caption{Estimation of bronchial tree for 23 generations. Surface reconstruction was performed for the first 7 generations}
    \label{fig:tree_final_23}
\end{figure}

Furthermore, distributions of airway lengths, branching angles and diameters were plotted for each generation, for \textbf{AVATREE}\cite{nousias2020avatree} and \textit{pone.0168026.s001}\cite{montesantos2016creation}.
Airway lengths maintain the same exponential decay pattern for both models. Differences appear in generations 1 to 4 distinctively defined by body size and anatomical features. The distribution of branching angles of our model is also confirmed by \textit{pone.0168026.s001}\cite{montesantos2016creation} maintaining a nearly linear decay with a meager rate. The distributions of diameters per generation are also observed to follow an exponential decay pattern. Both our model and \textit{pone.0168026.s001}\cite{montesantos2016creation} decay similarly after generation $4$, validating the morphometric characteristics of the airway trees generated by our approach. Figures \ref{fig:dist_lengths} to \ref{fig:dist_diameters} present the distribution of airway length, branching angle and diameter for each generation for AVATREE and for \textit{pone.0168026.s001}\cite{montesantos2016creation}. 
Table \ref{table:summary} presents an overview of quantitative macroscopic figures for AVATREE and relevant studies.
Branching ratios ($RB_H$,$RB_S$), diameter ratios ($RD_H$,$RD_S$) and length ratios $RL_H$,$RL_S$) were calculated for Strahler and Horsfield ratios denoted as $*_S$ and $*_H$ respectively. Specifically, $RB_H$,$RD_H$ and $RL_H$ were calculated equal to $RB_H=1.74$, $RD_H=1.259$ and $RL_H=1.26\pm 1.01$. Montesantos et al.\cite{montesantos2016creation} reported $RB_H=1.56$, $RD_H=1.115$ and $RL_H=1.13$ respectively. Additionally, $RB_S$,$RD_S$ and $RL_S$ were calculated equal to $RB_S=2.35$, $RD_S=1.25$ and $RL_S=1.23\pm 1.02$ and are close to the figures provides by relative studies \cite{horsfield1986morphometry,montesantos2016creation} as Table \ref{table:summary} reveals.  Likewise, rate of decline for diameters per generation $RD$ was calculated to $RD=0.83 \pm 0.21$, being in agreement to \cite{montesantos2016creation}. 
Average branching angle $\theta$ for our model was calculated to $32.44\pm 28.95$ comparable to \cite{montesantos2016creation} reporting a $\theta$ equal to $42.1\pm 21.4$. Finally, Figures \ref{fig:tree_final_12} and \ref{fig:tree_final_23} present bronchial tree 1-dimensional representations extended up to $12$ and $23$ generations respectively. Additionally, for both generated models the surface has been reconstructed for the first $7$ generations.

\begin{table}[t!]
\caption{Structural features comparison}
\resizebox{\textwidth}{!}{

\begin{tabular}{|p{2.2cm}|p{1.4cm}|p{1.8cm}|p{1.2cm}|p{1.2cm}|p{1.2cm}|p{1.2cm}|p{1.2cm}|p{1.2cm}|p{1.5cm}|}
\hline
\multirow{2}{*}{\textbf{}}         & \textbf{No of acini} & \textbf{Diameter Rate of Decline}                             & \multirow{2}{*}{\textbf{RB$_H$}} & \multirow{2}{*}{\textbf{RD$_H$}} & \multirow{2}{*}{\textbf{RL$_H$}} & \multirow{2}{*}{\textbf{RB$_S$}} & \multirow{2}{*}{\textbf{RD$_S$}} & \multirow{2}{*}{\textbf{RL$_S$}} & \multirow{2}{*}{\textbf{Mean $\theta$  }} \\ \cline{2-3}
                                   &   & \multicolumn{1}{l|}{} &                               &                               &                               &                               &                               &                               &                                  \\ \hline
\textbf{AVATREE (Nousias et al.\cite{nousias2020avatree})}                     & 31204          & $0.83 \pm 0.21$                                          & $1.74$                          & 1.259                         & $1.26 \pm 1.01$                    & 2.35                          & 1.25                          & $1.23 \pm 1.02$                    & $32.4488 \pm 28.95$                   \\ \hline
\textbf{Tawhai et al.\cite{tawhai2004ct}}      & 29445          &                                               & 1.47                          &                               & 0.13                          & 2.8                           & 1.41                          & 1.39                          &                                  \\ \hline
\textbf{Horsfield et al.\cite{horsfield1986morphometry}}    & 27992          &                                               &                               &                               &                               & 2.54-2.81                     & 1.5                           & 1.55                          & 37.28                            \\ \hline
\textbf{Bordas et al.\cite{bordas2015development}}       &                &                                               &                               &                               &                               &                               &                               &                               & $42.90\pm 0.10$                      \\ \hline
\textbf{Montesantos et al. \cite{montesantos2016creation}} & $27763 \pm 7118.5$ & $0.789\pm 0.16$                                   & 1.56                          & 1.116                         & 1.13                          & 2.49                          & 1.397                         & 1.392                         & $42.1 \pm 21.4$                       \\ \hline
\end{tabular}
}
\label{table:summary}
\end{table}

\begin{figure}[t!]
\includegraphics[width=\textwidth]{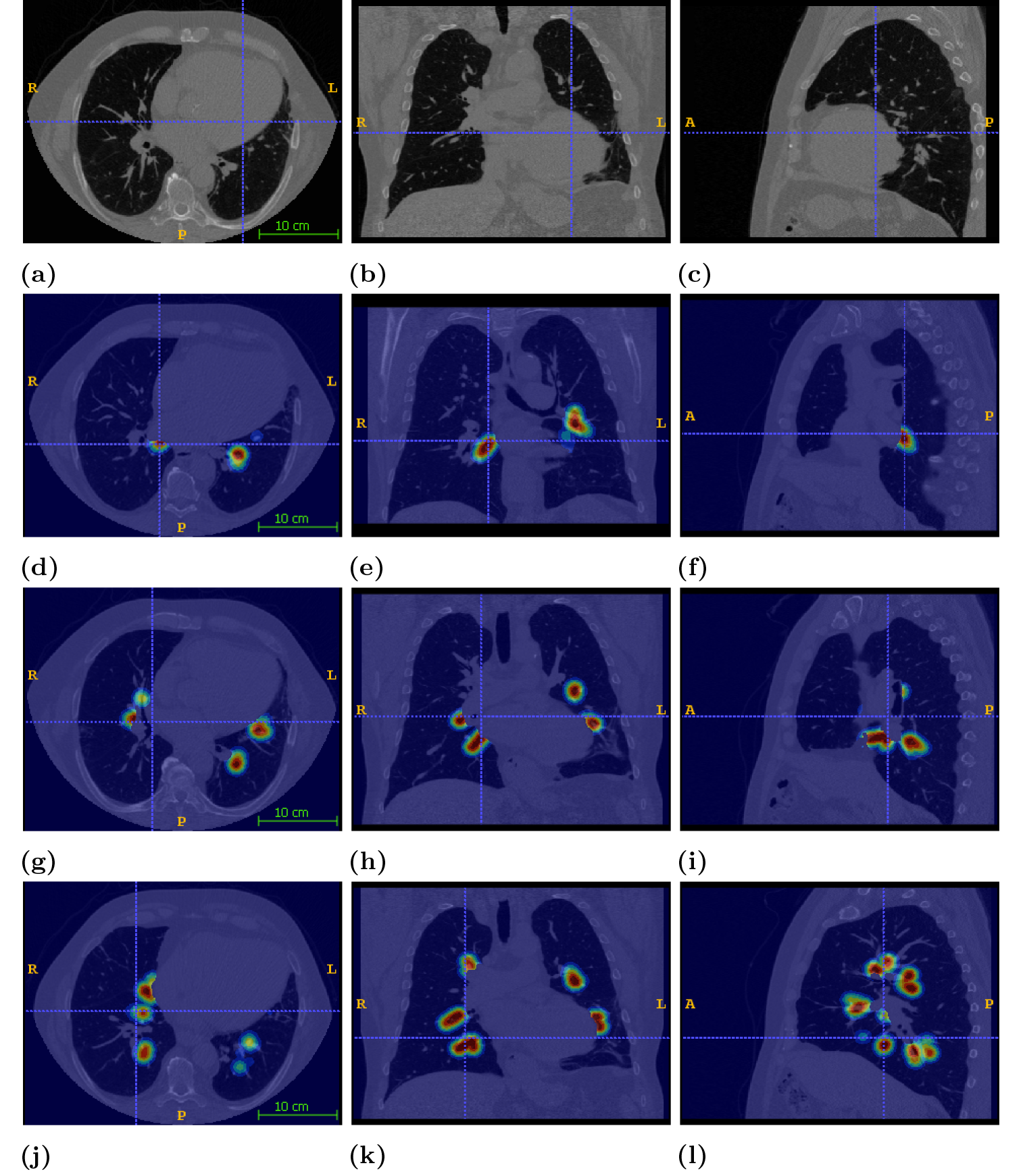}
\caption{Visualization of spatial likelihood for each branching generation. The extracted spatial maps are overlaid on the CT scans. Given a certain slice of the CT scan, the second row presents the probability to locate a branch of the second generation. Likewise in the third row the probability to locate a third generation airway is visualized and in the fourth row, the probability to locate a fourth generation airway is presented.}
\label{fig:visualizationProbabilityMap}
\end{figure}

\subsection{Spatial probability maps of branching properties}

The location of each new generation branch is calculated as explained before and provides a random sample out of all possible bronchial tree conformations. In this step of the proposed framework, we produce probabilistic maps for each generation branch that provide estimates of the spatial probability of encountering a particular generation at some point in the imaging data. Such a probabilistic model allows for optimising clinical decision-making by accounting for the branches' distributional uncertainty. 
Let's denote with 
\begin{equation}
W_g(\mathbf{x}),I:\Omega \xrightarrow{}\mathcal{R}
\end{equation}
the probability map for generation $g$ where 
\begin{equation}
\mathbf{x}=(x,y,z) , \mathbf{x} \in \Omega
\end{equation}
is a voxel in the spatial domain $\Omega \subset \mathcal{R}^3$ of the volumetric imaging data. Then
\begin{equation}
    W_g(\mathbf{x})=\frac{1}{{\sigma \sqrt {2\pi } }}e^{{{ - \left( {d } \right)^2 } \mathord{\left/ {\vphantom {{ - \left( {d } \right)^2 } {2\sigma ^2 }}} \right. \kern-\nulldelimiterspace} {2\sigma ^2 }}}
\end{equation}
where $d$ is the distance of voxel $\mathbf{x}$ to the closest edge of $\mathcal{G}$ labeled with generation $g$. Parameter $\sigma$ is set experimentally to $\sigma=1$.

Figure \ref{fig:visualizationProbabilityMap}
provides insightful visualization of the spatial likelihood for each branching generation. The extracted spatial map is overlaid on the CT scans
The first column corresponds to the axial view, the second column corresponds to the coronal view, and the third column corresponds to the sagittal view.
The first row depicts raw imaging data. The second row presents the probabilistic maps for the second generation of airways. The third row presents the probabilistic maps for the third generation of airways. The fourth row presents the probabilistic maps for the fourth generation of airways.

\subsection{Computational fluid dynamics to compute flow in normal and constricted lungs}

This work aims to develop precise models of airway constiction that correlate to respiratory maladies. Moreover, we provide an extensive assessment of the airflow features and clarification of the inhaled particle attributes as a next step towards attaining effective personalized treatment with appropriately tailored substances. To pursue these goals, a geometry processing methodology is established and employed to form various essential models of the pulmonary system. The proposed sequence mitigates several effects that degrade the quality of the surface models, thus facilitating the required simulations. Apart from the typical lung case, two obstructed cases are examined, i.e. a model with obstructions in both lungs and a model with obstructions in the right lung only. The partially obstructed model is evaluated only for the right lung to assess the effects of obstructions in the asymmetric behaviour of the typical model since most particles are deposited in the right section, as shown by our initial investigations. These particular cases are examined through the finite volume method (FVM) and specifically the CFD algorithms, which allow accurate deductions about airflow upon numerous parts of the lungs. In addition, a thorough parametric analysis is conducted to identify the effects of particle size and density on their deposition efficacy in various sections of the pulmonary system. In this way, the behaviour of various substances with diverse dosages can be assessed in terms of the deposition pattern they exhibit. The data demonstrate potential employment in personalized medical remedies for lung diseases by providing regional-oriented deposition information.

\begin{figure}[t]
\centering
\includegraphics[width=0.8\textwidth]{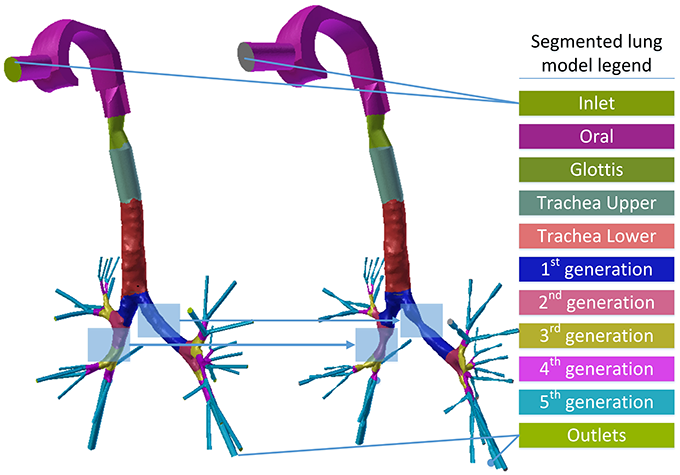}
\caption{Visualization of bronchocontriction simulation for the first five generations of a patient-specific reconstructed bronchial tree surface}
\label{fig:bmc-overview}
\end{figure}

\subsubsection{Simulation setup}

Various lung models related to diverse narrowing levels have been generated using the previously designated method. More specifically, and in addition to the unobstructed lung model, a group of obstructed models have been created. Each model is characterized by the lung generation index, after which narrowings with bottleneck structure is introduced. Furthermore, an asymmetric case where obstructions are introduced only in the right lung after the second generation is examined. An in-house established software based on the geometry described above processing sequences has been effectively utilized to develop these models. 

An open-source platform associated with CFD and FPT analyses, called OpenFoam\cite{jasak2007openfoam} is efficiently employed in order to implement the volume meshing of the lung by utilizing the snappyHexMesh algorithm. A very fine discretization near the walls of the pulmonary system is adopted by our analysis that involves hex-dominant cells. In this manner, the stimulating properties of turbulence can be discerned, while the positions of particles can be precisely determined. Specifically, the domain is discretized into 5,569,106 cells, which denotes a substantial computational burden. In order to effectively tackle this issue, ARIS, a high-performance computer (HPC), was utilized to conduct the simulations.

\begin{figure}[t]
\centering
\includegraphics[width=0.8\textwidth]{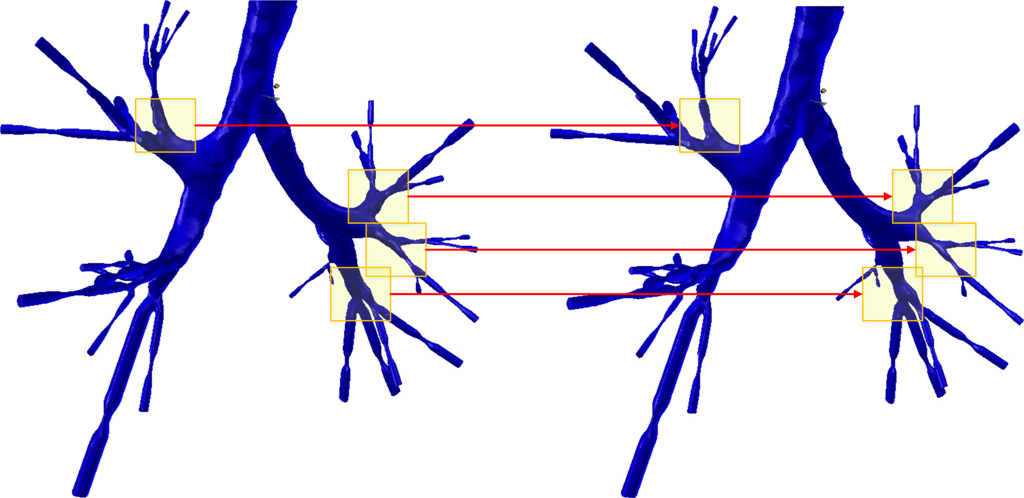}
\caption{Visualization of bronchocontriction simulation for the first 5 generations of a patient-specific reconstructed bronchial tree surface}
\label{fig:bmc-nar-3}
\end{figure}

The lung's activity can be computationally evaluated by employing various models, such as Reynolds-averaged Navier-Stokes (RANS) or Reynolds-averaged simulation (RAS), direct numerical simulation (DNS) and large eddy simulation (LES). Amid them, the RAS model exhibits acceptable accuracy levels and the minimum computational cost. 
In this context, RAS is effectively utilized to conduct a steady-state CFD analysis of the human respiratory system by solving the Navier-Stokes equations. Specifically, the airflow properties are clarified by employing a kOmegaSST turbulence model. The investigation is performed through the FVM algorithm, particularly the SIMPLE technique, part of the OpenFoam platform. Second-order schemes in time and space are applied to perform the discretization of the associated equations. 

A time step of $5 \times 10^{-4}s$ is employed to allow numerical stability. As designated in the CFD theoretical formulations, the Courant number, the air velocity, and the mesh discretization determine the value of the timestep mentioned above. In this manner, stability and convergence of the simulations are ensured. A hypothesis of a total pressure drop of -15 Pa is considered throughout the inspiratory states simulations. An additional evaluation of the inhaled particle attributes is allowed by the airflow velocity and pressure, assessed over the computational domain.

\begin{figure}[H]
\centering
\includegraphics[width=0.5\textwidth]{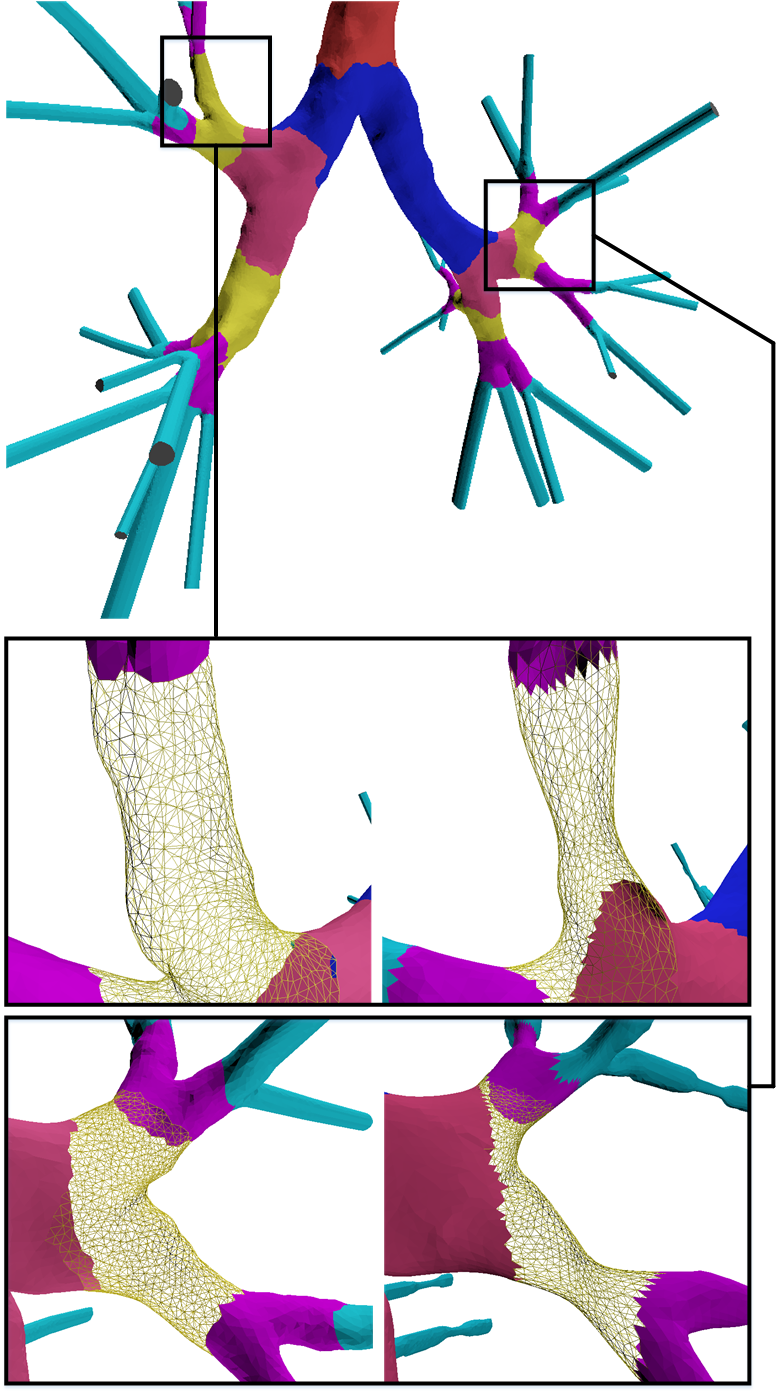}
\caption{Detailed visualization of bronchocontriction simulation for the first 5 generations of a patient-specific reconstructed bronchial tree surface}
\label{fig:bmc-narrowing-3rd-gen}
\end{figure}

\newpage
\subsubsection{Results}

\begin{figure}[H]
\centering
\ifdefined\showFigures
\includegraphics[width=0.6\textwidth]{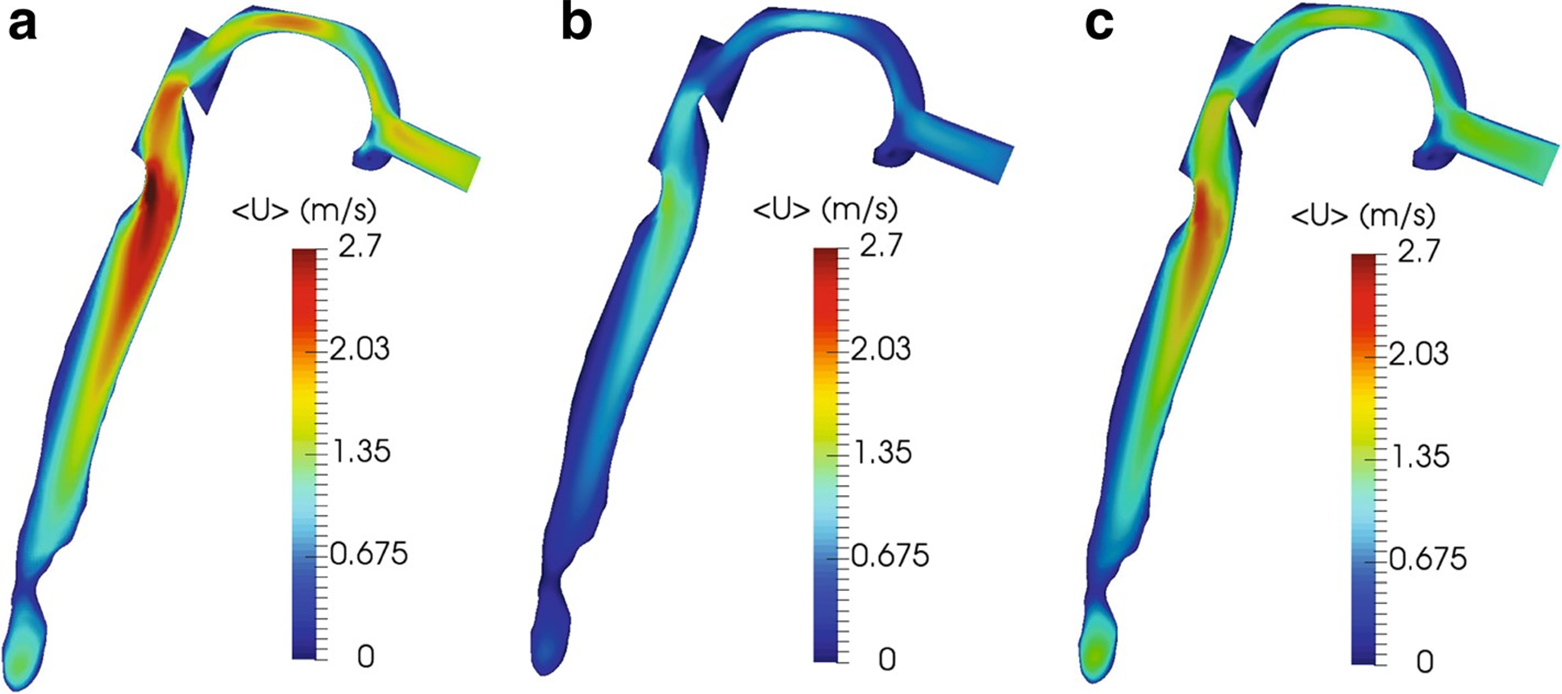}
\fi
\caption{Sagital cross-section of oral cavity, glottis and trachea in three cases. Visualization of the airflow velocity profiles for a) Normal case b) Narrowing is introduced in both left and right lungs and c) narrowing is introduced only in the left lung.}
\label{fig:sagital}
\end{figure}

\begin{figure}[H]
\centering
\ifdefined\showFigures
\includegraphics[width=0.6\textwidth]{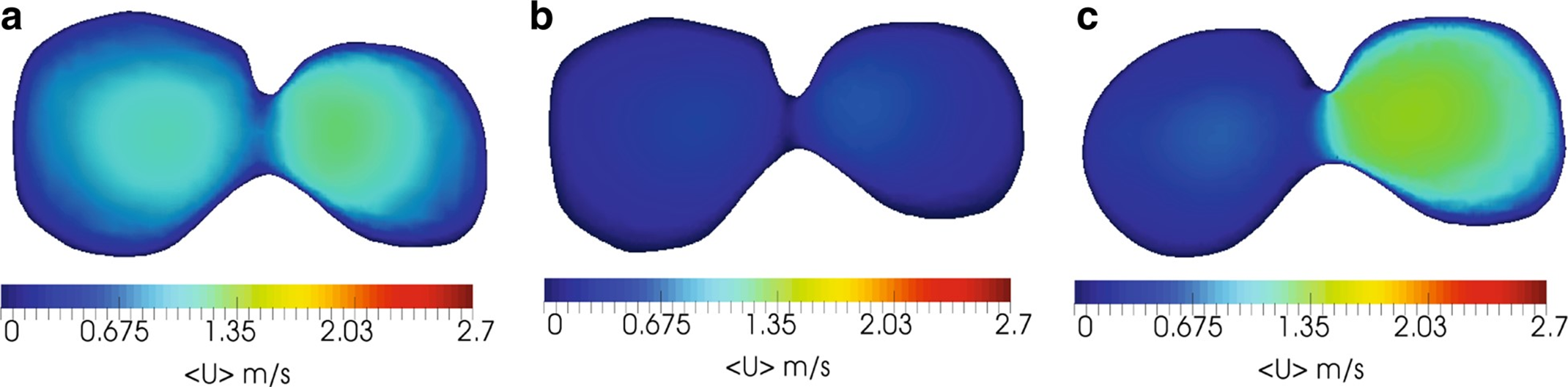}
\fi
\caption{Velocity profile on the cross-section of the first bifurcation.a) Normal case b) Narrowing is introduced in both left and right lungs and c) narrowing is introduced only in the left lung.}
\label{fig:transverse}
\end{figure}

\begin{figure}[H]
\centering
\ifdefined\showFigures
\includegraphics[width=0.7\textwidth]{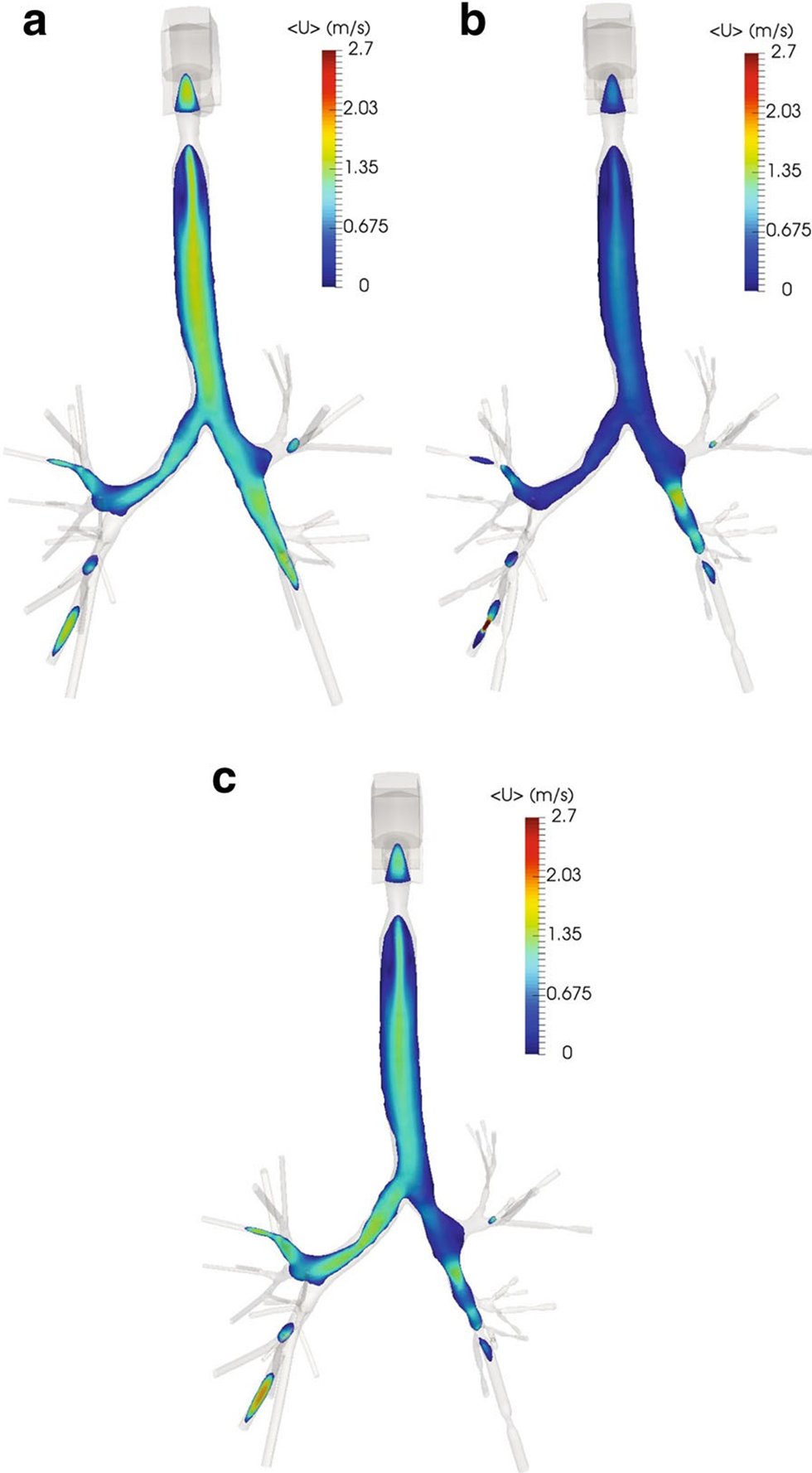}
\fi
\caption{Visualization of velocity distribution for coronal cross-section normal and constricted lung geometry. a) Normal lung geometry b) Constrictions are introduced in both left and right lung airways c) Constrictions are introduced only on the left lung.}
\label{fig:coronal}
\end{figure}

\begin{figure}[H]
\centering
\ifdefined\showFigures
\includegraphics[width=0.8\textwidth]{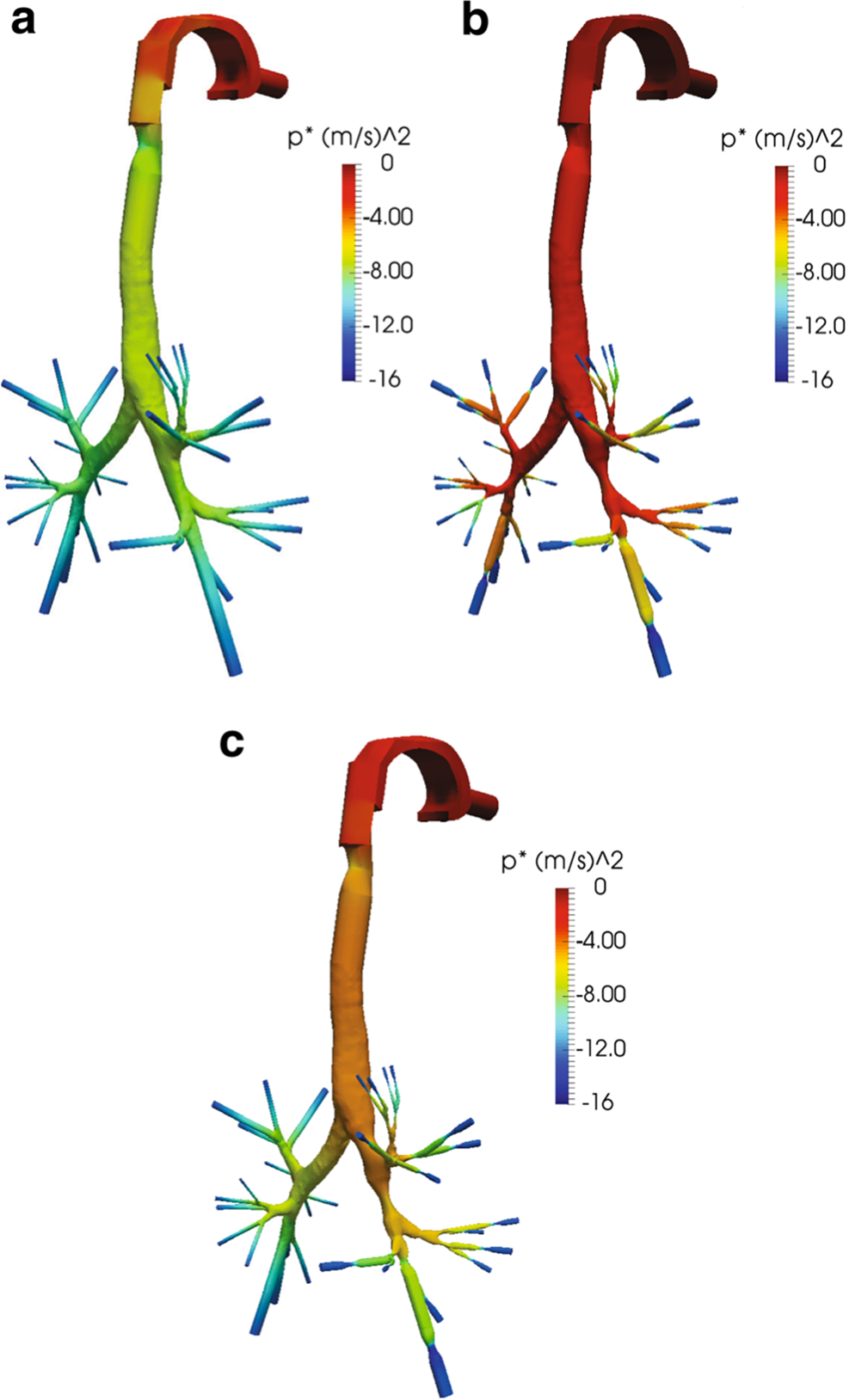}
\fi
\caption{Pressure distribution for the surface of the oral cavity, glottis trachea and the first five generations. a) Normal case. b) Narrowing is introduced in both left and right lungs and c) narrowing is introduced only in the left lung.}
\label{fig:pressuredrop}
\end{figure}

The normal lung demonstrates an inlet velocity of 1.475 m/s and a flowrate of 12.6 L/min, whereas the velocity of air when narrowings are detected after the second generation is 0.711 m/s and the flowrate is 6.06 L/min. Also, when obstructions occur in the right lung only, an inlet velocity of 1.311 m/s and a flowrate of 11.2 L/min are observed. Indicative results are illustrated in Figures \ref{fig:pressuredrop}, \ref{fig:sagital}, \ref{fig:transverse}, \ref{fig:coronal}, where the normalized pressure drop at the surface of the lung, the air velocity distribution cross-sections at the sagittal plane, the transverse plane, the trachea, and main bronchi, are presented. From the results depicted in Figure \ref{fig:pressuredrop}, higher values of pressure drop are observed on the trachea and the main bronchi when obstructions in both lungs occur. As observed from Figure \ref{fig:sagital}, when several narrowings are present, the airflow is decreased in the region of the trachea, thus diminishing the breathing capability. On the other hand, the airflow is fairly decreased if only one lung is obstructed. Similar conclusions can be derived from Fig.\ref{fig:transverse}. Furthermore, the airflow features at the region of the main bronchi are clarified in Figure \ref{fig:coronal}. The airflow is driven at the other functional part when only one lung is obstructed. This observation designates that any inhaled particles, ambient or medical oriented, will be directed mostly away from the side of the pulmonary system, where the inflammation is more severe.

\newpage

\newpage

\newpage

\section{Human-machine interfaces for patient-specific pulmonary modelling}

The presented components of AVATREE are provided as an open-source solution accompanied by a graphical user interface (GUI). The implemented application programming interface (API) includes the following modules, a) input-output functionalities, b) 1-dimensional representation tools including centerline extraction, graph generation, derivation of graph node properties, c) bronchial tree extension tools extending the 1-dimensional representation to the desired number of generations, d) 3D surface generation and processing tools and e) airway narrowing simulation tools.       

\begin{figure}[t]
    \centering
    \ifdefined\showFigures
    \includegraphics[width=\textwidth]{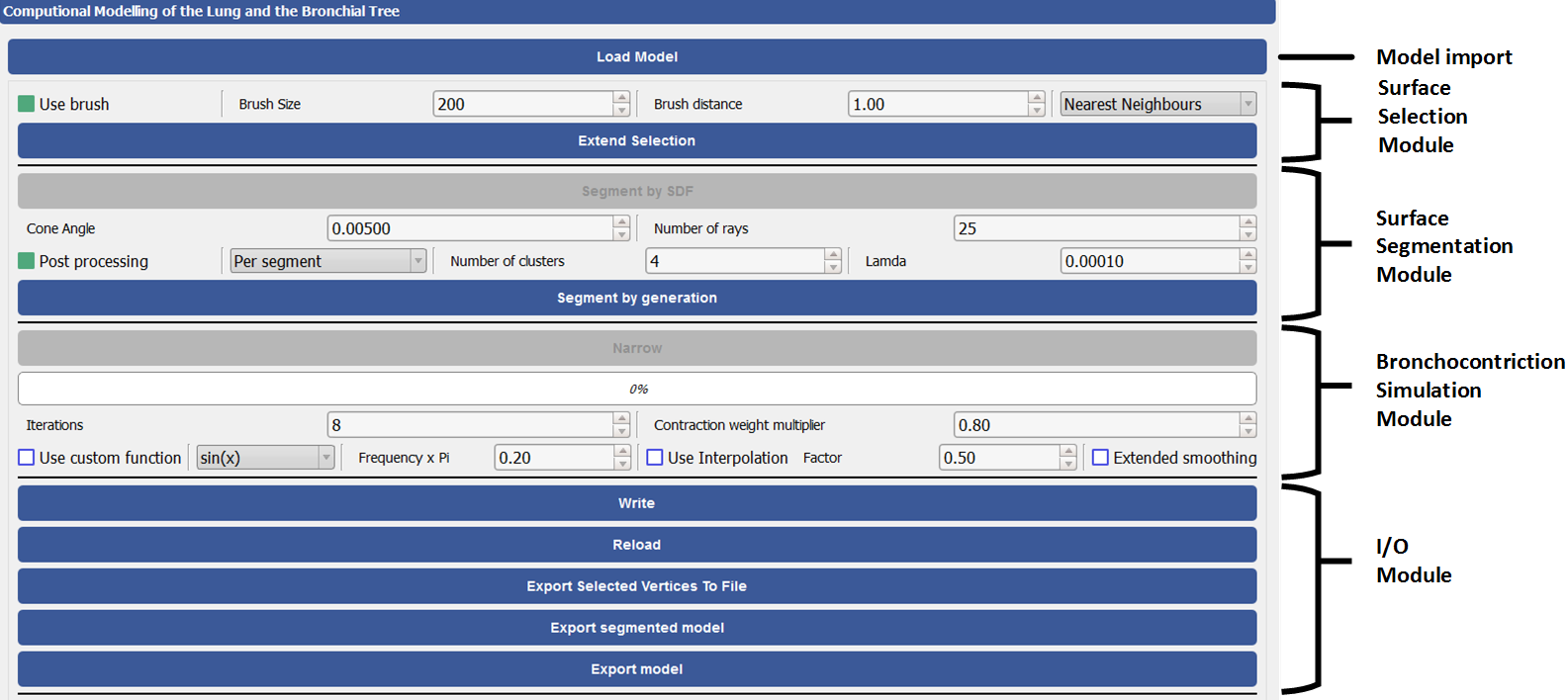}
    \fi
    \caption{User interface. The UI is comprised of four panels, namely the data input and output panel, area selection panel, segmentation panel and bronchoconstriction simulation panel}
  \label{fig:user-interface}
\end{figure}

\begin{figure}[t]
    \centering
    \ifdefined\showFigures
    \includegraphics[width=0.8\textwidth]{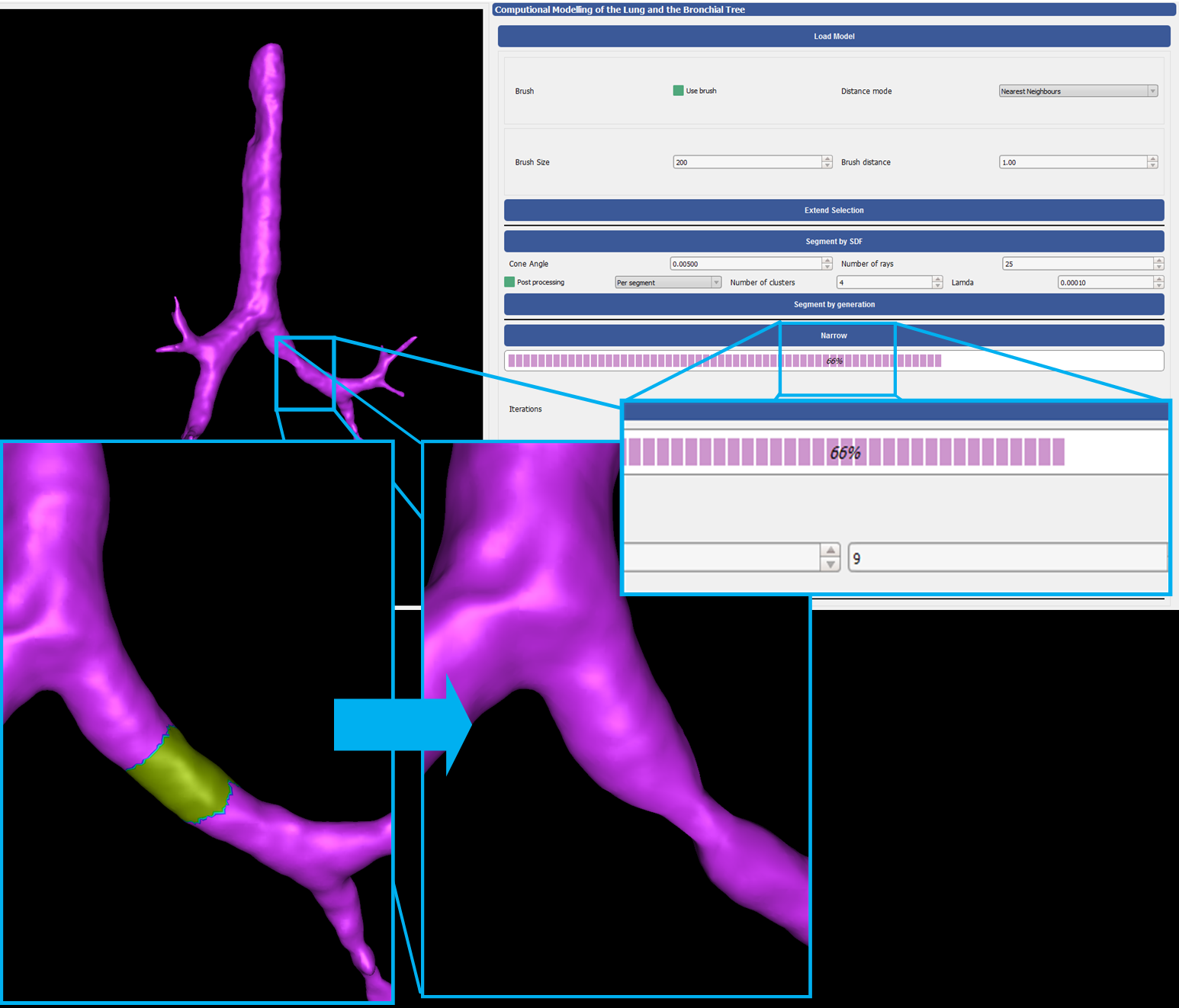}
    \fi
    \caption{Demonstration of broncho-constriction simulation. Airway of second generation narrowed at 34\% of original diameter.}
  \label{fig:ui-constriction}
\end{figure}

\begin{figure}[t]
 \centering
\ifdefined\showFigures
\includegraphics[width=0.8\textwidth]{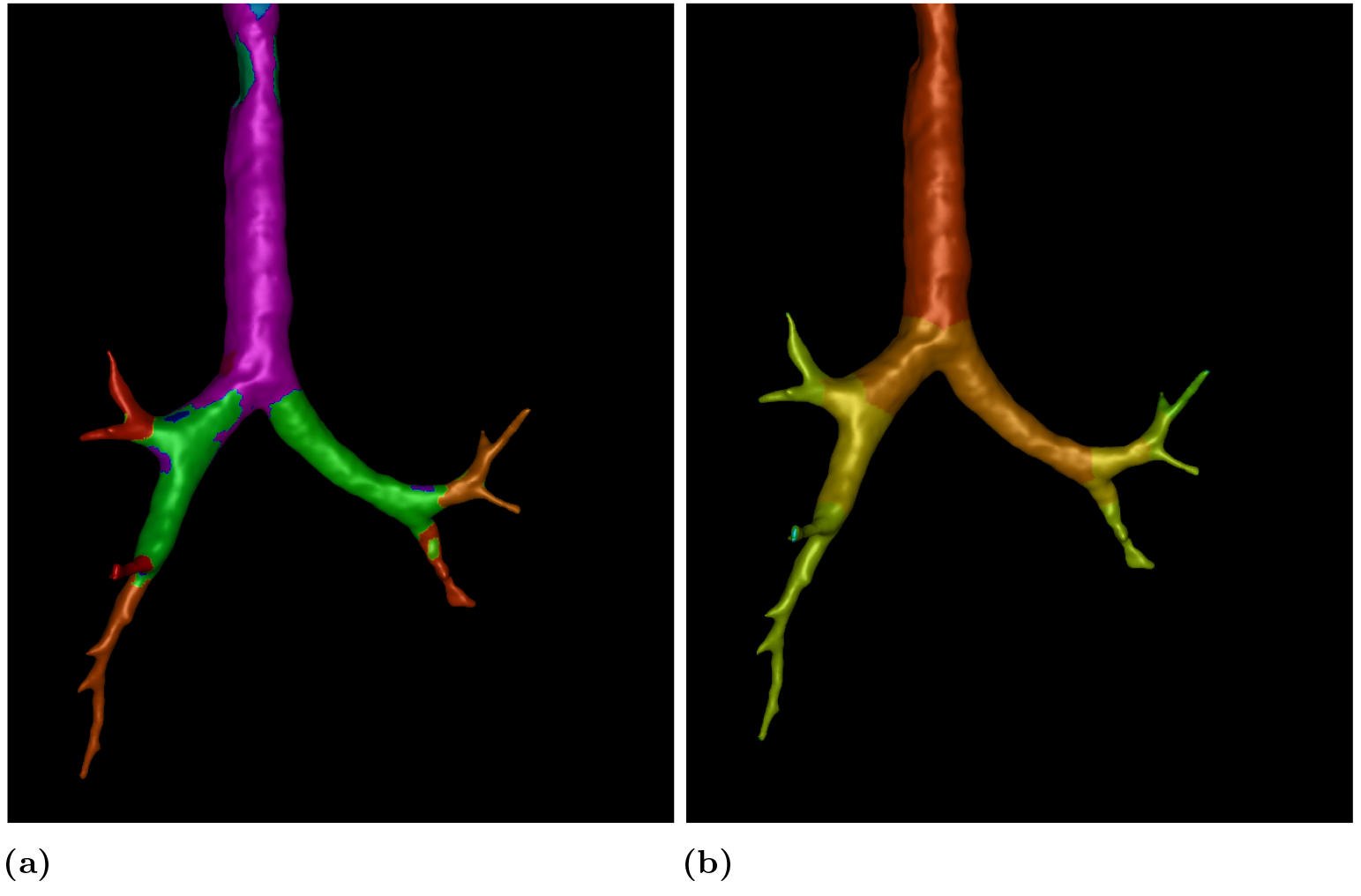}
\fi
\caption{%
Surface annotation with a) SDF function visualizing local diameter b) airway generation.}%
  \label{fig:ui-segmentation}
\end{figure}

The deformed surface introduced into computational fluid dynamics can provide insight into the breathing pattern and drug delivery in asthmatic lungs\cite{das2018targeting}. The user interface, presented in Figure \ref{fig:user-interface}, employs the set of functionalities defined by AVATREE  and is comprised of four panels, namely the data input and output panel, area selection panel, segmentation panel and bronchoconstriction simulation panel. Through the GUI, the user can load a 3D model, select the area to be processed, as Figure \ref{fig:ui-constriction} visualizes, and use the narrowing functionalities to reduce the airway diameter by the desired percentage. The amount of narrowing depends on the number of iterations and contraction weight multiplier. In Figure \ref{fig:ui-constriction} an airway of the first generation was constricted by 66\%.
The surface faces can be classified based on local properties in the segmentation panel. The one illustrates the shape diameter function (SDF)\cite{shapira2008consistent}, while the other one the 3D surface according to the generation number. The results are visualized in Figure \ref{fig:ui-segmentation}.

\chapter{Monitoring medication adherence in constrictive pulmonary diseases}
\label{chapter:monitoring}


The respiratory system is a vital structure vulnerable to airborne infection and injury. Respiratory diseases are the leading causes of death and disability worldwide. Specifically, 334 million people have asthma, the most common chronic disease of childhood, affecting 14\% of all children globally \cite{world2017global}. The effective management of chronic constrictive pulmonary conditions lies mainly in the proper and timely medication administration. However, as recently reported \cite{ngo2019inhaler}, a large proportion of patients misuse their inhalers. Studies have shown that possible technique errors can harm clinical outcomes for users of inhaler medication \cite{darcy2014method,jardim2019importance}. Incorrect inhaler usage and poor adherence were associated with high durations of hospitalization and high numbers of exacerbations.

Several methods have been introduced to monitor a patient's adherence to medication. As a series of studies indicate, effective medication adherence monitoring can be defined by successfully identifying actions performed by the patient during inhaler usage. Several inhaler types are available in the market, among which the pressurized metered-dose inhalers and dry powder inhalers are the most common. In any case, developing a smart inhaler setup that allows better monitoring and direct feedback to the user independently of the drug type is expected to lead to more efficient drug delivery, thereby becoming the main product used by patients.

The pMDI usage technique is characterized as successful if a specific sequence of actions is followed \cite{murphy2019help}.
Appropriate audio-based monitoring could help patients synchronize their breath with drug activation and remind them to keep their breath after inhalation for a sufficient amount of time. Several methodologies that engage electronic monitoring of medication adherence have been introduced in the past two decades \cite{aldeer2018review}, aiming to alter patient behavioural patterns \cite{heath2015theory,alquran2018smartphone}. In the field of inhaler-based health monitoring devices, a recent comprehensive review by Kikidis et al. \cite{kikidis2016digital} provides a comparative analysis of research and commercial attempts in this direction. Cloud-based self-management platforms and sensor networks constitute the next step towards adequate medication adherence and self-management of respiratory conditions \cite{khusial2019myaircoach,polychronidou2019systematic}.\par

It is crucial to successfully identify audio events related to medication adherence in all cases. In this direction, several approaches have been proposed in the literature, presenting mainly decision trees or state-of-the-art classifiers applied to a series of extracted features. However, the above-mentioned methodologies come with high computational costs, limiting the applicability of monitoring medication adherence to offline processing or online complex distributed cloud-based architectures that can handle the need for resources. Therefore, the demand for computationally fast yet highly accurate classification techniques remains.

Motivated by the aforementioned open issues, this study presents methodologies \cite{nousias2018mhealth,ntalianis2019assessment,pettas2019recognition} that recognize the respiration and drug delivery phases on acoustic signals derived from pMDI usage.

\section{Prescribed inhaler use}

\begin{figure}
    \centering
    \includegraphics[width=0.9\linewidth]{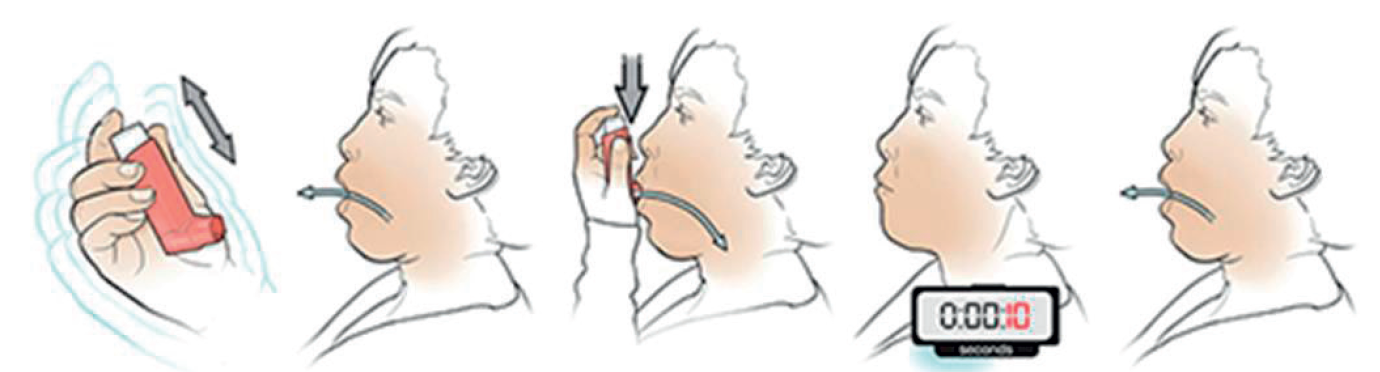}
    \caption{How to use a metered dose inhaler. A. The cap should be removed and the inhaler shaken  B. The patient should breathe out, away from the inhaler C. The patient should bring the inhaler to the mouth,  place it between the teeth and close the lips around it. Afterwards, the patient should start to breathe in slowly, press the top of the inhaler once, and keep breathing in slowly until a full breath is taken. D. The patient should remove the inhaler from the mouth, hold breath for about 10 seconds, and then breathe out. E. The patient should breathe out}
    \label{fig:technique}
\end{figure}

In clinical practice, specific instructions for correct usage of the pressurized Metered Dose Inhalers (pMDI) are provided to the patients based on four states, namely the pMDI inhaler actuation inhalation, exhalation, and background/environmental activity.
Figure \ref{fig:technique} demonstrates the correct usage of a Pressurized Metered Dose Inhaler (pMDI) according to clinical guidelines and forms the basis for the separation of the four classes of events that are assessed in the current study, namely: pMDI inhaler actuations, inhalation sounds, exhalation sounds, and background/environmental sounds. 
More specifically, proper inhaler use includes the following steps \cite{gillette2016inhaler}: 
\begin{enumerate}
\item  The cap should be removed and the inhaler shaken
\item  The patient should breathe out, away from the inhaler
\item  The patient should bring the inhaler to the mouth,  place it between the teeth and close the lips around it. Afterwards, the patient should start to breathe in slowly, press the top of the inhaler once, and keep breathing in slowly until a full breath is taken.
\item  The patient should remove the inhaler from the mouth, hold breath for about 10 seconds, and then breathe out.
\item The patient should breathe out
\end{enumerate}.
To automatically recognize those four different states, recent approaches apply machine learning for the classification of sound signals\cite{nousias2018mhealth,taylor2018estimation,taylor2018objective}. 

\section{System architecture}

\begin{figure}
    \centering
    \includegraphics[width=\textwidth]{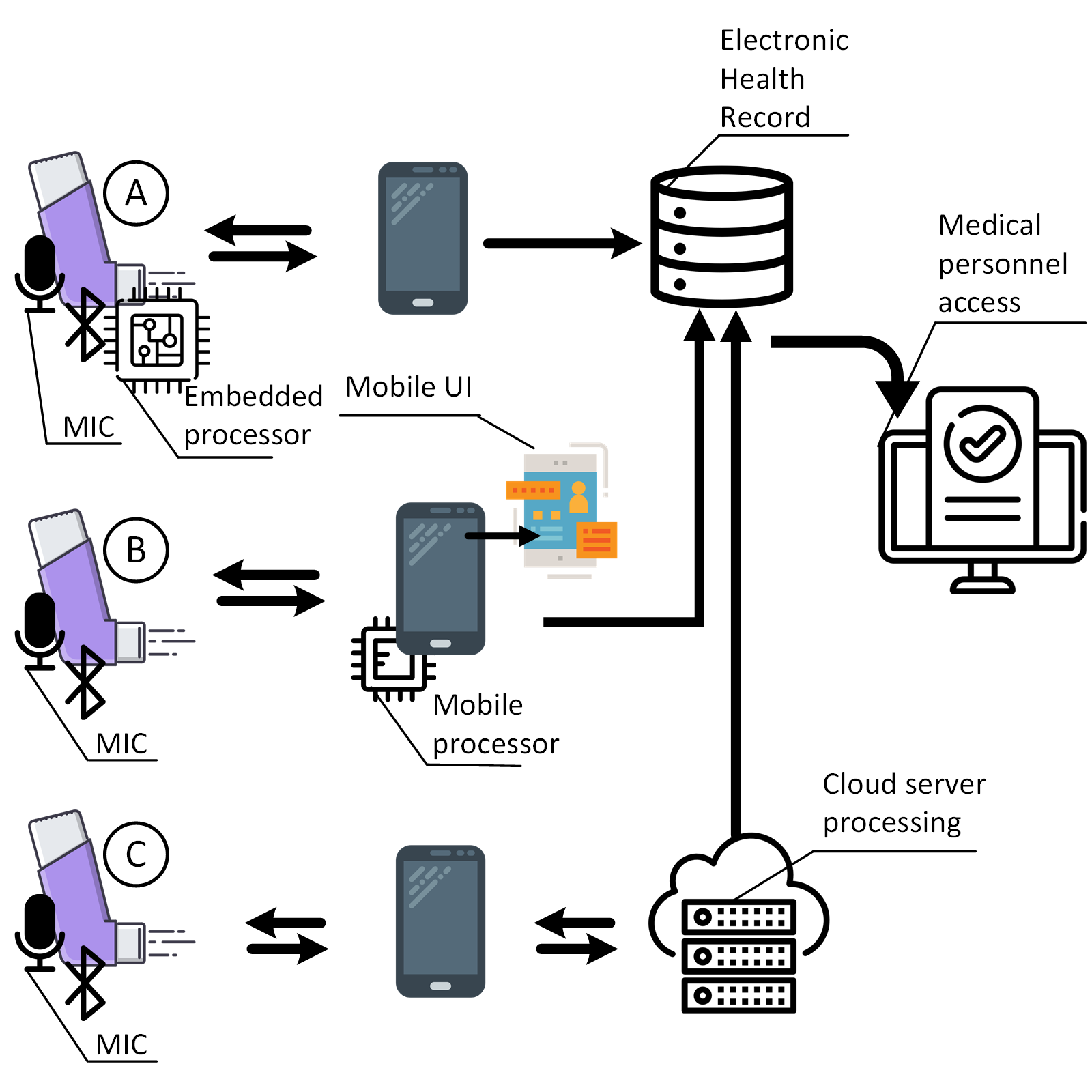}
    \caption{Monitoring system architecture. Three scenarios are taken into account. A) The inhaler device transmits the captured audio samples to the mobile device. The mobile device stores the audio files and sends them to a cloud processing server for differentiation. B) The audio device transmits the captured audio samples to the mobile device, and the differentiation occurs in the Mobile device processor. C) The captured audio is directly processed in the embedded processor.}
    \label{fig:systemarchitecture}
\end{figure}

Figure \ref{fig:systemarchitecture} presents a diagram of the overall architecture. The system consists of three parts. Three scenarios are taken into account. A) The inhaler device transmits the captured audio samples to the mobile device. The mobile device stores the audio files and sends them to a cloud processing server for differentiation. B) The audio device transmits the captured audio samples to the mobile device, and the differentiation occurs in the Mobile device processor. C) The captured audio is directly processed in the embedded processor. The algorithms presented in the following sections cover all of the cases mentioned above.

\section{Data collection}
 This section describes the data collection and annotation process. Two datasets were used, aiming to facilitate the training of audio analysis models toward the differentiation of inhaler related sounds.
 
\subsection{Dataset A}
\begin{figure}[H]
  \begin{subfigure}[b]{0.30\linewidth}
    \includegraphics[width=\textwidth]{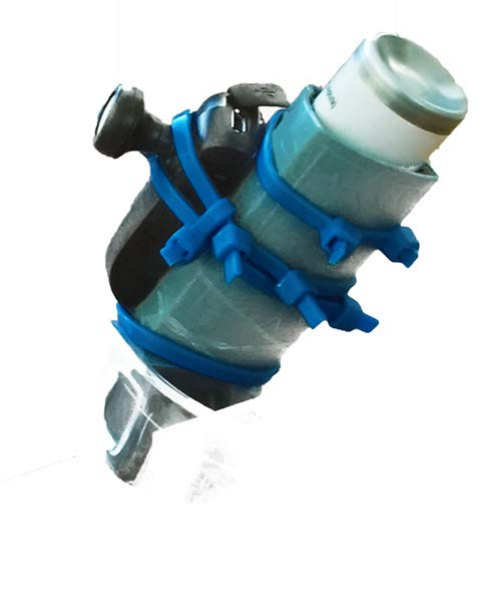}
    \caption{}
    \label{fig:exp_setup}
  \end{subfigure}
  \hfill
  \begin{subfigure}[b]{0.30\linewidth}
    \includegraphics[width=\textwidth]{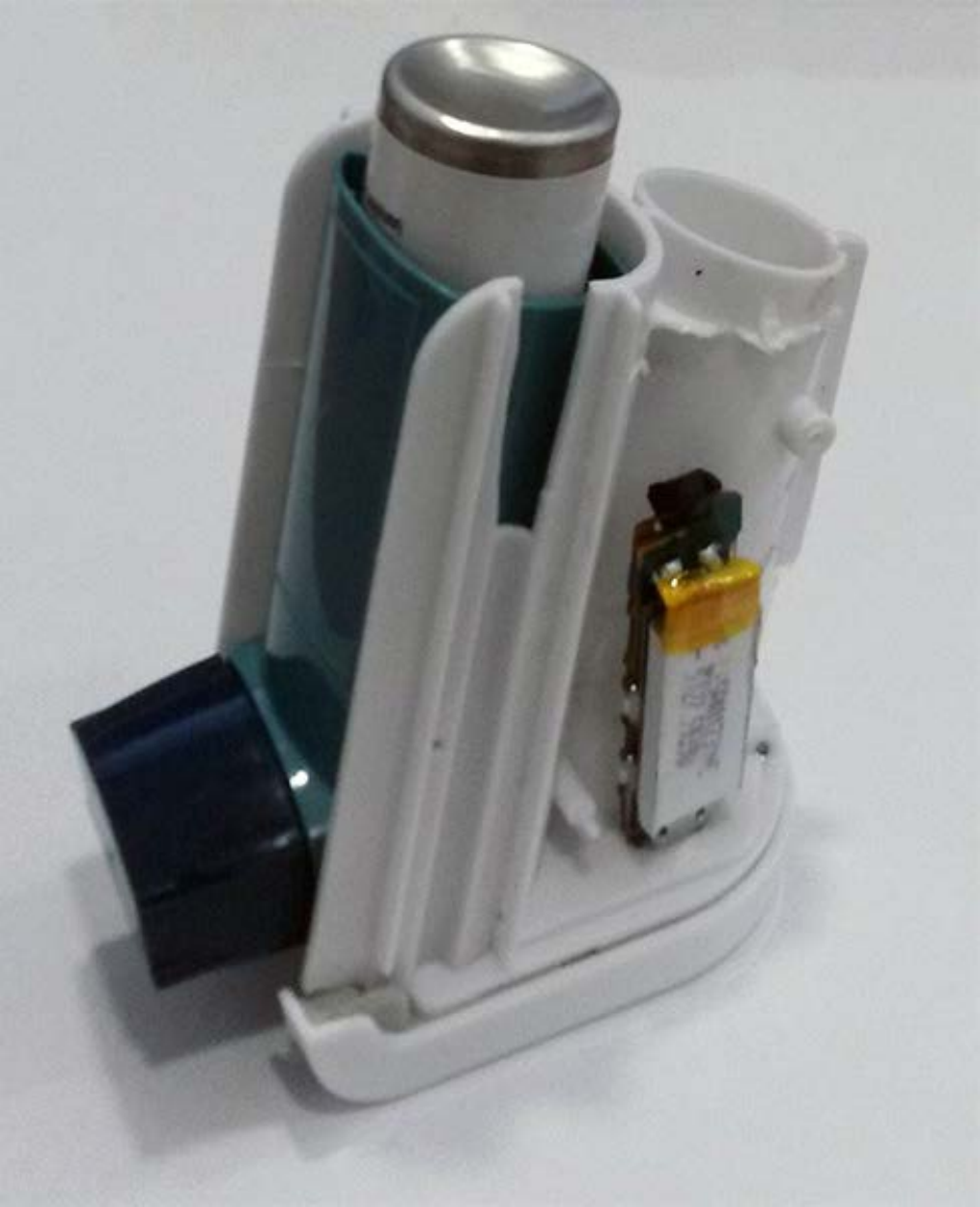}
    \caption{}
    \label{fig:no_case}
  \end{subfigure}
  \hfill
  \begin{subfigure}[b]{0.30\linewidth}
    \includegraphics[width=\textwidth]{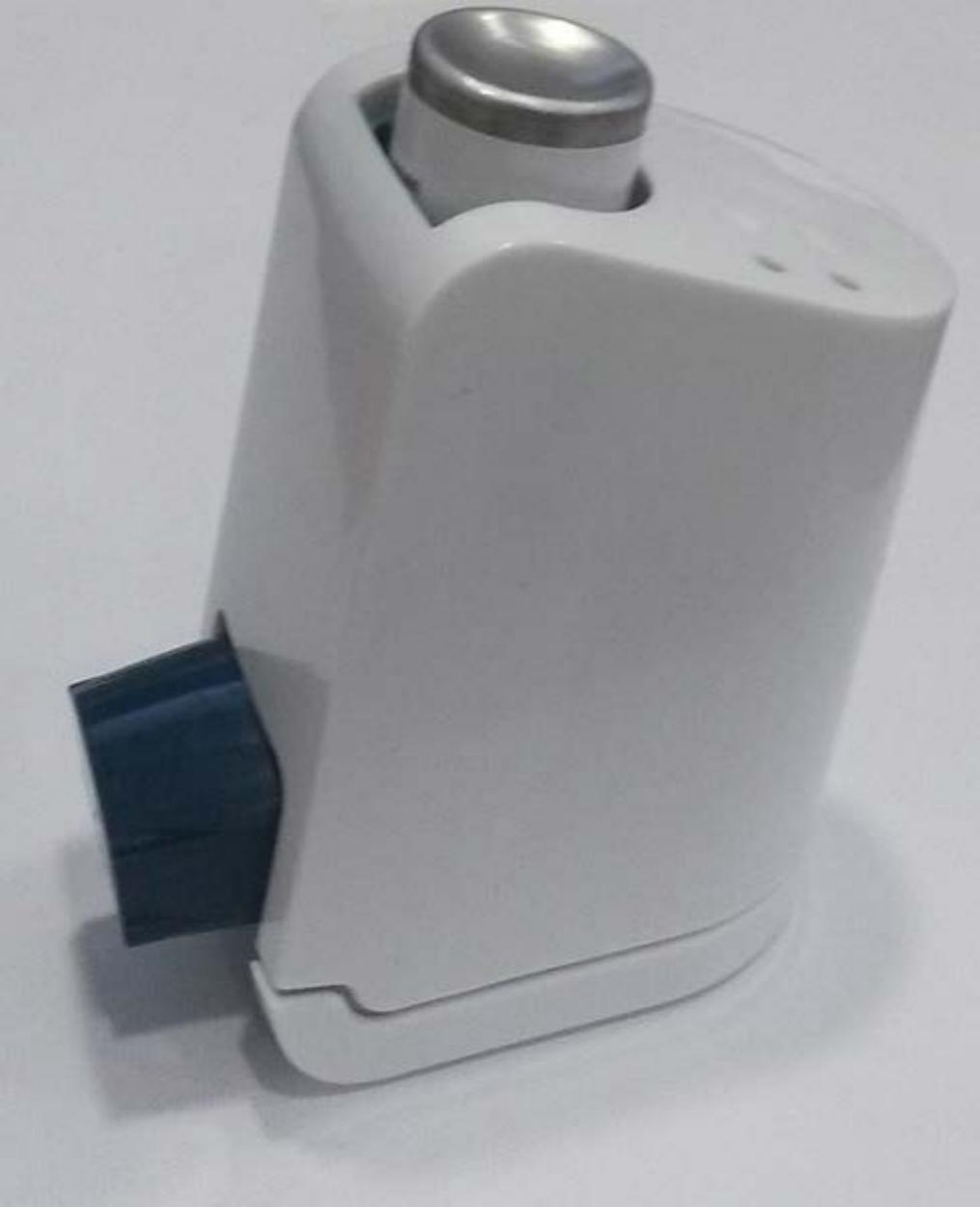}
    \caption{}
    \label{fig:with_case}
  \end{subfigure}
  \caption{a)Experimental setup of the pMDI. The Bluetooth microphone is firmly locked on the device. b) Inhaler prototype without casing. The pMDI is placed within a cavity. c) Inhaler prototype with casing.}
  \label{fig:inhaler-prototype3}
\end{figure}

Several inhaler sounds were recorded in indoor and outdoor environments for the first dataset at an 8 kHz sampling rate and 16-bit depth. The recording device was a Samsung HM1200 Bluetooth microphone mounted on an inhaler loaded with placebo canisters. For Dataset A, a mobile device application was developed that connected to the Bluetooth microphone and allowed the user to activate the microphone before using the inhaler. The sounds were categorized into inhaler actuations, exhalations, inhalations, and noise referring to environmental or other sounds. Several full inhalers use audio segments with a total duration of 12 seconds each were captured with the participation of 12 persons. Afterwards, audio segments were annotated and cropped to a duration of 0.5 seconds each. In total, 1980 segments of 0.5 seconds each were compiled in a balanced collection, where to each class, 495 segments were dedicated. 
The experimental inhaler device is described in Figure \ref{fig:inhaler-prototype3}

\begin{figure}[!tbp]
\begin{center}
\includegraphics[width =\linewidth]{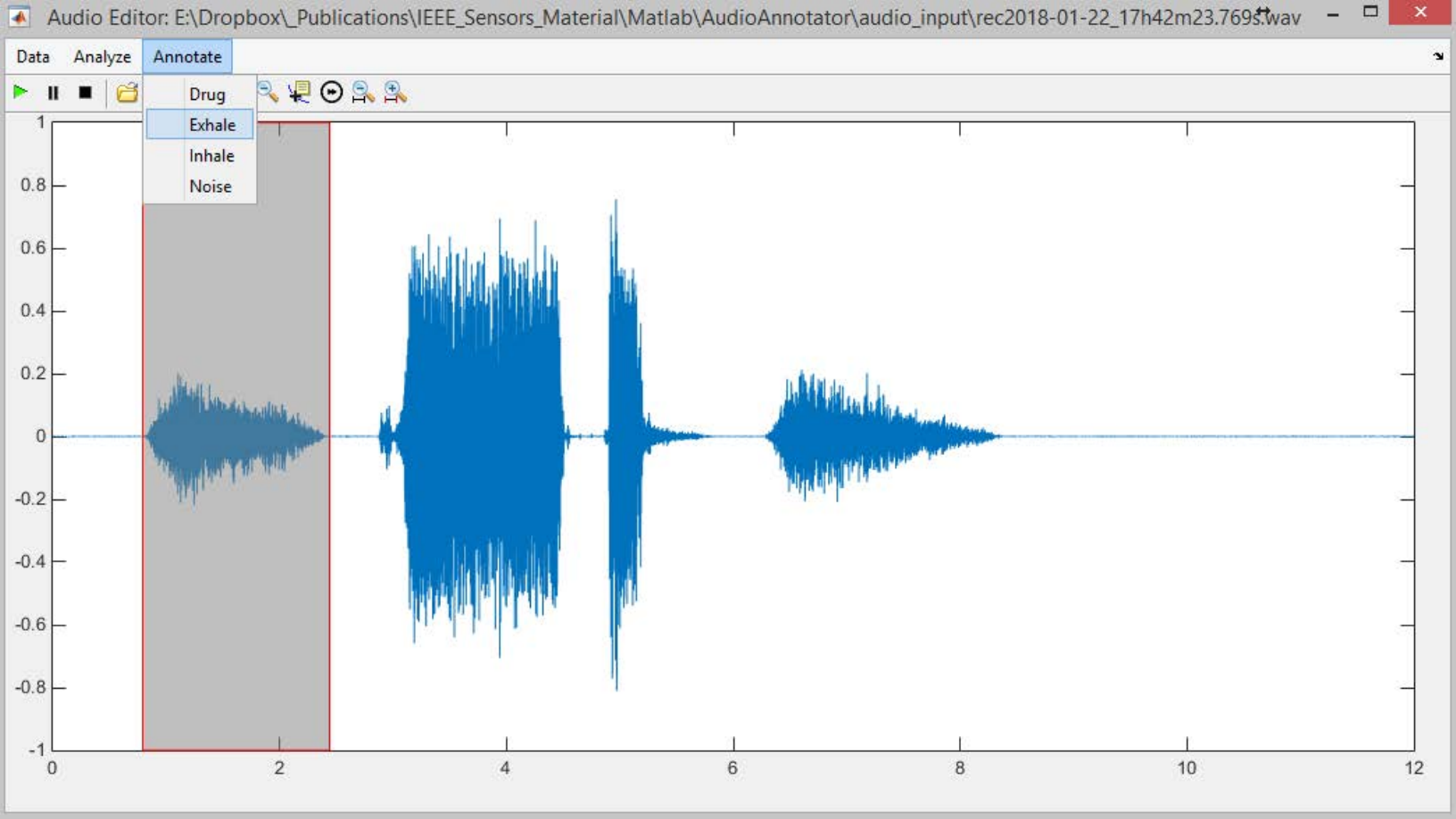}
\end{center}
\caption{Annotation toolkit UI. The user inspects the audio graph, selects a segment corresponding to a certain class, and attaches the proper annotation.}
\label{fig:annotation-tool}
\end{figure}

\begin{figure}[H]
    \centering
    \includegraphics[width=\linewidth]{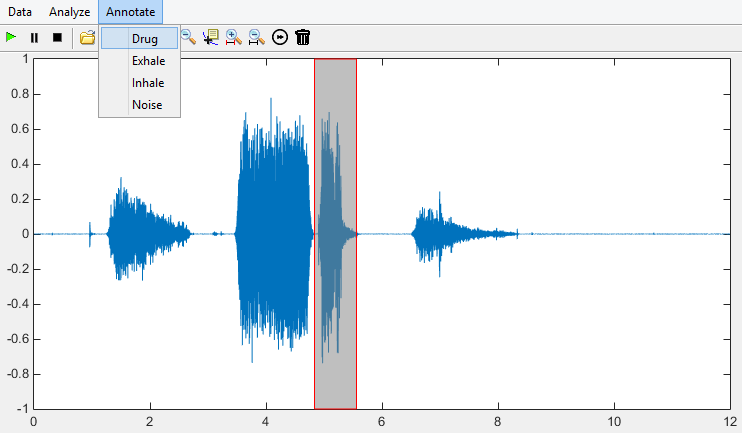}
    \caption{Manual segmented and annotation of audio recordings}
    \label{fig:annotation_process}
\end{figure}

To compile the dataset, an annotation toolkit was employed. The user interface visualizes the audio samples while the user selects parts of the audio files and assigns a class through the menu item "Annotate". The annotated part is stored in a separate audio file. Figure \ref{fig:annotation-tool} and \ref{fig:annotation_process} show the user interface of the annotation toolkit \cite{nousias2018mhealth}.

Even though "Dataset A" generated adequate results with high accuracy, there is a significant drawback. The precisely selected audio segments do not describe the reality of a patient's everyday life since real inhaler audio data encompass overlapping and mixed audio events. A second dataset was compiled to adequately describe the inhaler use, referred to as "Dataset B". 

\subsection{Dataset B}
\begin{figure}[H]
\centering
    \begin{subfigure}{\textwidth}
         \includegraphics[width=\linewidth]{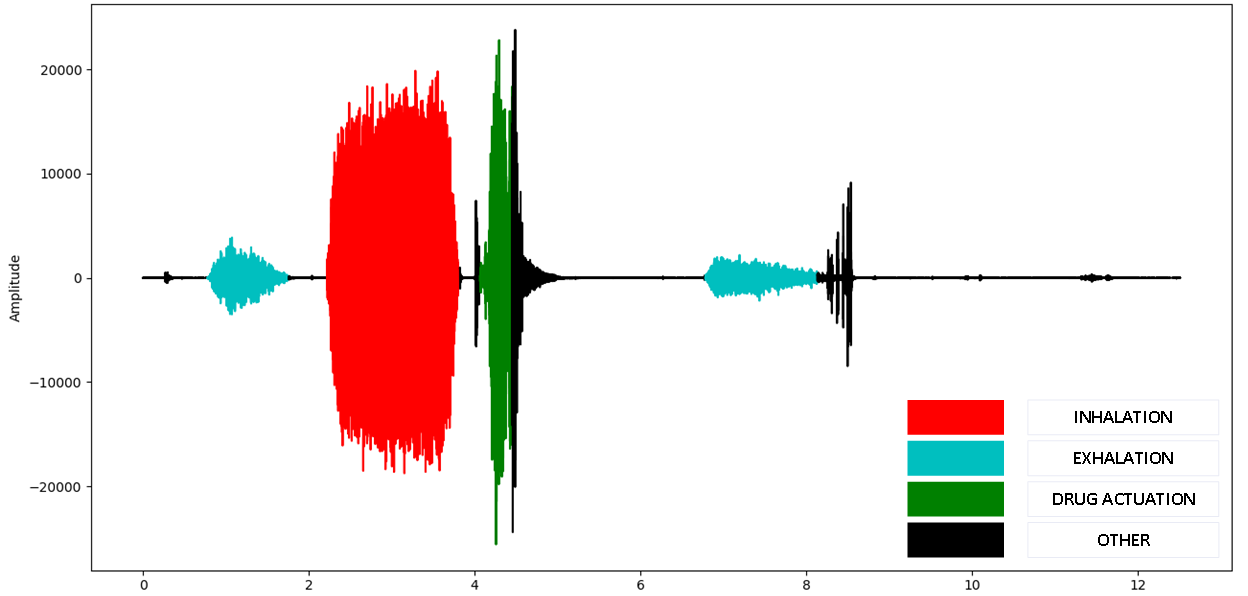}
         \caption{}
    \end{subfigure}
    \caption{Annotated audio file of 12 seconds. Red color corresponds to inhalation, cyan to exhalation, green to drug activation and black to other sounds}
    \label{fig:annotation}
\end{figure}

For the second dataset, referred to as "Dataset B", audio recordings from pMDI use were received using a standard checklist of steps recommended by National Institute of Health (NIH) guidelines. It was essential to ensure that the actuation sounds were used accurately recorded. The data were acquired from three subjects, between 28 and 34 years old, who used the same inhaler device loaded with placebo canisters. The first person (female)  committed 240 audio files, the second subject (male), 70 audio files and the third subject (female) 50 audio files. In total, 360 audio files were recorded for twelve seconds each, containing a full inhaler usage case. The recordings were performed in an acoustically controlled indoor environment, free of ambient noise, at the University of Patras to reflect possible use in real-life conditions and to ensure accurate data acquisition. The study supervisors were responsible for inhaler actuation sounds and respiratory sounds and followed a protocol that defined all the essential steps of the pMDI inhalation technique. Prior training of the participants on this procedure allowed to reduce the experimental variability and increase the consistency of action sequences. Each participant annotated the onset and duration of each respiratory phase in written form during the whole experiment. Also, the annotation of the different actions was subsequently verified and completed by a trained researcher and based on a visual inspection of the acquired temporal signal.
The acoustics of inhaler use was recorded as monophonic audio at a sampling rate of 8000Hz. The sensor's characteristics are 105dB-SPL sensitivity and 20Hz - 20kHz bandwidth. After quantization, the signal had a resolution of 16 bits/sample. Throughout the audio data processing, no further quantization of the data took place. A complete audio recording time-series example is presented in Figure \ref{fig:annotation}, coloured according to the annotated events. Any signal part that has not been annotated was considered noise during the validation stage. This dataset can allow in-depth analysis of patterns in sound classification and data analysis of inhaler use in clinical trial settings.

\section{Methodology}

\subsection{Audio feature extraction}
\label{section:audio-features-extraction}

As presented in Section \ref{section:monitoring-medication}, there are several feature extraction methods based either on well-known classical approaches. We mainly examine features based on Spectrogram, Cepstrogram and Mel-Frequency Cepstral Coefficients. The following paragraphs describe how the features are extracted so as to be utilized within machine learning algorithms in the following sections. Even though the theoretical approach for each algorithm is presented in Section \ref{section:monitoring-medication}, the formulation of the feature vector primarily defines the performance of the classification algorithm. 

\paragraph{Spectrogram}
In details, for a timeseries $x[n]$ with $n$ timepoints, where $n \in (1,\cdots,N)$, the spectrogram $\mathbf{S}$ is described as follows:
\begin{equation}
    \mathbf{S}=spectrogram\{x[n]\}(m,\omega)=\vert  X(m,\omega)\vert^2
\end{equation}
where $X(m,\omega)$ is the Fourier transform of $x[n]$,
\begin{equation}
\omega \in (1,\cdots,\frac{w}{2}+1)
\end{equation}
and 
\begin{equation}
   m \in (1,\cdots,\floor*{\nicefrac{N-(w-h)}{h}}) 
\end{equation}
$w$ is the window size and $h$ is the frame increment.
The latter is set to $h=\nicefrac{w}{4}$. 

\noindent Feature vector $\mathbf{v}_s$ 
\begin{equation}
\mathbf{v}_s=\sum_{m=1}^{M}S(m,k)
\end{equation}
$\mathbf{v}_s$ is subsequently downsampled to $\mathbf{v}'_s$ so that 
\begin{equation}
    \mathbf{v}_s\in \mathcal{R}^{[\nicefrac{w}{2}+1)]\times [\nicefrac{N-(w-h)}{h}]} \rightarrow \mathbf{v}'_s\in \mathcal{R}^{N_F \times [\nicefrac{N-(w-h)}{h}]} 
\end{equation}
where $N_F=32$ is the number of frequencies after downsampling.

\paragraph{Cepstrogram}
Cepstrogram $C(m,k)$ is formulated as
\begin{equation}
C(m,k)= \left|\sum_{n=0}^{N-1}log|X(m,n)|^2\cos(\frac{2\pi}{N}kn)\right|^{2}
\end{equation}
where $X(m,n)$ is the short time Fourier transform, $m$ denotes the $m-th$ temporal component and $k$ the $k-th$ cepstral coefficient and $n$ the $n-th$ frequency component. The audio feature vector $\mathbf{v}=[v_1 v_2 v_3 \ldots v_k]$ is derived by summing up the quefrency magnitude for every temporal window for each quefrency component. 
\noindent Feature vector $\mathbf{v}_c$ 
\begin{equation}
\mathbf{v}_c=\sum_{m=1}^{M}C(m,k)
\end{equation}
$\mathbf{v}_c$ is subsequently downsampled to $\mathbf{v}'_c$ so that 
\begin{equation}
    \mathbf{v}_c\in \mathcal{R}^{[\nicefrac{w}{2}+1)]\times [\nicefrac{N-(w-h)}{h}]} \rightarrow \mathbf{v}'_c\in \mathcal{R}^{N_F \times [\nicefrac{N-(w-h)}{h}]} 
\end{equation}
where $N_F=32$ is the number of frequencies after downsampling.

\paragraph{Mel-Frequency Cepstral Coefficients}
MFCC feature extraction is presented in \ref{section:preliminaries:mfcc}. $N_F=32$ ceptral coefficients are extracted.

\subsection{Supervised feature classification with Gaussian Mixture models}

\noindent A Gaussian Mixture Model (GMM) is defined with parameters 

\begin{equation}
    \{ a_i, \mathbf{\mu}_i, \mathbf{C}_i\}, i \in K
\end{equation}
where $K$ is the number of components, $a_i$ is the mixture weight of component $i$, $\mathbf{\mu}_i$ is the $d$-dimensional vector, containing the mean values for each feature, and $\mathbf{C}$ is the covariance matrix.

\noindent The Gaussian mixture density $p(\mathbf{v}|\lambda_n)$ is modeled as a linear combination of multivariate Gaussian PDFs, where $\mathbf{v}$ is a feature vector and  $\lambda_n$ is the mixture model corresponding to class $n$. 

\noindent The Gaussian mixture density of each feature vector $\mathbf{v}$ is modeled as a linear combination of multivariate Gaussian PDFs with the general form:
\begin{equation}
p(\mathbf{v}|\theta_i)=\frac{1}{(2\pi)^{\frac{d}{2}}|\mathbf{C_i}|^2}e^{[-\frac{1}{2}(\mathbf{v}-\mathbf{\mu}_i)^T\mathbf{C_i}^{-1}(\mathbf{v}-\mathbf{\mu}_i)]}
\end{equation}
where: 
\begin{itemize}
\item $\theta_i=(\mathbf{\mu_i},\mathbf{C_i}),$
\item $\mathbf{v}$ is the d-dimensional feature vector,
\item $\mathbf{\mu}$ is the d-dimensional vector, containing the mean values for each feature,
\item $\mathbf{C}$ is the $dxd$ covariance matrix and
\item $|\mathbf{C}|$ is the determinant
\end{itemize}
\noindent The complete set of parameters for a mixture model with $K$ components is 
\begin{equation}
    \Theta=\{ a_1,\cdots,a_K,\theta_1,\cdots,\theta_K \}
\end{equation}

\noindent Each GMM model $\lambda_n$ for class $n$ is parameterized as follows:
\begin{equation}
\label{eq:GMM_params}
\lambda_n=\{a_k^n,\mathbf{\mu}_k^n,\mathbf{C}_k^n \},k=1,\cdots,K
\end{equation}

\noindent An expectation maximization (EM) approach is utilized to derive the parameters $K_n$ , $\{a_i, \mathbf{\mu}_i, \mathbf{C}_i\}_n$ for the GMM $\lambda_n$ corresponding to class $n$ that best fit the input data.\\

\noindent At this point we analyze the expectation-maximization (EM) algorithm \cite{mclachlan2007algorithm} employed to compute the GMM parameters in eq.\eqref{eq:GMM_params}. 
In each iteration of the EM algorithm for Gaussian Mixtures, we deploy an E-step and an M-step:

\paragraph{E-step} 
We compute $w_{ik}$ for all feature vectors $\mathbf{v}_i$ and all mixture components $k$.
The membership weight of data point $\mathbf{v}$ in component $k$ given parameter $\Theta$ is defined as:

\begin{equation}
\label{eq:membership_weight}
w_{ik}=\frac{p_k(\mathbf{v}_i,\theta_k)\cdot a_k }{\sum_{m=1}^Kp_m(\mathbf{v}_i|\theta_m)\cdot a_m }
\end{equation}
for all components $k$ , $1 \leq k \leq K$ and all data samples $i$ , $1 \leq i \leq N$.

\paragraph{M-step} We calculate the new parameters. Given $N_k=\sum_{i=1}^Nw_{ik}$ the sum of membership weights for the $k-th$ component we get the mixture weights:
\begin{equation}
a_k^{new}=\frac{N_k}{N},1\leq k\leq K
\end{equation}
The updated mean:
\begin{equation}
\mathbf{\mu}_k^{new}=\frac{1}{N_k} \sum_{i=1}^N w_{ik} \cdot \mathbf{v}_i,1\leq k\leq K 
\end{equation}
and the updated covariance:
\begin{equation}
\mathbf{C}_k^{new}=\frac{1}{N_k} \sum_{i=1}^N w_{ik} (\mathbf{v}-\mathbf{\mu_i})^T(\mathbf{v}-\mathbf{\mu_i})
\end{equation}

\paragraph{Termination criteria}
The termination criteria for the EM is the following:
\begin{equation}
\log l(\Theta)_{t+1}-\log l(\Theta)_{t} \leq \epsilon
\end{equation}
where the log-likelihood, defined as $\log l(\Theta)=\sum_{i=1}^N\log p(\mathbf{v}_i|\Theta)$ and $\epsilon$ is a small user-defined scalar value.

\noindent In order to find the best fit for the data, we compute the Gaussian Mixture for $K=1$ to $K=32$ components iterating over full and diagonal covariance matrices for each class.
With the generation of each model, we estimate the Bayesian Information Criteria(BIC) \cite{schwarz1978estimating}. 
The model with the lowest BIC best fits the input data. For $n$ classes we get $n$ GMMs

\noindent After the optimal parameters for the GMMs have been computed and given $d$ the number of features, $K$ the number of components of the $i^{th}$ feature vector $\mathbf{v}_i$ , $\lambda_n$ the GMM of class $n$ we get:

\begin{equation}
P(\mathbf{v_i}|\lambda_n)=\sum_{i=1}^{K}a_i^n p_i^n(\mathbf{v})
\end{equation}
where  $a_i^n$ are the mixture weights to satisfy the constraint:
\begin{equation}
\sum_{i=1}^M a_i^n=1 , a_i^n>0
\end{equation}
Finally, and after the $P(\mathbf{v}| \lambda_n)$ for the test feature vector $\mathbf{v}$ and for each class $n$ is estimated, the test feature vector is assigned to the class with the greatest likelihood.


\begin{algorithm}
\caption{Supervised feature classification with Gaussian Mixture models via exhaustive search}
For each class of the data
\begin{enumerate}
    \item For $K=1 \cdots K_{max}$
    \begin{enumerate}
        \item For assumption that covariance matrix is either diagonal or full 
        \begin{enumerate}
        \item Compute Gaussian Mixture
        \item Compute BIC
        \end{enumerate}
    \end{enumerate}
    \item GMM with lowest BIC is assumed to best fit the data
\end{enumerate}
For test sample $\mathbf{v}$ and for each GMM model compute likelyhood. \\
The test feature vector is assigned to the class with the greatest likelihood.
\end{algorithm}

\subsection{Relevance Feedback}

\begin{figure}[t!]
  \begin{subfigure}[b]{0.35\linewidth}
    \includegraphics[width=\textwidth]{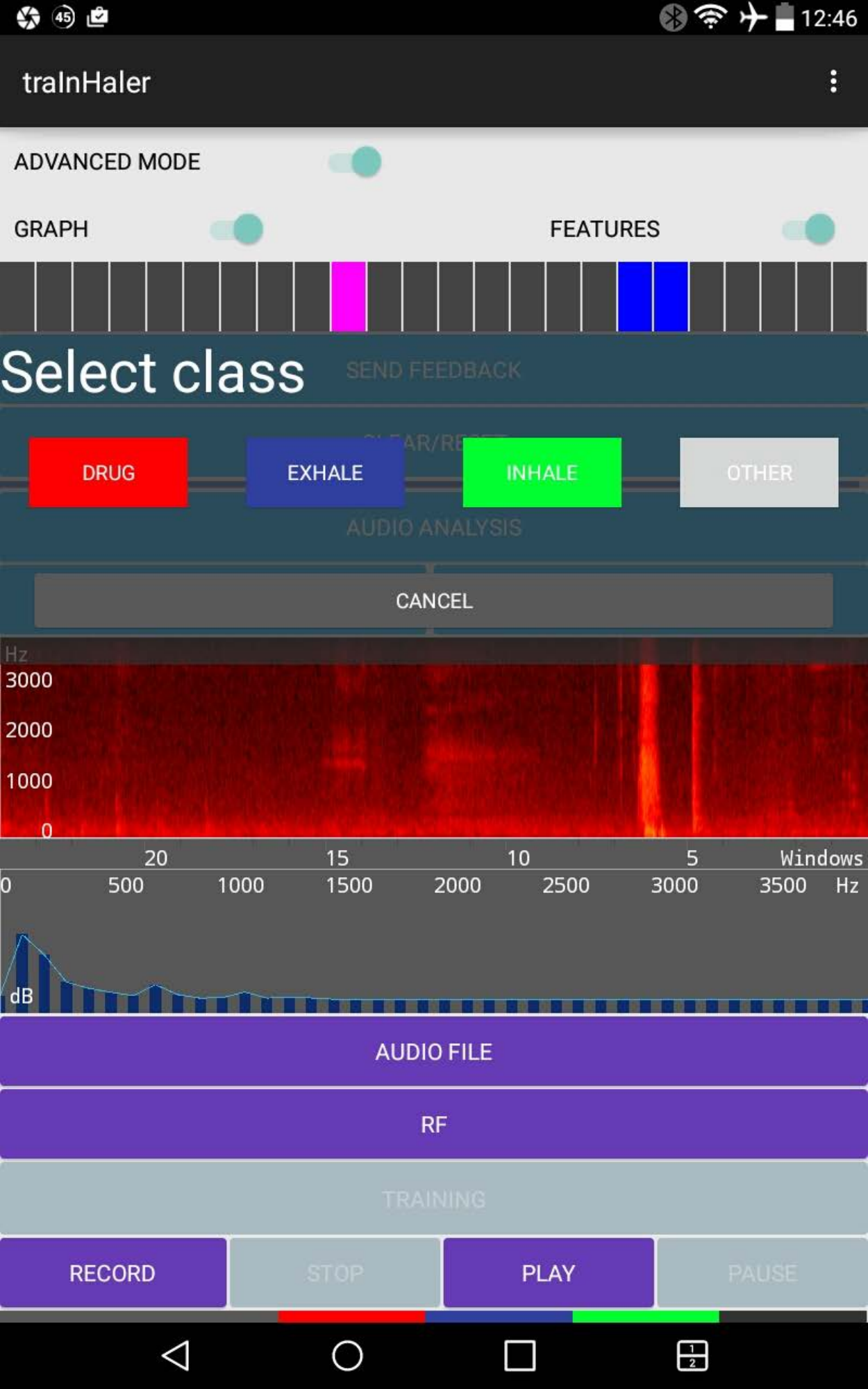}
    \caption{}
    \label{fig:rel-1}
  \end{subfigure}
  \hfill
  \begin{subfigure}[b]{0.35\linewidth}
    \includegraphics[width=\textwidth]{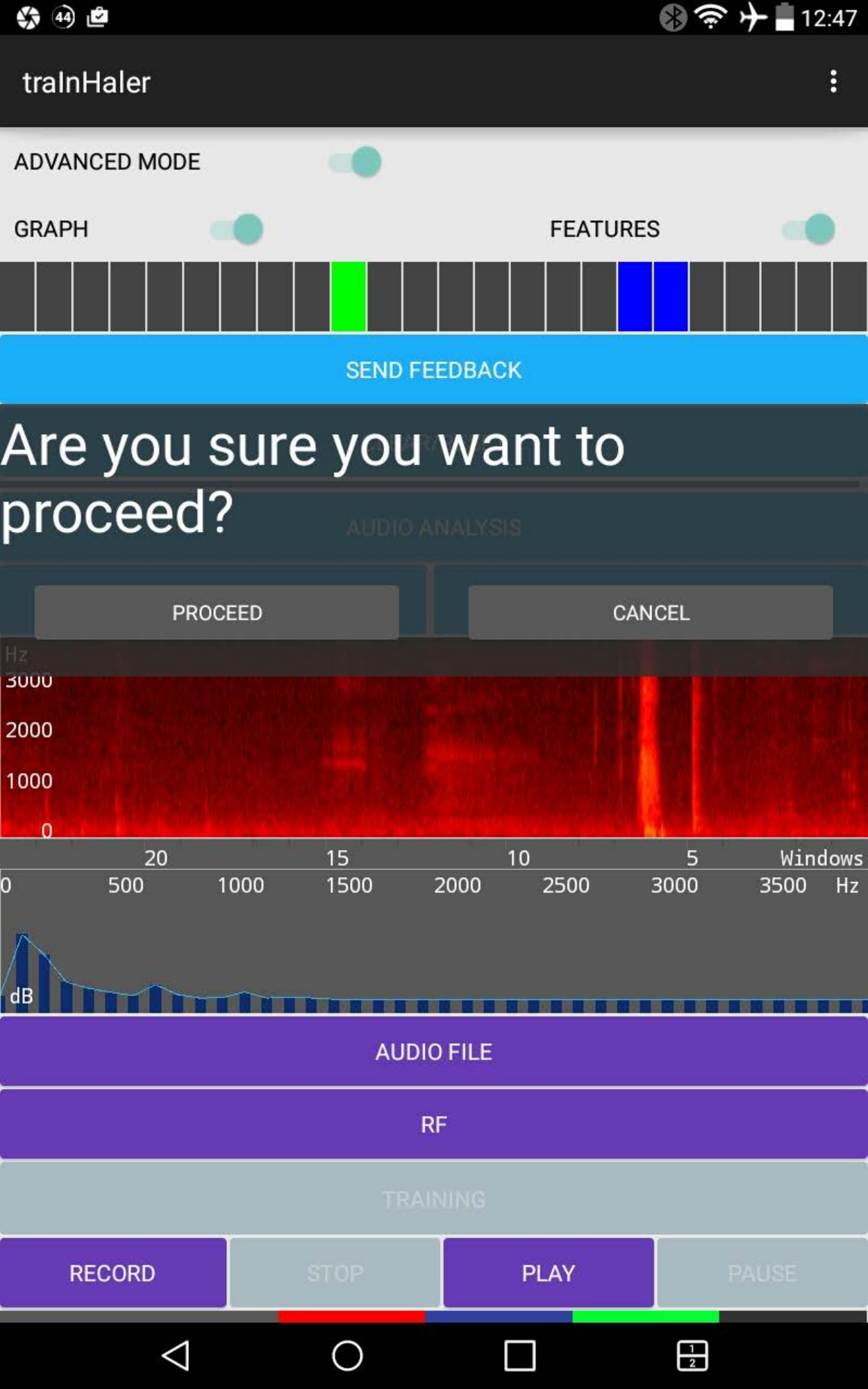}
    \caption{}
    \label{fig:app_relevance_feedback}
  \end{subfigure}
  \caption{a) Relevance feedback functionality selection menu. The user taps a classification result and activates the selection UI. b) After the result is corrected, it can be resubmitted.}
	\label{fig:user-interface-relevance-feedback}
\end{figure}

This section describes the proposed relevance feedback approach for the personalization of the trained models. The importance of the relevant feedback mechanism lies in the assumption that a small group of people compiled the initial dataset. This means that it may not contain the unique frequency patterns related to how different end-users exhale, inhale or activate the drug. Thus, a relevant feedback mechanism should allow the personalization of the trained models and the compilation of patient-specific datasets.
Initially, it is assumed that the patient has submitted a set of personal feature vectors annotated to the corresponding class using the relevance feedback functionality depicted in Figure \ref{fig:user-interface-relevance-feedback}. Each complete user submission includes $N=24$ feature vectors corresponding to 12-seconds of audio recording. The dataset used to train the models consists of $M=1980$ feature vectors, with 495 feature vectors per class.

Given the set of feature vectors $\mathcal{F}$ we perform kNN search with $k=1$  in the dataset $\mathcal{D}$ for each feature vector ${\mathbf{v}_{F}}_n$. The result is denoted as ${\mathcal{D}_{F}}_n$.
The new personalized dataset $\mathcal{D}_F$ is the union of $\mathcal{F}$ with each ${\mathcal{D}_{F}}_n$. At this point, it is important to remove all the duplicate vectors. Algorithm 1 presents a more detailed overview of the procedure.

\begin{algorithm}[t!]
\caption{Relevance feedback algorithm}
\begin{algorithmic}[1]
\Require  \\User defined entries $\mathcal{F}=\{ {\mathbf{v}_{F}}_1,\cdots,{\mathbf{v}_{F}}_n,\cdots,{\mathbf{v}_{F}}_N\}$
\\Dataset $\mathcal{D}=\{ {\mathbf{v}_D}_1,\cdots,{\mathbf{v}_D}_m,\cdots,{\mathbf{v}_D}_M \}$
\Initialize 
\State Personalized dataset $\mathcal{D}_F=\{ \}$ as an empty set

\ForEach {$\mathbf{v}_{F_n} \in \mathcal{F} $}
\State ${\mathcal{D}_{F}}_n \gets$ k nearest neighbors of ${\mathbf{v}_{F}}_n$ using $\mathcal{D}$
\State $\mathcal{D}_F = \mathcal{D}_F \cup {\mathcal{D}_{F}}_n$
\EndFor
\State $\mathcal{D}_F = \mathcal{D}_F \cup \mathcal{F}$
\Ensure Each element of $\mathcal{D}_F$ is unique.
\end{algorithmic}
\end{algorithm}

\subsection{Audio event classification and localization with long-short term memory recurrent neural networks}

\begin{figure}[t!]
    \centering
    \includegraphics[width=0.8\linewidth]{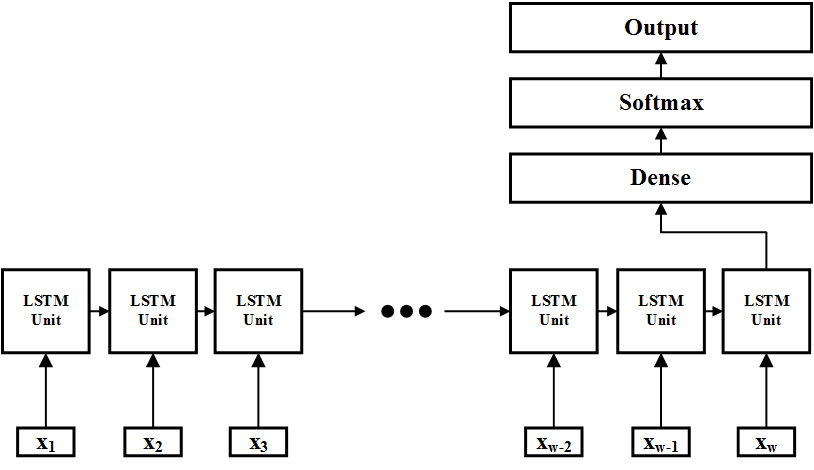}
    \caption{Deployed LSTM architecture}
    \label{fig:lstm_arch}
\end{figure}

This section describes the network architecture facilitating the audio event detection task.
The proposed sequential architecture is presented in Figure~\ref{fig:lstm_arch}. It consists, initially, of one layer of LSTM memory cells, with each one consisting of $h=64$ units. After the LSTM input layer, a dropout layer \cite{srivastava2014dropout} is introduced in order to reduce overfitted parts after training, with a dropout rate set to $0.3$, followed by a flatten layer and a dense output layer that returns a $4 \times 1$ vector. Finally, a softmax activation function is used. The model is optimized using binary cross-entropy loss \cite{liu2017learning} and the Adam optimizer \cite{kingma2014adam}.

\begin{figure}[t!]
    \centering
    \includegraphics[width=0.8\linewidth]{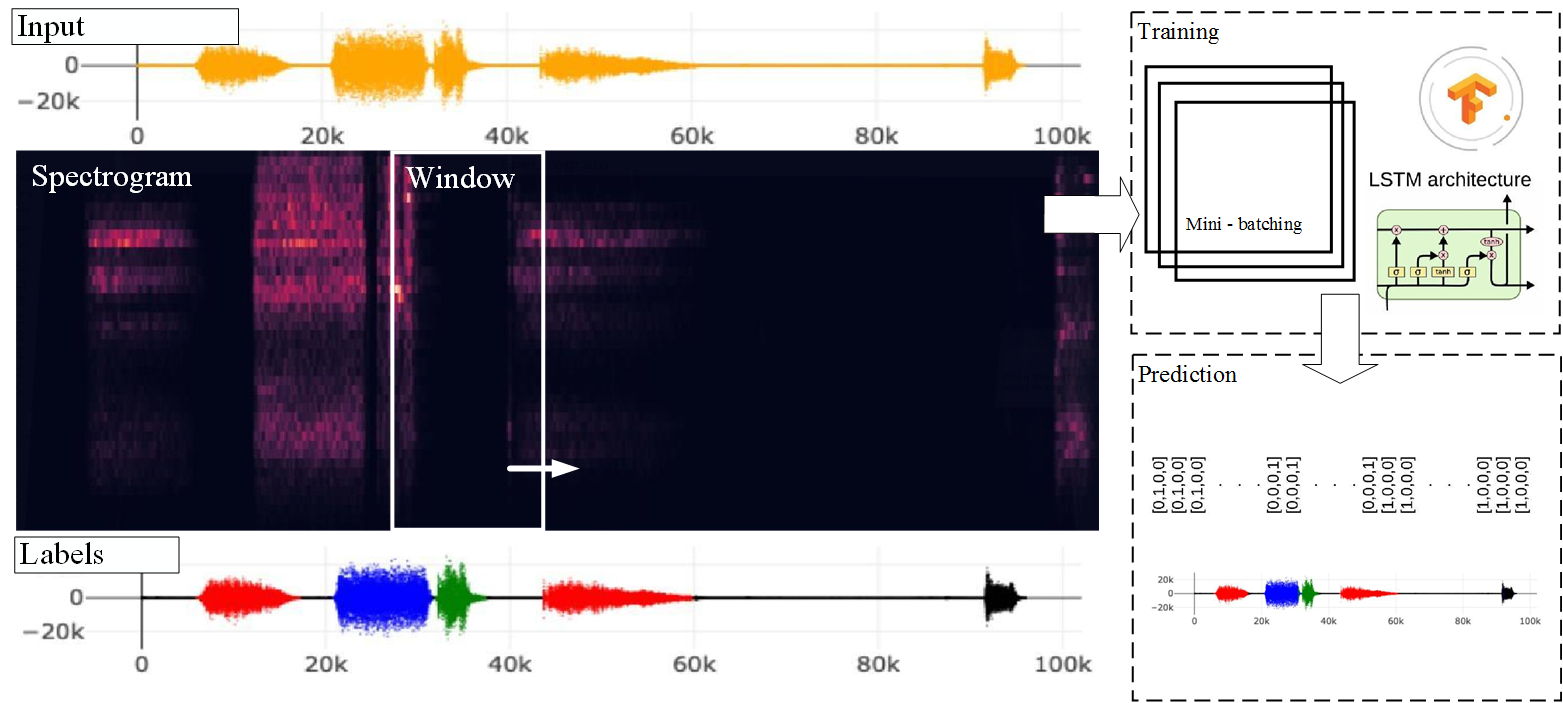}
    \caption{Deployed LSTM architecture}
    \label{fig:lstm_pipeline}
\end{figure}

The spectrogram was used as a tool to develop a classifier of inhaler sounds. It is swept across the temporal dimension, with a sliding window with length $w=15$, moving at a step size equal to $\mathbf{1}$ window. In order to form the training instances, we segment $\mathbf{S}$ into time windows of size $N_f\times T$ and assign a class to each one of them, according to the class of the central point of the window, as presented in Figure~\ref{fig:lstm_pipeline}. Each training example $\mathbf{W}_k$ is defined as:

\begin{equation}
  \mathbf{W}_k \in \mathcal{R}^{ N_F \times T}= (\mathbf{S}_{i j})_{\substack{1 \le i \le N_F\\ k-w \le j \le k}}
\end{equation}
The training set is organized in microbatches $\mathbf{B}$, so that
\begin{equation}
    \mathbf{B} \in \mathcal{R}^{b\times (T\times N_F)}
\end{equation}
with $b=25$ in our experiments. The input tensor can be defined as
\begin{equation}
   \mathbf{X} \in \mathcal{R}^{n \times w \times F} 
\end{equation}
where $n=25$ is the minibatch size, $w=25$ is the window size and $F=42$ is the dimension of spectrogram feature vector.

Extensive hyper-parameter optimization took place to define the number of hidden units, the number of dropout rates and the minibatch size. By observing the network's performance on the validation set, we stopped training at 70 epochs to avoid over-fitting. The testing loss increases, and the average testing accuracy stabilizes after around 70 epochs.

\subsection{Monitoring medication adherence through convolutional neural networks}
\label{section:monitoring-medication-cnn}
In previous sections, a series of features, namely STFT features, Cepstrum Features, and MFCC features, was employed with the state of the art classifiers, namely random forest (RF) ADABoost, support vector machines (SVM) and Gaussian mixture models (GMM). The classification accuracy reported is in the range of 97\% to 98\%. However, the feature extraction process is tenuous and computationally complex, especially in the case of embedded processors, thus limiting the applicability of monitoring medication adherence to offline processing or online complex distributed cloud-based architectures to handle the need for resources. 
However, the need remains for highly accurate and fast classification methods. Motivated by the aforementioned open issues, we present a fast, data-driven approach based on convolutional neural networks allowing us to differentiate the four audio events, namely, drug actuation, inhalation, exhalation and other sounds. The benefits can be summarized in the following points. i) The presented approach is applied directly to the time domain without requiring time-consuming and computationally expensive feature extraction approaches such as Spectrogram or Mel-frequencies.ii) An extensive execution time evaluation study demonstrates that the presented approach is faster than state-of-the-art methods. iii) Convolutional deep sparse coding speeds up the computational graph aiming to allow the real-time implementation. iv) Finally, the achieved classification accuracy is around 95\%, allowing for real-time implementations.

The CNN architecture consists of three convolutional layers with a max-pooling layer, a dropout function \cite{wu2015max} and four fully connected layers. For the convolutional kernels, the stride is set equal to one, with zero padding to keep the shape of the output of each filter constant and equal to its input's dimensionality. Table \ref{table:audio-cnn-architecture} presents the stacked layers for each model, the values of dropout layers, the number of filters in each convolutional layer, and the number of neurons in fully connected layers and the activation function.

\begin{table}[t!]
\caption{CNN Parameters}
\label{table:audio-cnn-architecture}
\centering
\begin{tabular}{|p{5cm}|p{5cm}|p{3cm}|}
\hline
Layers                               & Layer Parameters    & Model        \\ \hline
\multirow{3}{*}{Convolutional Layer} & Filters             & 16           \\ \cline{2-3} 
                                     & Kernel Size         & $4 \times 4$ \\ \cline{2-3} 
                                     & Activation Function & ReLu         \\ \hline
Max Pooling                          & Kernel Size         & $2 \times 2$ \\ \hline
Dropout                              &                     & 0.2          \\ \hline
\multirow{3}{*}{Convolutional Layer} & Filters             & 16           \\ \cline{2-3} 
                                     & Kernel Size         & $5 \times 5$ \\ \cline{2-3} 
                                     & Activation Function & ReLu         \\ \hline
Max Pooling                          & Kernel Size         & $2 \times 2$ \\ \hline
Dropout                              &                     & 0.1          \\ \hline
\multirow{3}{*}{Convolutional Layer} & Filters             & 16           \\ \cline{2-3} 
                                     & Kernel Size         & $6 \times 6$ \\ \cline{2-3} 
                                     & Activation Function & ReLu         \\ \hline
Max Pooling                          & Kernel Size         & $2 \times 2$ \\ \hline
\multirow{2}{*}{Dense}               & Neurons             & 64           \\ \cline{2-3} 
                                     & Activation Function & ReLu         \\ \hline
\multirow{2}{*}{Dense}               & Neurons             & 128          \\ \cline{2-3} 
                                     & Activation Function & ReLu         \\ \hline
\multirow{2}{*}{Dense}               & Neurons             & 64           \\ \cline{2-3} 
                                     & Activation Function & ReLu         \\ \hline
\multirow{2}{*}{Dense}               & Neurons             & 4            \\ \cline{2-3} 
                                     & Activation Function & ReLu       \\
                                     \hline
\end{tabular}
\end{table}

The audio files used for training and testing were loaded through appropriate libraries in a vector of $4000 \times 1$ dimension. Then, reshaping was performed to employ two-dimensional convolutions. In particular, the first 16 samples are placed in the first row of the matrix, the subsequent 16 samples in the second, until a $250 \times 16$ matrix is constructed. An example of the reshaping procedure is given in Figure \ref{fig:reshaping}, while Figure \ref{fig:reshaping_4_class} visualizes examples of sounds per class after this reshaping procedure.

\begin{figure}[t!]
  \centering
  \includegraphics[width=\linewidth]{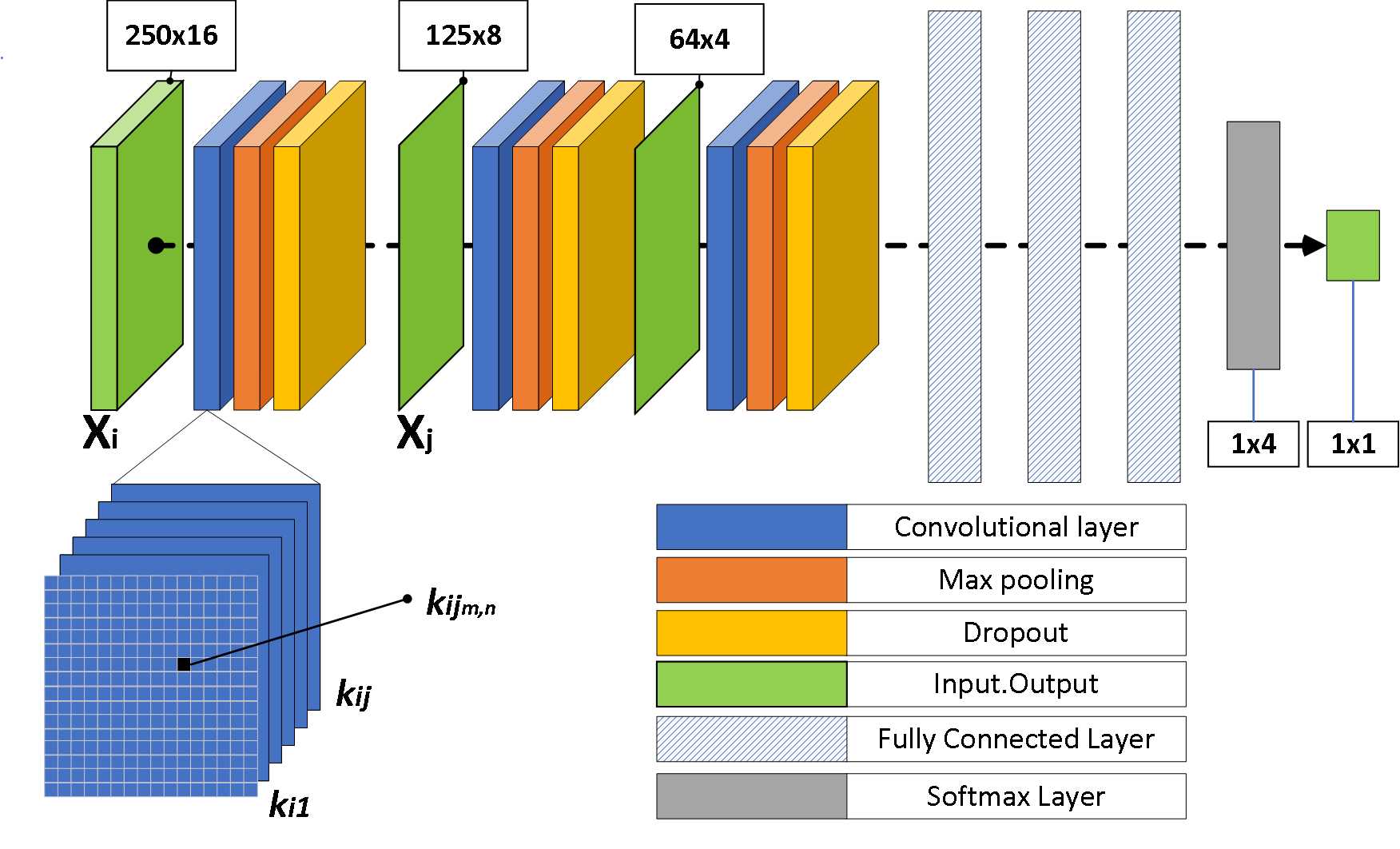}
  \caption{CNN Architecture. The CNN architecture consists of three convolutional layers with a max-pooling layer, a dropout function \cite{wu2015max} and four fully connected layers. For the convolutional kernels, the stride is set equal to one, with zero padding to keep the shape of the output of each filter constant and equal to its input's dimensionality.}
  \label{fig:CNN-architecture}
\end{figure}

\begin{figure}[t!]
    \centering
\includegraphics[width=\linewidth]{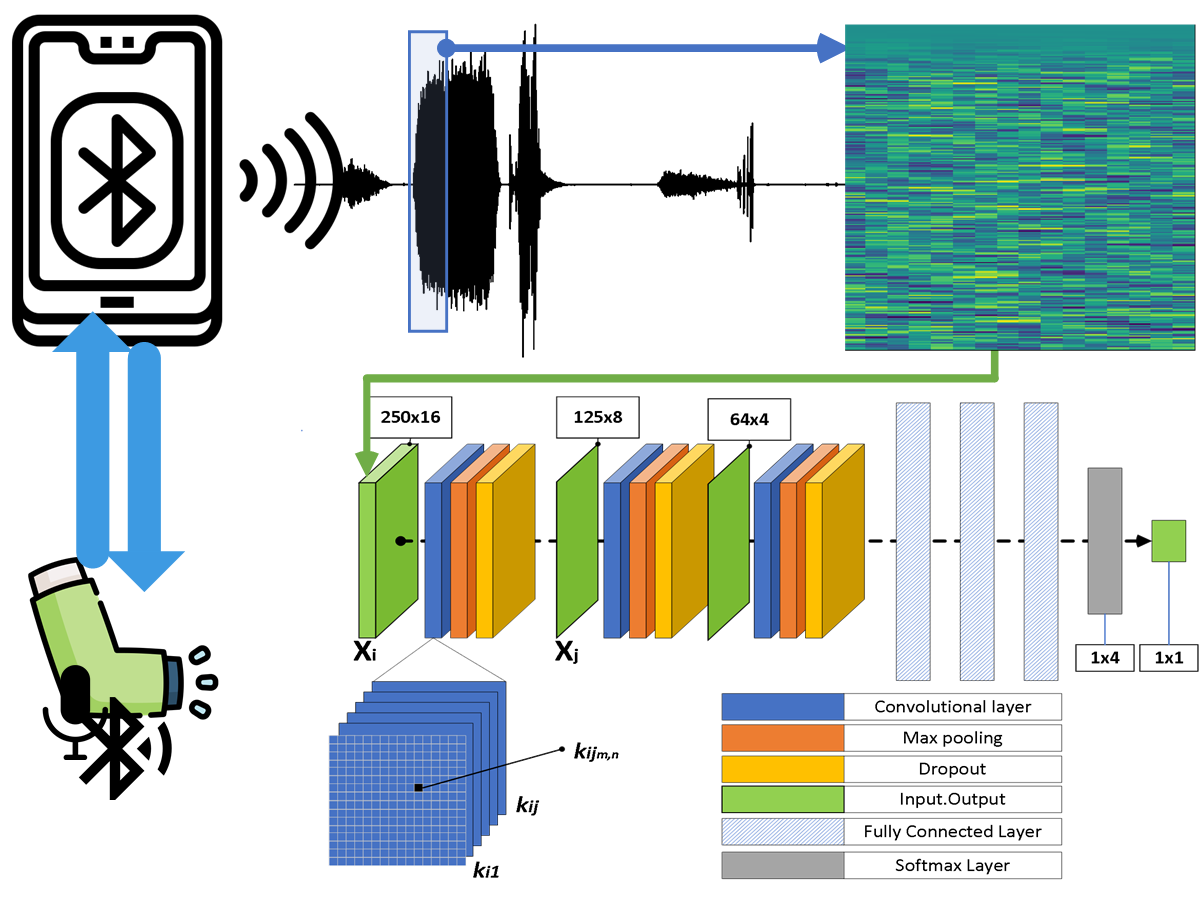}
         \caption{Overview of the processing pipeline}
         \label{fig:cnnpipeline} 
\end{figure}

\begin{figure}[t!]
    \centering
     \includegraphics[width=0.6\linewidth]{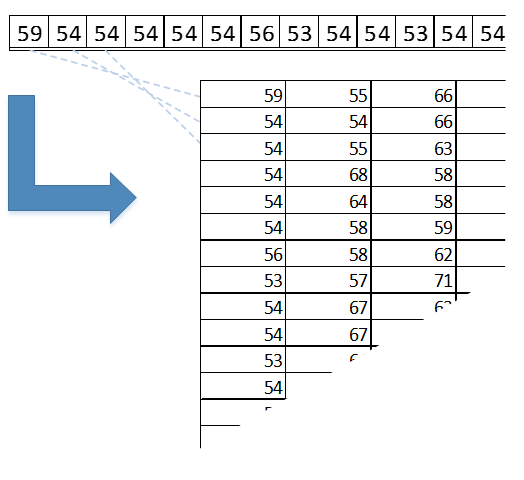}
          \caption{Illustration of reshaping of a vector into a two-dimensional matrix.}
          \label{fig:reshaping}
\end{figure}

\begin{figure}[t!]
\centering
\begin{subfigure}{0.48\textwidth}
         \includegraphics[width=\linewidth]{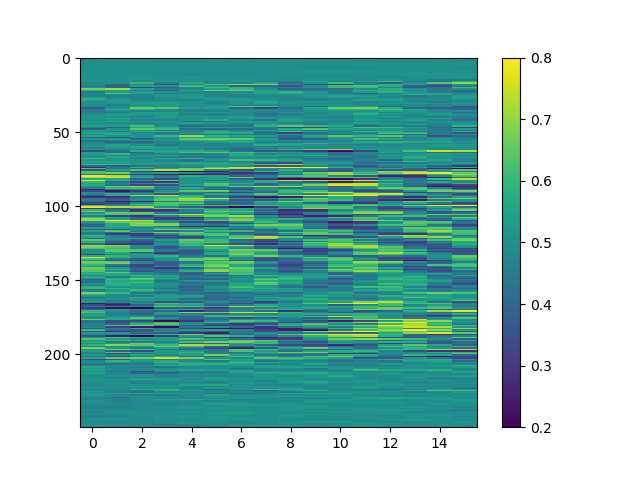}
         \caption{Sound of drug class after reshaping}
         \label{fig:Drug_class}
    \end{subfigure}
    \begin{subfigure}{0.48\textwidth}
         \includegraphics[width=\linewidth]{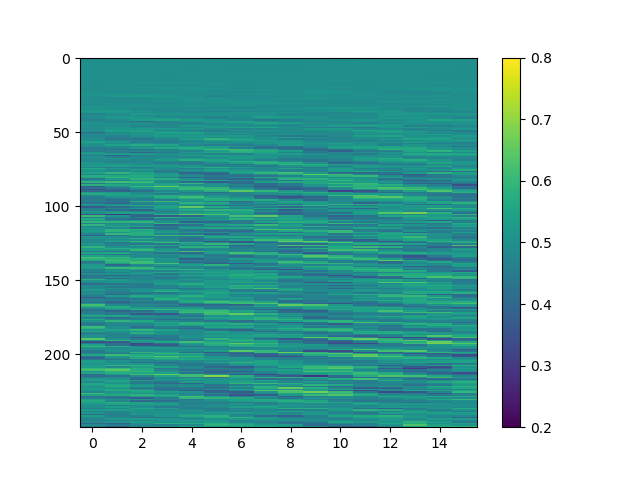}
         \caption{Sound of exhale class after reshaping}
         \label{fig:Exhale_class}
    \end{subfigure}\\
    \begin{subfigure}{0.48\textwidth}
         \includegraphics[width=\linewidth]{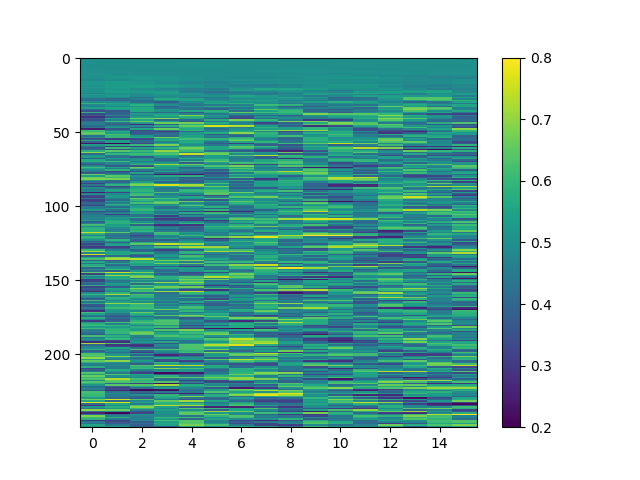}
         \caption{Sound of inhale class after reshaping}
         \label{fig:Inhale_class}
    \end{subfigure}
    \begin{subfigure}{0.48\textwidth}
         \includegraphics[width=\linewidth]{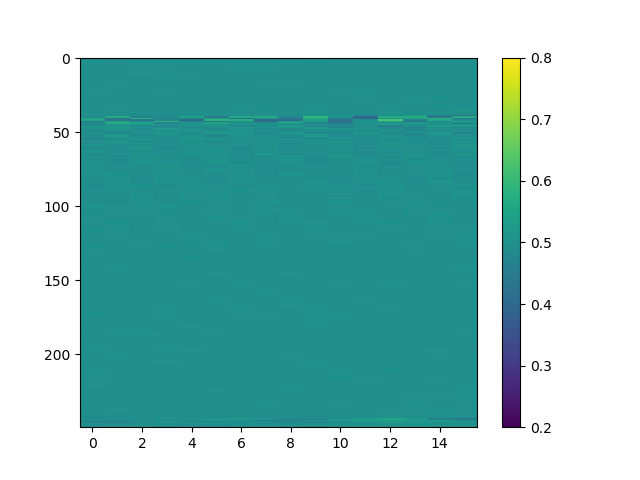}
         \caption{Sound of noise class after reshaping}
         \label{fig:Noise_class}
    \end{subfigure}
    
    \caption{Visualization of the segmented audio files for each respiratory phase after the reshaping procedure.}
    \label{fig:reshaping_4_class}
\end{figure}


\section{Experimental evaluation}

This section elaborates on the experimental evaluation of the approaches mentioned above. Two evaluation studies take place, one for each dataset. The motivation behind using two datasets for the evaluations aims to adequately provide answers on how trustworthy is the differentiation of inhaler audio events in actual conditions based on a comparison of traditional ML algorithms and deep neural networks.

\subsection{Validation settings} 

During validation, the main differentiation parameter comes from the availability of previous recordings from a specific individual. It is expected that preliminary information can increase classification accuracy. However, it puts an additional burden on the usage of the monitoring system since it requires the collection of data every time a new patient wants to test the framework. 

Firstly, we consider the \textit{Multi Subject} modeling approach, denoted as \textit{MultiSubj}. In this case, the recordings of all subjects are used to form a large dataset, divided into five equal parts used to perform five-fold cross-validation, thereby allowing different samples from the same subject to be used in the training and test set, respectively. This validation scheme was followed in previous work \cite{ntalianis2019assessment} and thus performed, also here for comparison purposes.\par

The second case includes the \textit{Single Subject} setting, in which the performance of the classifier is validated through training and testing within each subject's recordings. Let's denote such models as \textit{SingleSubj}. Specifically, the recordings of each subject are split into five equal parts to perform cross-validation. The accuracy is assessed for each subject separately, and then the classifier's overall performance is calculated by averaging the three individual results.\par

The third evaluation setting refers to the case when no previous recordings for the testing subject are available. Thus samples from other subjects should be used. This is the \textit{leave-one-subject-out (LOSO)} approach that illustrates how well the trained network can generalize to individuals that it never saw before during training. \textit{LOSO} models facilitate the use of the monitoring system since they do not require a data pre-collection phase, and they also have the lowest risk of over-fitting. However, if the inter-subject variability is high, they might not adapt well, especially if the number of training subjects is small, as in our case. With this approach, we use the recordings of two subjects for training and the recordings of the third subject for testing. This procedure is completed when all subjects have been used for testing, and the accuracy is averaged to obtain the overall performance of the classifiers.

\subsection{Dataset A}
This section presents the adherence monitoring accuracy for Dataset A compared with respect to well-established classification algorithms, namely GMMs, SVMs \cite{burges1998tutorial}, Random Forests \cite{breiman2001random}, ADABoost \cite{zhu2009multi} for Spectrogram, Cepstrogram and MFCC features. An additional comparison entails Fisher Kernel Representations\cite{jaakkola1999exploiting,maaten2011learning} of the aforementioned features. In order to compare the proposed method with the FKL approach, we utilized the publicly available implementation of \cite{maaten2011learning}. FKL receives time series as input and generates as output new vectors. Thus, we utilize two approaches: In the first approach, we use the extracted features of MFCC, Spectrogram and Cepstrogram as input time series. This way, the FKL is used as preprocessing step. The results of this approach are presented in the blocks named MFCC FKL, SPECT FKL, and CEPST FKL. In the second approach, we use the output of the GMM probability function as input time series. The results are presented in the blocks named MFCC GMM FKL, SPECT GMM FKL, and CEPST GMM FKL blocks of Table \ref{table:classification_accuracy}. Furthermore, to provide a comparison with Continuous Wavelet Transform (CWT) based approaches, we utilized the CWT with Morlet wavelet as a feature extraction method\cite{taylor2014acoustic}.

\subsubsection{Feature separability} 
\begin{figure}[p]
  \begin{subfigure}[b]{0.42\linewidth}
    \includegraphics[width=\textwidth]{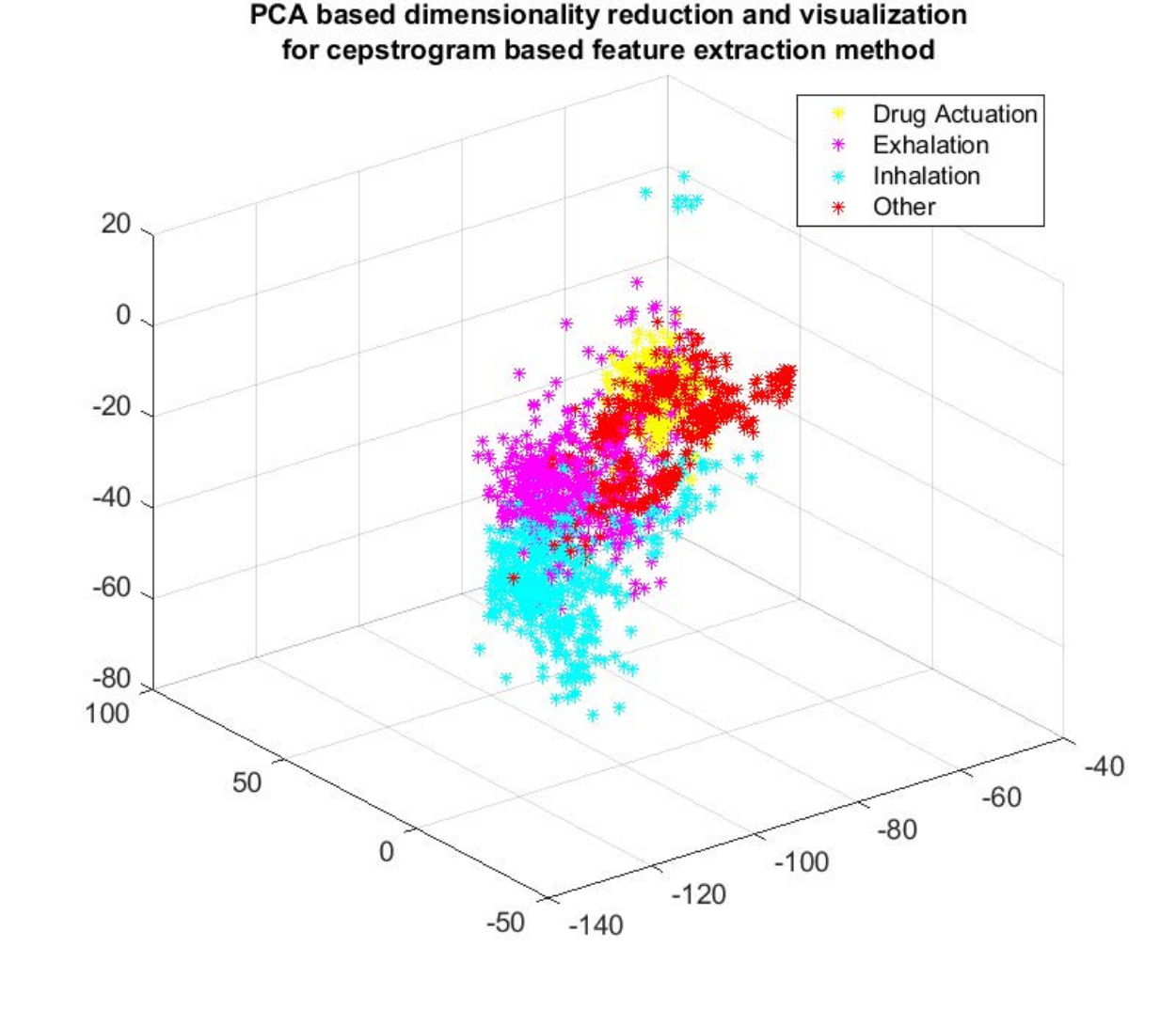}
    \caption{PCA based}
    \label{fig:dimensionality-3}
  \end{subfigure}
  \hfill
  \begin{subfigure}[b]{0.42\linewidth}
    \includegraphics[width=\textwidth]{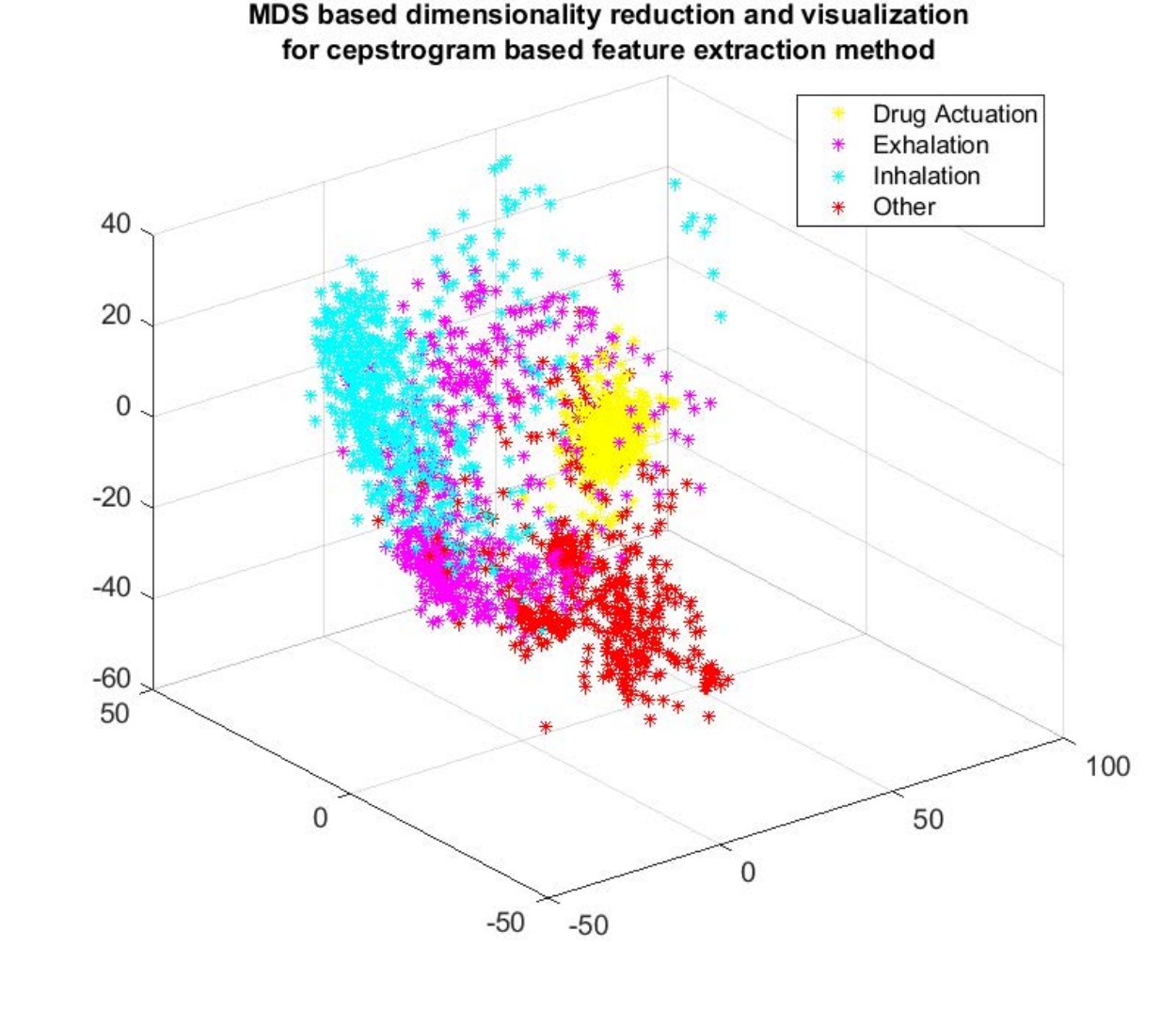}
    \caption{MDS based}
    \label{fig:dimensionality-4}
  \end{subfigure}
    \begin{subfigure}[b]{0.42\linewidth}
    \includegraphics[width=\textwidth]{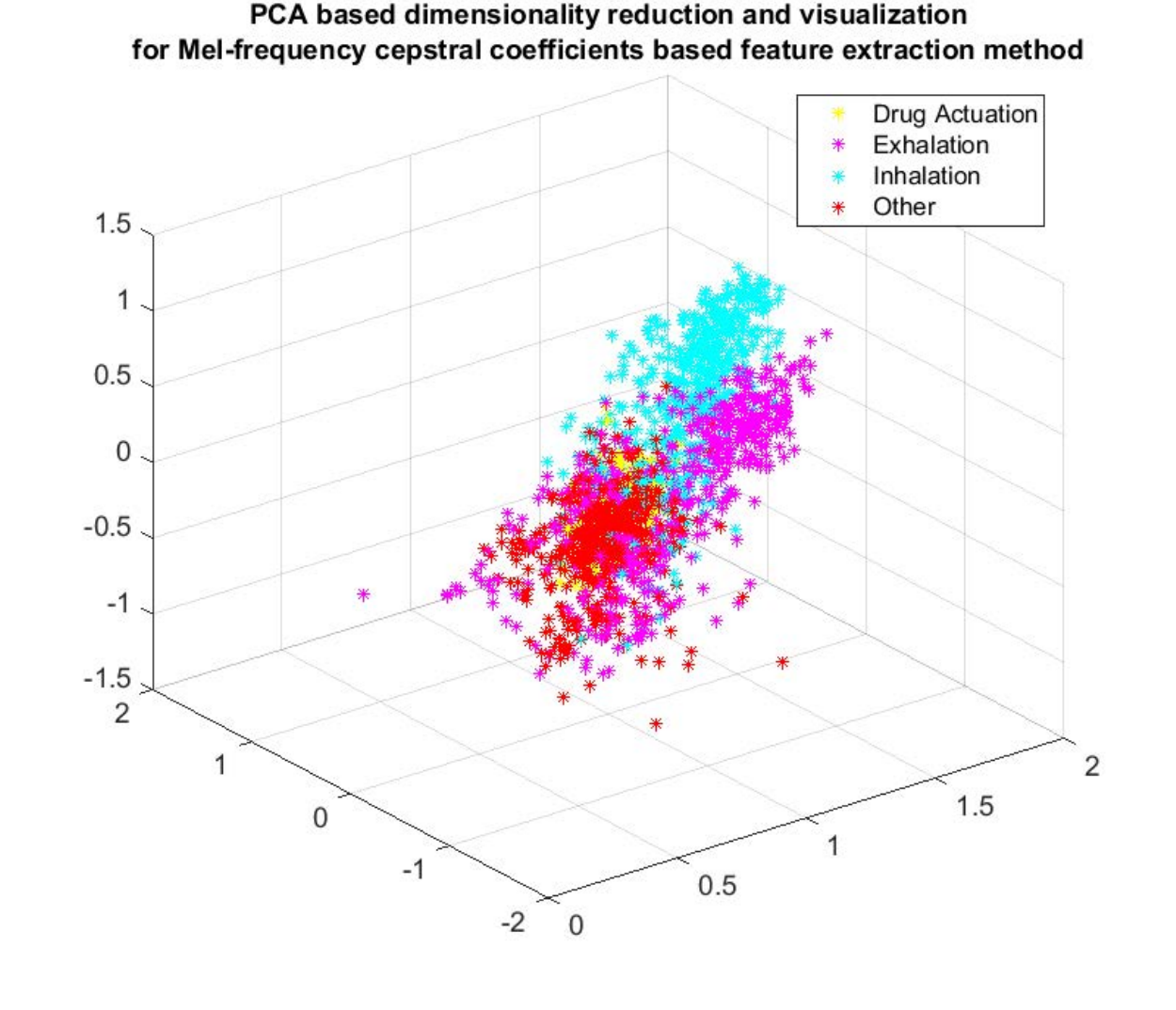}
     \caption{PCA based}
    \label{fig:dimensionality-5}
  \end{subfigure}
  \hfill
  \begin{subfigure}[b]{0.48\linewidth}
    \includegraphics[width=\textwidth]{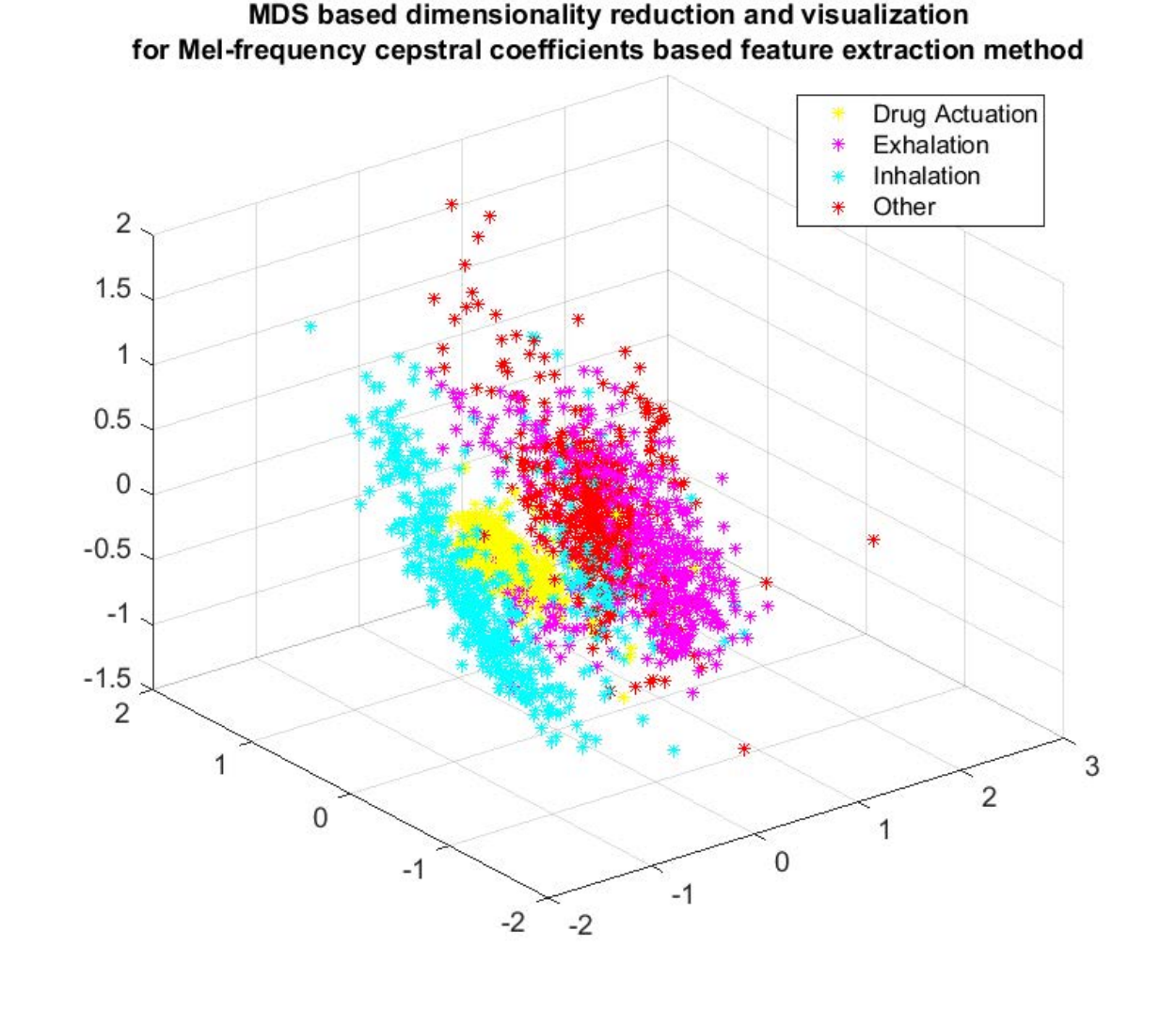}
    \caption{MDS based}
    \label{fig:dimensionality-6}
  \end{subfigure}
    \begin{subfigure}[b]{0.48\linewidth}
    \includegraphics[width=\textwidth]{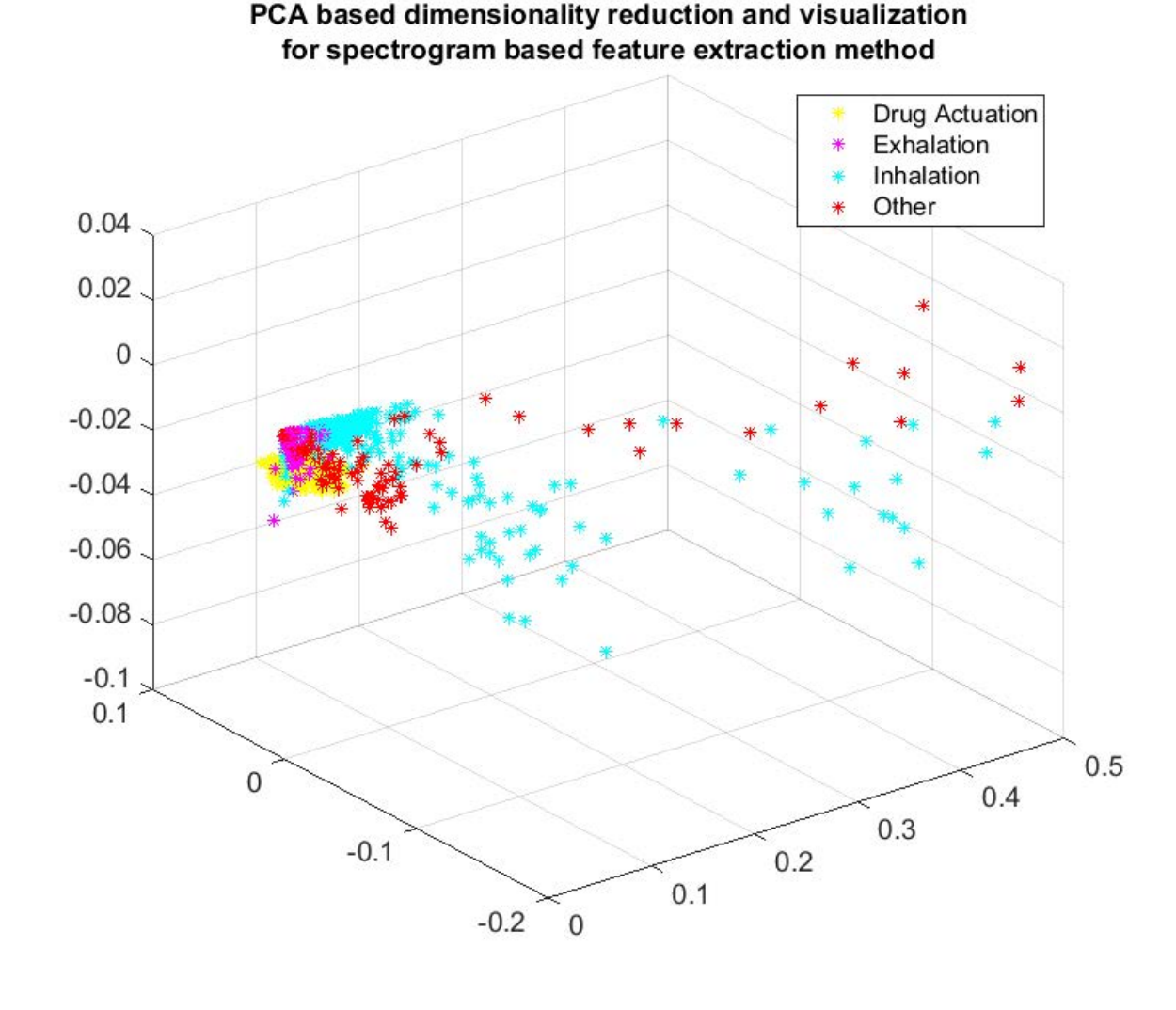}
    \caption{PCA based}
    \label{fig:dimentsonality-1}
  \end{subfigure}
  \hfill
  \begin{subfigure}[b]{0.48\linewidth}
    \includegraphics[width=\textwidth]{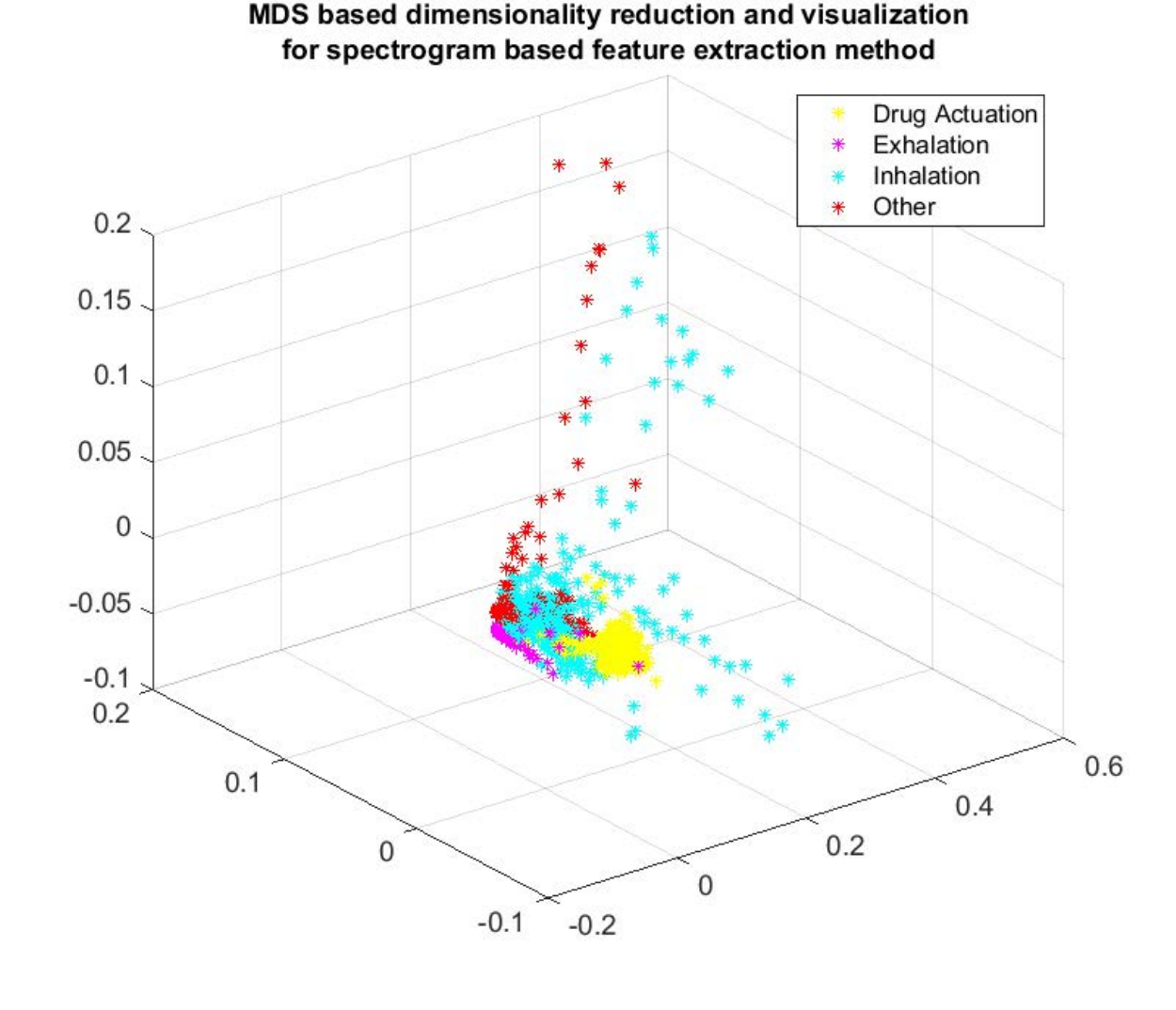}
   \caption{MDS based}
    \label{fig:dimentsonality-2}
  \end{subfigure}

 \caption{Dimensionality reduction and visualization for a,b) Cepstrogram based feature extraction, c,d)Mel-frequency cepstral coefficients based feature extraction and e,f) for spectrogram based feature extraction. The colours correspond to yellow for drug actuation, magenta for exhalations, cyan for inhalations and red for other types of noise.}
	\label{fig:dimensionality}
\end{figure}

In order to examine the feature separability for the approaches mentioned above, we visualize the feature space by employing two different methods: a) principal component analysis (PCA) method and b) multi-dimensional scaling (MDS) \cite{Cox2001}. Figure \ref{fig:dimensionality} depicts the visualization of the feature vectors in the three-dimensional feature space for each of the dimensionality reduction methods. As it can be observed, the Cepstrogram based features demonstrate higher separability than the other approaches, which is later verified by the classification accuracy results.
Given spectrogram  $S(m,\omega)$, cepstrogram $C(m,k)$ and MFCC $c(n)$ we employ a post-processing step to generate a one-dimensional feature vector

\subsubsection{Results}

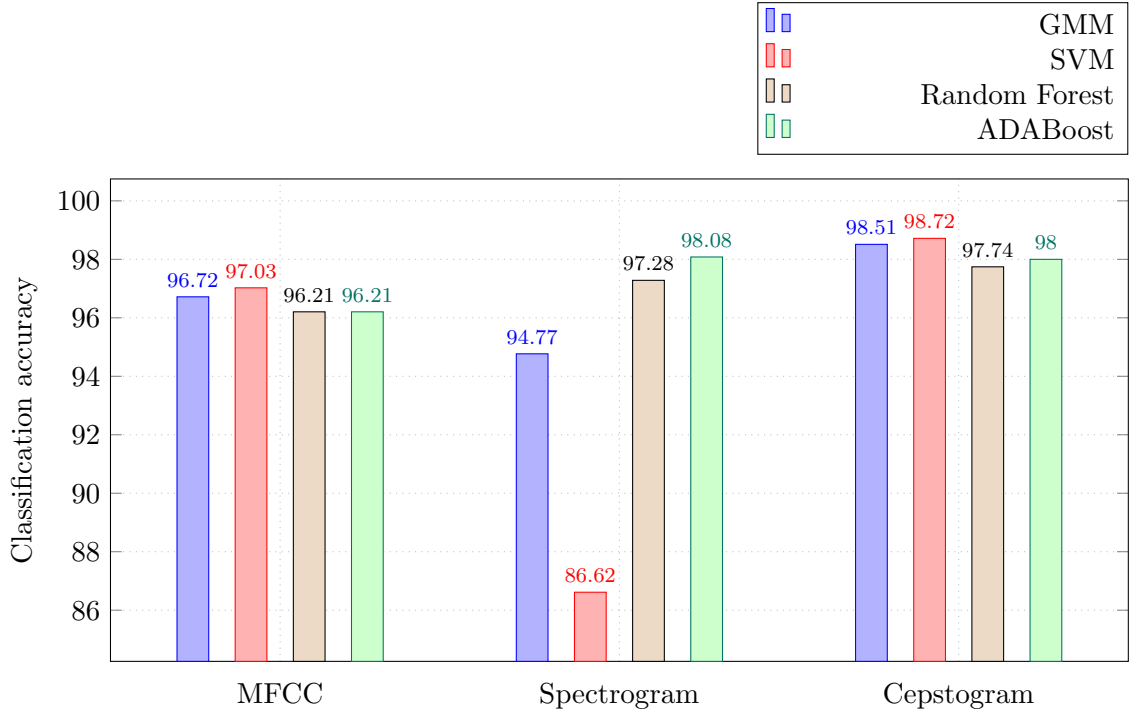
\begin{figure}[H]
\centering
\begin{tikzpicture}
    \begin{axis}[
        width  = \linewidth,
        height = 8cm,
        scaled y ticks=true,
        ymin=85,
		ymax=100,
        major x tick style = transparent,
        ybar=10pt,
        bar width=12pt,
        ymajorgrids = true,
        ylabel = {Classification accuracy},
        enlargelimits=0.05,
        nodes near coords,
        every node near coord/.append style={font=\scriptsize},
  		 nodes near coords align={vertical},
  	    symbolic x coords={MFCC,Spectrogram,Cepstogram},
        xtick = data,
        x tick label style={font=\normalsize,text width=3in,align=center},
        scaled y ticks = true,
        enlarge x limits=0.25,
        legend cell align=right,
        legend style={
                at={(1,1.05)},
                anchor=south east,
                column sep=10ex
        }
    ]
        
         \addplot[style={blue,fill=LightSteelBlue,mark=none}]
             coordinates {(MFCC,96.718) (Spectrogram,94.769) (Cepstogram,98.513)};
        
        \addplot[style={red,fill=LightPink,mark=none}]
            coordinates {(MFCC, 97.026) (Spectrogram,86.615) (Cepstogram,98.718)};

        \addplot[style={black,fill=Wheat,mark=none}]
             coordinates {(MFCC,96.205) (Spectrogram,97.282) (Cepstogram,97.744)};

        \addplot[style={JungleGreen,fill=LightGreen,mark=none}]
             coordinates {(MFCC,96.205) (Spectrogram,98.08081) (Cepstogram,98)};
            
        \legend{GMM,SVM,Random Forest,ADABoost}
    \end{axis}
		
\end{tikzpicture}
\caption{Classification accuracy for the 4-class problem. Three feature extraction algorithms are compared: i) MFCC ii) Spectrogram iii) Cepstogram}
\label{fig:classification_accuracy}
\end{figure}

The classification results are presented in Table \ref{table:classification_accuracy} and in Figure \ref{fig:classification_accuracy}. Table \ref{table:confusion-matrix-cwt} presents the confusion matrices for SVM, Random Forest and AdaBoost classifiers for the 4-class problem and for the Drug vs other sounds 2-class problem. Our results agree with the results provided in \cite{taylor2014acoustic} yielding a 99.18\% sensitivity, 99.73\% specificity and 99.45\% accuracy in the identification of drug actuation sounds.

As observed, the classification accuracy reaches 97\% for all feature extraction methods.GMMs yield the best results reaching 98\% in the case of Cepstogram features, while SVM seems to be a good classifier for MFCC and Cepstogram but not for Spectrogram features. Considering the MFCC based feature extraction, GMM reaches 96\%, SVM reaches 97\%, Random Forest reaches 96\%, and ADABoost 96\%. For the spectrogram-based feature extraction method, the classification accuracy is 94.7\% for GMM, 86\% for SVM, 97\% for Random Forest, and 98\% for ADABoost. Finally, for the Cepstrogram, the classification accuracy reaches 98\% in the case of the GMM classifier, 98\% in the case of the SVM classifier, 97\% for the Random Forest approach 97\% for ADABoost. The utilization of the FKL preprocessing step does not provide better results than the corresponding features used as input time series. e.g. CEPST FKL with RF has yielded worse results than CEPST RF.

\begin{table}[H]
\caption{Classification accuracy (\%) for the 4-class problem}
\label{table:classification_accuracy}
\centering
\begin{tabular}{|l|l|l|l|}
\hline
& \textbf{MFCC}         
& \textbf{Spectrogram}   
& \textbf{Cepstogram}    
\\ 
\hline
\textbf{SVM} 
& 97.026               
& 86.615                
& \textbf{98.718}       
\\ 
\hline
\textbf{RF}  
& 96.205               
& 97.282                
& \textbf{97.744}                
\\ 
\hline
\textbf{ADA} 
& 96.205               
& \textbf{98}                   
& \textbf{98}                   
\\ \hline
\textbf{GMM} 
& 96.718               
& 94.769                
& \textbf{98.513}       
\\ 
\hline
& \textbf{MFCC FKL}     
& \textbf{SPECT FKL}     
& \textbf{CEPST FKL}     
\\ 
\hline
\textbf{SVM} 
& 93.128               
& 86.308                
& \textbf{95.744}                
\\ 
\hline
\textbf{RF}  
& 92.41                
& \textbf{97.179}                
& 94.615                
\\ 
\hline
\textbf{ADA} 
& 92.821               
& \textbf{97.333}                
& 96                   
\\ 
\hline
& \textbf{MFCC GMM FKL} 
& \textbf{SPECT GMM FKL} 
& \textbf{CEPST GMM FKL} 
\\ 
\hline
\textbf{SVM} 
& 96.718               
& 94.821                
& \textbf{98.513}       
\\ 
\hline
\textbf{RF}  
& 96.308               
& 95.026                
& \textbf{98.513}       
\\ 
\hline
\textbf{ADA} 
& 95.487               
& 95.333                
& \textbf{98.205}       
\\ 
\hline
\end{tabular}
\end{table}

\subsubsection{Noise robustness assessment}
To assess the robustness of noise and other sounds, assuming that the initial dataset was created under ideal conditions, we compiled noisy datasets by adding background and environmental sounds \cite{kikidis2015utilizing}, collected from freesound.org \cite{fonseca2017freesound}, by superposing dataset audio segments $\mathbf{x}$ and noise $\mathbf{n}$ in the following manner:
\begin{equation}
\mathbf{x}'=\mathbf{x}+k*\mathbf{n}
\end{equation}
The classification accuracy for different values of factor $k$ is shown in table \ref{table:classification_accuracy_noise}. 
As it is made obvious, the classification accuracy of each method drops below 85\% as added noise factor reaches $0.5$. Table \ref{table:classification_accuracy_noise} demonstrates that in noisy conditions, Cepstrogram based GMM approach yields the best results.

\begin{table}[H]
\caption{Classification accuracy (\%) vs added noise factor}
\label{table:classification_accuracy_noise}
\resizebox{\textwidth}{!}{
\begin{tabular}{ll|l|l|l|l|}
\cline{3-6}
                                                  &              & \multicolumn{4}{l|}{\textbf{Added noise and environmental sounds factor}}  \\ \cline{3-6} 
                                                      &              
 & \textbf{0.00}  
 & \textbf{0.10}  
 & \textbf{0.20}   
 & \textbf{0.50}   \\ \hline
\multicolumn{1}{|l|}{\multirow{4}{*}{\textbf{MFCC}}}  & \textbf{GMM} 
 & 96.718          
 & 93.13          
 & 87.778          
 & 79.899          \\ \cline{2-6} 
\multicolumn{1}{|l|}{}                                
& \textbf{SVM} 
& 97.026       
& 93.59          
& 87.0202         
& 79.4444         \\ \cline{2-6} 
\multicolumn{1}{|l|}{}                                & \textbf{RF}  
& 96.205        
& 93.23          
& 86.2626         
& 78.333          \\ \cline{2-6} 
\multicolumn{1}{|l|}{}                                & \textbf{ADA} 
& 96.205        
& 92.93          
& 85.9091         
& 72              \\ \hline
\multicolumn{1}{|l|}{\multirow{4}{*}{\textbf{SPECT}}} & \textbf{GMM} 
& 94.768           
& 92.83          
& 88.384          
& 78.485          \\ \cline{2-6} 
\multicolumn{1}{|l|}{}                               
& \textbf{SVM} 
& 86.615       
& 85.35          
& 83.2323         
& 78.68687        \\ \cline{2-6} 
\multicolumn{1}{|l|}{}                                & \textbf{RF}  
& 97.282       
& 95.66          
& 91.8182         
& 83.28283        \\ \cline{2-6} 
\multicolumn{1}{|l|}{}                                & \textbf{ADA} 
& 98       
& 95.15          
& 92.0202         
& 81.71717        \\ \hline
\multicolumn{1}{|l|}{\multirow{4}{*}{\textbf{CEPST}}} & \textbf{GMM} 
& 98.513 & \textbf{96.47} 
& \textbf{96.414} 
& 82.879 \\ \cline{2-6} 
\multicolumn{1}{|l|}{}                                & \textbf{SVM} 
& \textbf{98.718}      
& 95.81          
& 91.9192         
& \textbf{83.93939}        \\ \cline{2-6} 
\multicolumn{1}{|l|}{}                                & \textbf{RF}  
& 97.744       
& 95.91          
& 92.7778         
& 83.8889         \\ \cline{2-6} 
\multicolumn{1}{|l|}{}                                & \textbf{ADA} 
& 98        
& 96.11          
& 91.6667         
& 81.9697         \\ \hline
\multicolumn{1}{|l|}{\textbf{MFCC GMM FKL}}           & \textbf{ADA} 
& 96.718         
& 91.364         
& 87.172          
& 78.131          \\ \hline
\multicolumn{1}{|l|}{\textbf{SPECT GMM FKL}}          & \textbf{ADA} 
& 94.821         
& 93.384         
& 87.727          
& 77.121          \\ \hline
\multicolumn{1}{|l|}{\textbf{CEPST GMM FKL}}          & \textbf{ADA} 
& 98.513         
& 94.242         
& 94.444          
& 81.162          \\ \hline
\end{tabular}}
\end{table}

\subsubsection{Confusion matrices}
Table \ref{table:confusion-matrix} presents the confusion matrices of each feature extraction method for all classification algorithms performed in the current study. As observed, the classification accuracy reaches 99\% in some cases. However, it is as low as 58\% in the case of support vector machines when accessing exhalations in spectrogram-based feature extraction. Finally, an important observation is that the Cepstrogram feature-based extraction method demonstrates the lowest misclassification rate in comparison to other approaches, which supports our initial assumption that this feature extraction approach yields the most separable feature representation.

\begin{table}[H]
\centering
\caption{Normalized \% confusion matrix for MFCC, spectrogram and cepstrogram feature extraction approaches}
\label{table:confusion-matrix}
\resizebox{\textwidth}{!}{
\begin{tabular}{cll|rrrrrrrrrrrr|}
\cline{4-15}
\multicolumn{1}{l}{}                                       
& 
&                 
& \multicolumn{12}{c|}{
\textbf{Reference}} \\ 
\cline{4-15} 
\multicolumn{1}{l}{}
&                                                    
&                 
& \multicolumn{4}{c|}{\textbf{MFCC}}                                  
& \multicolumn{4}{c|}{\textbf{SPECT}}                                 
& \multicolumn{4}{c|}{\textbf{CEPST}}
\\ 
\cline{4-15} 
\multicolumn{1}{l}{}  
&                                                    
&                 
&\multicolumn{1}{l|}{\textbf{Drug}}   
& \multicolumn{1}{l|}{\textbf{Exhale}} 
& \multicolumn{1}{l|}{\textbf{Inhale}}
& \multicolumn{1}{l|}{\textbf{Noise}}
& \multicolumn{1}{l|}{\textbf{Drug}}
& \multicolumn{1}{l|}{\textbf{Exhale}} 
& \multicolumn{1}{l|}{\textbf{Inhale}} 
& \multicolumn{1}{l|}{\textbf{Noise}}
& \multicolumn{1}{l|}{\textbf{Drug}}
& \multicolumn{1}{l|}{\textbf{Exhale}}
& \multicolumn{1}{l|}{\textbf{Inhale}}
& \multicolumn{1}{l|}{\textbf{Noise}}
\\ 
\hline
\hline
\multicolumn{1}{|c|}{
\multirow{16}{*}{
\rotatebox[origin=c]{90}{
\textbf{Prediction}}}} 
& \multicolumn{1}{l|}{\multirow{4}{*}{\textbf{SVM}}} 
& \textbf{Drug}   
& \cellcolor{color2}\textbf{97.54} 
& \cellcolor{color2}0.00            
& \cellcolor{color2}0.21            
& \cellcolor{color2}0.21           
& \cellcolor{color1}\textbf{97.54} 
& \cellcolor{color1}0.00            
& \cellcolor{color1}0.00            
& \cellcolor{color1}0.00           
& \cellcolor{color2}\textbf{99.38} 
& \cellcolor{color2}0.41            
& \cellcolor{color2}0.00            
& \cellcolor{color2}0.41   
\\ 
\cline{3-15} 
\multicolumn{1}{|c|}{}                                     
& \multicolumn{1}{l|}{}                              
& \textbf{Exhale} 
& \cellcolor{color2}0.41           
& \cellcolor{color2}\textbf{96.14}  
& \cellcolor{color2}1.23            
& \cellcolor{color2}2.89           
& \cellcolor{color1}2.46           
& \cellcolor{color1}\textbf{58.94}  
& \cellcolor{color1}1.43            
& \cellcolor{color1}3.73           
& \cellcolor{color2}0.41           
& \cellcolor{color2}\textbf{98.16}  
& \cellcolor{color2}1.43            
& \cellcolor{color2}0.41
\\ 
\cline{3-15} 
\multicolumn{1}{|c|}{}                                     
& \multicolumn{1}{l|}{}                              
& \textbf{Inhale} 
& \cellcolor{color2}0.00           
& \cellcolor{color2}0.00            
& \cellcolor{color2}\textbf{97.74}  
& \cellcolor{color2}0.20           
& \cellcolor{color1}0.00           
& \cellcolor{color1}0.41            
& \cellcolor{color1}\textbf{94.88}  
& \cellcolor{color1}0.83           
& \cellcolor{color2}0.00          
& \cellcolor{color2}0.20            
& \cellcolor{color2}\textbf{98.57}  
& \cellcolor{color2}0.41
\\ 
\cline{3-15} 
\multicolumn{1}{|c|}{}                                     
& \multicolumn{1}{l|}{}                              
& \textbf{Other}  
& \cellcolor{color2}2.05           
& \cellcolor{color2}3.86            
& \cellcolor{color2}0.82            
& \cellcolor{color2}\textbf{96.70} 
& \cellcolor{color1}0.00           
& \cellcolor{color1}40.65           
& \cellcolor{color1}3.69            
& \cellcolor{color1}\textbf{95.44} 
& \cellcolor{color2}0.21           
& \cellcolor{color2}1.23            
& \cellcolor{color2}0.00            
& \cellcolor{color2}\textbf{98.77}
\\ \cline{2-15} 
\multicolumn{1}{|c|}{}                                     
& \multicolumn{1}{l|}{\multirow{4}{*}{\textbf{RF}}}  
& \textbf{Drug}   
& \cellcolor{color1}\textbf{97.13} 
& \cellcolor{color1}0.20            
& \cellcolor{color1}0.21            
& \cellcolor{color1}0.00           
& \cellcolor{color2}\textbf{97.74} 
& \cellcolor{color2}0.20            
& \cellcolor{color2}0.00            
& \cellcolor{color2}0.00           
& \cellcolor{color1}\textbf{98.97} 
& \cellcolor{color1}0.41            
& \cellcolor{color1}0.00            
& \cellcolor{color1}0.62 
\\ \cline{3-15} 
\multicolumn{1}{|c|}{}                                     
& \multicolumn{1}{l|}{}                              
& \textbf{Exhale} 
& \cellcolor{color1}1.44           
& \cellcolor{color1}\textbf{95.93}  
& \cellcolor{color1}2.26            
& \cellcolor{color1}4.33           
& \cellcolor{color2}0.00           
& \cellcolor{color2}\textbf{96.95}  
& \cellcolor{color2}2.05            
& \cellcolor{color2}2.69           
& \cellcolor{color1}0.82           
& \cellcolor{color1}\textbf{97.35}  
& \cellcolor{color1}1.64            
& \cellcolor{color1}1.65
\\ 
\cline{3-15} 
\multicolumn{1}{|c|}{}                                     
& \multicolumn{1}{l|}{}                              
& \textbf{Inhale} 
& \cellcolor{color1}0.41           
& \cellcolor{color1}0.00            
& \cellcolor{color1}\textbf{96.71}  
& \cellcolor{color1}0.62           
& \cellcolor{color2}0.62           
& \cellcolor{color2}0.61            
& \cellcolor{color2}\textbf{97.95}  
& \cellcolor{color2}0.83           
& \cellcolor{color1}0.00           
& \cellcolor{color1}0.61            
& \cellcolor{color1}\textbf{98.16}  
& \cellcolor{color1}1.24
\\ \cline{3-15} 
\multicolumn{1}{|c|}{}                                     
& \multicolumn{1}{l|}{}                              
& \textbf{Other}  
& \cellcolor{color1}1.02           
& \cellcolor{color1}3.87            
& \cellcolor{color1}0.82            
& \cellcolor{color1}\textbf{95.05} 
& \cellcolor{color2}1.64           
& \cellcolor{color2}2.24            
& \cellcolor{color2}0.00            
& \cellcolor{color2}\textbf{96.48} 
& \cellcolor{color1}0.21           
& \cellcolor{color1}1.63            
& \cellcolor{color1}0.20            
& \cellcolor{color1}\textbf{96.49}
\\ \cline{2-15} 
\multicolumn{1}{|c|}{}                                     
& \multicolumn{1}{l|}{\multirow{4}{*}{\textbf{ADA}}} 
& \textbf{Drug}   
& \cellcolor{color2}\textbf{97.54} 
& \cellcolor{color2}0.00            
& \cellcolor{color2}0.00            
& \cellcolor{color2}0.41           
& \cellcolor{color1}\textbf{98.77} 
& \cellcolor{color1}0.41            
& \cellcolor{color1}0.00            
& \cellcolor{color1}0.00           
& \cellcolor{color2}\textbf{99.18} 
& \cellcolor{color2}0.20            
& \cellcolor{color2}0.20            
& \cellcolor{color2}0.21
\\ 
\cline{3-15} 
\multicolumn{1}{|c|}{}                                     
& \multicolumn{1}{l|}{}                              
& \textbf{Exhale} 
& \cellcolor{color2}1.03           
& \cellcolor{color2}\textbf{96.75}  
& \cellcolor{color2}1.65            
& \cellcolor{color2}4.95           
& \cellcolor{color1}0.41           
& \cellcolor{color1}\textbf{96.95}  
& \cellcolor{color1}1.43            
& \cellcolor{color1}1.45           
& \cellcolor{color2}0.62           
& \cellcolor{color2}\textbf{97.35}  
& \cellcolor{color2}1.84            
& \cellcolor{color2}1.03
\\ 
\cline{3-15} 
\multicolumn{1}{|c|}{}                                     
& \multicolumn{1}{l|}{}                              
& \textbf{Inhale} 
& \cellcolor{color2}0.00           
& \cellcolor{color2}0.20            
& \cellcolor{color2}\textbf{96.91}  
& \cellcolor{color2}1.03           
& \cellcolor{color1}0.00           
& \cellcolor{color1}0.20            
& \cellcolor{color1}\textbf{98.36}  
& \cellcolor{color1}0.62           
& \cellcolor{color2}0.00           
& \cellcolor{color2}0.61            
& \cellcolor{color2}\textbf{97.54}  
& \cellcolor{color2}0.82
\\ 
\cline{3-15} 
\multicolumn{1}{|c|}{}                                    
& \multicolumn{1}{l|}{}                              
& \textbf{Other}  
& \cellcolor{color2}1.43           
& \cellcolor{color2}3.05            
& \cellcolor{color2}1.44            
& \cellcolor{color2}\textbf{93.61} 
& \cellcolor{color1}0.82           
& \cellcolor{color1}2.44            
& \cellcolor{color1}0.21            
& \cellcolor{color1}\textbf{97.93} 
& \cellcolor{color2}0.20           
& \cellcolor{color2}1.84            
& \cellcolor{color2}0.42            
& \cellcolor{color2}\textbf{97.94}
\\ 
\cline{2-15} 
\multicolumn{1}{|c|}{}                                     
& \multicolumn{1}{l|}{\multirow{4}{*}{\textbf{GMM}}} 
& \textbf{Drug}   
& \cellcolor{color1}\textbf{96.71} 
& \cellcolor{color1}0.00            
& \cellcolor{color1}0.00            
& \cellcolor{color1}0.00           
& \cellcolor{color2}\textbf{99.18} 
& \cellcolor{color2}0.41            
& \cellcolor{color2}0.00            
& \cellcolor{color2}2.69           
& \cellcolor{color1}\textbf{99.38} 
& \cellcolor{color1}0.00            
& \cellcolor{color1}0.00            
& \cellcolor{color1}0.62  
\\ 
\cline{3-15} 
\multicolumn{1}{|c|}{}                                     
& \multicolumn{1}{l|}{}                              
& \textbf{Exhale} 
& \cellcolor{color1}0.82           
& \cellcolor{color1}\textbf{96.14}  
& \cellcolor{color1}1.23            
& \cellcolor{color1}2.68           
& \cellcolor{color2}0.62           
& \cellcolor{color2}\textbf{93.29}  
& \cellcolor{color2}1.64            
& \cellcolor{color2}6.00           
& \cellcolor{color1}0.41           
& \cellcolor{color1}\textbf{99.18}  
& \cellcolor{color1}1.43            
& \cellcolor{color1}2.27           
\\ 
\cline{3-15} 
\multicolumn{1}{|c|}{}                                     
& \multicolumn{1}{l|}{}                              
& \textbf{Inhale} 
& \cellcolor{color1}0.21           
& \cellcolor{color1}0.00            
& \cellcolor{color1}\textbf{97.74}  
& \cellcolor{color1}1.03           
& \cellcolor{color2}0.20           
& \cellcolor{color2}3.46            
& \cellcolor{color2}\textbf{98.16}  
& \cellcolor{color2}2.90           
& \cellcolor{color1}0.00           
& \cellcolor{color1}0.41            
& \cellcolor{color1}\textbf{98.57}  
& \cellcolor{color1}0.21           
\\ 
\cline{3-15} 
\multicolumn{1}{|c|}{}                                     
& \multicolumn{1}{l|}{}                             
& \textbf{Other}  
& \cellcolor{color1}2.26           
& \cellcolor{color1}3.86            
& \cellcolor{color1}1.03            
& \cellcolor{color1}\textbf{96.29} 
& \cellcolor{color2}0.00           
& \cellcolor{color2}2.84            
& \cellcolor{color2}0.20            
& \cellcolor{color2}\textbf{88.41} 
& \cellcolor{color1}0.21           
& \cellcolor{color1}0.41            
& \cellcolor{color1}0.00            
& \cellcolor{color1}\textbf{96.90} 
\\ 
\cline{1-15} 
\end{tabular}
}
\end{table}

\begin{table}[H]
\centering
\caption{Normalized \% confusion matrix for continuous wavelet transform.}
\label{table:confusion-matrix-cwt}
\resizebox{\textwidth}{!}{
\begin{tabular}{cll|p{3cm}|p{3cm}|p{3cm}|p{3cm}|}
\cline{4-7}
\multicolumn{1}{l}{}                           
& 
&                 
& \multicolumn{4}{c|}{
\textbf{Reference}} \\ 
\cline{4-7} 
\multicolumn{1}{l}{}
&                                             
&                     
& \multicolumn{4}{c|}{\textbf{CWT}}  
\\ 
\cline{4-7} 
\multicolumn{1}{l}{}                           
&                                             
&                 
& \textbf{Drug}  
& \textbf{Exhale} 
& \textbf{Inhale} 
& \textbf{Noise}
\\ 
\hline
\hline
\multicolumn{1}{|c|}{
\multirow{12}{*}{
\rotatebox[origin=c]{90}{
\textbf{Prediction}}}} 
& \multicolumn{1}{l|}{\multirow{4}{*}{\textbf{SVM}}} 
& \textbf{Drug}   
& \cellcolor{color1}\textbf{97.11} 
& \cellcolor{color1}0.20            
& \cellcolor{color1}0.20          
& \cellcolor{color1}0.00  
\\ 
\cline{3-7} 
\multicolumn{1}{|c|}{}                         
& \multicolumn{1}{l|}{}                       
& \textbf{Exhale} 
& \cellcolor{color1}1.44    
& \cellcolor{color1}\textbf{98.16}  
& \cellcolor{color1}27.72          
& \cellcolor{color1}85.66
\\ 
\cline{3-7} 
\multicolumn{1}{|c|}{}                         
& \multicolumn{1}{l|}{}                       
& \textbf{Inhale}  
& \cellcolor{color1}1.03          
& \cellcolor{color1}0.82      
& \cellcolor{color1}\textbf{71.66}  
& \cellcolor{color1}6.56
\\ 
\cline{3-7} 
\multicolumn{1}{|c|}{}                         
& \multicolumn{1}{l|}{}                       
& \textbf{Other}   
& \cellcolor{color1}0.42        
& \cellcolor{color1}0.82        
& \cellcolor{color1}0.42      
& \cellcolor{color1}\textbf{7.78} 
\\ \cline{2-7} 
\multicolumn{1}{|c|}{}                         
& \multicolumn{1}{l|}{\multirow{4}{*}{\textbf{RF}}}  
& \textbf{Drug}   
& \cellcolor{color2}\textbf{98.97} 
& \cellcolor{color2}0.40      
& \cellcolor{color2}0.20        
& \cellcolor{color2}0.20
\\ \cline{3-7} 
\multicolumn{1}{|c|}{}                         
& \multicolumn{1}{l|}{}                       
& \textbf{Exhale} 
& \cellcolor{color2}0.00      
& \cellcolor{color2}\textbf{94.10}  
& \cellcolor{color2}1.64      
& \cellcolor{color2}8.60
\\ 
\cline{3-7} 
\multicolumn{1}{|c|}{}                         
& \multicolumn{1}{l|}{}                       
& \textbf{Inhale} 
& \cellcolor{color2}1.03          
& \cellcolor{color2}0.81         
& \cellcolor{color2}\textbf{95.70}  
& \cellcolor{color2}2.67
\\ 
\cline{3-7} 
\multicolumn{1}{|c|}{}                         
& \multicolumn{1}{l|}{}                       
& \textbf{Other}  
& \cellcolor{color2}0.00       
& \cellcolor{color2}4.69          
& \cellcolor{color2}2.46         
& \cellcolor{color2}\textbf{88.53} 
\\ 
\cline{2-7} 
\multicolumn{1}{|c|}{}                         
& \multicolumn{1}{l|}{\multirow{4}{*}{\textbf{ADA}}} 
& \textbf{Drug}   
& \cellcolor{color1}\textbf{99.18} 
& \cellcolor{color1}0.00          
& \cellcolor{color1}0.41           
& \cellcolor{color1}0.41
\\ 
\cline{3-7} 
\multicolumn{1}{|c|}{}                        
& \multicolumn{1}{l|}{}                       
& \textbf{Exhale}     
& \cellcolor{color1}0.00
& \cellcolor{color1}\textbf{95.91}           
& \cellcolor{color1}1.64          
& \cellcolor{color1}6.56 
\\ 
\cline{3-7} 
\multicolumn{1}{|c|}{}                         
& \multicolumn{1}{l|}{}                       
& \textbf{Inhale} 
& \cellcolor{color1}0.20
& \cellcolor{color1}0.61         
& \cellcolor{color1}\textbf{96.31}            
& \cellcolor{color1}1.64
\\ 
\cline{3-7} 
\multicolumn{1}{|c|}{}                         
& \multicolumn{1}{l|}{}                       
& \textbf{Other}  
& \cellcolor{color1}0.62        
& \cellcolor{color1}3.48           
& \cellcolor{color1}1.64          
& \cellcolor{color1}\textbf{91.39} 
\\ 
\cline{2-7}
\cline{2-7}
\multicolumn{1}{|c|}{}                         
& \multicolumn{1}{l|}{\multirow{2}{*}{\textbf{ADA}}} 
& \textbf{}   
& \multicolumn{2}{c|}{\textbf{Drug}}
& \multicolumn{2}{c|}{\textbf{Other}}
\\ 
\cline{3-7} 
\multicolumn{1}{|c|}{}                         
& \multicolumn{1}{l|}{\multirow{2}{*}{\textbf{2-Class}}}   
& \textbf{Drug}   
& \multicolumn{2}{c|}{\textbf{99.18}}
& \multicolumn{2}{c|}{0.27}       
\\ 
\cline{3-7} 
\multicolumn{1}{|c|}{}                         
& \multicolumn{1}{l|}{}   
& \textbf{Other}   
& \multicolumn{2}{c|}{0.82}
& \multicolumn{2}{c|}{\textbf{99.73}}       
\\ 
\cline{1-7} 
\end{tabular}
}
\end{table}

\subsubsection{Employing relevance feedback to improve classification accuracy}
\label{subsection:relevance_feedback_results}
\begin{figure}[t!] 
\centering
\includegraphics[width=0.9\textwidth]{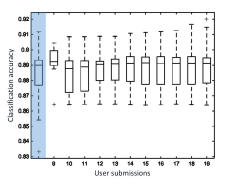}
     \begin{center}(a)\end{center}
\includegraphics[width=0.8\textwidth]{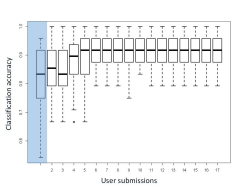}
     \begin{center}(b)\end{center}

  \caption{Relevance feedback of Cepstrogram based features.Classification accuracy box plot as a function of included user submissions for the \textbf{CEPST-GMM} approach and \textbf{CEPST-SVM} approach.}
  \label{fig:rfig1}
\end{figure}

In order to validate the relevance feedback functionality, we employed the relevance feedback of a second group consisting of five subjects. It is important to note that subjects of the first group that provided the audio samples were not included in the second group. Each person provided $20$ sets of annotated submissions, with each submission containing $24$ annotated feature vectors. In the validation process, we assume that one of the $20$ sets is not annotated and derive the classification accuracy for this set by employing the following cross-validation approach: At first, we include only two annotated sets for the compilation of the relevant dataset and derive the classification accuracy. Then, one more set is included in each iteration, the relevant dataset is recompiled, and the classification accuracy is recalculated. The process is repeated until all remaining $19$ sets are included. 
The process evaluates the improvement of the classification accuracy score relative to the number of user submissions. 
The results of the evaluation process are presented in Figure \ref{fig:rfig1}. The first column represents the classification accuracy result without relevant feedback, while the following columns represent the classification accuracy for the corresponding number of user submissions. CEPST-SVM shows low classification accuracy without relevant feedback but significantly improves as user submissions increase. CEPST-GMM shows the slowest improvement rate but appears more robust since the first column representing the classification results without relevant feedback is concentrated around 89\%. 

\subsection{Dataset B}

Dataset B is tested against a series of algorithms and settings. Specifically,  SVM, random forests, Adaboost, Gaussian Mixture Models, LSTM and CNN based approaches are compared for Spectrogram, Ceptrogram, MFCC and time-domain features where applicable, across three different setups, Single Subject, Multi-Subject, and LOSO. Furthermore, an additional feature is tested, the mixed vs the unmixed events in a window. To provide a thorough description of the mixing feature, in a non-mixed setup, each test sample contains only one type of audio event. In contrast, each test sample may contain multiple audio events in a mixed setup. In the latter, the testing set is formulated by a sliding window, and the class of the central sample defines the class index. The following tables provide an exhaustive analysis of the benefits and drawbacks of each algorithm.

\subsubsection{Simulation setup and validation settings}

The Respiratory and Drug Actuation Benchmark is a publicly available benchmark suite\footnote{https://gitlab.com/vvr/monitoring-medication-adherence/rda-benchmark} employed to carry out the following study. We compare the classification performance of those mentioned above essential and widely used machine learning and deep learning algorithms, namely the Random Forests (RF), Support Vector Machines (SVM), Adaboost, Convolutional Neural Networks (CNNs) and Recurrent Neural Networks (LSTMs). They were evaluated with spectrogram, cepstrogram and MFCC features. CNNs were directly applied to time-series values to demonstrate their capability to provide dependable solutions at lower execution times.\par

During validation, the main differentiation parameter comes from the availability of previous recordings from a specific individual. It is expected that preliminary information can increase classification accuracy. However, it puts an additional burden on the usage of the monitoring system since it requires the collection of data every time a new patient wants to test the framework.\par

Firstly, we consider the \textit{Multi Subject} modeling approach, denoted as \textit{MultiSubj}. In this case, the recordings of all subjects are used to form a large dataset, divided into five equal parts used to perform five-fold cross-validation, thereby allowing different samples from the same subject to be used in the training and test set respectively. This validation scheme was followed in previous work \cite{ntalianis2019assessment} and thus performed, also here for comparison purposes.\par

The second case includes the \textit{Single Subject} setting, in which the performance of the classifier is validated through training and testing within each subject's recordings. We denote such models as \textit{SingleSubj}. Specifically, the recordings of each subject are split into five equal parts to perform cross-validation. The accuracy is assessed for each subject separately, and then the classifier's overall performance is calculated by averaging the three individual results.\par

The third evaluation setting refers to the case when no previous recordings for the testing subject are available. Thus, samples from other subjects are used. This is the \textit{leave-one-subject-out (LOSO)} approach that illustrates how well the trained network can generalize to individuals that it never saw before during training. \textit{LOSO} models facilitate the use of the monitoring system since they do not require a data pre-collection phase, and they also have the lowest risk of over-fitting. However, if the inter-subject variability is high, they might not adapt well, especially if the number of training subjects is small, as in our case. With this approach, we use the recordings of two subjects for training and the recordings of the third subject for testing. This procedure is complete when all subjects have been used for testing, and the accuracy is averaged to obtain the overall performance of the classifiers.\par

\textbf{Furthermore, the testing dataset mixing brought a new parameter into the comparison. A non-mixed setup means that each audio segment in the testing set consists of a single class. However, in a mixed setup, each testing set sample emanates from a sliding window that naturally includes parts belonging to multiple classes. In this case, the class of an audio segment is the class of the central sample.}\par

\subsubsection{Key performance indicators}

Table~\ref{table:results:Accuracy} summarizes the classification accuracy for drug, exhalations and inhalations across all validation setups. Out of a superficial examination, no method is better, but key performance indicators need to be established. Multi-subject and single-subject settings indicate the classifier's success if the user has previously participated in the sampling process. LOSO setting is the closest to a real situation setting since a commercial classifier would not assume that a new user had already submitted samples to the training process. Even though a feedback loop can improve accuracy \cite{nousias2018mhealth} such a process cannot be a prerequisite. As a result, LOSO performance is the most representative. Furthermore, between mixed and non-mixed, the first is closer to the real situation since the non-mixed setup assumes that the position of audio segments containing a certain audio event is known before the classifier is applied. Furthermore, tables \ref{table:results:recall} and \ref{table:results:precision} present the performance of classifiers in drug detection and exhalation-inhalation differentiation. Again, the LOSO-mixed setup is the most representative of reality. In short, the key performance indicators can be summarized below:
\begin{itemize}
    \item The three highest non-mixed LOSO accuracies should be highlighted (Table~\ref{table:results:Accuracy})
    \item The three highest mixed multisubject accuracies should be highlighted (Table~\ref{table:results:Accuracy})
    \item The three highest mixed LOSO accuracies should be highlighted (Table~\ref{table:results:Accuracy})
    \item The three highest drug precision mixed and non-mixed sensitivity should be highlighted (Table~\ref{table:results:Accuracy})
\end{itemize}
Metrics to measure the performance of the compared classifiers are accuracy , sensitivity and specificity. For the sake of self completeness

\begin{figure}[H]
\centering
  \includegraphics[width=\textwidth]{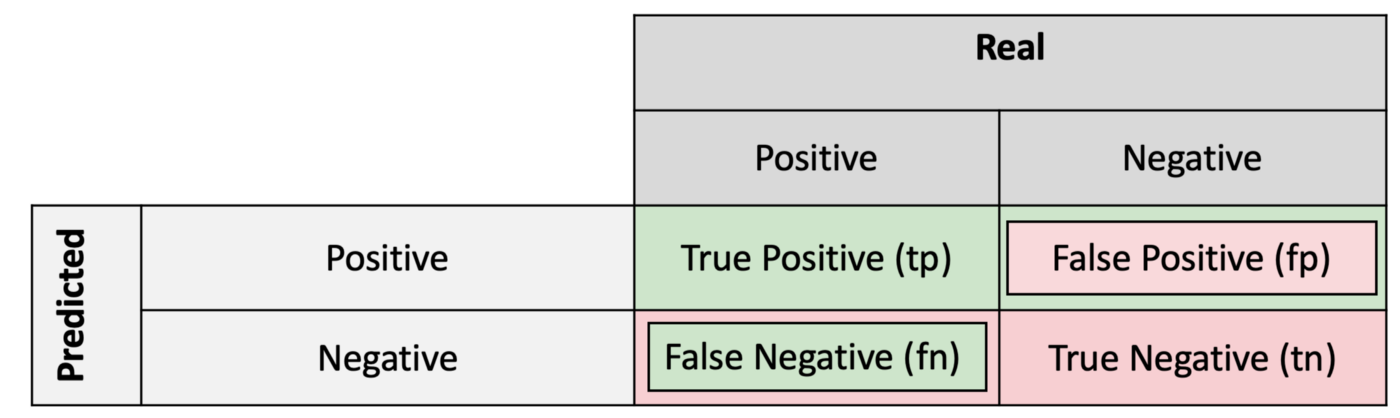}
\caption{Accuracy matrix}
\label{fig:accuracyMatrix}
\end{figure}

\begin{equation}
    \text{accuracy}=\frac{(TP+TN)}{(TP+FP+TN+FN)}
\end{equation}
\begin{equation}
    \text{precision}=\frac{(TP)}{(TP+FP)}
\end{equation}
\begin{equation}
   \text{recall} =\frac{TN}{(TN+FP)}
\end{equation}
where TP is The number of positive correct identifications, TN the number of negative correct identifications, FP is the number of positive incorrect identifications and FN the number of negative incorrect identifications.

\subsubsection{Results}

Close inspection of Tables ~\ref{table:results:Accuracy} and ~\ref{table:results:f1} reveals that no method clearly outperforms the others. The performance of data-driven approaches largely depends on the dataset and the preprocessing steps. However, since all the approaches were trained with the same dataset, we can quantify which can capture the individual characteristics more efficiently. For all methodologies, the multisubject approach yields the highest score. This means that if a user's data have been included in the training process, the success probability climbs to a level near 96\%. 

Given the KPIs defined in the previous section, in the non-mixed setup, SVM-MFCC, ADA-MFCC, and CNN-TIME yield the highest accuracy when patient data are already included in the dataset, corresponding to Multi and Single cases. However, in the "Leave One Subject Out" Case, which also corresponds to a real-life scenario CNN-TIME, ADA-CEPST, and RF-CEPST yield the best results. In the mixed setup, the results are similar with LSTM-SPECT to replace CNN-TIME only in the Multi-Subject setup.

Further insight is provided by Table~\ref{table:results:f1}. Measuring F1 score is highly important since it consists of the harmonic mean of precision and recall of a candidate detector. Close inspection reveals that drug detection has much more different characteristics than exhalations and inhalations. 
For the Exhale/LOSO case, CNN-TIME has the best performance that also concurs with the results presented in the Accuracy table. For detecting inhalations in LOSO setup, CNN-TIME also yields the best performance. However, in the Multisubject setup, MFCC methods-based features demonstrate more stable performance than the CNN-TIME method. 

In the drug classification case, the F1 score reveals much more different results. Close examination reveals that mixed setup demonstrates lower F1 Scores than non-mixed setup, meaning that "Drug" detection is highly affected by the surrounding audio events. ADA-SPECT, GMM-SPECT and RF-SPECT have the higher accuracy in MultiSubject mixed setup, reaching 73\%, while for the non-mixed case, the accuracy is 96\%. Likewise, LOSO/Non-mixed reaches with Random Forest-based methods 84\%, but in the mixed setup, the corresponding performance is 32\%. \textbf{Such outcome confirms and quantifies the assumption that Relevance Feedback, where the user submits their samples, can significantly improve performance.\cite{nousias2018mhealth}}



\begin{table}
\centering
\caption{Accuracy}
\label{table:results:Accuracy}
\resizebox{0.6\textwidth}{!}{
\begin{tabular}{|l|l|l|r|r|r|}
\hline
{\color[HTML]{000000} }                             & {\color[HTML]{000000} }     & {\color[HTML]{000000} }      & {\color[HTML]{000000} Multi}                         & {\color[HTML]{000000} LOSO}                          & {\color[HTML]{000000} Single}                        \\ \hline
{\color[HTML]{000000} }                             & {\color[HTML]{000000} ada}  & {\color[HTML]{000000} cepst} & \cellcolor[HTML]{CFCFCF}{\color[HTML]{000000} 92.11} & \cellcolor[HTML]{CFCFCF}{\color[HTML]{000000} 90.35} & {\color[HTML]{000000} 90.67}                         \\ \cline{2-6} 
{\color[HTML]{000000} }                             & {\color[HTML]{000000} ada}  & {\color[HTML]{000000} mfcc}  & {\color[HTML]{000000} 91.76}                         & {\color[HTML]{000000} 88.31}                         & {\color[HTML]{000000} 89.43}                         \\ \cline{2-6} 
{\color[HTML]{000000} }                             & {\color[HTML]{000000} ada}  & {\color[HTML]{000000} spect} & {\color[HTML]{000000} 89.8}                          & {\color[HTML]{000000} 84.55}                         & {\color[HTML]{000000} 88.79}                         \\ \cline{2-6} 
{\color[HTML]{000000} }                             & {\color[HTML]{000000} cnn}  & {\color[HTML]{000000} time}  & {\color[HTML]{000000} 91.71}                         & \cellcolor[HTML]{CFCFCF}{\color[HTML]{000000} 92.94} & {\color[HTML]{000000} 90.61}                         \\ \cline{2-6} 
{\color[HTML]{000000} }                             & {\color[HTML]{000000} gmm}  & {\color[HTML]{000000} cepst} & {\color[HTML]{000000} 89.81}                         & {\color[HTML]{000000} 80.54}                         & {\color[HTML]{000000} 89.33}                         \\ \cline{2-6} 
{\color[HTML]{000000} }                             & {\color[HTML]{000000} gmm}  & {\color[HTML]{000000} mfcc}  & {\color[HTML]{000000} 82.83}                         & {\color[HTML]{000000} 82.12}                         & {\color[HTML]{000000} 78.49}                         \\ \cline{2-6} 
{\color[HTML]{000000} }                             & {\color[HTML]{000000} gmm}  & {\color[HTML]{000000} spect} & {\color[HTML]{000000} 86.45}                         & {\color[HTML]{000000} 83.58}                         & {\color[HTML]{000000} 86.88}                         \\ \cline{2-6} 
{\color[HTML]{000000} }                             & {\color[HTML]{000000} lstm} & {\color[HTML]{000000} spect} & \cellcolor[HTML]{CFCFCF}{\color[HTML]{000000} 92.16} & {\color[HTML]{000000} 87.73}                         & \cellcolor[HTML]{CFCFCF}{\color[HTML]{000000} 91.1}  \\ \cline{2-6} 
{\color[HTML]{000000} }                             & {\color[HTML]{000000} rf}   & {\color[HTML]{000000} cepst} & {\color[HTML]{000000} 91.69}                         & \cellcolor[HTML]{CFCFCF}{\color[HTML]{000000} 90.28} & \cellcolor[HTML]{CFCFCF}{\color[HTML]{000000} 91.05} \\ \cline{2-6} 
{\color[HTML]{000000} }                             & {\color[HTML]{000000} rf}   & {\color[HTML]{000000} mfcc}  & {\color[HTML]{000000} 91.61}                         & {\color[HTML]{000000} 89.36}                         & {\color[HTML]{000000} 90.86}                         \\ \cline{2-6} 
{\color[HTML]{000000} }                             & {\color[HTML]{000000} rf}   & {\color[HTML]{000000} spect} & {\color[HTML]{000000} 89.47}                         & {\color[HTML]{000000} 85.7}                          & {\color[HTML]{000000} 89.85}                         \\ \cline{2-6} 
{\color[HTML]{000000} }                             & {\color[HTML]{000000} svm}  & {\color[HTML]{000000} cepst} & {\color[HTML]{000000} 91.67}                         & {\color[HTML]{000000} 82.34}                         & {\color[HTML]{000000} 90.89}                         \\ \cline{2-6} 
{\color[HTML]{000000} }                             & {\color[HTML]{000000} svm}  & {\color[HTML]{000000} mfcc}  & \cellcolor[HTML]{CFCFCF}{\color[HTML]{000000} 92.49} & {\color[HTML]{000000} 87.29}                         & \cellcolor[HTML]{CFCFCF}{\color[HTML]{000000} 92.14} \\ \cline{2-6} 
\multirow{-14}{*}{{\color[HTML]{000000} Mixed}}     & {\color[HTML]{000000} svm}  & {\color[HTML]{000000} spect} & {\color[HTML]{000000} 34.03}                         & {\color[HTML]{000000} 34}                            & {\color[HTML]{000000} 33.53}                         \\ \hline
{\color[HTML]{000000} }                             & {\color[HTML]{000000} ada}  & {\color[HTML]{000000} cepst} & {\color[HTML]{000000} 95.02}                         & \cellcolor[HTML]{CFCFCF}{\color[HTML]{000000} 93.32} & {\color[HTML]{000000} 92.57}                         \\ \cline{2-6} 
{\color[HTML]{000000} }                             & {\color[HTML]{000000} ada}  & {\color[HTML]{000000} mfcc}  & \cellcolor[HTML]{CFCFCF}{\color[HTML]{000000} 95.91} & {\color[HTML]{000000} 87.02}                         & {\color[HTML]{000000} 93.8}                          \\ \cline{2-6} 
{\color[HTML]{000000} }                             & {\color[HTML]{000000} ada}  & {\color[HTML]{000000} spect} & {\color[HTML]{000000} 94.01}                         & {\color[HTML]{000000} 84.92}                         & {\color[HTML]{000000} 92.99}                         \\ \cline{2-6} 
{\color[HTML]{000000} }                             & {\color[HTML]{000000} cnn}  & {\color[HTML]{000000} time}  & \cellcolor[HTML]{CFCFCF}{\color[HTML]{000000} 95.29} & \cellcolor[HTML]{CFCFCF}{\color[HTML]{000000} 94.12} & {\color[HTML]{000000} 92.69}                         \\ \cline{2-6} 
{\color[HTML]{000000} }                             & {\color[HTML]{000000} gmm}  & {\color[HTML]{000000} cepst} & {\color[HTML]{000000} 94.1}                          & {\color[HTML]{000000} 80.92}                         & \cellcolor[HTML]{CFCFCF}{\color[HTML]{000000} 94.02} \\ \cline{2-6} 
{\color[HTML]{000000} }                             & {\color[HTML]{000000} gmm}  & {\color[HTML]{000000} mfcc}  & {\color[HTML]{000000} 76.48}                         & {\color[HTML]{000000} 79.39}                         & {\color[HTML]{000000} 68.82}                         \\ \cline{2-6} 
{\color[HTML]{000000} }                             & {\color[HTML]{000000} gmm}  & {\color[HTML]{000000} spect} & {\color[HTML]{000000} 86.81}                         & {\color[HTML]{000000} 81.49}                         & {\color[HTML]{000000} 86.52}                         \\ \cline{2-6} 
{\color[HTML]{000000} }                             & {\color[HTML]{000000} lstm} & {\color[HTML]{000000} spect} & {\color[HTML]{000000} 92.93}                         & {\color[HTML]{000000} 88.89}                         & {\color[HTML]{000000} 92.54}                         \\ \cline{2-6} 
{\color[HTML]{000000} }                             & {\color[HTML]{000000} rf}   & {\color[HTML]{000000} cepst} & {\color[HTML]{000000} 93.92}                         & \cellcolor[HTML]{CFCFCF}{\color[HTML]{000000} 91.98} & {\color[HTML]{000000} 93.3}                          \\ \cline{2-6} 
{\color[HTML]{000000} }                             & {\color[HTML]{000000} rf}   & {\color[HTML]{000000} mfcc}  & {\color[HTML]{000000} 94.87}                         & {\color[HTML]{000000} 90.08}                         & {\color[HTML]{000000} 93.26}                         \\ \cline{2-6} 
{\color[HTML]{000000} }                             & {\color[HTML]{000000} rf}   & {\color[HTML]{000000} spect} & {\color[HTML]{000000} 92.79}                         & {\color[HTML]{000000} 85.5}                          & {\color[HTML]{000000} 93.26}                         \\ \cline{2-6} 
{\color[HTML]{000000} }                             & {\color[HTML]{000000} svm}  & {\color[HTML]{000000} cepst} & {\color[HTML]{000000} 95.15}                         & {\color[HTML]{000000} 77.67}                         & \cellcolor[HTML]{CFCFCF}{\color[HTML]{000000} 94.57} \\ \cline{2-6} 
{\color[HTML]{000000} }                             & {\color[HTML]{000000} svm}  & {\color[HTML]{000000} mfcc}  & \cellcolor[HTML]{CFCFCF}{\color[HTML]{000000} 96.21} & {\color[HTML]{000000} 83.97}                         & \cellcolor[HTML]{CFCFCF}{\color[HTML]{000000} 96.23} \\ \cline{2-6} 
\multirow{-14}{*}{{\color[HTML]{000000} Non-mixed}} & {\color[HTML]{000000} svm}  & {\color[HTML]{000000} spect} & {\color[HTML]{000000} 75.32}                         & {\color[HTML]{000000} 69.85}                         & {\color[HTML]{000000} 77.09}                         \\ \hline
\end{tabular}
}
\end{table}

\begin{table}
\centering
\caption{F1 Score summarization}
\label{table:results:f1}
\resizebox{\textwidth}{!}{
\begin{tabular}{@{}|l|l|l||lll||lll||lll|@{}}
\toprule
                             &      &       & \multicolumn{3}{l|}{Drug}                                                                                                                                      & \multicolumn{3}{l|}{Exhale}                                                                                                                                    & \multicolumn{3}{l|}{Inhale}                                                                                                                                    \\ \midrule \midrule
                             &      &       & \multicolumn{1}{l|}{LOSO}                                                 & \multicolumn{1}{l|}{Multi}                                                & Single & \multicolumn{1}{l|}{LOSO}                                                 & \multicolumn{1}{l|}{Multi}                                                & Single & \multicolumn{1}{l|}{LOSO}                                                 & \multicolumn{1}{l|}{Multi}                                                & Single \\ \midrule
                             & ada  & cepst & \multicolumn{1}{l|}{17.58}                                                & \multicolumn{1}{l|}{64.84}                                                & 63.28  & \multicolumn{1}{l|}{\cellcolor[HTML]{CFCFCF}{\color[HTML]{000000} 83.57}} & \multicolumn{1}{l|}{\cellcolor[HTML]{CFCFCF}{\color[HTML]{000000} 87.27}} & 84.45  & \multicolumn{1}{l|}{86.6}                                                 & \multicolumn{1}{l|}{89.24}                                                & 86.79  \\ \cmidrule(l){2-12} 
                             & ada  & mfcc  & \multicolumn{1}{l|}{11.87}                                                & \multicolumn{1}{l|}{64.1}                                                 & 65.51  & \multicolumn{1}{l|}{82.46}                                                & \multicolumn{1}{l|}{86.67}                                                & 81.03  & \multicolumn{1}{l|}{80.55}                                                & \multicolumn{1}{l|}{89.46}                                                & 85.42  \\ \cmidrule(l){2-12} 
                             & ada  & spect & \multicolumn{1}{l|}{\cellcolor[HTML]{CFCFCF}{\color[HTML]{000000} 25.89}} & \multicolumn{1}{l|}{\cellcolor[HTML]{CFCFCF}{\color[HTML]{000000} 74.85}} & 74.5   & \multicolumn{1}{l|}{72.77}                                                & \multicolumn{1}{l|}{81.48}                                                & 78.43  & \multicolumn{1}{l|}{84.73}                                                & \multicolumn{1}{l|}{\cellcolor[HTML]{CFCFCF}{\color[HTML]{000000} 91.19}} & 88.94  \\ \cmidrule(l){2-12} 
                             & cnn  & time  & \multicolumn{1}{l|}{21.85}                                                & \multicolumn{1}{l|}{58.1}                                                 & 63.41  & \multicolumn{1}{l|}{\cellcolor[HTML]{CFCFCF}{\color[HTML]{000000} 89.08}} & \multicolumn{1}{l|}{86.81}                                                & 83.06  & \multicolumn{1}{l|}{\cellcolor[HTML]{CFCFCF}{\color[HTML]{000000} 94.94}} & \multicolumn{1}{l|}{89.95}                                                & 89.05  \\ \cmidrule(l){2-12} 
                             & gmm  & cepst & \multicolumn{1}{l|}{25.81}                                                & \multicolumn{1}{l|}{63.96}                                                & 67.46  & \multicolumn{1}{l|}{72.42}                                                & \multicolumn{1}{l|}{83.73}                                                & 81.02  & \multicolumn{1}{l|}{48.3}                                                 & \multicolumn{1}{l|}{84.74}                                                & 85.21  \\ \cmidrule(l){2-12} 
                             & gmm  & mfcc  & \multicolumn{1}{l|}{}                                                     & \multicolumn{1}{l|}{17.13}                                                & 0.47   & \multicolumn{1}{l|}{72.12}                                                & \multicolumn{1}{l|}{73.85}                                                & 65.76  & \multicolumn{1}{l|}{41.62}                                                & \multicolumn{1}{l|}{53.86}                                                & 53.32  \\ \cmidrule(l){2-12} 
                             & gmm  & spect & \multicolumn{1}{l|}{\cellcolor[HTML]{CFCFCF}{\color[HTML]{000000} 32.78}} & \multicolumn{1}{l|}{\cellcolor[HTML]{CFCFCF}{\color[HTML]{000000} 73.59}} & 74.27  & \multicolumn{1}{l|}{67.42}                                                & \multicolumn{1}{l|}{72.78}                                                & 74.44  & \multicolumn{1}{l|}{\cellcolor[HTML]{CFCFCF}{\color[HTML]{000000} 89.34}} & \multicolumn{1}{l|}{89.13}                                                & 85.89  \\ \cmidrule(l){2-12} 
                             & lstm & spect & \multicolumn{1}{l|}{23.1}                                                 & \multicolumn{1}{l|}{66.61}                                                & 70.67  & \multicolumn{1}{l|}{75.5}                                                 & \multicolumn{1}{l|}{86.4}                                                 & 83.21  & \multicolumn{1}{l|}{86.04}                                                & \multicolumn{1}{l|}{90}                                                   & 88.87  \\ \cmidrule(l){2-12} 
                             & rf   & cepst & \multicolumn{1}{l|}{25.41}                                                & \multicolumn{1}{l|}{63.22}                                                & 64.73  & \multicolumn{1}{l|}{\cellcolor[HTML]{CFCFCF}{\color[HTML]{000000} 83.01}} & \multicolumn{1}{l|}{86.55}                                                & 84.77  & \multicolumn{1}{l|}{86.6}                                                 & \multicolumn{1}{l|}{89.19}                                                & 89.38  \\ \cmidrule(l){2-12} 
                             & rf   & mfcc  & \multicolumn{1}{l|}{15.87}                                                & \multicolumn{1}{l|}{59.78}                                                & 63.84  & \multicolumn{1}{l|}{82.57}                                                & \multicolumn{1}{l|}{\cellcolor[HTML]{CFCFCF}{\color[HTML]{000000} 86.97}} & 84.88  & \multicolumn{1}{l|}{84.69}                                                & \multicolumn{1}{l|}{87.57}                                                & 87.01  \\ \cmidrule(l){2-12} 
                             & rf   & spect & \multicolumn{1}{l|}{\cellcolor[HTML]{CFCFCF}{\color[HTML]{000000} 30.98}} & \multicolumn{1}{l|}{\cellcolor[HTML]{CFCFCF}{\color[HTML]{000000} 73.69}} & 75.91  & \multicolumn{1}{l|}{74.26}                                                & \multicolumn{1}{l|}{80.86}                                                & 80.83  & \multicolumn{1}{l|}{88.41}                                                & \multicolumn{1}{l|}{\cellcolor[HTML]{CFCFCF}{\color[HTML]{000000} 90.65}} & 90.27  \\ \cmidrule(l){2-12} 
                             & svm  & cepst & \multicolumn{1}{l|}{11.42}                                                & \multicolumn{1}{l|}{62.82}                                                & 65.09  & \multicolumn{1}{l|}{68.1}                                                 & \multicolumn{1}{l|}{86.9}                                                 & 84.8   & \multicolumn{1}{l|}{68.26}                                                & \multicolumn{1}{l|}{88.92}                                                & 88.77  \\ \cmidrule(l){2-12} 
                             & svm  & mfcc  & \multicolumn{1}{l|}{12.83}                                                & \multicolumn{1}{l|}{59.93}                                                & 66.03  & \multicolumn{1}{l|}{79.33}                                                & \multicolumn{1}{l|}{\cellcolor[HTML]{CFCFCF}{\color[HTML]{000000} 88.91}} & 87.66  & \multicolumn{1}{l|}{80.38}                                                & \multicolumn{1}{l|}{88.93}                                                & 88.66  \\ \cmidrule(l){2-12} 
\multirow{-14}{*}{Mixed}     & svm  & spect & \multicolumn{1}{l|}{25.59}                                                & \multicolumn{1}{l|}{71.28}                                                & 73.71  & \multicolumn{1}{l|}{40.93}                                                & \multicolumn{1}{l|}{37.5}                                                 & 35.19  & \multicolumn{1}{l|}{\cellcolor[HTML]{CFCFCF}{\color[HTML]{000000} 89.51}} & \multicolumn{1}{l|}{\cellcolor[HTML]{CFCFCF}{\color[HTML]{000000} 90.98}} & 89.73  \\ \midrule
                             & ada  & cepst & \multicolumn{1}{l|}{63.64}                                                & \multicolumn{1}{l|}{93.02}                                                & 92.93  & \multicolumn{1}{l|}{\cellcolor[HTML]{CFCFCF}{\color[HTML]{000000} 94.25}} & \multicolumn{1}{l|}{95.98}                                                & 93.18  & \multicolumn{1}{l|}{\cellcolor[HTML]{CFCFCF}{\color[HTML]{000000} 96.17}} & \multicolumn{1}{l|}{\cellcolor[HTML]{CFCFCF}{\color[HTML]{000000} 96.41}} & 93.77  \\ \cmidrule(l){2-12} 
                             & ada  & mfcc  & \multicolumn{1}{l|}{40}                                                   & \multicolumn{1}{l|}{\cellcolor[HTML]{CFCFCF}{\color[HTML]{000000} 96.43}} & 94.77  & \multicolumn{1}{l|}{91.4}                                                 & \multicolumn{1}{l|}{\cellcolor[HTML]{CFCFCF}{\color[HTML]{000000} 96.15}} & 94.09  & \multicolumn{1}{l|}{84.76}                                                & \multicolumn{1}{l|}{\cellcolor[HTML]{CFCFCF}{\color[HTML]{000000} 96.72}} & 94.03  \\ \cmidrule(l){2-12} 
                             & ada  & spect & \multicolumn{1}{l|}{56}                                                   & \multicolumn{1}{l|}{95.38}                                                & 94.91  & \multicolumn{1}{l|}{85.98}                                                & \multicolumn{1}{l|}{93.89}                                                & 92.7   & \multicolumn{1}{l|}{90.75}                                                & \multicolumn{1}{l|}{95.79}                                                & 93.43  \\ \cmidrule(l){2-12} 
                             & cnn  & time  & \multicolumn{1}{l|}{70}                                                   & \multicolumn{1}{l|}{82.8}                                                 & 85.84  & \multicolumn{1}{l|}{\cellcolor[HTML]{CFCFCF}{\color[HTML]{000000} 95.23}} & \multicolumn{1}{l|}{\cellcolor[HTML]{CFCFCF}{\color[HTML]{000000} 96.3}}  & 93.58  & \multicolumn{1}{l|}{\cellcolor[HTML]{CFCFCF}{\color[HTML]{000000} 98.24}} & \multicolumn{1}{l|}{\cellcolor[HTML]{CFCFCF}{\color[HTML]{000000} 96.77}} & 96.2   \\ \cmidrule(l){2-12} 
                             & gmm  & cepst & \multicolumn{1}{l|}{62.5}                                                 & \multicolumn{1}{l|}{93.16}                                                & 94.85  & \multicolumn{1}{l|}{88.37}                                                & \multicolumn{1}{l|}{94.92}                                                & 94.26  & \multicolumn{1}{l|}{74.23}                                                & \multicolumn{1}{l|}{96.38}                                                & 96     \\ \cmidrule(l){2-12} 
                             & gmm  & mfcc  & \multicolumn{1}{l|}{}                                                     & \multicolumn{1}{l|}{10.78}                                                & 18.43  & \multicolumn{1}{l|}{81.72}                                                & \multicolumn{1}{l|}{83.02}                                                & 76.11  & \multicolumn{1}{l|}{48.1}                                                 & \multicolumn{1}{l|}{66.32}                                                & 75.89  \\ \cmidrule(l){2-12} 
                             & gmm  & spect & \multicolumn{1}{l|}{47.06}                                                & \multicolumn{1}{l|}{94.36}                                                & 95.14  & \multicolumn{1}{l|}{81.6}                                                 & \multicolumn{1}{l|}{85.48}                                                & 83.73  & \multicolumn{1}{l|}{92.95}                                                & \multicolumn{1}{l|}{93.13}                                                & 88.58  \\ \cmidrule(l){2-12} 
                             & lstm & spect & \multicolumn{1}{l|}{69.8}                                                 & \multicolumn{1}{l|}{93.24}                                                & 92.97  & \multicolumn{1}{l|}{89.9}                                                 & \multicolumn{1}{l|}{93.55}                                                & 93.19  & \multicolumn{1}{l|}{92.93}                                                & \multicolumn{1}{l|}{94.08}                                                & 94.03  \\ \cmidrule(l){2-12} 
                             & rf   & cepst & \multicolumn{1}{l|}{\cellcolor[HTML]{CFCFCF}{\color[HTML]{000000} 84.21}} & \multicolumn{1}{l|}{92.35}                                                & 92.9   & \multicolumn{1}{l|}{\cellcolor[HTML]{CFCFCF}{\color[HTML]{000000} 91.61}} & \multicolumn{1}{l|}{94.57}                                                & 94.24  & \multicolumn{1}{l|}{\cellcolor[HTML]{CFCFCF}{\color[HTML]{000000} 97.07}} & \multicolumn{1}{l|}{96.1}                                                 & 94.47  \\ \cmidrule(l){2-12} 
                             & rf   & mfcc  & \multicolumn{1}{l|}{\cellcolor[HTML]{CFCFCF}{\color[HTML]{000000} 74.07}} & \multicolumn{1}{l|}{93.61}                                                & 93.8   & \multicolumn{1}{l|}{89.95}                                                & \multicolumn{1}{l|}{95.68}                                                & 93.58  & \multicolumn{1}{l|}{95.24}                                                & \multicolumn{1}{l|}{95.48}                                                & 95.26  \\ \cmidrule(l){2-12} 
                             & rf   & spect & \multicolumn{1}{l|}{66.67}                                                & \multicolumn{1}{l|}{94.9}                                                 & 95.72  & \multicolumn{1}{l|}{86.03}                                                & \multicolumn{1}{l|}{92.57}                                                & 92.73  & \multicolumn{1}{l|}{93.51}                                                & \multicolumn{1}{l|}{95.15}                                                & 94.47  \\ \cmidrule(l){2-12} 
                             & svm  & cepst & \multicolumn{1}{l|}{43.9}                                                 & \multicolumn{1}{l|}{93.81}                                                & 94.59  & \multicolumn{1}{l|}{78.28}                                                & \multicolumn{1}{l|}{96.05}                                                & 95.3   & \multicolumn{1}{l|}{83.25}                                                & \multicolumn{1}{l|}{95.79}                                                & 96.04  \\ \cmidrule(l){2-12} 
                             & svm  & mfcc  & \multicolumn{1}{l|}{47.62}                                                & \multicolumn{1}{l|}{\cellcolor[HTML]{CFCFCF}{\color[HTML]{000000} 96.18}} & 96.76  & \multicolumn{1}{l|}{86.85}                                                & \multicolumn{1}{l|}{\cellcolor[HTML]{CFCFCF}{\color[HTML]{000000} 96.54}} & 96.62  & \multicolumn{1}{l|}{85.31}                                                & \multicolumn{1}{l|}{96.24}                                                & 95.76  \\ \cmidrule(l){2-12} 
\multirow{-14}{*}{Non-mixed} & svm  & spect & \multicolumn{1}{l|}{\cellcolor[HTML]{CFCFCF}{\color[HTML]{000000} 81.82}} & \multicolumn{1}{l|}{\cellcolor[HTML]{CFCFCF}{\color[HTML]{000000} 95.96}} & 96.81  & \multicolumn{1}{l|}{75.37}                                                & \multicolumn{1}{l|}{76.28}                                                & 75.89  & \multicolumn{1}{l|}{95.36}                                                & \multicolumn{1}{l|}{94.98}                                                & 94.24  \\ \bottomrule
\end{tabular}
}
\end{table}

\begin{table}
\centering
\caption{Precision summarization}
\label{table:results:precision}
\resizebox{\textwidth}{!}{

\begin{tabular}{@{}|l|l|l||lll||lll||lll|@{}}
\toprule
                             &      &       & \multicolumn{3}{l||}{Drug}                                                                                                                                      & \multicolumn{3}{l||}{Exhale}                                                                                                                                    & \multicolumn{3}{l|}{Inhale}                                                                                                                                    \\ \midrule \midrule
                             &      &       & \multicolumn{1}{l|}{LOSO}                                                 & \multicolumn{1}{l|}{Multi}                                                & Single & \multicolumn{1}{l|}{LOSO}                                                 & \multicolumn{1}{l|}{Multi}                                                & Single & \multicolumn{1}{l|}{LOSO}                                                 & \multicolumn{1}{l|}{Multi}                                                & Single \\ \midrule \midrule
                             & ada  & cepst & \multicolumn{1}{l|}{41.77}                                                & \multicolumn{1}{l|}{66.48}                                                & 67.08  & \multicolumn{1}{l|}{77.74}                                                & \multicolumn{1}{l|}{92.15}                                                & 89.46  & \multicolumn{1}{l|}{77.64}                                                & \multicolumn{1}{l|}{83.86}                                                & 81.52  \\ \cmidrule(l){2-12} 
                             & ada  & mfcc  & \multicolumn{1}{l|}{69.21}                                                & \multicolumn{1}{l|}{80.93}                                                & 75.81  & \multicolumn{1}{l|}{78.28}                                                & \multicolumn{1}{l|}{\cellcolor[HTML]{CFCFCF}{\color[HTML]{000000} 93.77}} & 88.73  & \multicolumn{1}{l|}{68.99}                                                & \multicolumn{1}{l|}{84.18}                                                & 79.55  \\ \cmidrule(l){2-12} 
                             & ada  & spect & \multicolumn{1}{l|}{\cellcolor[HTML]{CFCFCF}{\color[HTML]{000000} 89.02}} & \multicolumn{1}{l|}{84.4}                                                 & 80.76  & \multicolumn{1}{l|}{71.28}                                                & \multicolumn{1}{l|}{89.81}                                                & 87.47  & \multicolumn{1}{l|}{76.94}                                                & \multicolumn{1}{l|}{\cellcolor[HTML]{CFCFCF}{\color[HTML]{000000} 90.66}} & 86.23  \\ \cmidrule(l){2-12} 
                             & cnn  & time  & \multicolumn{1}{l|}{66.77}                                                & \multicolumn{1}{l|}{65.32}                                                & 65.06  & \multicolumn{1}{l|}{\cellcolor[HTML]{CFCFCF}{\color[HTML]{000000} 86.71}} & \multicolumn{1}{l|}{88.83}                                                & 83.73  & \multicolumn{1}{l|}{\cellcolor[HTML]{CFCFCF}{\color[HTML]{000000} 93.48}} & \multicolumn{1}{l|}{85.85}                                                & 87.79  \\ \cmidrule(l){2-12} 
                             & gmm  & cepst & \multicolumn{1}{l|}{42.68}                                                & \multicolumn{1}{l|}{72.98}                                                & 72.17  & \multicolumn{1}{l|}{64.03}                                                & \multicolumn{1}{l|}{90.18}                                                & 87.91  & \multicolumn{1}{l|}{32.01}                                                & \multicolumn{1}{l|}{75.49}                                                & 76.57  \\ \cmidrule(l){2-12} 
                             & gmm  & mfcc  & \multicolumn{1}{l|}{0}                                                    & \multicolumn{1}{l|}{11.02}                                                & 0.24   & \multicolumn{1}{l|}{\cellcolor[HTML]{CFCFCF}{\color[HTML]{000000} 94.45}} & \multicolumn{1}{l|}{\cellcolor[HTML]{CFCFCF}{\color[HTML]{000000} 97.04}} & 97.15  & \multicolumn{1}{l|}{26.55}                                                & \multicolumn{1}{l|}{39.33}                                                & 47.21  \\ \cmidrule(l){2-12} 
                             & gmm  & spect & \multicolumn{1}{l|}{72.87}                                                & \multicolumn{1}{l|}{\cellcolor[HTML]{CFCFCF}{\color[HTML]{000000} 85.8}}  & 86.02  & \multicolumn{1}{l|}{63.38}                                                & \multicolumn{1}{l|}{74.29}                                                & 78.71  & \multicolumn{1}{l|}{\cellcolor[HTML]{CFCFCF}{\color[HTML]{000000} 87}}    & \multicolumn{1}{l|}{\cellcolor[HTML]{CFCFCF}{\color[HTML]{000000} 93.08}} & 92.38  \\ \cmidrule(l){2-12} 
                             & lstm & spect & \multicolumn{1}{l|}{\cellcolor[HTML]{CFCFCF}{\color[HTML]{000000} 86.36}} & \multicolumn{1}{l|}{79.72}                                                & 79.24  & \multicolumn{1}{l|}{69.09}                                                & \multicolumn{1}{l|}{89.64}                                                & 87.39  & \multicolumn{1}{l|}{79.28}                                                & \multicolumn{1}{l|}{\cellcolor[HTML]{CFCFCF}{\color[HTML]{000000} 90.5}}  & 89.34  \\ \cmidrule(l){2-12} 
                             & rf   & cepst & \multicolumn{1}{l|}{50}                                                   & \multicolumn{1}{l|}{61.91}                                                & 63.75  & \multicolumn{1}{l|}{77.3}                                                 & \multicolumn{1}{l|}{92.36}                                                & 90.53  & \multicolumn{1}{l|}{78}                                                   & \multicolumn{1}{l|}{83.66}                                                & 84.26  \\ \cmidrule(l){2-12} 
                             & rf   & mfcc  & \multicolumn{1}{l|}{74.09}                                                & \multicolumn{1}{l|}{77.08}                                                & 77.09  & \multicolumn{1}{l|}{76.06}                                                & \multicolumn{1}{l|}{92.95}                                                & 91.7   & \multicolumn{1}{l|}{74.97}                                                & \multicolumn{1}{l|}{80.89}                                                & 80.06  \\ \cmidrule(l){2-12} 
                             & rf   & spect & \multicolumn{1}{l|}{85.06}                                                & \multicolumn{1}{l|}{\cellcolor[HTML]{CFCFCF}{\color[HTML]{000000} 84.69}} & 83.82  & \multicolumn{1}{l|}{73.42}                                                & \multicolumn{1}{l|}{87.85}                                                & 89.28  & \multicolumn{1}{l|}{82.89}                                                & \multicolumn{1}{l|}{89.4}                                                 & 88.49  \\ \cmidrule(l){2-12} 
                             & svm  & cepst & \multicolumn{1}{l|}{57.62}                                                & \multicolumn{1}{l|}{72.18}                                                & 73.97  & \multicolumn{1}{l|}{56.12}                                                & \multicolumn{1}{l|}{91.84}                                                & 91.58  & \multicolumn{1}{l|}{52.78}                                                & \multicolumn{1}{l|}{83.75}                                                & 84.42  \\ \cmidrule(l){2-12} 
                             & svm  & mfcc  & \multicolumn{1}{l|}{76.83}                                                & \multicolumn{1}{l|}{83.96}                                                & 83.73  & \multicolumn{1}{l|}{70.42}                                                & \multicolumn{1}{l|}{93.08}                                                & 93.26  & \multicolumn{1}{l|}{68.55}                                                & \multicolumn{1}{l|}{83.03}                                                & 82.73  \\ \cmidrule(l){2-12} 
\multirow{-14}{*}{Mixed}     & svm  & spect & \multicolumn{1}{l|}{\cellcolor[HTML]{CFCFCF}{\color[HTML]{000000} 90.85}} & \multicolumn{1}{l|}{\cellcolor[HTML]{CFCFCF}{\color[HTML]{000000} 85.63}} & 85.57  & \multicolumn{1}{l|}{\cellcolor[HTML]{CFCFCF}{\color[HTML]{000000} 97.24}} & \multicolumn{1}{l|}{\cellcolor[HTML]{CFCFCF}{\color[HTML]{000000} 98.49}} & 97.5   & \multicolumn{1}{l|}{\cellcolor[HTML]{CFCFCF}{\color[HTML]{000000} 84.68}} & \multicolumn{1}{l|}{89.17}                                                & 87.1   \\ \midrule \midrule
                             & ada  & cepst & \multicolumn{1}{l|}{70}                                                   & \multicolumn{1}{l|}{93.26}                                                & 93.44  & \multicolumn{1}{l|}{89.87}                                                & \multicolumn{1}{l|}{\cellcolor[HTML]{CFCFCF}{\color[HTML]{000000} 96.92}} & 93.67  & \multicolumn{1}{l|}{\cellcolor[HTML]{CFCFCF}{\color[HTML]{000000} 94.96}} & \multicolumn{1}{l|}{\cellcolor[HTML]{CFCFCF}{\color[HTML]{000000} 96.87}} & 94     \\ \cmidrule(l){2-12} 
                             & ada  & mfcc  & \multicolumn{1}{l|}{90}                                                   & \multicolumn{1}{l|}{\cellcolor[HTML]{CFCFCF}{\color[HTML]{000000} 97.93}} & 93.99  & \multicolumn{1}{l|}{85.23}                                                & \multicolumn{1}{l|}{96.77}                                                & 93.47  & \multicolumn{1}{l|}{74.79}                                                & \multicolumn{1}{l|}{\cellcolor[HTML]{CFCFCF}{\color[HTML]{000000} 97.18}} & 94.5   \\ \cmidrule(l){2-12} 
                             & ada  & spect & \multicolumn{1}{l|}{70}                                                   & \multicolumn{1}{l|}{96.37}                                                & 96.72  & \multicolumn{1}{l|}{78.9}                                                 & \multicolumn{1}{l|}{94.19}                                                & 94.52  & \multicolumn{1}{l|}{86.55}                                                & \multicolumn{1}{l|}{96.24}                                                & 92.5   \\ \cmidrule(l){2-12} 
                             & cnn  & time  & \multicolumn{1}{l|}{63.64}                                                & \multicolumn{1}{l|}{83.75}                                                & 88.06  & \multicolumn{1}{l|}{\cellcolor[HTML]{CFCFCF}{\color[HTML]{000000} 91.83}} & \multicolumn{1}{l|}{96.54}                                                & 92.26  & \multicolumn{1}{l|}{\cellcolor[HTML]{CFCFCF}{\color[HTML]{000000} 98.07}} & \multicolumn{1}{l|}{\cellcolor[HTML]{CFCFCF}{\color[HTML]{000000} 97.19}} & 96.32  \\ \cmidrule(l){2-12} 
                             & gmm  & cepst & \multicolumn{1}{l|}{50}                                                   & \multicolumn{1}{l|}{95.34}                                                & 95.63  & \multicolumn{1}{l|}{80.17}                                                & \multicolumn{1}{l|}{95.62}                                                & 95.25  & \multicolumn{1}{l|}{60.5}                                                 & \multicolumn{1}{l|}{95.92}                                                & 96     \\ \cmidrule(l){2-12} 
                             & gmm  & mfcc  & \multicolumn{1}{l|}{0}                                                    & \multicolumn{1}{l|}{5.7}                                                  & 10.93  & \multicolumn{1}{l|}{\cellcolor[HTML]{CFCFCF}{\color[HTML]{000000} 96.2}}  & \multicolumn{1}{l|}{\cellcolor[HTML]{CFCFCF}{\color[HTML]{000000} 99.35}} & 98.17  & \multicolumn{1}{l|}{31.93}                                                & \multicolumn{1}{l|}{60.82}                                                & 85     \\ \cmidrule(l){2-12} 
                             & gmm  & spect & \multicolumn{1}{l|}{40}                                                   & \multicolumn{1}{l|}{95.34}                                                & 96.17  & \multicolumn{1}{l|}{77.64}                                                & \multicolumn{1}{l|}{84.52}                                                & 83.29  & \multicolumn{1}{l|}{94.12}                                                & \multicolumn{1}{l|}{95.61}                                                & 95     \\ \cmidrule(l){2-12} 
                             & lstm & spect & \multicolumn{1}{l|}{\cellcolor[HTML]{CFCFCF}{\color[HTML]{000000} 98.11}} & \multicolumn{1}{l|}{95.27}                                                & 95.15  & \multicolumn{1}{l|}{86.32}                                                & \multicolumn{1}{l|}{93.27}                                                & 93.74  & \multicolumn{1}{l|}{88.42}                                                & \multicolumn{1}{l|}{95.26}                                                & 94.94  \\ \cmidrule(l){2-12} 
                             & rf   & cepst & \multicolumn{1}{l|}{80}                                                   & \multicolumn{1}{l|}{90.67}                                                & 92.9   & \multicolumn{1}{l|}{85.23}                                                & \multicolumn{1}{l|}{96.1}                                                 & 94.99  & \multicolumn{1}{l|}{\cellcolor[HTML]{CFCFCF}{\color[HTML]{000000} 97.48}} & \multicolumn{1}{l|}{96.55}                                                & 94     \\ \cmidrule(l){2-12} 
                             & rf   & mfcc  & \multicolumn{1}{l|}{\cellcolor[HTML]{CFCFCF}{\color[HTML]{000000} 100}}   & \multicolumn{1}{l|}{94.82}                                                & 95.08  & \multicolumn{1}{l|}{83.12}                                                & \multicolumn{1}{l|}{96.45}                                                & 93.21  & \multicolumn{1}{l|}{92.44}                                                & \multicolumn{1}{l|}{95.92}                                                & 95.5   \\ \cmidrule(l){2-12} 
                             & rf   & spect & \multicolumn{1}{l|}{80}                                                   & \multicolumn{1}{l|}{96.37}                                                & 97.81  & \multicolumn{1}{l|}{83.12}                                                & \multicolumn{1}{l|}{92.42}                                                & 91.64  & \multicolumn{1}{l|}{90.76}                                                & \multicolumn{1}{l|}{95.3}                                                 & 94     \\ \cmidrule(l){2-12} 
                             & svm  & cepst & \multicolumn{1}{l|}{90}                                                   & \multicolumn{1}{l|}{94.3}                                                 & 95.63  & \multicolumn{1}{l|}{65.4}                                                 & \multicolumn{1}{l|}{96.75}                                                & 96.31  & \multicolumn{1}{l|}{73.11}                                                & \multicolumn{1}{l|}{96.24}                                                & 97     \\ \cmidrule(l){2-12} 
                             & svm  & mfcc  & \multicolumn{1}{l|}{\cellcolor[HTML]{CFCFCF}{\color[HTML]{000000} 100}}   & \multicolumn{1}{l|}{\cellcolor[HTML]{CFCFCF}{\color[HTML]{000000} 97.93}} & 97.81  & \multicolumn{1}{l|}{78.06}                                                & \multicolumn{1}{l|}{96.77}                                                & 97.13  & \multicolumn{1}{l|}{75.63}                                                & \multicolumn{1}{l|}{96.24}                                                & 96     \\ \cmidrule(l){2-12} 
\multirow{-14}{*}{Non-mixed} & svm  & spect & \multicolumn{1}{l|}{90}                                                   & \multicolumn{1}{l|}{\cellcolor[HTML]{CFCFCF}{\color[HTML]{000000} 98.45}} & 99.45  & \multicolumn{1}{l|}{\cellcolor[HTML]{CFCFCF}{\color[HTML]{000000} 97.47}} & \multicolumn{1}{l|}{\cellcolor[HTML]{CFCFCF}{\color[HTML]{000000} 98.55}} & 97.39  & \multicolumn{1}{l|}{\cellcolor[HTML]{CFCFCF}{\color[HTML]{000000} 94.96}} & \multicolumn{1}{l|}{94.98}                                                & 94     \\ \bottomrule
\end{tabular}

}
\end{table}

\begin{table}
\centering
\caption{Recall summarization}
\label{table:results:recall}
\resizebox{\textwidth}{!}{
\begin{tabular}{@{}|l|l|l||lll||lll||lll|@{}}
\toprule
                             &      &       & \multicolumn{3}{l||}{Drug}                                                                                                                                      & \multicolumn{3}{l||}{Exhale}                                                                                                                                    & \multicolumn{3}{l|}{Inhale}                                                                                                                                    \\ \midrule \midrule
                             &      &       & \multicolumn{1}{l|}{LOSO}                                                 & \multicolumn{1}{l|}{Multi}                                                & Single & \multicolumn{1}{l|}{LOSO}                                                 & \multicolumn{1}{l|}{Multi}                                                & Single & \multicolumn{1}{l|}{LOSO}                                                 & \multicolumn{1}{l|}{Multi}                                                & Single \\ \midrule \midrule
                             & ada  & cepst & \multicolumn{1}{l|}{41.77}                                                & \multicolumn{1}{l|}{66.48}                                                & 67.08  & \multicolumn{1}{l|}{77.74}                                                & \multicolumn{1}{l|}{92.15}                                                & 89.46  & \multicolumn{1}{l|}{77.64}                                                & \multicolumn{1}{l|}{83.86}                                                & 81.52  \\ \cmidrule(l){2-12} 
                             & ada  & mfcc  & \multicolumn{1}{l|}{69.21}                                                & \multicolumn{1}{l|}{80.93}                                                & 75.81  & \multicolumn{1}{l|}{78.28}                                                & \multicolumn{1}{l|}{\cellcolor[HTML]{CFCFCF}{\color[HTML]{000000} 93.77}} & 88.73  & \multicolumn{1}{l|}{68.99}                                                & \multicolumn{1}{l|}{84.18}                                                & 79.55  \\ \cmidrule(l){2-12} 
                             & ada  & spect & \multicolumn{1}{l|}{\cellcolor[HTML]{CFCFCF}{\color[HTML]{000000} 89.02}} & \multicolumn{1}{l|}{84.4}                                                 & 80.76  & \multicolumn{1}{l|}{71.28}                                                & \multicolumn{1}{l|}{89.81}                                                & 87.47  & \multicolumn{1}{l|}{76.94}                                                & \multicolumn{1}{l|}{\cellcolor[HTML]{CFCFCF}{\color[HTML]{000000} 90.66}} & 86.23  \\ \cmidrule(l){2-12} 
                             & cnn  & time  & \multicolumn{1}{l|}{66.77}                                                & \multicolumn{1}{l|}{65.32}                                                & 65.06  & \multicolumn{1}{l|}{\cellcolor[HTML]{CFCFCF}{\color[HTML]{000000} 86.71}} & \multicolumn{1}{l|}{88.83}                                                & 83.73  & \multicolumn{1}{l|}{\cellcolor[HTML]{CFCFCF}{\color[HTML]{000000} 93.48}} & \multicolumn{1}{l|}{85.85}                                                & 87.79  \\ \cmidrule(l){2-12} 
                             & gmm  & cepst & \multicolumn{1}{l|}{42.68}                                                & \multicolumn{1}{l|}{72.98}                                                & 72.17  & \multicolumn{1}{l|}{64.03}                                                & \multicolumn{1}{l|}{90.18}                                                & 87.91  & \multicolumn{1}{l|}{32.01}                                                & \multicolumn{1}{l|}{75.49}                                                & 76.57  \\ \cmidrule(l){2-12} 
                             & gmm  & mfcc  & \multicolumn{1}{l|}{0}                                                    & \multicolumn{1}{l|}{11.02}                                                & 0.24   & \multicolumn{1}{l|}{\cellcolor[HTML]{CFCFCF}{\color[HTML]{000000} 94.45}} & \multicolumn{1}{l|}{\cellcolor[HTML]{CFCFCF}{\color[HTML]{000000} 97.04}} & 97.15  & \multicolumn{1}{l|}{26.55}                                                & \multicolumn{1}{l|}{39.33}                                                & 47.21  \\ \cmidrule(l){2-12} 
                             & gmm  & spect & \multicolumn{1}{l|}{72.87}                                                & \multicolumn{1}{l|}{\cellcolor[HTML]{CFCFCF}{\color[HTML]{000000} 85.8}}  & 86.02  & \multicolumn{1}{l|}{63.38}                                                & \multicolumn{1}{l|}{74.29}                                                & 78.71  & \multicolumn{1}{l|}{\cellcolor[HTML]{CFCFCF}{\color[HTML]{000000} 87}}    & \multicolumn{1}{l|}{\cellcolor[HTML]{CFCFCF}{\color[HTML]{000000} 93.08}} & 92.38  \\ \cmidrule(l){2-12} 
                             & lstm & spect & \multicolumn{1}{l|}{\cellcolor[HTML]{CFCFCF}{\color[HTML]{000000} 86.36}} & \multicolumn{1}{l|}{79.72}                                                & 79.24  & \multicolumn{1}{l|}{69.09}                                                & \multicolumn{1}{l|}{89.64}                                                & 87.39  & \multicolumn{1}{l|}{79.28}                                                & \multicolumn{1}{l|}{\cellcolor[HTML]{CFCFCF}{\color[HTML]{000000} 90.5}}  & 89.34  \\ \cmidrule(l){2-12} 
                             & rf   & cepst & \multicolumn{1}{l|}{50}                                                   & \multicolumn{1}{l|}{61.91}                                                & 63.75  & \multicolumn{1}{l|}{77.3}                                                 & \multicolumn{1}{l|}{92.36}                                                & 90.53  & \multicolumn{1}{l|}{78}                                                   & \multicolumn{1}{l|}{83.66}                                                & 84.26  \\ \cmidrule(l){2-12} 
                             & rf   & mfcc  & \multicolumn{1}{l|}{74.09}                                                & \multicolumn{1}{l|}{77.08}                                                & 77.09  & \multicolumn{1}{l|}{76.06}                                                & \multicolumn{1}{l|}{92.95}                                                & 91.7   & \multicolumn{1}{l|}{74.97}                                                & \multicolumn{1}{l|}{80.89}                                                & 80.06  \\ \cmidrule(l){2-12} 
                             & rf   & spect & \multicolumn{1}{l|}{85.06}                                                & \multicolumn{1}{l|}{\cellcolor[HTML]{CFCFCF}{\color[HTML]{000000} 84.69}} & 83.82  & \multicolumn{1}{l|}{73.42}                                                & \multicolumn{1}{l|}{87.85}                                                & 89.28  & \multicolumn{1}{l|}{82.89}                                                & \multicolumn{1}{l|}{89.4}                                                 & 88.49  \\ \cmidrule(l){2-12} 
                             & svm  & cepst & \multicolumn{1}{l|}{57.62}                                                & \multicolumn{1}{l|}{72.18}                                                & 73.97  & \multicolumn{1}{l|}{56.12}                                                & \multicolumn{1}{l|}{91.84}                                                & 91.58  & \multicolumn{1}{l|}{52.78}                                                & \multicolumn{1}{l|}{83.75}                                                & 84.42  \\ \cmidrule(l){2-12} 
                             & svm  & mfcc  & \multicolumn{1}{l|}{76.83}                                                & \multicolumn{1}{l|}{83.96}                                                & 83.73  & \multicolumn{1}{l|}{70.42}                                                & \multicolumn{1}{l|}{93.08}                                                & 93.26  & \multicolumn{1}{l|}{68.55}                                                & \multicolumn{1}{l|}{83.03}                                                & 82.73  \\ \cmidrule(l){2-12} 
\multirow{-14}{*}{Mixed}     & svm  & spect & \multicolumn{1}{l|}{\cellcolor[HTML]{CFCFCF}{\color[HTML]{000000} 90.85}} & \multicolumn{1}{l|}{\cellcolor[HTML]{CFCFCF}{\color[HTML]{000000} 85.63}} & 85.57  & \multicolumn{1}{l|}{\cellcolor[HTML]{CFCFCF}{\color[HTML]{000000} 97.24}} & \multicolumn{1}{l|}{\cellcolor[HTML]{CFCFCF}{\color[HTML]{000000} 98.49}} & 97.5   & \multicolumn{1}{l|}{\cellcolor[HTML]{CFCFCF}{\color[HTML]{000000} 84.68}} & \multicolumn{1}{l|}{89.17}                                                & 87.1   \\ \midrule \midrule
                             & ada  & cepst & \multicolumn{1}{l|}{70}                                                   & \multicolumn{1}{l|}{93.26}                                                & 93.44  & \multicolumn{1}{l|}{89.87}                                                & \multicolumn{1}{l|}{\cellcolor[HTML]{CFCFCF}{\color[HTML]{000000} 96.92}} & 93.67  & \multicolumn{1}{l|}{\cellcolor[HTML]{CFCFCF}{\color[HTML]{000000} 94.96}} & \multicolumn{1}{l|}{\cellcolor[HTML]{CFCFCF}{\color[HTML]{000000} 96.87}} & 94     \\ \cmidrule(l){2-12} 
                             & ada  & mfcc  & \multicolumn{1}{l|}{90}                                                   & \multicolumn{1}{l|}{\cellcolor[HTML]{CFCFCF}{\color[HTML]{000000} 97.93}} & 93.99  & \multicolumn{1}{l|}{85.23}                                                & \multicolumn{1}{l|}{96.77}                                                & 93.47  & \multicolumn{1}{l|}{74.79}                                                & \multicolumn{1}{l|}{\cellcolor[HTML]{CFCFCF}{\color[HTML]{000000} 97.18}} & 94.5   \\ \cmidrule(l){2-12} 
                             & ada  & spect & \multicolumn{1}{l|}{70}                                                   & \multicolumn{1}{l|}{96.37}                                                & 96.72  & \multicolumn{1}{l|}{78.9}                                                 & \multicolumn{1}{l|}{94.19}                                                & 94.52  & \multicolumn{1}{l|}{86.55}                                                & \multicolumn{1}{l|}{96.24}                                                & 92.5   \\ \cmidrule(l){2-12} 
                             & cnn  & time  & \multicolumn{1}{l|}{63.64}                                                & \multicolumn{1}{l|}{83.75}                                                & 88.06  & \multicolumn{1}{l|}{\cellcolor[HTML]{CFCFCF}{\color[HTML]{000000} 91.83}} & \multicolumn{1}{l|}{96.54}                                                & 92.26  & \multicolumn{1}{l|}{\cellcolor[HTML]{CFCFCF}{\color[HTML]{000000} 98.07}} & \multicolumn{1}{l|}{\cellcolor[HTML]{CFCFCF}{\color[HTML]{000000} 97.19}} & 96.32  \\ \cmidrule(l){2-12} 
                             & gmm  & cepst & \multicolumn{1}{l|}{50}                                                   & \multicolumn{1}{l|}{95.34}                                                & 95.63  & \multicolumn{1}{l|}{80.17}                                                & \multicolumn{1}{l|}{95.62}                                                & 95.25  & \multicolumn{1}{l|}{60.5}                                                 & \multicolumn{1}{l|}{95.92}                                                & 96     \\ \cmidrule(l){2-12} 
                             & gmm  & mfcc  & \multicolumn{1}{l|}{0}                                                    & \multicolumn{1}{l|}{5.7}                                                  & 10.93  & \multicolumn{1}{l|}{\cellcolor[HTML]{CFCFCF}{\color[HTML]{000000} 96.2}}  & \multicolumn{1}{l|}{\cellcolor[HTML]{CFCFCF}{\color[HTML]{000000} 99.35}} & 98.17  & \multicolumn{1}{l|}{31.93}                                                & \multicolumn{1}{l|}{60.82}                                                & 85     \\ \cmidrule(l){2-12} 
                             & gmm  & spect & \multicolumn{1}{l|}{40}                                                   & \multicolumn{1}{l|}{95.34}                                                & 96.17  & \multicolumn{1}{l|}{77.64}                                                & \multicolumn{1}{l|}{84.52}                                                & 83.29  & \multicolumn{1}{l|}{94.12}                                                & \multicolumn{1}{l|}{95.61}                                                & 95     \\ \cmidrule(l){2-12} 
                             & lstm & spect & \multicolumn{1}{l|}{\cellcolor[HTML]{CFCFCF}{\color[HTML]{000000} 98.11}} & \multicolumn{1}{l|}{95.27}                                                & 95.15  & \multicolumn{1}{l|}{86.32}                                                & \multicolumn{1}{l|}{93.27}                                                & 93.74  & \multicolumn{1}{l|}{88.42}                                                & \multicolumn{1}{l|}{95.26}                                                & 94.94  \\ \cmidrule(l){2-12} 
                             & rf   & cepst & \multicolumn{1}{l|}{80}                                                   & \multicolumn{1}{l|}{90.67}                                                & 92.9   & \multicolumn{1}{l|}{85.23}                                                & \multicolumn{1}{l|}{96.1}                                                 & 94.99  & \multicolumn{1}{l|}{\cellcolor[HTML]{CFCFCF}{\color[HTML]{000000} 97.48}} & \multicolumn{1}{l|}{96.55}                                                & 94     \\ \cmidrule(l){2-12} 
                             & rf   & mfcc  & \multicolumn{1}{l|}{\cellcolor[HTML]{CFCFCF}{\color[HTML]{000000} 100}}   & \multicolumn{1}{l|}{94.82}                                                & 95.08  & \multicolumn{1}{l|}{83.12}                                                & \multicolumn{1}{l|}{96.45}                                                & 93.21  & \multicolumn{1}{l|}{92.44}                                                & \multicolumn{1}{l|}{95.92}                                                & 95.5   \\ \cmidrule(l){2-12} 
                             & rf   & spect & \multicolumn{1}{l|}{80}                                                   & \multicolumn{1}{l|}{96.37}                                                & 97.81  & \multicolumn{1}{l|}{83.12}                                                & \multicolumn{1}{l|}{92.42}                                                & 91.64  & \multicolumn{1}{l|}{90.76}                                                & \multicolumn{1}{l|}{95.3}                                                 & 94     \\ \cmidrule(l){2-12} 
                             & svm  & cepst & \multicolumn{1}{l|}{90}                                                   & \multicolumn{1}{l|}{94.3}                                                 & 95.63  & \multicolumn{1}{l|}{65.4}                                                 & \multicolumn{1}{l|}{96.75}                                                & 96.31  & \multicolumn{1}{l|}{73.11}                                                & \multicolumn{1}{l|}{96.24}                                                & 97     \\ \cmidrule(l){2-12} 
                             & svm  & mfcc  & \multicolumn{1}{l|}{\cellcolor[HTML]{CFCFCF}{\color[HTML]{000000} 100}}   & \multicolumn{1}{l|}{\cellcolor[HTML]{CFCFCF}{\color[HTML]{000000} 97.93}} & 97.81  & \multicolumn{1}{l|}{78.06}                                                & \multicolumn{1}{l|}{96.77}                                                & 97.13  & \multicolumn{1}{l|}{75.63}                                                & \multicolumn{1}{l|}{96.24}                                                & 96     \\ \cmidrule(l){2-12} 
\multirow{-14}{*}{Non-mixed} & svm  & spect & \multicolumn{1}{l|}{90}                                                   & \multicolumn{1}{l|}{\cellcolor[HTML]{CFCFCF}{\color[HTML]{000000} 98.45}} & 99.45  & \multicolumn{1}{l|}{\cellcolor[HTML]{CFCFCF}{\color[HTML]{000000} 97.47}} & \multicolumn{1}{l|}{\cellcolor[HTML]{CFCFCF}{\color[HTML]{000000} 98.55}} & 97.39  & \multicolumn{1}{l|}{\cellcolor[HTML]{CFCFCF}{\color[HTML]{000000} 94.96}} & \multicolumn{1}{l|}{94.98}                                                & 94     \\ \bottomrule
\end{tabular}
}
\end{table}

\begin{table}
\centering
\caption{Classification accuracy in relevant state of the art}
\label{table:result:relevant_soa}
\resizebox{\textwidth}{!}{
\begin{tabular}{@{}|p{10cm}|p{3cm}|p{3cm}|p{3cm}|@{}}
\toprule
                    \textbf{Method} & \textbf{Drug} & \textbf{Inhale} & \textbf{Exhale} \\ \midrule
Holmes et al. (2012) \cite{holmes2012automatic}      & 89            & -               & -               \\ \midrule
Holmes et al. (2013-14) \cite{holmes2013acoustic,holmes2014acoustic}    & 92.1          & 91.7            & 93.7            \\ \midrule
Taylor et al. (2017) / QDA \cite{taylor2018advances} & 88.2          & -               & -               \\ \midrule
Taylor et al. (2017) / ANN \cite{taylor2018advances} & 65.6          & -               & -               \\ \bottomrule
\end{tabular}
}
\end{table}

In order to better assess the contribution of the proposed approach, we first summarize in Table \ref{table:result:relevant_soa} the classification performance of state-of-the-art algorithms. In more detail, Holmes et al. \cite{holmes2012automatic} presented, in 2012, a method that differentiates blister and non-blister events with an accuracy of 89.0\%. A year later, Holmes et al. \cite{holmes2013acoustic, holmes2014acoustic} also developed an algorithm that recognizes blister events and breath events (with an accuracy of 92.1\%) and separates inhalations from exhalations (with an accuracy of more than 90\%). Later, Taylor et al. developed two main algorithms for blister detection \cite{taylor2018advances, taylor2018objective} based on Quadratic Discriminant Analysis and ANN and achieved an accuracy of 88.2\% and 65.6\%, respectively. 
From Tables \ref{table:result:relevant_soa}, \ref{table:results:f1}, \ref{table:results:Accuracy} it is apparent that the classification accuracy achieved by our approach does not exceed the performance of the relevant state of the art approaches. Our approach performs similarly to the methods developed by Holmes et al. \cite{holmes2012automatic, holmes2013acoustic}, Taylor et al. \cite{taylor2014acoustic}, and Pettas et al. \cite{pettas2019recognition}. However, the approach of Nousias et al. \cite{nousias2018mhealth} outperforms our algorithm, mainly for the drug and environmental noise classes. However, the utilization of a CNN architecture in the time domain allows for an implicit signal representation that circumvents the need for additional feature extraction (e.g. in the spectral domain) and results in significantly lower execution times.

\subsubsection{Performance evaluation}

\begin{figure}[H]
\centering
 \includegraphics[width=0.8\linewidth]{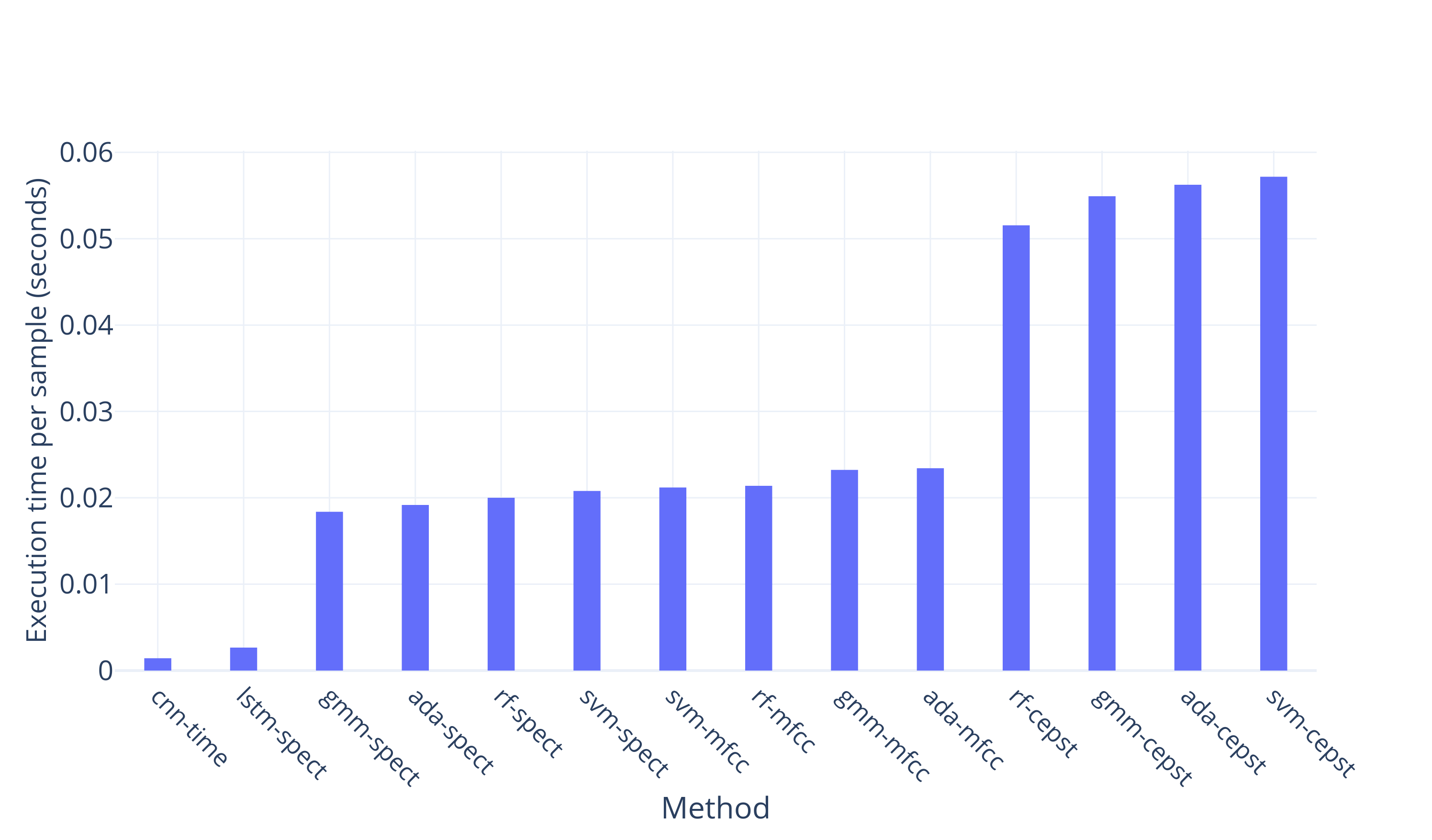}
\caption{Comparison of the computational cost of CNN based approach  approach and other studies. }
\label{fig:execution-time}  
 \end{figure}

We compare the computational cost of the CNN Model with other approaches executed in the same machine (Intel(R) Core(TM) i5-5250U CPU @ 2.7GHz). The results are summarized in Figure \ref{fig:execution-time}. 
This figure highlights our approach's gain in computational speedup compared to the time-consuming Spectrogram and similar feature-based algorithms. Specifically, Figure \ref{fig:execution-time} shows that classification by CNN is 40 times faster than the slowest Cepstrogram-based methods and fifteen times faster than Spectrogram and MFCC-based methods. Using multiple features certainly has a multiplying effect.     

\begin{table}[t!]
\centering
\caption{Comparison of the computational cost of CNN based approach  approach and other studies. }
\label{table:result:performance_comparison}
\resizebox{\textwidth}{!}{
\begin{tabular}{@{}|p{4cm}|l|l|p{4cm}|@{}}
\toprule
Method     & Classification execution time & Feature extraction Time & Sum    \\ \midrule
cnn-time   & 0.000460227                   & 0.000968868             & 0.0014 \\ \midrule
lstm-spect & 0.000700771                   & 0.001955032             & 0.0027 \\ \midrule
gmm-spect  & 1.82257E-05                   & 0.018367229             & 0.0184 \\ \midrule
ada-spect  & 0.000773556                   & 0.018398599             & 0.0192 \\ \midrule
rf-spect   & 0.000278274                   & 0.019732354             & 0.02   \\ \midrule
svm-spect  & 4.70897E-05                   & 0.02074642              & 0.0208 \\ \midrule
svm-mfcc   & 2.1926E-05                    & 0.021171822             & 0.0212 \\ \midrule
rf-mfcc    & 0.000275223                   & 0.021114178             & 0.0214 \\ \midrule
gmm-mfcc   & 2.35955E-05                   & 0.023210378             & 0.0232 \\ \midrule
ada-mfcc   & 0.000852494                   & 0.022573248             & 0.0234 \\ \midrule
rf-cepst   & 0.000272068                   & 0.051282082             & 0.0516 \\ \midrule
gmm-cepst  & 1.19811E-05                   & 0.054902331             & 0.0549 \\ \midrule
ada-cepst  & 0.0008844                     & 0.055359981             & 0.0562 \\ \midrule
svm-cepst  & 3.10628E-05                   & 0.057136457             & 0.0572 \\ \bottomrule
\end{tabular}
}
\end{table}

\section{User interfaces for medication adherence monitoring and relevance feedback}

In order to facilitate monitoring data collection and experimental evaluation, an application for mobile devices was developed. The application connects to the microphone mounted on the inhaler case using Bluetooth. Initially, the user is presented with guidelines on how to use the application visualized in Figure \ref{fig:mobileUI_3}A. In the next view, \ref{fig:mobileUI_3}B, the main interface appears where the user needs to activate the inhaler device recording through the record button. The visualization consists of 24 colour-coded rectangles corresponding to $0.5$ seconds. At any given time, the user can also stop and restart the application through the Clear/Reset functionality. As soon as the recording initializes, the Spectrogram also appears for cross-checking and debugging purposes. After twelve seconds, the user receives a notification that the audio capture is complete. Afterwards, the user can activate the "Audio analysis" functionality that initializes communication with a cloud server where the processing and the classification take place. The outcome is sent back to the mobile device and visualized to the user.

A \textit{relevance feedback} mechanism can be afterwards activated. The user interacts with the decision with the 24 colour-coded rectangles mentioned above and can press a particular rectangle and alter the outcome. Afterwards, the user can upload the annotation for the relevant audio segments to the server to be included in the following training sessions. Including user-defined input within the training dataset significantly increases accuracy, as experimental evaluation in Table \ref{table:results:f1} reveals, since the used accuracy will correspond to the "Multi-Subject" case instead of the "LOSO" case.

\begin{figure}[t!]
\centering
\includegraphics[width=\linewidth]{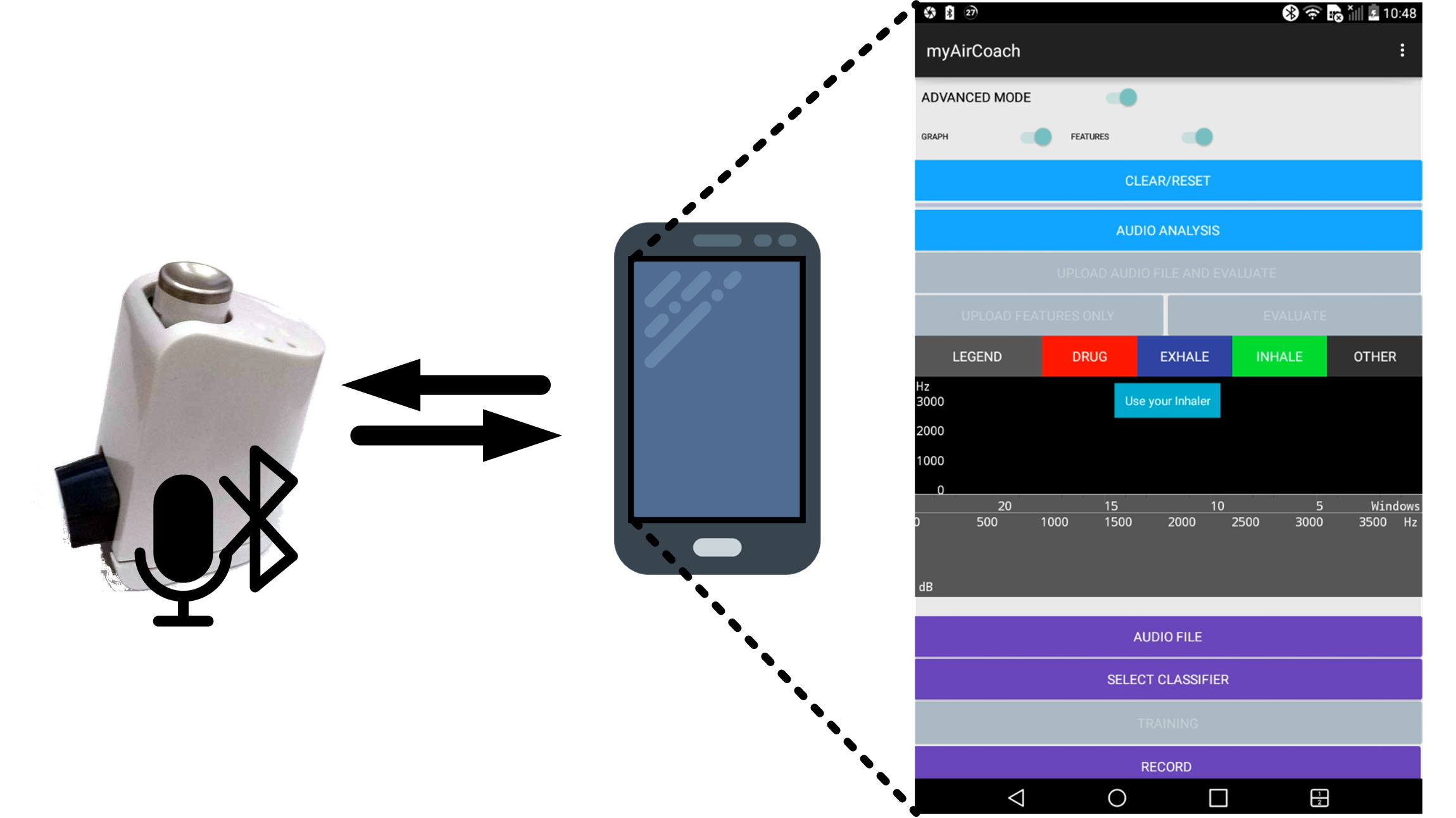}
\caption{Visualization of the relation between the inhaler device, the mobile device and the application.}
\label{fig:mobileUI_2}
 \end{figure}
 
\begin{figure}[t!]
\centering
\includegraphics[width=\linewidth]{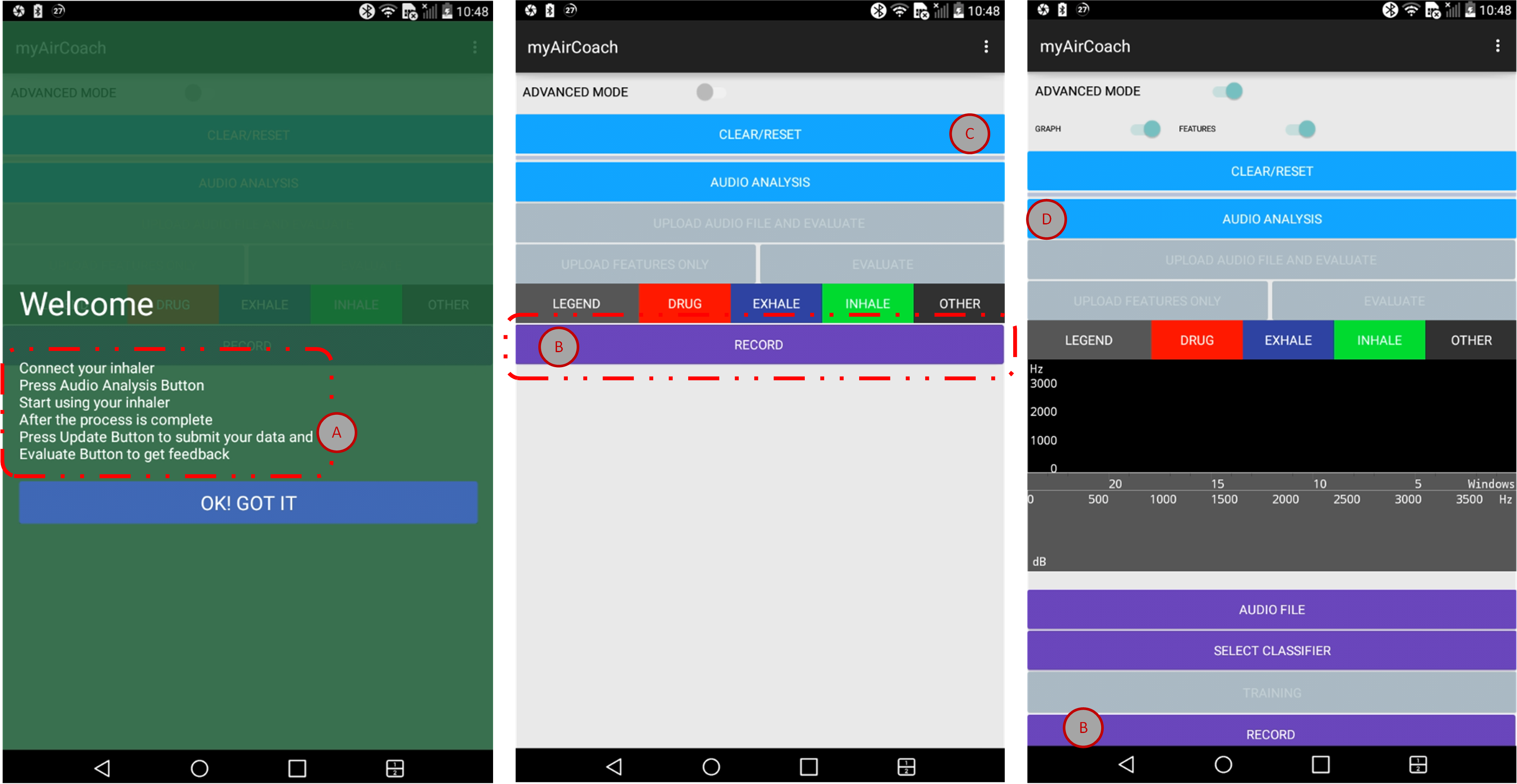}
\caption{Basic features and views of the application. A) Welcoming screen and guidelines B) Record button to start audio capturing from the inhaler device. C) Clear/Reset functionality to restart the application. D) Audio analysis to classify the audio data.}
\label{fig:mobileUI_3}
 \end{figure}

\begin{figure}[t!]
\centering
\includegraphics[width=\linewidth]{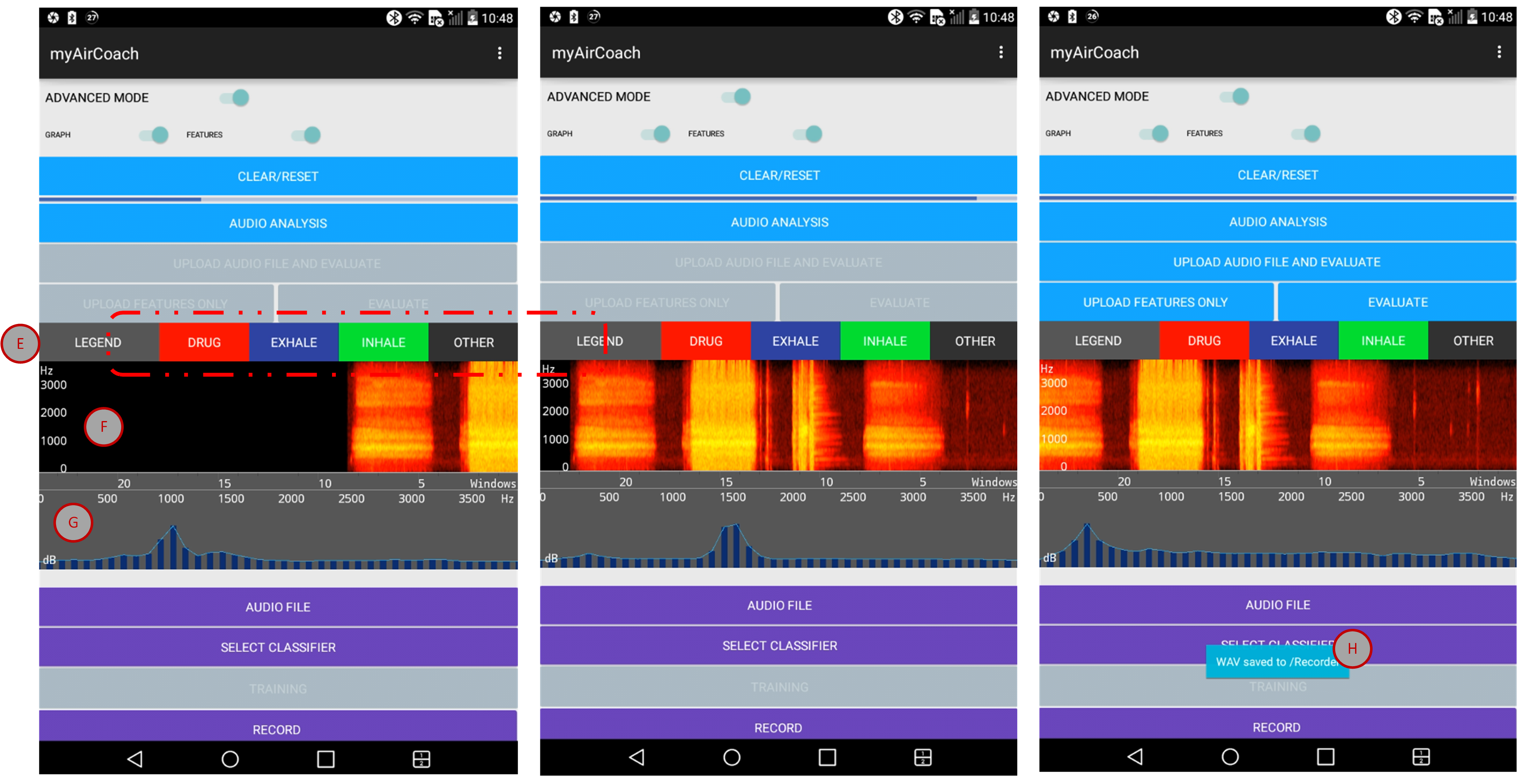}
\caption{E) Color legend. F) Spectrogram visualization. G) Spectrogram audio features for the given window under processing. H) "WAV saved" signal to notify the user that the audio file has been captured.}
\label{fig:mobileUI_4}
 \end{figure}

 \begin{figure}[t!]
\centering
 \includegraphics[width=0.5\linewidth]{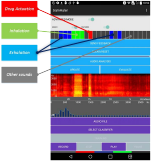}
\caption{Classification outcome for a certain inhaler use. The color coding is explained in the legend at the left side. Each colored rectangle corresponds to one of the classes, namely drug actuation, inhalation, exhalation and other sounds.}
\label{fig:app_classification_result}
 \end{figure}


\chapter{Conclusion}
\label{chapter:conclusion}

This dissertation lies in the family of approaches that constitute the virtual physiological human and aims to provide novel methods that facilitate patient-specific computational modelling of the pulmonary system within the scope of Asthma. Additional data-driven machine and deep learning methodologies are discussed that contribute to the improvement of self-management of Asthma.

Chapter \ref{chapter:novel-patient-specific} focused on the patient-specific modelling of the pulmonary system. Coupled structural and functional respiratory system models allow to investigate of personalized ventilation patterns for personalized Asthma and COPD management, create personalized pulmonary medicine, and predict the severity of one's condition. However, current imaging techniques do not reconstruct the bronchial tree structure beyond the sixth generation. Thus, novel methods are required to estimate the possible structure given the imaging data. The presented methodologies are a modified version of existing volume filling algorithms that emphasize available volume maximization. They were validated with datasets available in the literature. Furthermore, the generated data were utilized to introduce a probabilistic visualization of the presence of certain generations on the CT data as an extra overlay. Additional broncho-constriction simulation tools were designed to create digital twins of constricted lungs and perform computational fluid dynamics simulations.

Chapter \ref{chapter:monitoring} presented data-driven machine and deep learning approaches that facilitate monitoring of medication adherence through the audio-based identification of inhaler events, namely, inhalation, exhalation, medicine actuation and environmental sounds. Accurate differentiation of medicine-related actions allows for determining a patient's adherence to medical guidance by identifying the sequence and duration of each event. By determining the quality of each inhaler session combined with environmental data, self-assessment questionnaires and spirometry measurements can facilitate the prediction of a patient's condition, i.e. exacerbations and hospitalizations. We aimed to investigate the effect on accuracy and computational complexity towards efficiently identifying inhaler events. Specifically, we compare a series of shallow machine learning models, namely support vector machines, random forests, boosting algorithms with long-term memory recurrent neural networks and convolutional networks.

Our ambition is to combine functional pulmonary models with spirometry data and medication adherence data to provide the tools for evaluating and predicting a patient's condition while allowing proper self-management and better medical oversight. Such a strategy would set the prerequisites for reducing the societal burden of constrictive pulmonary diseases.

\section{Publications}
The following list contains papers published by the author of this dissertation, mainly as a leading and contributing author, until May of the Year 2022. Publications that are presented, analyzed and discussed  within this dissertation are listed in Subsections \ref{subsection:relevant_journal_publications} and \ref{subsection:relevant_conference_publications}. Furthermore, subsection \ref{subsection:other_publications} presents a list of research papers that are not relevant to the core idea of this dissertation but were curated in the years 2016-2022.

\subsection{Journal article related to this dissertation}
\label{subsection:relevant_journal_publications}
\begin{enumerate}
\item \textbf{Nousias S}, Zacharaki EI, Moustakas K. AVATREE: An open-source computational modelling framework modelling Anatomically Valid Airway TREE conformations. PloS one. 2020 Apr 3;15(4):e0230259.
\item \textbf{Nousias S}, Lalos AS, Arvanitis G, Moustakas K, Tsirelis T, Kikidis D, Votis K, Tzovaras D. An mHealth system for monitoring medication adherence in obstructive respiratory diseases using content based audio classification. IEEE Access. 2018 Feb 26;6:11871-82.
\item Ntalianis V, Fakotakis ND, \textbf{Nousias S}, Lalos AS, Birbas M, Zacharaki EI, Moustakas K. Deep CNN Sparse Coding for Real Time Inhaler Sounds Classification. Sensors. 2020 Jan;20(8):2363.
\item Lalas A, \textbf{Nousias S}, Kikidis D, Lalos A, Arvanitis G, Sougles C, Moustakas K, Votis K, Verbanck S, Usmani O, Tzovaras D. Substance deposition assessment in obstructed pulmonary system through numerical characterization of airflow and inhaled particles attributes. BMC medical informatics and decision making. 2017 Dec;17(3):25-44.
\item Fakotakis ND, \textbf{Nousias S}, Arvanitis G, Zacharaki EI, Moustakas K. Revisiting Audio Pattern Recognition for Asthma Medication Adherence: Evaluation with the RDA Benchmark Suite.  Under review.
\end{enumerate}

\subsection{Conference publications related to this dissertation}
\label{subsection:relevant_conference_publications}

\begin{enumerate}
    \item Ntalianis V, Nousias S, Lalos AS, Birbas M, Tsafas N, Moustakas K. Assessment of medication adherence in respiratory diseases through deep sparse convolutional coding. In: IEEE Conference on Emerging Technologies and Factory Automation. Vol 2019-Septe. Zaragoza, Spain: IEEE; 2019: 1657-1660. doi:10.1109/ ETFA.2019.8869054
\item Pettas D, Nousias S, Zacharaki EI, Moustakas K. Recognition of Breathing Activity and Medication Adherence using LSTM Neural Networks. In: 2019 IEEE 19th International Conference on Bioinformatics and Bioengineering (BIBE). IEEE; 2019:941-946. doi:10.1109/BIBE.2019.00176
\item Arvanitis G, Kocsis O, Lalos AS, Nousias S, Moustakas K, Fakotakis N. 3-class prediction of asthma control status using a Gaussian mixture model approach. In: ACM International Conference Proceeding Series. New York, New York, USA: ACM Press; 2018:1-2. doi:10.1145/3200947.3201056
\item Lalas A, Kikidis D, Votis K, et al. Numerical assessment of airflow and inhaled particles attributes in the obstructed pulmonary system. In: Proceedings - 2016 IEEE International Conference on Bioinformatics and Biomedicine, BIBM 2016. IEEE; 2017:606-612. doi:10.1109/BIBM.2016.7822588
\item Nousias S, Lakoumentas J, Lalos A, et al. Monitoring asthma medication adherence through content-based audio classification. In: 2016 IEEE Symposium Series on Computational Intelligence, SSCI 2016. IEEE; 2017:1-5. doi:10.1109/ SSCI.2016.7849898
\item Nousias S, Lalos AS, Moustakas K. Computational modeling for simulating obstructive lung diseases based on geometry processing methods. In: International Conference on Digital Human Modeling and Applications in Health, Safety, Ergonomics and Risk Management. Springer, Cham; 2016:100-109.
\item Nousias S, Lalos A, Moustakas K, et al. Computational modeling methods for simulating obstructive human lung diseases. In: 9.1 Respiratory Function Technologists/Scientists. Vol 48. London: European Respiratory Society; 2016:PA4401. doi:10.1183/13993003.congress-2016.PA4401
\end{enumerate}

\subsection{Other publications}
\label{subsection:other_publications}
\begin{enumerate}
 \item Nousias S, Pikoulis E, Mavrokefalidis C, Lalos AS, Moustakas K. Accelerating 3D scene analysis for autonomous driving systems. In: IEEE International Conference on Very Large Scale Integration. ; 2021.
\item Nousias S, Arvanitis G, Lalos AS, Moustakas K. Fast Mesh Denoising with Data Driven Normal Filtering Using Deep Variational Autoencoders. IEEE Trans Ind Informatics. 2021;17(2):980-990. doi:10.1109/TII.2020.3000491
\item Nousias S, Arvanitis G, Lalos AS, Moustakas K. Mesh Saliency Detection Using Convolutional Neural Networks. In: 2020 IEEE International Conference on Multimedia and Expo (ICME). IEEE; 2020:1-6. doi:10.1109/ICME46284.2020.9102796
\item Nousias S, Pikoulis EV, Mavrokefalidis C, Lalos AS. Accelerating deep neural networks for efficient scene understanding in automotive cyber-physical systems. IEEE Int Conf Ind Cyber-Physical Syst. 2021.
\item Ntalianis V, Fakotakis ND, Nousias S, et al. Deep CNN Sparse Coding for Real Time Inhaler Sounds Classification. Sensors. 2020;20(8):2363. doi:10.3390/s20082363
\item Nousias S, Arvanitis G, Lalos AS, et al. A Saliency Aware CNN-Based 3D Model Simplification and Compression Framework for Remote Inspection of Heritage Sites. IEEE Access. 2020;8(1):169982-170001. doi:10.1109/access.2020.3023167
\item Bitzas D, Tselios C, Nousias S, et al. GamECAR : A gamified educational system for supporting the adoption of eco-driving. Univers Access Inf Soc. 2020:1-23.
\item Tselios C, Nousias S, Bitzas D, et al. Enhancing an eco-driving gamification platform through wearable and vehicle sensor data integration. In: AMI 2019 : European Conference on Ambient Intelligence 2019. ; 2019:1-6.
\item Ortmann S, Maye O, Zinke B, et al. Wearable sensing technologies for smart asthma guidance. In: Current Directions in Biomedical Engineering. ; 2019:643607.
\item Nousias S, Lalos AS, Kalogeras A, Alexakos C, Koulamas C, Moustakas K. Sparse modelling and optimization tools for energy-efficient and reliable IoT. In: 2019 1st International Conference on Societal Automation, SA 2019. Krakow, Poland; 2019. doi:10.1109/SA47457.2019.8938029
\item Papoulias G, Nousias S, Moustakas K. Simulation framework for fluid-solid interaction of cerebral aneurysm wall deformation. In: International Conference on Information, Intelligence, Systems and Applications. Vol 0. ; 2019:0-0.
\item Nousias S, Tselios C, Bitzas D, et al. Exploiting Gamification to Improve Eco-driving Behaviour: The GamECAR Approach. Electron Notes Theor Comput Sci. 2019;343:103-116. doi:10.1016/j.entcs.2019.04.013
\item Nousias S, Tselios C, Bitzas D, Lalos AS, Moustakas K, Chatzigiannakis I. Uncertainty Management for Wearable IoT Wristband Sensors Using Laplacian-Based Matrix Completion. In: 2018 IEEE 23rd International Workshop on Computer-Aided Modeling and Design of Communication Links and Networks (CAMAD). Vol 2018-Septe. IEEE; 2018:1-6. doi:10.1109/CAMAD.2018.8515001
\item Nousias S, Tseliosl C, Uitzasl D, et al. Managing nonuniformities and uncertainties in vehicle-oriented sensor data over next-generation networks. In: 2018 IEEE International Conference on Pervasive Computing and Communications Workshops (PerCom Workshops). IEEE; 2018:272-277. doi:10.1109/PERCOMW.2018.8480342
\item Lalos AS, Nousias S, Moustakas K. GamECAR: Gamifying Self Management of Eco-driving. ERCIM News Spec Theme Digit Twins. 2018.
\item Nousias S, Lalos AS, Tselios C, et al. Gamification of EcoDriving Behaviours through Intelligent Management of dynamic car and driver information. In: European Project Space. SCITEPRESS; 2018.
\end{enumerate}

\bibliographystyle{unsrturl-custom}

\appendix

\part*{Appendix}

\chapter{Medication adherence}

\section{Raw Medication adherence data for Dataset B}

Tables \ref{table:results:svm} to \ref{table:results:lstm_cnn:raw} present in detail the classification outcomes for all methods, evaluation setups and classes. The aforementioned tables were used to derive tables in the results section. Tables \ref{table:results:svm} to \ref{table:results:lstm_cnn} present the normalized confusion matrices for Support Vector Machines, Random Forests, ADABoost,GMM, LSTM, CNNs for spectrogram, ceptrogram and MFCC features on the basis or MultiSubject, SingleSubject and LOSO evaluation setups under mixed and non-mixed testing configurations. Furthermore, Tables  \ref{table:results:svm:raw} to \ref{table:results:lstm_cnn:raw} present the non-normalized confusion matrices for Support Vector Machines, Random Forests, ADABoost,GMM, LSTM, CNNs for spectrogram, ceptrogram and MFCC features on the basis or MultiSubject, SingleSubject and LOSO evaluation setups under mixed and non-mixed testing configurations. Tables \ref{table:results:svm} to \ref{table:results:lstm_cnn} have each case highlighted with pale blue while performance higher than 90\% is marked with bold. On the other hand Tables \ref{table:results:svm:raw} to \ref{table:results:lstm_cnn:raw} are not annotated since they are only cited to allow the reader to vaidate the outcomes of Chapter 5 and the accuracy metrics presented in Tables \ref{table:results:svm} to \ref{table:results:lstm_cnn}. 

\noindent \textbf{Reference values are on the Y axis and prediction on the X axis}

\begin{table}
\centering
\caption{Experimental evaluation with Random Forest algorithm\\ Normalized confusion matrices.}
\label{table:results:rf}
\resizebox{\textwidth}{!}{
\begin{tabular}{@{}|l|l|l|l|llll|llll|llll|@{}}
\toprule
                      &                         &                             &        & \multicolumn{4}{l|}{Multi}                                                                                                                                                                   & \multicolumn{4}{l|}{Single}                                                                                                                                                                  & \multicolumn{4}{l|}{LOSO}                                                                                                                                                                   \\ \midrule
                      &                         &                             &        & \multicolumn{1}{l|}{Drug}                          & \multicolumn{1}{l|}{Exhale}                        & \multicolumn{1}{l|}{Inhale}                        & Noise                         & \multicolumn{1}{l|}{Drug}                          & \multicolumn{1}{l|}{Exhale}                        & \multicolumn{1}{l|}{Inhale}                        & Noise                         & \multicolumn{1}{l|}{Drug}                         & \multicolumn{1}{l|}{Exhale}                        & \multicolumn{1}{l|}{Inhale}                        & Noise                         \\
\bottomrule
\end{tabular}
}
\end{table}

\begin{table}
\centering
\caption{Experimental evaluation with Gaussian Mixture Models\\ Normalized confusion matrices.}
\label{table:results:gmm}
\resizebox{\textwidth}{!}{
\begin{tabular}{@{}|l|l|l|l|llll|llll|llll|@{}}
\toprule
                      &                         &                             &        & \multicolumn{4}{l|}{Multi}                                                                                                                                                                   & \multicolumn{4}{l|}{Single}                                                                                                                                                                  & \multicolumn{4}{l|}{LOSO}                                                                                                                                                                   \\ \midrule
                      &                         &                             &        & \multicolumn{1}{l|}{Drug}                          & \multicolumn{1}{l|}{Exhale}                        & \multicolumn{1}{l|}{Inhale}                        & Noise                         & \multicolumn{1}{l|}{Drug}                          & \multicolumn{1}{l|}{Exhale}                        & \multicolumn{1}{l|}{Inhale}                        & Noise                         & \multicolumn{1}{l|}{Drug}                         & \multicolumn{1}{l|}{Exhale}                        & \multicolumn{1}{l|}{Inhale}                        & Noise                         \\
\bottomrule
\end{tabular}
}
\end{table}

\begin{table}
\centering
\caption{Experimental evaluation with ADABoost\\ Normalized confusion matrices.}
\label{table:results:ada}
\resizebox{\textwidth}{!}{
\begin{tabular}{@{}|l|l|l|l|llll|llll|llll|@{}}
\toprule
                       &                         &                             &        & \multicolumn{4}{l|}{Multi}                                                                                                                                                                   & \multicolumn{4}{l|}{Single}                                                                                                                                                                  & \multicolumn{4}{l|}{LOSO}                                                                                                                                                                    \\ \midrule
                       &                         &                             &        & \multicolumn{1}{l|}{Drug}                          & \multicolumn{1}{l|}{Exhale}                        & \multicolumn{1}{l|}{Inhale}                        & Noise                         & \multicolumn{1}{l|}{Drug}                          & \multicolumn{1}{l|}{Exhale}                        & \multicolumn{1}{l|}{Inhale}                        & Noise                         & \multicolumn{1}{l|}{Drug}                          & \multicolumn{1}{l|}{Exhale}                        & \multicolumn{1}{l|}{Inhale}                        & Noise                         \\
\bottomrule
\end{tabular}
}
\end{table}

\begin{table}
\centering
\caption{Experimental evaluation with SVM\\ Normalized confusion matrices.}
\label{table:results:svm}
\resizebox{\textwidth}{!}{
\begin{tabular}{@{}|l|l|l|l|llll|llll|llll|@{}}
\toprule
                       &                         &                             &        & \multicolumn{4}{l|}{Multi}                                                                                                                                                                   & \multicolumn{4}{l|}{Single}                                                                                                                                                                  & \multicolumn{4}{l|}{LOSO}                                                                                                                                                                    \\ \midrule
                       &                         &                             &        & \multicolumn{1}{l|}{Drug}                          & \multicolumn{1}{l|}{Exhale}                        & \multicolumn{1}{l|}{Inhale}                        & Noise                         & \multicolumn{1}{l|}{Drug}                          & \multicolumn{1}{l|}{Exhale}                        & \multicolumn{1}{l|}{Inhale}                        & Noise                         & \multicolumn{1}{l|}{Drug}                          & \multicolumn{1}{l|}{Exhale}                        & \multicolumn{1}{l|}{Inhale}                        & Noise                         \\
\bottomrule
\end{tabular}
}
\end{table}

\begin{table}
\centering
\caption{Experimental evaluation with LSTM, CNN\\ Normalized confusion matrices.}
\label{table:results:lstm_cnn}
\resizebox{\textwidth}{!}{
\begin{tabular}{@{}|l|l|l|l|llll|llll|llll|@{}}
\toprule
                       &                         &                             &        & \multicolumn{4}{l|}{Multi}                                                                                                                                                                  & \multicolumn{4}{l|}{Single}                                                                                                                                                                  & \multicolumn{4}{l|}{LOSO}                                                                                                                                                                    \\ \midrule
                       &                         &                             &        & \multicolumn{1}{l|}{Drug}                          & \multicolumn{1}{l|}{Exhale}                        & \multicolumn{1}{l|}{Inhale}                        & Noise                        & \multicolumn{1}{l|}{Drug}                          & \multicolumn{1}{l|}{Exhale}                        & \multicolumn{1}{l|}{Inhale}                        & Noise                         & \multicolumn{1}{l|}{Drug}                          & \multicolumn{1}{l|}{Exhale}                        & \multicolumn{1}{l|}{Inhale}                        & Noise                         \\
\bottomrule
\end{tabular}
}
\end{table}

\begin{table}
\centering
\caption{Experimental evaluation with Random forests\\ Non-normalized confusion matrices.}
\label{table:results:rf:raw}
\resizebox{\textwidth}{!}{
\begin{tabular}{@{}|l|l|l|l|llll|llll|llll|@{}}
\toprule
                      &                         &                             &        & \multicolumn{4}{l|}{Multi}                                                                                                                                                           & \multicolumn{4}{l|}{Single}                                                                                                                                                                  & \multicolumn{4}{l|}{LOSO}                                                                                                                                                           \\ \midrule
                      &                         &                             &        & \multicolumn{1}{l|}{Drug}                        & \multicolumn{1}{l|}{Exhale}                      & \multicolumn{1}{l|}{Inhale}                      & Noise                       & \multicolumn{1}{l|}{Drug}                         & \multicolumn{1}{l|}{Exhale}                        & \multicolumn{1}{l|}{Inhale}                        & Noise                          & \multicolumn{1}{l|}{Drug}                       & \multicolumn{1}{l|}{Exhale}                      & \multicolumn{1}{l|}{Inhale}                      & Noise                       \\ \midrule
                      &                         &                             & Drug   & \multicolumn{1}{l|}{\cellcolor[HTML]{D9D9D9}186} & \multicolumn{1}{l|}{\cellcolor[HTML]{D9D9D9}0}   & \multicolumn{1}{l|}{\cellcolor[HTML]{D9D9D9}6}   & \cellcolor[HTML]{D9D9D9}1   & \multicolumn{1}{l|}{179}                          & \multicolumn{1}{l|}{0}                             & \multicolumn{1}{l|}{3}                             & 1                              & \multicolumn{1}{l|}{\cellcolor[HTML]{D9D9D9}8}  & \multicolumn{1}{l|}{\cellcolor[HTML]{D9D9D9}0}   & \multicolumn{1}{l|}{\cellcolor[HTML]{D9D9D9}2}   & \cellcolor[HTML]{D9D9D9}0   \\
\bottomrule
\end{tabular}
}
\end{table}

\begin{table}
\centering
\caption{Experimental evaluation with Gaussian Mixture Models\\ Non-normalized confusion matrices.}
\label{table:results:gmm:raw}
\resizebox{\textwidth}{!}{
\begin{tabular}{@{}|l|l|l|l|llll|llll|llll|@{}}
\toprule
                       &                         &                             &        & \multicolumn{4}{l|}{Multi}                                                                                                                                                           & \multicolumn{4}{l|}{Single}                                                                                                                                                                  & \multicolumn{4}{l|}{LOSO}                                                                                                                                                          \\ \midrule
                       &                         &                             &        & \multicolumn{1}{l|}{Drug}                        & \multicolumn{1}{l|}{Exhale}                      & \multicolumn{1}{l|}{Inhale}                      & Noise                       & \multicolumn{1}{l|}{Drug}                         & \multicolumn{1}{l|}{Exhale}                        & \multicolumn{1}{l|}{Inhale}                        & Noise                          & \multicolumn{1}{l|}{Drug}                      & \multicolumn{1}{l|}{Exhale}                      & \multicolumn{1}{l|}{Inhale}                      & Noise                       \\ \midrule
                       &                         &                             & Drug   & \multicolumn{1}{l|}{\cellcolor[HTML]{D9D9D9}184} & \multicolumn{1}{l|}{\cellcolor[HTML]{D9D9D9}0}   & \multicolumn{1}{l|}{\cellcolor[HTML]{D9D9D9}8}   & \cellcolor[HTML]{D9D9D9}1   & \multicolumn{1}{l|}{176}                          & \multicolumn{1}{l|}{0}                             & \multicolumn{1}{l|}{7}                             & 0                              & \multicolumn{1}{l|}{\cellcolor[HTML]{D9D9D9}4} & \multicolumn{1}{l|}{\cellcolor[HTML]{D9D9D9}0}   & \multicolumn{1}{l|}{\cellcolor[HTML]{D9D9D9}6}   & \cellcolor[HTML]{D9D9D9}0   \\
\bottomrule
\end{tabular}
}
\end{table}

\begin{table}
\centering
\caption{Experimental evaluation with ADABoost\\ Non-normalized confusion matrices.}
\label{table:results:ada:raw}
\resizebox{\textwidth}{!}{
\begin{tabular}{@{}|l|l|l|l|llll|llll|llll|@{}}
\toprule
                       &                         &                             &        & \multicolumn{4}{l|}{Multi}                                                                                                                                                           & \multicolumn{4}{l|}{Single}                                                                                                                                                                  & \multicolumn{4}{l|}{LOSO}                                                                                                                                                           \\ \midrule
                       &                         &                             &        & \multicolumn{1}{l|}{Drug}                        & \multicolumn{1}{l|}{Exhale}                      & \multicolumn{1}{l|}{Inhale}                      & Noise                       & \multicolumn{1}{l|}{Drug}                         & \multicolumn{1}{l|}{Exhale}                        & \multicolumn{1}{l|}{Inhale}                        & Noise                          & \multicolumn{1}{l|}{Drug}                       & \multicolumn{1}{l|}{Exhale}                      & \multicolumn{1}{l|}{Inhale}                      & Noise                       \\ \midrule
                       &                         &                             & Drug   & \multicolumn{1}{l|}{\cellcolor[HTML]{D9D9D9}186} & \multicolumn{1}{l|}{\cellcolor[HTML]{D9D9D9}0}   & \multicolumn{1}{l|}{\cellcolor[HTML]{D9D9D9}6}   & \cellcolor[HTML]{D9D9D9}1   & \multicolumn{1}{l|}{177}                          & \multicolumn{1}{l|}{0}                             & \multicolumn{1}{l|}{5}                             & 1                              & \multicolumn{1}{l|}{\cellcolor[HTML]{D9D9D9}7}  & \multicolumn{1}{l|}{\cellcolor[HTML]{D9D9D9}0}   & \multicolumn{1}{l|}{\cellcolor[HTML]{D9D9D9}3}   & \cellcolor[HTML]{D9D9D9}0   \\
\bottomrule
\end{tabular}
}
\end{table}

\begin{table}
\centering
\caption{Experimental evaluation with SVM\\ Non-normalized confusion matrices.}
\label{table:results:svm:raw}
\resizebox{\textwidth}{!}{
\begin{tabular}{@{}|l|l|l|l|llll|llll|llll|@{}}
\toprule
                       &                         &                             &        & \multicolumn{4}{l|}{Multi}                                                                                                                                                           & \multicolumn{4}{l|}{Single}                                                                                                                                                                   & \multicolumn{4}{l|}{LOSO}                                                                                                                                                           \\ \midrule
                       &                         &                             &        & \multicolumn{1}{l|}{Drug}                        & \multicolumn{1}{l|}{Exhale}                      & \multicolumn{1}{l|}{Inhale}                      & Noise                       & \multicolumn{1}{l|}{Drug}                         & \multicolumn{1}{l|}{Exhale}                         & \multicolumn{1}{l|}{Inhale}                        & Noise                          & \multicolumn{1}{l|}{Drug}                       & \multicolumn{1}{l|}{Exhale}                      & \multicolumn{1}{l|}{Inhale}                      & Noise                       \\
\bottomrule
\end{tabular}
}
\end{table}

\begin{table}
\centering
\caption{Experimental evaluation with LSTM,CNN\\ Non-normalized confusion matrices.}
\label{table:results:lstm_cnn:raw}
\resizebox{\textwidth}{!}{
\begin{tabular}{@{}|l|l|l|l|llll|llll|llll|@{}}
\toprule
                       &                         &                             &        & \multicolumn{4}{l|}{Multi}                                                                                                                                                                  & \multicolumn{4}{l|}{Single}                                                                                                                                                                  & \multicolumn{4}{l|}{LOSO}                                                                                                                                                                \\ \midrule
                       &                         &                             &        & \multicolumn{1}{l|}{Drug}                         & \multicolumn{1}{l|}{Exhale}                        & \multicolumn{1}{l|}{Inhale}                        & Noise                         & \multicolumn{1}{l|}{Drug}                         & \multicolumn{1}{l|}{Exhale}                        & \multicolumn{1}{l|}{Inhale}                        & Noise                          & \multicolumn{1}{l|}{Drug}                       & \multicolumn{1}{l|}{Exhale}                        & \multicolumn{1}{l|}{Inhale}                        & Noise                        \\
\bottomrule
\end{tabular}
}
\end{table}

\end{document}